\documentclass[10pt]{iopartmod}
\usepackage{amssymb}
\usepackage{amsmath}
\usepackage{makeidx}
\usepackage[dvips]{graphicx}

\input{tcilatex}

\begin{document}

\title{Slow light in photonic crystals}
\author{Alex Figotin and Ilya Vitebskiy}

\begin{abstract}
The problem of slowing down light by orders of magnitude has been
extensively discussed in the literature. Such a possibility can be useful in
a variety of optical and microwave applications. Many qualitatively
different approaches have been explored. Here we discuss how this goal can
be achieved in linear dispersive media, such as photonic crystals. The
existence of slowly propagating electromagnetic waves in photonic crystals
is quite obvious and well known. The main problem, though, has been how to
convert the input radiation into the slow mode without losing a significant
portion of the incident light energy to absorption, reflection, etc. We show
that the so-called frozen mode regime offers a unique solution to the above
problem. Under the frozen mode regime, the incident light enters the
photonic crystal with little reflection and, subsequently, is completely
converted into the frozen mode with huge amplitude and almost zero group
velocity. The linearity of the above effect allows the slowing of light
regardless of its intensity. An additional advantage of photonic crystals
over other methods of slowing down light is that photonic crystals can
preserve both time and space coherence of the input electromagnetic wave.
\end{abstract}


\address{Department of Mathematics, University of California at Irvine, CA
92697}

\section{Introduction}

\subsection{What is slow light?}

It is common knowledge that, in vacuum, light propagates with constant
velocity $c\approx 3\times 10^{8}$m/sec. In optically transparent
nondispersive media, the speed of light propagation is different 
\begin{equation}
v=\omega /k=c/n,  \label{v=c/n}
\end{equation}%
where $k$ is the wave number, $\omega $ is the respective frequency, and $n$
is the refractive index of the medium. At optical frequencies, the
refractive index $n$ of transparent materials usually does not exceed
several units, and the speed of light propagation is of the same order of
magnitude as in vacuum.

The situation can change dramatically in strongly dispersive media. Although
the phase velocity of light is still determined by the same expression (\ref%
{v=c/n}), the speed of electromagnetic pulse propagation is different from $%
v $ and is determined by the group velocity \cite{Brill,LLEM,Yariv}%
\begin{equation}
u=\frac{\partial\omega}{\partial k}=c\left( n+\omega\frac{dn}{d\omega }%
\right) ^{-1},  \label{u(n)}
\end{equation}
which is one of the most important electromagnetic characteristics of the
medium. With certain reservations, the group velocity $u$ coincides with the
electromagnetic energy velocity and is usually referred to simply as the
propagation speed of light in the medium. Hereinafter, the speed of light
propagation means the group velocity (\ref{u(n)}), rather then the phase
velocity (\ref{v=c/n}).

Strong dispersion means that the group velocity $u$ strongly depends on the
frequency and can be substantially different from $c$. In the slow light
case, which is the subject of our interest, the electromagnetic pulse
propagates through the dispersive medium at the speed $u\ll c$, regardless
of the respective value of the phase velocity (\ref{v=c/n}). In some cases, $%
u$ can even become vanishingly small implying that the propagating
electromagnetic mode at the respective frequency does not transfer energy.
In another extreme case, the group velocity $u$ can exceed $c$ (the
so-called case of superluminal pulse propagation), without contradicting the
causality principle \cite{Brill,Caus Somm,Caus Chiao,Caus Boyd,Caus Milo}.
In yet another case of a left-handed medium, the group velocity $u$ can have
the opposite sign to that of the phase velocity $v$ \cite{Veselago}. But
again, in this paper we will focus exclusively on the slow light and related
phenomena.

Slow and ultraslow light have numerous and diverse practical applications.
The related phenomena include dramatic enhancement of various light-matter
interactions such as nonlinear effects (higher harmonic generation, wave
mixing, etc.), magnetic Faraday rotation, as well as many other important
electromagnetic properties of the optical media. Such an enhancement can
facilitate design of controllable optical delay lines, phase shifters,
miniature and efficient optical amplifiers and lasers, etc. In addition,
ultraslow light might allow nonlinear interactions down to a single photon
level, which could significantly benefit the design of ultrasensitive
optical switches, quantum all-optical data storage and data processing
devices. Ultraslow light can also be used in quantum communication and
design of novel acousto-optical devices. This list can be continued. For
more detailed information on the prospective practical applications of slow
light phenomena see, for example, \cite{EIT 1,EIT 2,EIT 3,EIT 4,EIT 5,EIT
6,Boyd 1,Boyd 2,Fan,CR Boyd1,CR Boyd2,CR Meloni 03,CR Yariv04,OPN
05,Khurgin1,Khurgin2,LDSL,SL Scal1,SL Scal2,SL Joann,PRB03,PRE03,PRE05} and
references therein.

\subsection{Temporal dispersion versus spatial dispersion}

In recent years, several different approaches have been pursued in order to
slow down or even completely stop light. These approaches can be grouped
into two major categories:

\begin{itemize}
\item[-] those where the low group velocity results from strong temporal
dispersion $dn/d\omega$ of optical media;

\item[-] those where the low speed of pulse propagation is a result of
coherent interference in spatially periodic heterogeneous media, such as
photonic crystals.
\end{itemize}

Let us start with a brief discussion of slow light phenomena in media with
strong temporal dispersion.

\subsubsection{Slow light in media with strong temporal dispersion}

Assuming that the refractive index $n$ in (\ref{u(n)}) is of the order of
unity, which is usually the case at optical frequencies, one can state that
a very low group velocity can only occur if $n$ varies strongly as a
function of $\omega$%
\begin{equation}
u\approx c\left( \omega\frac{dn}{d\omega}\right) ^{-1}\ll c\text{ \ only if }%
\omega\frac{dn}{d\omega}\gg1.  \label{u<<c}
\end{equation}
Strong frequency dependence of the refractive index $n$ can be a result of
excitation of electronic or some other intrinsic resonances of the medium,
which are normally accompanied by strong absorption of light. Recently,
though, several techniques have been developed that allow to significantly
suppress the absorption of light at the frequency where the derivative $%
dn/d\omega$ peaks.

One of the most successful ways to suppress light absorption is based on the
effect of electromagnetically induced transparency (EIT) \cite{EIT}. In such
a case, the incident light interacts with atomic spin excitations forming
combined excitations of photons and spins, called dark-state polaritons.
These polaritons propagate slowly through the medium in the form of a
sharply compressed pulse, the energy of which is much smaller than that of
the incident light pulse. Most of the incident light energy is expended to
create the coherent state of the atomic spins, the rest is carried away by
the control electromagnetic field. The pulse delay inside the medium is
limited by the bandwidth of the transparency window, which decreases with
propagation distance. At higher propagation distances the medium becomes
increasingly opaque at frequencies other than the line center, further
reducing the available transparency window \cite{EIT 4,EIT 5}. Specific
physical mechanisms of such transformations are very diverse. The detailed
description of EIT and related phenomena can be found in the extensive
literature on the subject (see, for example, \cite{EIT 1,EIT 2,EIT 3,EIT
4,EIT 5,EIT 6}, and references therein). The techniques based on EIT have
already produced some amazing results, such as reduction of the speed of
pulse propagation by 7 -- 8 orders of magnitude compared to the speed of
light in vacuum, while providing a huge and controllable pulse delay.

Another method to create a transparency window in otherwise opaque substance
was used in \cite{Boyd 1,Boyd 2}. This method involves the creation of a
spectral hole by the periodic modulation of the ground state population at
the beat frequency between the pump and the probe fields applied to the
material sample. It can produce slow light in a solid-state material at room
temperature. The spectral hole created by this technique can be extremely
narrow (36 Hz in the experiment \cite{Boyd 1,Boyd 2}), and leads to a rapid
spectral variation of refractive index. It allowed to reduce the light group
velocity in a ruby crystal down to 57 m/s.

Physically, the above approaches to slowing down the light can be viewed as
a reversible transformation of the input nearly monochromatic light into
some kind of coherent atomic excitations (e.g., dark-state polaritons) with
very low relaxation rate and low group velocity. In other words, ultraslow
pulse propagating through such a medium is, in fact, an intrinsic coherent
excitation triggered by the input light, rather than a light pulse per se.
This process always involves some kind of a delicate resonant light-matter
interaction with extremely small bandwidth. Indeed, the relation (\ref{u<<c}%
) yields the following limitation on the slow pulse bandwidth 
\begin{equation}
\frac{\Delta \omega }{\omega }<\frac{u}{c},  \label{Delta om}
\end{equation}%
where the assumption is made that the refractive index $n$ within the
transparency window is of the order of unity. The condition (\ref{Delta om})
can also be viewed as a constraint on the minimal propagation speed of a
light pulse with a given bandwidth $\Delta \omega $. On the positive side,
the approach based on EIT or its modifications does produce an exceptionally
low speed of pulse propagation, which can have some very important practical
implications.

In the rest of the paper we focus exclusively on those techniques which do
not involve any intrinsic resonant excitations of the medium and, therefore,
do not essentially rely on strong temporal dispersion. Instead, we will
focus on spatially periodic dielectric arrays, in which low group velocity
results solely from spatial inhomogeneity of the optical medium.

\subsubsection{Slow light in spatially periodic arrays}

Well-known examples of optical periodic dielectric structures include
photonic crystals \cite{Joann}, periodic arrays of coupled optical
resonators \cite{CR Boyd1,CR Boyd2,CR Meloni 03,CR Yariv04,OPN
05,Khurgin1,Khurgin2}, and line-defect waveguides \cite{LDSL}. Generally, a
periodic heterogeneous medium can be assigned a meaningful refractive index $%
n$ only if the structural period $L$ is much smaller than the light
wavelength $\lambda $%
\begin{equation}
L\ll \lambda .  \label{L<<lambda}
\end{equation}%
On the other hand, a significant spatial dispersion associated with
heterogeneity of the medium can occur only when $L$ and $\lambda $ are
comparable in value%
\begin{equation}
L\sim \lambda .  \label{L=lambda}
\end{equation}%
In particular, the relation (\ref{L=lambda}) defines a necessary condition
under which heterogeneity of the medium can lead to low speed of
electromagnetic pulse propagation. Hence, in the cases where low speed of
pulse propagation is a result of strong spatial dispersion, one cannot
assign a meaningful refractive index to the composite medium, and the
expression (\ref{u<<c}) for the group velocity of light does not apply.

At optical frequencies, the speed of pulse propagation in periodic
dielectric arrays \ can be reduced by two or three orders of magnitude. This
is not a fundamental restriction, but rather a technological limitation
related to the difficulty of building flawless periodic arrays at
nanoscales. On the positive side, the dielectric components of the periodic
array are not required to display strong temporal dispersion and, hence,
absorption of light is not an essential and unavoidable problem in this
case. In addition, the photonic crystal based approach is much more
versatile in terms of the input light intensity. It allows the same photonic
device to operate both at high and low intensity of the input light. By
contrast, utilizing strong temporal dispersion always involves significant
nonlinearity and usually is limited to a certain amplitude of the input
light.

There is a natural bandwidth limitation on the slowed pulse in periodic
dielectric arrays, which is similar to the case of slow light in
time-dispersive media. Indeed, let $\Delta\omega$ be the frequency bandwidth
of a pulse and $\Delta k$ -- the respective range of the Bloch wave number.
The average group velocity $\left\langle u\right\rangle $ of the pulse is
defined as%
\begin{equation}
\left\langle u\right\rangle \approx\frac{\Delta\omega}{\Delta k}.
\label{<u>}
\end{equation}
Let us make the following natural assumptions.

\begin{enumerate}
\item The pulse propagating inside the periodic medium is composed of the
Bloch eigenmodes belonging to the same spectral branch of the dispersion
relation $\omega\left( k\right) $. This assumption implies that $\Delta k$
cannot exceed the size $2\pi/L$ of the Brillouin zone%
\begin{equation}
\Delta k<2\pi/L,  \label{Dk < 1/L}
\end{equation}
where $L$ is the unit cell length of the periodic array.

\item The refractive index of the constitutive components of the periodic
array is of the order of unity and, therefore,%
\begin{equation}
L\sim\lambda_{0}=2\pi c/\omega,  \label{L=c/om}
\end{equation}
where $\lambda_{0}$ is the light wavelength in vacuum.
\end{enumerate}

The relations (\ref{<u>}-\ref{L=c/om}) yield the following limitation on the
minimal speed of pulse propagation for a pulse with a given bandwidth $%
\Delta\omega$%
\begin{equation}
\left\langle u\right\rangle >\frac{L}{2\pi}\Delta\omega\sim c\frac {%
\Delta\omega}{\omega}.  \label{u min}
\end{equation}
The restriction (\ref{u min}) is similar to that defined by the inequality (%
\ref{Delta om}) and related to the case of slow light in a uniform medium
with strong temporal dispersion. In either case, a higher refractive index
would lower the minimal speed $\left\langle u\right\rangle $ of pulse
propagation for a given pulse bandwidth $\Delta\omega$.

Any attempt to circumvent the restriction (\ref{u min}) would involve some
kind of pulse compression techniques \cite{Fan}.

\subsubsection{Examples of periodic arrays supporting slow light}

\paragraph{Coupled resonator optical waveguide.}

During the last several years, a tremendous progress has been made in theory
and applications of periodic arrays of coupled optical resonators.
Generally, if the coupling between adjacent resonators in a periodic chain
is weak, the group velocity of Bloch excitations supported by such a
periodic array is low. This is true regardless of the nature of individual
resonators. The above simple idea forms the basis for one of the most
popular approaches to slowing down the light. An extensive discussion on the
subject and numerous examples and references can be found in \cite{CR
Boyd1,CR Boyd2,CR Meloni 03,CR Yariv04,OPN 05,Khurgin1,Khurgin2,CR MV}.

A qualitatively similar situation occurs in line-defect waveguides in a
photonic crystal slab, where a periodic array of structural defects plays
the role of weakly coupled optical resonators. Following \cite{LDSL},
consider a dielectric slab with a two-dimensional periodic array of holes in
it. A row of missing holes in this periodic array forms a line defect, which
supports a waveguiding mode with two types of cutoff within the photonic
band gap. These characteristics can be tuned by controlling the defect
width. Theoretical calculations supported by interference measurements show
that the single waveguiding mode of the line-defect waveguide displays
extraordinarily large group dispersion. In some instances, the corresponding
traveling speed is 2 orders of magnitude slower than that in air. According
to \cite{LDSL}, one of the major limiting factor here is structural
imperfection of the array.

Slow light phenomena in periodic arrays of weakly coupled resonators have
been the subject of a great number of recent publications, including some
excellent review articles cited above. For this reason, further in this
paper we will not discuss this subject any more.

\paragraph{Photonic crystals.}

Photonic crystals are spatially periodic structures composed of usually two
different transparent dielectric materials \cite{Joann}. Similarly to
periodic arrays of coupled resonators, in photonic crystals, a low group
velocity of light can result from multiple scattering of individual photons
by periodic spatial inhomogeneities, rather than from temporal dispersion of
the substance \cite{OPN 05,SL Scal1,SL Scal2,SL Joann,PRB03,PRE03,PRE05}.
The lowest group velocity achievable in photonic crystals for a given pulse
bandwidth can be close to that defined by the fundamental restriction (\ref%
{u min}). For example, if we want a pulse to propagate undistorted at speed
as low as $10^{-3}c$, its bandwidth $\Delta \omega $ should be less than $%
10^{-3}\omega $, which at optical frequencies is of the order of $10$ GHz.
In this respect, the situation in photonic crystals is as good as it can
possibly be in any other linear passive media with limited refractive index.

Unlike the case of optical waveguides and linear arrays of coupled
resonators, in photonic crystals we have bulk electromagnetic waves capable
of propagating in any direction through the periodic heterogeneous
structure. This results in much greater density of modes, compared to that
of the above-mentioned arrays of coupled resonators. In addition,
electromagnetic waves in photonic crystals can remain coherent in all three
dimensions, which is also essential for a variety of practical applications.

A major problem with slow light in photonic crystals is the efficiency of
conversion of the incident light into the slow mode inside the heterogeneous
medium. We shall see in the next section that in most cases an incident
electromagnetic wave with the frequency of one of the slow modes is simply
reflected back to space, without creating the slow mode inside the photonic
crystal. How to overcome this fundamental problem and, thereby, how to
transform a significant fraction of the incident light energy into a slow
mode with drastically enhanced amplitude, is one of the primary subjects of
this paper.\bigskip

The paper is organized as follows. In Section 2 we describe, in general
terms, what kind of slow modes can exist in photonic crystals and under what
circumstances some of these modes can be effectively excited by incident
light. We show, that there is a unique situation, which we call the frozen
mode regime, in which the incident light can enter the photonic crystal with
little reflection and be completely converted into a slow mode with nearly
zero group velocity and drastically enhanced amplitude.

Section 3 gives an overall picture of the frozen mode regime in periodic
layered media, without going into the detailed analysis based on the Maxwell
equations. All the statements made in this section are later proven in
Sections 5 through 11.

In section 4 we define the physical conditions under which a periodic
layered array can support the frozen mode regime. These conditions boil down
to whether or not the electromagnetic dispersion relation of the periodic
array can develop a stationary inflection point (\ref{SIP}). This
requirement imposes quite severe restrictions on composition and geometry of
the periodic layered medium. We show, in particular, that in the case of
light propagating normally to the layers, the frozen mode regime can only
occur if some of the layers are magnetic with significant nonreciprocal
Faraday rotation. In the case of oblique light propagation, the presence of
magnetic layers is not required, which makes it possible to realize the
frozen mode regime at any frequency range, including optical and UV. A
trade-off though is that at least some of the layers of a non-magnetic stack
must display significant dielectric anisotropy with tilted orientation of
the anisotropy axis.

Section 5 is devoted to electrodynamics of periodic layered media.
Particular attention is given to the cases where some of the layers display
dielectric and/or magnetic anisotropy, because otherwise, the
electromagnetic dispersion relation $\omega \left( k\right) $ of the
periodic array cannot develop a stationary inflection point (\ref{SIP}) and,
therefore, such a structure cannot support the frozen mode regime.

Sections 6 through 12 constitute the analytical basis for the entire
investigation. There we present a rigorous and systematic analysis of the
scattering problem for a semi-infinite periodic array of anisotropic
dielectric layers. The emphasis is on the vicinity of stationary points (\ref%
{SP}) of the electromagnetic dispersion relation, where the slow
electromagnetic modes can be excited. The comparative analysis of all
possible stationary points shows that only a stationary inflection point (%
\ref{SIP}) can provide necessary conditions for slowing down and freezing a
significant fraction of incoming radiation. In all other cases, the incident
wave is either reflected back to space, or gets converted into a fast
propagating mode with low amplitude. The exact analytical results of these
sections are supported by a number of numerical simulations.

\section{Stationary points of dispersion relations and slow modes}

\begin{figure}[tbph]
\scalebox{0.8}{\includegraphics[viewport=0 0 400 400,clip]{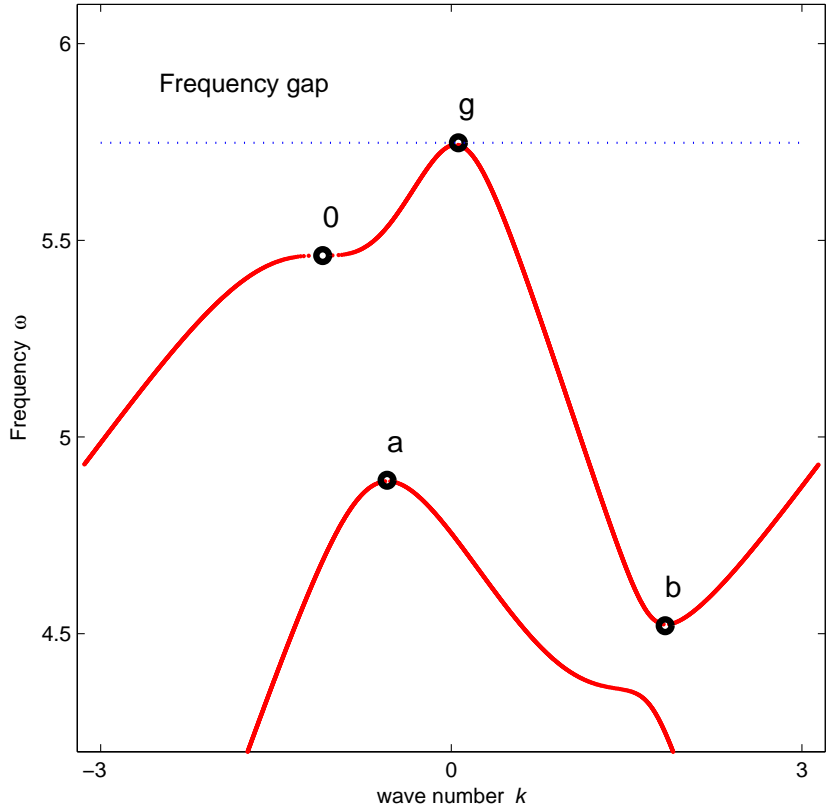}}
\caption{An example of electromagnetic dispersion relation $\protect\omega%
(k) $ with various stationary points: (i) extreme points $a$ and $b$ of the
respective spectral branches, (ii) a photonic band edge $g$, (iii) a
stationary inflection point $0$. Each stationary point is associated with
slow light.}
\label{DR_PRE03}
\end{figure}

In periodic heterogeneous media, such as photonic crystals, the velocity of
light is defined as the wave group velocity%
\begin{equation}
\vec{u}=\partial \omega /\partial \vec{k},  \label{u}
\end{equation}%
where $\vec{k}$ is the Bloch wave vector and $\omega =\omega \left( \vec{k}%
\right) $ is the respective frequency. At some frequencies, the dispersion
relation $\omega \left( \vec{k}\right) $ can develop stationary points%
\begin{equation}
\partial \omega /\partial \vec{k}=0,  \label{SP}
\end{equation}%
where the group velocity $\vec{u}$ vanishes. Zero group velocity usually
implies that the respective Bloch eigenmode does not transfer
electromagnetic energy. Indeed, with certain reservations, the energy flux $%
\vec{S}$ of a propagating Bloch mode is%
\begin{equation}
\vec{S}=W\vec{u},  \label{S=Wu}
\end{equation}%
where $W$ is the electromagnetic energy density associated with this mode.
If $W$ is bounded, then the group velocity $\vec{u}$ and the energy flux $%
\vec{S}$ vanish simultaneously at the respective stationary point (\ref{SP})
of the dispersion relation. Such modes are referred to as slow modes, or
slow light. Some examples of stationary points (\ref{SP}) are shown in Fig. %
\ref{DR_PRE03}, where each of the frequencies $\omega _{a}$, $\omega _{b}$, $%
\omega _{g}$, $\omega _{0}$ is associated with a slow mode.

The electromagnetic dispersion relation of any photonic crystal displays an
infinite number of stationary points like those shown in Fig. \ref{DR_PRE03}%
. But, a common problem with almost all of them is that the respective slow
modes cannot be excited in a semi-infinite photonic crystal by incident
light. This explains why there have been only a few attempts to exploit the
photonic crystals for slowing down the light. Let us take a closer look at
this problem.

\begin{figure}[tbph]
\begin{center}
\scalebox{0.8}{\includegraphics[viewport=0 0 300 250,clip]{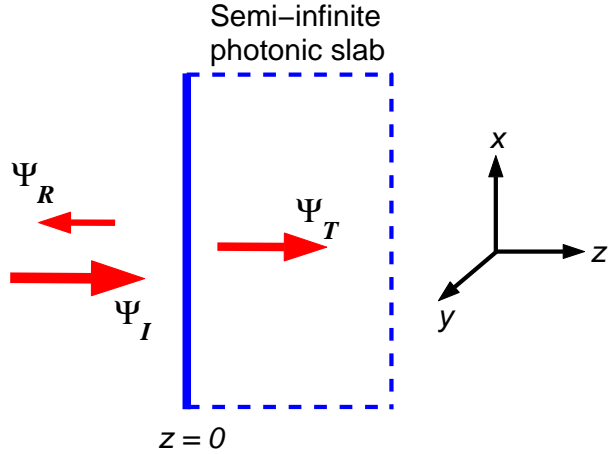}}
\end{center}
\caption{Plane wave normally incident on a lossless semi-infinite photonic
slab. The subscripts $I$, $R$, and $T$ refer to the incident, reflected and
transmitted waves, respectively.}
\label{SISn}
\end{figure}

Consider a scattering problem of a plane monochromatic wave normally
incident on a lossless semi-infinite photonic slab with the electromagnetic
dispersion relation shown in Fig. \ref{DR_PRE03}. The symbol $k$ in Fig. \ref%
{DR_PRE03} denotes the normal component of the Bloch wave number $\vec{k}$
in the periodic structure, which in the case of a normal incidence is the
only nonzero component of $\vec{k}$. The symbols $\Psi _{I}$, $\Psi _{R}$,
and $\Psi _{T}$ \ in Fig. \ref{SISn} denote the incident, reflected, and
transmitted waves, respectively. The transmittance $\tau $ and reflectance $%
\rho $ of the semi-infinite slab are defined as%
\begin{equation}
\tau =\frac{S_{T}}{S_{I}},\ \rho =-\frac{S_{R}}{S_{I}}=1-\tau .
\label{tau(n)}
\end{equation}%
where $S_{I}$, $S_{R}$ and $S_{T}$ are the normal energy fluxes of the
respective waves.

\begin{figure}[tbph]
\begin{center}
\scalebox{0.8}{\includegraphics[viewport=0 0 500 400,clip]{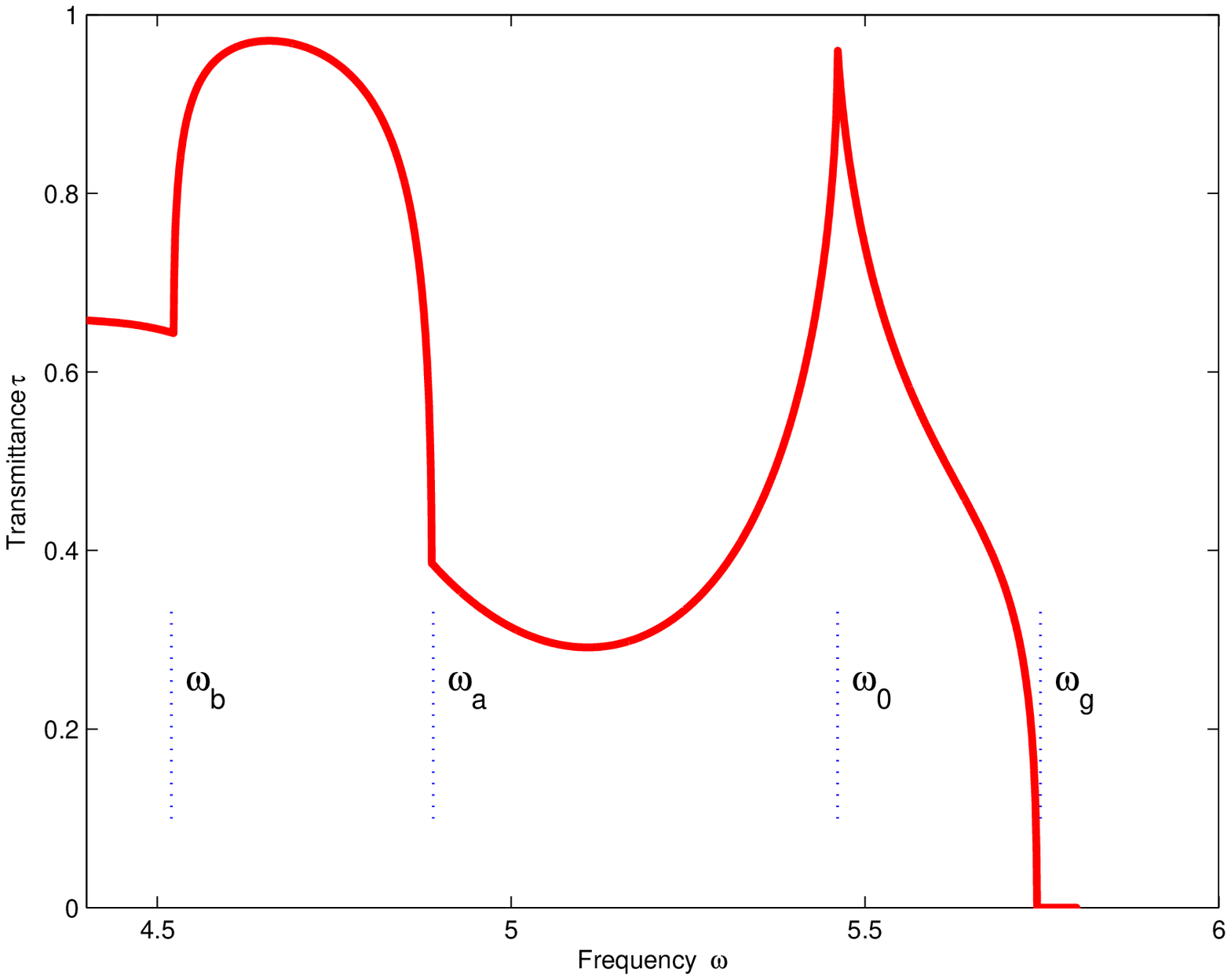}}
\end{center}
\caption{Transmittance $\protect\tau$ of the semi-infinite photonic slab as
a function of incident light frequency $\protect\omega$ for the
semi-infinite photonic slab with the dispersion relation presented in Fig.
1. The characteristic frequencies $\protect\omega_{a}$, $\protect\omega_{b}$%
, $\protect\omega_{0}$, and $\protect\omega_{g}$ are associated with the
respective stationary points in Fig. 1. Within the photonic band gap at $%
\protect\omega\geq\protect\omega_{g}$ the incident light is totally
reflected by the slab.}
\label{tE_PRE03}
\end{figure}

If the frequency $\omega$ is close to the band edge frequency $\omega_{g}$
in Fig. \ref{DR_PRE03}, then the incident wave will be totally reflected
back into space, as illustrated in Fig. \ref{tE_PRE03}. This implies that
the fraction of the incident wave energy converted into the slow mode
corresponding to the point $g$ in Fig. \ref{DR_PRE03} vanishes as $%
\omega\rightarrow\omega_{g}$.

In another case, where the incident wave frequency is close to either of the
characteristic values $\omega _{a}$ or $\omega _{b}$ in Fig. \ref{DR_PRE03},
the slab transmittance remains finite, as seen in Fig. \ref{tE_PRE03}. This
implies that the incident wave will be partially transmitted into the
semi-infinite photonic slab. The problem, though, is that none of the
transmitted light will propagate inside the slab in the form of the slow
mode corresponding to the respective stationary point $a$ or $b$. For
example, at frequency $\omega _{a}$, all the transmitted light corresponds
to a fast propagating mode with positive and large group velocity and the
wave number different from that corresponding to the point $a$ in Fig. \ref%
{DR_PRE03}. A similar situation takes place at $\omega =\omega _{b}$: the
fraction of the transmitted light that is converted into the respective slow
mode vanishes as $\omega \rightarrow \omega _{b}$.

Let us turn now to the stationary inflection point $0$ in Fig. \ref{DR_PRE03}%
, where both the first and the second derivatives of the frequency $\omega $
with respect to $k$ vanish, while the third derivative is finite%
\begin{equation}
\text{at }\omega =\omega _{0}\text{ \ and \ }k=k_{0}\text{ : \ }\frac{%
\partial \omega }{\partial k}=0;\;\frac{\partial ^{2}\omega }{\partial k^{2}}%
=0;\;\frac{\partial ^{3}\omega }{\partial k^{3}}>0.  \label{SIP}
\end{equation}%
In such a case, a plane wave with $\omega =\omega _{0}$ incident from the
left can be transmitted into the semi-infinite photonic crystal with little
reflection, as demonstrated in Fig. \ref{tE_PRE03}. But most remarkably,
having entered the photonic slab, the light is \emph{completely} converted
into the slow mode with infinitesimal group velocity and drastically
enhanced amplitude. Such a behavior is uniquely associated with stationary
inflection point (\ref{SIP}) of the dispersion relation and constitutes the 
\emph{frozen mode regime} \cite{PRB03,PRE03,PRE05}. In the frozen mode
regime, the vanishingly small group velocity $u$ in Eq. (\ref{S=Wu}) is
offset by the diverging value of the energy density $W$%
\begin{equation}
\text{as }\omega \rightarrow \omega _{0}\text{: \ \ \ }u\sim \left\vert
\omega -\omega _{0}\right\vert ^{2/3}\rightarrow 0,~~W\sim \left\vert \omega
-\omega _{0}\right\vert ^{-2/3}\rightarrow \infty ,  \label{u=0, W = inf}
\end{equation}%
As a result, the energy flux (\ref{S=Wu}) associated with the transmitted
frozen mode remains finite and comparable with that of the incident wave
even at the frozen mode frequency $\omega _{0}$ corresponding to the point $%
0 $ of the dispersion relation in Fig. \ref{DR_PRE03}. Such a spectacular
behavior is uniquely attributed to a stationary inflection point (\ref{SIP})
of the electromagnetic dispersion relation. Of course, in reality, the
electromagnetic energy density $W$ of the frozen mode will be limited by
such factors as absorption, nonlinear effects, imperfection of the periodic
dielectric array, deviation of the incident radiation from a perfect plane
monochromatic wave, finiteness of the photonic slab dimensions, etc. Still,
with all these limitations in place, the frozen mode regime can be very
attractive for a variety of practical applications.

In the following sections we present a detailed analysis of the frozen mode
regime associated with stationary inflection point (\ref{SIP}). In the rest
of this section we briefly discuss the effect of photonic crystal boundaries
on slow light phenomena.

\subsection{Slow light in a finite photonic slab}

Up to this point we have considered light incident on the surface of a
semi-infinite photonic crystal. Since real photonic crystals are always
bounded, the question arises whether and how the photonic crystal boundaries
affect the conditions of slow mode excitation and propagation.

\begin{figure}[tbph]
\begin{center}
\scalebox{0.8}{\includegraphics[viewport=0 0 400 150,clip]{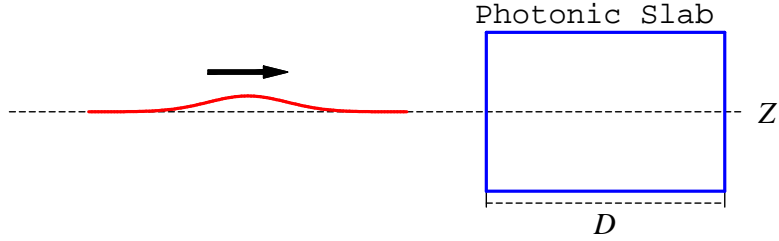}}
\end{center}
\caption{A pulse of length $l_{0}$ approaching a photonic slab of thickness $%
D$. The arrow shows the direction of pulse propagation. What happens after
the pulse hits the slab boundary is shown in Fig. 5.}
\label{PulseI}
\end{figure}

\begin{figure}[tbph]
\begin{center}
\scalebox{0.8}{\includegraphics[viewport=0 0 400 150,clip]{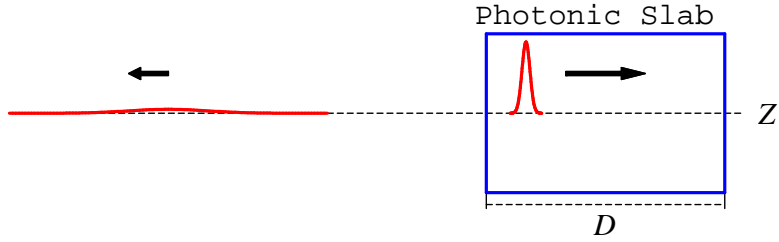}}
\end{center}
\caption{After hitting the slab, the pulse splits into the reflected and
transmitted pulses. In a slow light situation, the transmitted pulse gets
compressed in space.}
\label{PulseT}
\end{figure}

To start with, let us recall that in an unbounded (infinite) photonic
crystal, the speed of light propagation is defined as its group velocity (%
\ref{u}), which determines the speed of pulse propagation in the medium. The
spatial length $l$ of a pulse inside the unbounded periodic medium is%
\begin{equation}
l\sim l_{0}\frac{u}{c}  \label{l=l0 u/c}
\end{equation}%
where $l_{0}$ is the spatial length of the same pulse in vacuum. The
quantity $l_{0}$ is directly related to the pulse bandwidth $\Delta \omega $%
\begin{equation}
\frac{\Delta \omega }{\omega }\sim \frac{\lambda _{0}}{l_{0}}=\frac{2\pi }{%
\omega }\frac{c}{l_{0}},  \label{D om}
\end{equation}%
where%
\begin{equation*}
\lambda _{0}=\frac{2\pi }{\omega }c
\end{equation*}%
is the light wavelength in vacuum.

If instead of an infinite photonic crystal we have a bounded photonic slab
of thickness $D$, as shown in Fig. \ref{PulseI} and \ref{PulseT}, the simple
interpretation of the group velocity $u$ as the speed of pulse propagation
can still apply, provided that the pulse length $l$ inside the photonic slab
is much smaller than the slab itself%
\begin{equation}
l\ll D.  \label{l<<D}
\end{equation}%
In other words, one can introduce the speed of pulse propagation inside the
slab only if the entire pulse can fit inside the slab, as in the situation
shown in Fig. \ref{PulseT}. In the slow light case, the group velocity $u$
decreases sharply, and so does the pulse length $l$ in (\ref{l=l0 u/c}).
Therefore, a slow pulse with a fixed bandwidth $\Delta \omega $ is more
likely to fit inside the photonic slab than a fast pulse with the same
bandwidth. The slower the pulse is, the better the condition (\ref{l<<D}) is
satisfied. Taking into account the relations (\ref{l=l0 u/c}) and (\ref{D om}%
), the condition (\ref{l<<D}) can also be recast as a lower limit on the
pulse bandwidth%
\begin{equation}
\Delta \omega \gg \frac{2\pi u}{D},  \label{Dom>>u/D}
\end{equation}%
implying that in order to fit inside the slab, the pulse bandwidth should
not be too narrow.

If a pulse satisfying the condition (\ref{l<<D}) or, equivalently, (\ref%
{Dom>>u/D}) is incident on a finite photonic slab, the slab can be treated
as a semi-infinite medium until the pulse actually hits the opposite
boundary of the slab. Except for the next subsection, all the results
discussed in this paper relate to the case (\ref{l<<D}), where we can
explicitly and literally talk about pulse propagation inside the medium and
where the group velocity $u$ in (\ref{u}) does have the meaning of the speed
of pulse propagation.

\subsection{Resonance effects in a finite photonic slab}

A qualitatively different picture emerges if the pulse length $l$ defined in
Eq. (\ref{l=l0 u/c}) is comparable in magnitude or exceeds the slab
thickness $D$. In such a case, the slab is too thin to accommodate the
entire pulse and the electromagnetic field $\Psi _{T}$ inside the slab
becomes a superposition of forward and backward propagating waves undergoing
multiple reflections from two opposite boundaries of the slab. This
situation by no means can be interpreted as an individual pulse propagating
through the periodic medium, because at any moment of time the
electromagnetic field inside the slab cannot be viewed as a wave packet
built around a single propagating mode. The term slow light does not
literary apply here and, therefore, this case goes beyond the scope of this
paper. Yet, it would be appropriate to discuss briefly what happens if the
photonic slab becomes too thin to be treated as semi-infinite.

\begin{figure}[tbph]
\begin{center}
\scalebox{0.8}{\includegraphics[viewport=0 0 300 250,clip]{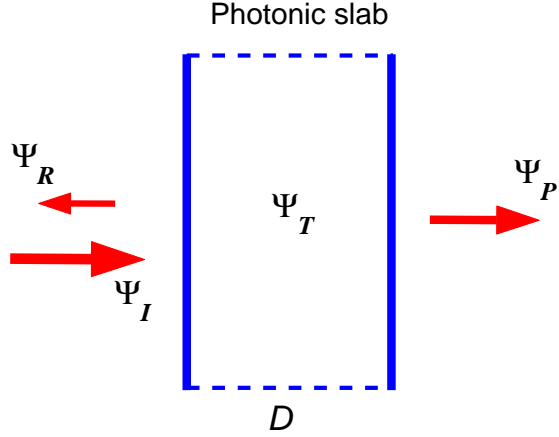}}
\end{center}
\caption{Light incident on a finite photonic slab of the thickness $D$. The
subscripts $I$, $R$, and $P$ refer to the incident, reflected, and passed
waves, respectively. The transmitted wave $\Psi_{T}$ inside the slab may
have Bloch componets propagating in either direction.}
\label{FSn}
\end{figure}

Assume that the photonic slab is thin enough to satisfy the inequality%
\begin{equation}
l\gg D,  \label{l>>D}
\end{equation}%
which is opposite to (\ref{l<<D}). The condition (\ref{l>>D}) establishes an
upper limit on the incident pulse bandwidth%
\begin{equation}
\Delta \omega \ll \frac{2\pi u}{D}.  \label{Dom<<u/D}
\end{equation}%
Consider a plane monochromatic wave incident on a finite photonic slab in
Fig. \ref{FSn}. Since a monochromatic wave packet has $l\rightarrow \infty $%
, the relations (\ref{l>>D}) and (\ref{Dom<<u/D}) are perfectly satisfied.
If the photonic slab is lossless, its steady-state transmittance and
reflectance are defined by the following expressions%
\begin{equation}
\tau =\frac{S_{P}}{S_{I}}=\frac{S_{T}}{S_{I}},\ \rho =-\frac{S_{R}}{S_{I}}%
=1-\tau ,  \label{tau(D)}
\end{equation}%
similar to those in (\ref{tau(n)}) related to the semi-infinite slab. The
Eqs. (\ref{tau(D)}) immediately follow from energy conservation
considerations.

\begin{figure}[tbph]
\begin{center}
\scalebox{0.8}{\includegraphics[viewport=0 0 500 400,clip]{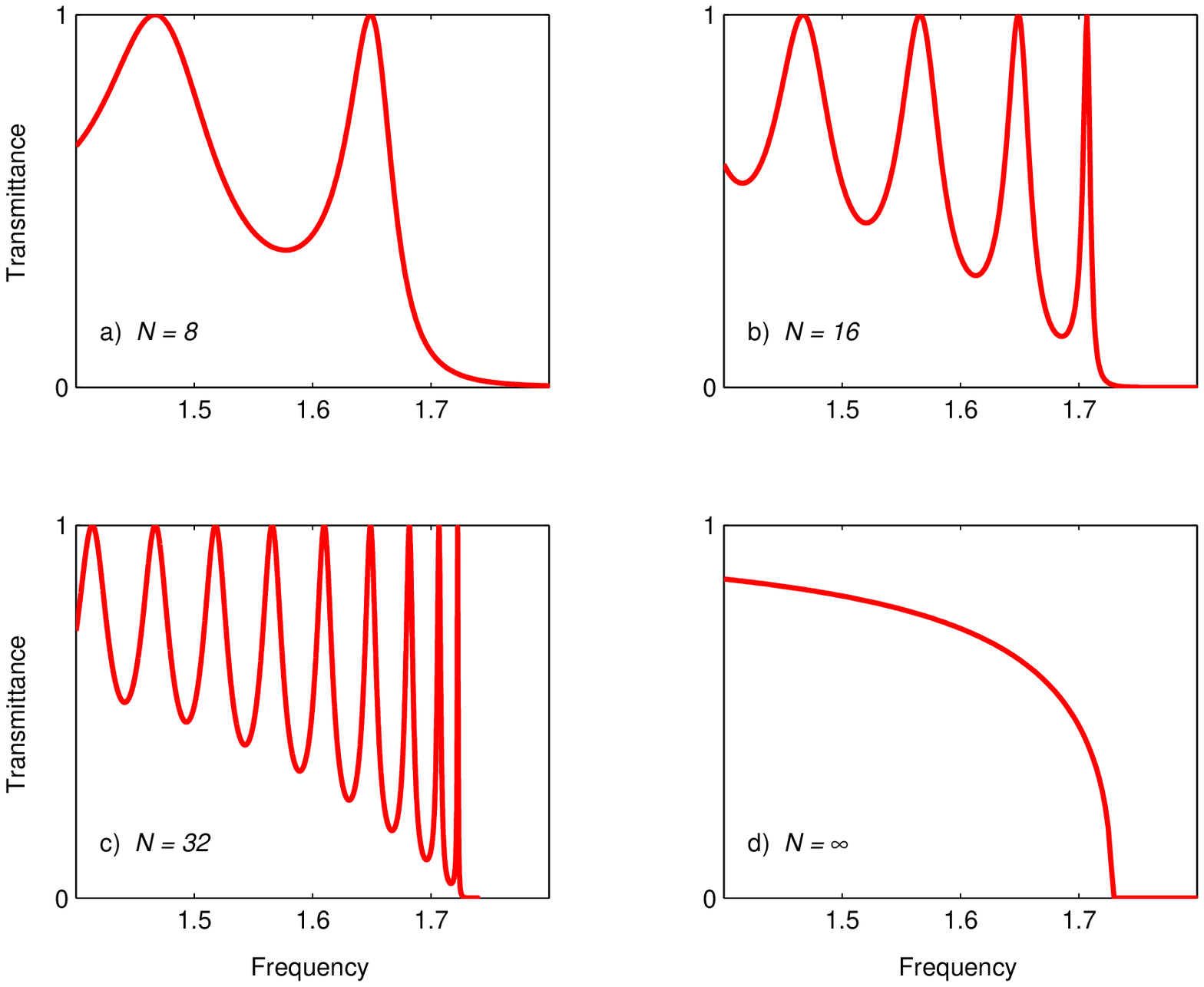}}
\end{center}
\caption{Typical plots of transmittance versus frequency of lossless
periodic stacks composed of different number $N$ of unit cells. The
frequency range shown includes a photonic band edge. The sharp transmission
peaks near the band edge frequency are associated with Fabry-Perot cavity
resonance. The case $N=\infty$ corresponds to a semi-infinite photonic slab
and is similar to that shown in Fig. 3.}
\label{tfN}
\end{figure}

A typical frequency dependence of finite photonic slab transmittance (\ref%
{tau(D)}) is shown in Figs. \ref{tfN} \emph{a}, \emph{b}, and \emph{c}. For
comparison, Fig. \ref{tfN} \emph{d} shows the transmittance (\ref{tau(n)})
of a semi-infinite photonic slab having the same periodic structure. The
sharp peaks in transmittance in the vicinity of photonic band edge at larger 
$N$ correspond to Fabry-Perot cavity resonances. At resonance, the
electromagnetic field $\Psi _{T}$ inside the slab is close to a standing
wave composed of one forward and one backward propagating Bloch eigenmodes
with large and nearly equal amplitudes. The slab boundaries coincide with
standing wave nodes, which determines the Bloch wavenumbers of the forward
and backward components%
\begin{equation}
k_{s}\approx k_{g}\pm \frac{\pi }{D}s,\ \ s=1,2,...  \label{k_s}
\end{equation}%
where $k_{g}$ corresponds to the photonic band edge. Usually, but not
necessarily, $k_{g}$ equals $0$ or $\pi /L$. The approximation (\ref{k_s})
is valid if $N\gg 1$ and only applies to the resonances close enough to the
photonic band edge.

The dispersion function $\omega \left( k\right) $ in the vicinity of a
photonic band edge can be approximated as follows 
\begin{equation}
\omega \approx \omega _{g}-\frac{\omega _{g}^{\prime \prime }}{2}\left(
k-k_{g}\right) ^{2},\text{ \ where }  \label{w = k^2}
\end{equation}%
where%
\begin{equation*}
\omega _{g}^{\prime \prime }=\left( \frac{\partial ^{2}\omega }{\partial
k^{2}}\right) _{k=k_{g}}.
\end{equation*}%
The propagating mode group velocity $u$ vanishes as $\omega \rightarrow
\omega _{g}$%
\begin{equation}
u=\frac{\partial \omega }{\partial k}\approx \omega _{g}^{\prime \prime
}\left( k_{g}-k\right) \approx \pm \sqrt{2\omega _{g}^{\prime \prime }}%
\left( \omega _{g}-\omega \right) ^{1/2},  \label{u = w^1/2}
\end{equation}%
where $\pm $ corresponds to the forward and backward propagating waves,
respectively. Inserting the resonance values (\ref{k_s}) of the Bloch
wavenumber into the dispersion relation Eq. (\ref{w = k^2}) yields the
resonance frequencies $\omega _{s}$ as 
\begin{equation}
\omega _{s}\approx \omega _{g}-\frac{\omega _{g}^{\prime \prime }}{2}\left( 
\frac{\pi }{D}s\right) ^{2},\ \ s=1,2,...  \label{w_s}
\end{equation}%
where $\omega _{g}=\omega \left( k_{g}\right) $ is the band edge. The
dependence (\ref{w_s}) is illustrated in Fig. \ref{tfN}\emph{c}.

Let us focus on the Fabry-Perot cavity resonance closest to the photonic
band edge. The respective frequency is%
\begin{equation}
\omega_{1}\approx\omega_{g}-\frac{\omega_{g}^{\prime\prime}}{2}\left( \frac{%
\pi}{D}\right) ^{2}.  \label{w_1}
\end{equation}
At frequency $\omega_{1}$, the group velocities of the forward and backward
propagating modes are%
\begin{equation}
u_{1}\approx\pm\omega_{g}^{\prime\prime}\frac{\pi}{D}=\pm\frac{\pi\omega
_{g}^{\prime\prime}}{LN},  \label{u_1}
\end{equation}
that is inversely proportional to the number $N$ of the unit cells $L$ in
the slab. The resonance field amplitude inside the slab is proportional to
the slab thickness%
\begin{equation}
\Psi_{T}\left( z\right) \approx N\Psi_{0}\sin\left( \pi zD^{-1}\right)
+\Psi_{1},\ 0\leq z\leq D,  \label{PsiT FP}
\end{equation}
where $\Psi_{0}$ and $\Psi_{1}$ are periodic functions of $z$ comparable in
magnitude with the incident wave $\Psi_{I}$. So, the maximum field amplitude
is reached in the middle of the slab and is proportional to the slab
thickness. The bandwidth $\Delta_{1}$ of the Fabry-Perot cavity resonance
decreases sharply, as the number of unit cells increases 
\begin{equation}
\Delta_{1}\propto\frac{c}{LN^{3}}\propto\frac{\omega_{1}}{N^{3}}.
\label{Delta1}
\end{equation}
This is clearly seen in Fig. \ref{tfN}.

At this point we would like to compare the frozen mode regime introduced in
the previous subsection and the Fabry-Perot cavity resonance. Both effects
result from coherent interference of light and can be thought of as photons
trapped inside the periodic medium. Both effects are accompanied by a huge
surge in electromagnetic field amplitude inside the photonic crystal. But
that is where their similarity ends. Indeed, in the case of a Fabry-Perot
cavity, the entire periodic stack (photonic slab) works as a resonator in
which the trapped photons are spread all over the place. For this reason,
all the major characteristics of Fabry-Perot cavity resonance are
essentially dependent on the slab thickness. If the slab thickness $D=NL$ is
too large, then even small absorption or structural irregularity will
completely smooth out the resonances. So, on the one hand, the slab should
have enough layers to support distinct Fabry-Perot cavity resonances. But on
the other hand, the number of layers should not be too large so that the
losses and structural irregularities would not wipe out the effect. In
addition, the number of layers essentially affects the resonance bandwidth.
By contrast, the frozen mode regime is not a resonance in a usual sense of
this word. Each trapped photon is now localized within certain small number
of unit cells depending on the pulse bandwidth, while the slab size is not
essential at all. Even if $N\rightarrow \infty $, it does not affect any
basic characteristics of the frozen mode regime, such as the bandwidth or
the frozen mode amplitude.\newline

\subsubsection{Photonic slab as a delay line}

In the case (\ref{l<<D}) of a thin slab, the idea of a distinct pulse slowly
propagating through the slab does not apply. On the other hand, one might be
interested in the relation between the input and the output pulses, rather
than in what is going on inside the photonic slab. This is the case, for
example, if the photonic slab is used as a delay line. Let $\Psi _{I}$ and $%
\Psi _{P}$ be the input and output pulses, respectively, as illustrated in
Fig. \ref{FSn}. The shape of the output pulse $\Psi _{P}$ can be close to
that of the input pulse $\Psi _{I}$ regardless of whether or not the
condition (\ref{l<<D}) is met. If the shape of the pulse is indeed
preserved, one can define the effective speed $\tilde{u}$ of pulse
propagation through the slab as%
\begin{equation}
\tilde{u}=\frac{D}{\tilde{t}},  \label{u_tilde}
\end{equation}%
where $\tilde{t}$ is the transit time of the pulse passed through the slab.
The quantity $\tilde{u}$ is referred to as the group delay. The transit time
determines the pulse delay due to the presence of the slab. Of course, in
the case (\ref{l<<D}) of a thick slab, the effective speed (\ref{u_tilde})
coincides with the pulse group velocity $u$. But now we consider the
opposite situation (\ref{l>>D}). It turns out that under the resonance
conditions, the transit time $\tilde{t}$ of a thin photonic slab increases
sharply, and the respective group delay $\tilde{u}$\ can be as low as $%
10^{-2}c$, while the pulse passes through the slab with little reflection
(see, for example, \cite{SL Scal1}, and references therein). In this sense,
the pulse delay can be classified as a slow light effect, although the
quantity $\tilde{u}$\ does not relate to the speed of any real pulse inside
the photonic slab. In the rest of this section we briefly discuss this well
known phenomenon.

Let us estimate the group delay associated with Fabry-Perot cavity
resonance. According to Eq. (\ref{PsiT FP}), the electromagnetic energy $%
\mathcal{H}$ stored in the entire slab at the resonance is%
\begin{equation}
\mathcal{H}\propto\left\vert \Psi_{T}\right\vert ^{2}D\propto\left\vert
\Psi_{I}\right\vert ^{2}LN^{3}.  \label{WD FP}
\end{equation}
This leads to the following rough estimate for the transit time $\tilde{t}$
in (\ref{u_tilde}) 
\begin{equation}
\tilde{t}_{1}\propto\frac{\mathcal{H}}{S_{I}}\propto c^{-1}LN^{3}.
\label{t1_tilde}
\end{equation}
The respective group delay (\ref{u_tilde}) is%
\begin{equation}
\tilde{u}_{1}=\frac{D}{\tilde{t}_{1}}\propto\frac{c}{N^{2}}.
\label{u1_tilde}
\end{equation}

Note that if the number $N$ is large, the value (\ref{u1_tilde}) of the
group delay is much lower than the group velocity (\ref{u_1}) of the
propagating Bloch mode at the same frequency $\omega _{1}$. The drawback,
though, is that the bandwidth (\ref{Delta1}) of the Fabry-Perot cavity
resonance shrinks even faster as the number $N$ of unit cells increases.
Eqs. (\ref{u1_tilde}) and (\ref{Delta1}) yield the following relation
between the bandwidth $\Delta _{1}$ and the group delay $\tilde{u}_{1}$%
\begin{equation}
\frac{\Delta _{1}}{\omega _{1}}\propto \frac{1}{N}\frac{\tilde{u}_{1}}{c}.
\label{Delta1/w}
\end{equation}%
Comparison of the slow light bandwidth (\ref{Delta1/w}) with its ideal value
(\ref{Delta om}) shows that the Fabry-Perot cavity resonance in a finite
periodic photonic slab has a fundamental bandwidth disadvantage, if used as
a delay line.

Note that real optical delay lines are commonly based on periodic arrays of
weakly coupled resonators, such as Fabry-Perot cavities, rather than on
individual Fabry-Perot cavities (see, for example, \cite{CR Boyd1,CR
Boyd2,CR Meloni 03,CR Yariv04,OPN 05,Khurgin1,Khurgin2} and references
therein).\bigskip

In conclusion, let us reiterate that in the cases other than (\ref{l<<D}),
there is no distinct pulse propagating inside the periodic medium and,
therefore, the notion of slow light does not literally apply there. Further
in this paper we assume that the condition (\ref{l<<D}) is satisfied,
warranting the approximation of a semi-infinite photonic crystal. This
allows us to investigate the slow light phenomenon in its pure form, when it
is directly related to the speed of electromagnetic pulse propagation
through the medium. In this case, the frozen mode regime associated with a
stationary inflection point (\ref{SIP}) provides a unique possibility of
converting a significant fraction of the incident light into a coherent mode
with extremely low group velocity and drastically enhanced amplitude.

\section{Slow light in periodic layered media}

From now on we restrict ourselves to stratified media, which are periodic
stacks of dielectric layers. Such systems are also referred to as photonic
crystals with one-dimensional periodicity. A major reason for such a choice
is that the electrodynamics of stratified media can be done within the
framework of a rigorous analytical approach. This is particularly important
since the frozen mode regime involves a unique and spectacular behavior, so
it would be desirable to be sure that such a behavior is not a numerical
artifact. As soon as we assume that the semi-infinite photonic slab in Fig. %
\ref{SISn} is a periodic array of plane-parallel uniform layers, we can give
a much more detailed and meaningful description of the frozen mode regime.

We start with some general remarks about electromagnetic eigenmodes in
periodic layered media. Then we proceed to a semi-qualitative description of
the situation taking place at different stationary points of the dispersion
relations. A consistent and complete analysis based on the Maxwell equations
will be presented in Sections 5 through 12.

\subsection{Propagating and evanescent eigenmodes in periodic stacks of
anisotropic layers}

Let $\Psi _{I}\left( z\right) $, $\Psi _{R}\left( z\right) $ and $\Psi
_{T}\left( z\right) $\ denote the incident, reflected and transmitted waves,
respectively, as shown in Fig. \ref{SISn}. In the frequency domain, each of
these waves can be explicitly represented by a column vector%
\begin{equation}
\Psi \left( z\right) =\left[ 
\begin{array}{c}
E_{x}\left( z\right) \\ 
E_{y}\left( z\right) \\ 
H_{x}\left( z\right) \\ 
H_{y}\left( z\right)%
\end{array}%
\right] ,  \label{Psi}
\end{equation}%
where $E_{x}\left( z\right) ,E_{y}\left( z\right) ,H_{x}\left( z\right)
,H_{y}\left( z\right) \mathbf{\ }$are the transverse components of
electromagnetic field. The exact definition of $\Psi \left( z\right) $ is
given in (\ref{LEM}) and (\ref{ME4}). The incident and reflected beams are
plane monochromatic waves propagating in vacuum, while the transmitted
electromagnetic field $\Psi _{T}\left( z\right) $ inside the periodic
layered medium is not a single Bloch eigenmode. At the slab boundary at $z=0$%
, the three waves satisfy the standard boundary condition%
\begin{equation}
\Psi _{I}\left( 0\right) +\Psi _{R}\left( 0\right) =\Psi _{T}\left( 0\right)
,  \label{BC}
\end{equation}%
implying continuity of the tangential field components (\ref{Psi}). Note
that periodic stacks capable of supporting the frozen mode regime must
include anisotropic layers with misaligned and/or oblique orientation of the
principal axes. As a consequence, the reflected and transmitted waves in
Fig. \ref{SISn} will have an elliptic polarization even if the incident wave
is linearly polarized.

In the setting of Fig. \ref{SISn} where the semi-infinite periodic layered
array occupies the half-space $z\geq0$, the transmitted wave $\Psi_{T}\left(
z\right) $ is a superposition of two Bloch components (Bloch eigenmodes)
with different polarizations and different values of the Bloch wave number $%
k $. There are three possibilities.

\begin{enumerate}
\item Both Bloch components of the transmitted wave $\Psi_{T}$ are
propagating modes%
\begin{equation}
\Psi_{T}\left( z\right) =\Psi_{pr1}\left( z\right) +\Psi_{pr2}\left(
z\right) ,\ \;z\geq0,  \label{PsiT=pr+pr}
\end{equation}
which means that the two respective values of $k$ are real. For example, at $%
\omega_{b}<\omega<\omega_{a}$ in Fig. \ref{DR_PRE03}, the transmitted wave $%
\Psi_{T}$ is composed of two Bloch eigenmodes with two different real wave
numbers $k_{1}$ and $k_{2}$ and two different group velocities $u_{1}>0$ and 
$u_{2}>0$. This constitutes the phenomenon of double refraction.

\item Both Bloch components of $\Psi_{T}$ are evanescent%
\begin{equation}
\Psi_{T}\left( z\right) =\Psi_{ev1}\left( z\right) +\Psi_{ev2}\left(
z\right) ,\ \;z\geq0,  \label{PsiT=ev+ev}
\end{equation}
which implies that the two respective values of $k$ are complex with $\func{%
Im}k>0$. For example, this is the case when the frequency $\omega$ falls
into the photonic band gap at $\omega>\omega_{g}$ in Fig. \ref{DR_PRE03}.
The fact that $\func{Im}k>0$ implies that the wave amplitude decays as the
distance $z$ from the semi-infinite slab surface increases. In the case (\ref%
{PsiT=ev+ev}), the incident wave is totally reflected back to space by the
semi-infinite slab, as seen in Fig. \ref{tE_PRE03}.

\item Of particular interest is the case where one of the Bloch components
of the transmitted wave $\Psi_{T}$ is a propagating mode with $u>0$, while
the other is an evanescent mode with $\func{Im}k>0$%
\begin{equation}
\Psi_{T}\left( z\right) =\Psi_{pr}\left( z\right) +\Psi_{ev}\left( z\right)
,\ \;z\geq0.  \label{PsiT=pr+ev}
\end{equation}
For example, this is the case at the frequency range%
\begin{equation}
\omega_{a}<\omega<\omega_{g}  \label{wa<w0<wb}
\end{equation}
in Fig. \ref{DR_PRE03}. As the distance $z$ from the slab/vacuum interface
increases, the evanescent contribution $\Psi_{ev}$ in (\ref{PsiT=pr+ev})
decays as $\exp\left( -z\func{Im}k\right) $, and the resulting transmitted
wave $\Psi_{T}$ turns into a single propagating Bloch mode $\Psi_{pr}$.
\end{enumerate}

Propagating modes with $u<0$, as well as evanescent modes with $\func{Im}k<0$%
, never contribute to the transmitted wave $\Psi_{T}$ inside the
semi-infinite stack in Fig. \ref{SISn}. This fact is based on the following
two assumptions:

\begin{itemize}
\item[-] The transmitted wave $\Psi_{T}$ and the reflected wave $\Psi_{R}$
are originated from the plane monochromatic wave $\Psi_{I}$ incident on the
semi-infinite photonic slab from the left, as shown in Fig. \ref{SISn}.

\item[-] The layered array in Fig. \ref{SISn} occupies the entire half-space
and is perfectly periodic at $z>0$.
\end{itemize}

If either of the above conditions is violated, the field $\Psi_{T}$ inside
the periodic stack can be a superposition of four Bloch eigenmodes with
either sign of the group velocity $u$ of propagating contributions, or
either sign of $\func{Im}k$\ of evanescent contributions. For example, this
would be the case if the periodic layered array in Fig. \ref{SISn} had some
kind of structural defects or a finite thickness like that presented in Fig. %
\ref{FSn}.

The propagating modes with $u>0$ and evanescent modes with $\func{Im}k>0$
are referred to as the \emph{forward} waves. Only forward modes contribute
to the transmitted wave $\Psi _{T}\left( z\right) $ in the case of a
periodic semi-infinite stack. The propagating modes with $u<0$ and
evanescent modes with $\func{Im}k<0$ are referred to as \emph{backward}
waves. Since the backward Bloch waves are not excited in the setting of Fig. %
\ref{SISn}, they play no role in further consideration.

In all three cases (\ref{PsiT=pr+pr} -- \ref{PsiT=ev+ev}), the contribution
of a particular Bloch eigenmode to the transmitted wave $\Psi _{T}$ depends
on the polarization $\Psi _{I}$ of the incident wave. One can always choose
the incident wave polarization so that only one Bloch component is excited.
In such a case, $\Psi _{T}$ is a single Bloch eigenmode.

Only propagating modes contribute to the normal component $S_{T}$ of the
energy flux inside a periodic semi-infinite slab. Evanescent modes do not
participate in energy transfer in such a case. In the important particular
case of a single propagating mode ($\Psi _{T}=\Psi _{pr}$), we have from (%
\ref{S=Wu}) and (\ref{tau(n)})%
\begin{equation}
S_{T}=Wu=\tau S_{I},  \label{ST=Wu}
\end{equation}%
where $W\sim \left\vert \Psi _{pr}\right\vert ^{2}$ is the energy density
associated with the transmitted propagating mode.

The assumption that the transmitted wave $\Psi _{T}\left( z\right) $ is a
superposition of propagating and/or evanescent Bloch eigenmodes may not be
valid at stationary points (\ref{SP}) of electromagnetic dispersion function 
$\omega \left( k\right) $, because each stationary point is a degeneracy
point of the frequency spectrum. For example, if the frequency $\omega $
exactly coincides with stationary inflection point defined by (\ref{SIP}),
the transmitted wave $\Psi _{T}\left( z\right) $ is dominated by a
(non-Bloch) Floquet eigenmode linearly growing with $z$, which constitutes
the frozen mode regime \cite{PRB03,PRE03}. At all other frequencies, the
transmitted wave is a superposition of two forward Bloch modes, each of
which can be either propagating or evanescent. A detailed analysis of this
and related phenomena is presented further in this paper.

Knowing the eigenmode composition of the transmitted wave $\Psi_{T}\left(
z\right) $ we can give a semi-qualitative description of what happens when
the frequency $\omega$ of the incident wave approaches one of the stationary
points (\ref{SP}) in Fig. \ref{DR_PRE03}. A complete analysis based on the
Maxwell equations will be presented later in the paper.

\subsection{Photonic band edge}

We start with the simplest case of a photonic band edge. Just below the band
edge frequency $\omega _{g}$ in Fig. \ref{DR_PRE03}, the transmitted field $%
\Psi _{T}\left( z\right) $ is a superposition (\ref{PsiT=pr+ev}) of one
propagating and one evanescent Bloch component. Due to the boundary
condition (\ref{BC}) at the slab/vacuum interface, the amplitude of the
transmitted wave at $z=0$ is comparable with that of the incident wave. In
the case of a generic polarization of the incident light, the amplitudes of
the propagating and evanescent Bloch components at $z=0$ are also comparable
to each other and to the amplitude of the incident light%
\begin{equation}
\left\vert \Psi _{pr}\left( 0\right) \right\vert \sim \left\vert \Psi
_{ev}\left( 0\right) \right\vert \sim \left\vert \Psi _{I}\right\vert ,\text{
at }\ \omega \leq \omega _{g}.  \label{pr = ev = I}
\end{equation}%
As the distance $z$ from the slab surface increases, the evanescent
component $\Psi _{ev}\left( z\right) $ decays rapidly, while the amplitude
of the propagating component remains constant. Eventually, at a certain
distance from the slab surface, the transmitted wave $\Psi _{T}\left(
z\right) $ becomes very close to its propagating Bloch component%
\begin{equation}
\Psi _{T}\left( z\right) \approx \Psi _{pr}\left( z\right) ,\text{ at }z\gg
L,\ \omega \leq \omega _{g}.  \label{PsiT=Psipr}
\end{equation}

The evanescent component $\Psi _{ev}$ of the transmitted wave does not
display any singularity in the vicinity of $\omega _{g}$. By contrast, the
propagating mode $\Psi _{pr}$ develops a singularity as $\omega \rightarrow
\omega _{g}-0$, which is associated with vanishing group velocity as in (\ref%
{SP}). At $\omega >\omega _{g}$, the propagating mode turns into another
evanescent mode in (\ref{PsiT=ev+ev}).

The dispersion relation in the vicinity of the band edge $g$ in Fig. \ref%
{DR_PRE03} can be approximated as%
\begin{equation*}
\omega_{g}-\omega\approx\frac{\omega_{g}^{\prime\prime}}{2}\left(
k-k_{g}\right) ^{2},\ \ \omega\lessapprox\omega_{g}.
\end{equation*}
This yields the following frequency dependence of the propagating mode group
velocity $u$ below the photonic band edge 
\begin{equation}
u=\frac{\partial\omega}{\partial k}\approx\omega_{g}^{\prime\prime}\left(
k_{g}-k\right) \approx\sqrt{2\omega_{g}^{\prime\prime}}\left( \omega
_{g}-\omega\right) ^{1/2},\ \ \omega\lessapprox\omega_{g}.  \label{u=w^1/2}
\end{equation}
The energy flux (\ref{ST=Wu}) associated with the slow propagating mode is%
\begin{equation}
S_{T}\approx\left\{ 
\begin{array}{c}
W\sqrt{2\omega_{g}^{\prime\prime}}\left( \omega_{g}-\omega\right) ^{1/2},%
\text{ at }\omega\lessapprox\omega_{g} \\ 
0,\text{ at }\omega\geq\omega_{g}%
\end{array}
\right. .  \label{ST(g)}
\end{equation}
where%
\begin{equation}
W\sim\left\vert \Psi_{pr}\right\vert ^{2}\sim\left\vert \Psi_{I}\right\vert
^{2}.  \label{W = pr = I}
\end{equation}
The latter estimation follows from (\ref{pr = ev = I}) and applies to the
case of a generic polarization of the incident wave. The semi-infinite slab
transmittance (\ref{tau(n)}) in the vicinity of $\omega_{g}$ is 
\begin{equation}
\tau=\frac{S_{T}}{S_{I}}\approx\left\{ 
\begin{array}{c}
\frac{W}{S_{I}}\sqrt{2\omega_{g}^{\prime\prime}}\left(
\omega_{g}-\omega\right) ^{1/2},\text{ at }\omega\lessapprox\omega_{g} \\ 
0,\text{ at }\omega\geq\omega_{g}%
\end{array}
\right. ,  \label{tau(g)}
\end{equation}
where according to (\ref{W = pr = I}) 
\begin{equation*}
\frac{W}{S_{I}}\sim\frac{\left\vert \Psi_{I}\right\vert ^{2}}{S_{I}}\sim c.
\end{equation*}
The relation (\ref{tau(g)}) is illustrated by the numerical example in Fig. %
\ref{tE_PRE03}.

Equation (\ref{tau(g)}) expresses the well-known fact that in the vicinity
of an electromagnetic band edge, the semi-infinite photonic crystal becomes
totally reflective, as illustrated in Fig. \ref{tE_PRE03}. This implies that
as $\omega \rightarrow \omega _{g}$, only an infinitesimal fraction of the
incident light energy is converted into the slow mode.

\subsection{Other extreme points of spectral branches}

For specificity, let us consider the stationary point $a$ of the dispersion
relation in Fig. \ref{DR_PRE03}, which qualitatively is not different from
the point $b$. At frequencies right below $\omega_{a}$ , the transmitted
wave $\Psi_{T}$ is a superposition (\ref{PsiT=pr+pr}) of two propagating
eigenmodes, one of which is the slow mode and the other is a regular forward
propagating mode. The slow mode develops a singularity at $\omega=\omega_{a}$
similar to that of the respective slow mode in the vicinity of the band edge
frequency $\omega_{g}$, while the other propagating mode (the fast mode)
remains regular in the vicinity of $\omega_{a}$ and does not produce any
anomaly. The two forward modes contribute additively to the energy flux $%
S_{T}$, but the contribution of the fast mode remains regular in the
vicinity of $\omega_{a}$, while the contribution of the slow mode shows the
same singular behavior as that described by Eqs. (\ref{ST(g)}) and (\ref%
{tau(g)}). Fig. \ref{tE_PRE03} provides a graphic illustration of such a
behavior.

The important point is that similar to the situation in the vicinity of a
photonic band edge, at $\omega=\omega_{a}$ and $\omega=\omega_{b}$ the
contribution of the respective slow mode to the transmitted wave $\Psi_{T}$
vanishes. In other words, in terms of slow mode excitation, the stationary
points $a$ and $b$ in Fig. \ref{DR_PRE03} are no different from the band
edge $g$.

\subsection{Stationary inflection point: the frozen mode regime}

A sharply different situation develops in the vicinity of a stationary
inflection point (\ref{SIP}) of the dispersion relation (point $0$ in Fig. %
\ref{DR_PRE03}). According to (\ref{SIP}), the dispersion relation in the
vicinity of $\omega_{0}$ can be approximated as follows%
\begin{equation}
\omega-\omega_{0}\approx\frac{\omega_{0}^{\prime\prime\prime}}{6}\left(
k-k_{0}\right) ^{3},  \label{w = k^3}
\end{equation}
where%
\begin{equation*}
\omega_{0}^{\prime\prime\prime}=\left( \frac{\partial^{3}\omega}{\partial
k^{3}}\right) _{k=k_{0}}.
\end{equation*}
The propagating mode group velocity $u$ vanishes as $\omega$ approaches $%
\omega_{0}$%
\begin{equation}
u=\frac{\partial\omega}{\partial k}\approx\frac{1}{2}\omega_{0}^{\prime
\prime\prime}\left( k-k_{0}\right) ^{2}\approx\frac{6^{2/3}}{2}\left(
\omega_{0}^{\prime\prime\prime}\right) ^{1/3}\left( \omega-\omega
_{0}\right) ^{2/3}.  \label{u=w^2/3}
\end{equation}
But remarkably, the electromagnetic energy density $W$ associated with the
transmitted frozen mode diverges as $\omega\rightarrow\omega_{0}$%
\begin{equation}
W\approx\frac{2\tau S_{I}}{6^{2/3}}\left( \omega_{0}^{\prime\prime\prime
}\right) ^{-1/3}\left( \omega-\omega_{0}\right) ^{-2/3},  \label{W(SIP)}
\end{equation}
where $S_{I}$ is the fixed energy flux of the incident wave. The slab
transmittance $\tau$ remains finite even at $\omega=\omega_{0}$, as
illustrated in Fig. \ref{tE_PRE03}. As a result, the energy flux (\ref{ST=Wu}%
) associated with the transmitted frozen mode also remains finite and can
even be close to unity in the vicinity of $\omega_{0}$. The latter implies
that the incident light is completely converted to the frozen mode with
infinitesimal group velocity (\ref{u=w^2/3}) and diverging energy density (%
\ref{W(SIP)}).

\begin{figure}[tbph]
\begin{center}
\scalebox{0.8}{\includegraphics[viewport=0 0 500 200,clip]{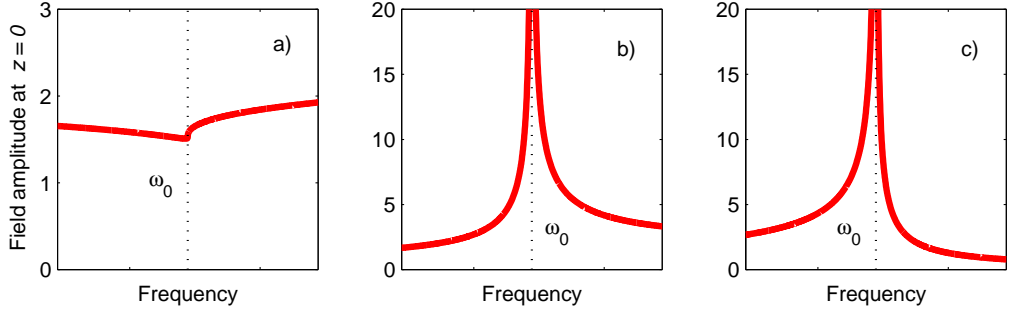}}
\end{center}
\caption{Destructive interference of the propagating and evanescent
contributions to the resulting field $\Psi_{T}$ at the slab/vacuum interface
under the frozen mode regime: a) resulting field amplitude $\left\vert
\Psi_{T}\left( 0\right) \right\vert ^{2}$, b) amplitude $\left\vert
\Psi_{pr}\left(0\right)\right\vert ^{2}$ of the propagating component, c)
amplitude $\left\vert \Psi_{ev}\left(0\right) \right\vert ^{2}$ of the
evanescent component.}
\label{AM0_PRE05}
\end{figure}

\begin{figure}[tbph]
\begin{center}
\scalebox{0.8}{\includegraphics[viewport=0 0 500 200,clip]{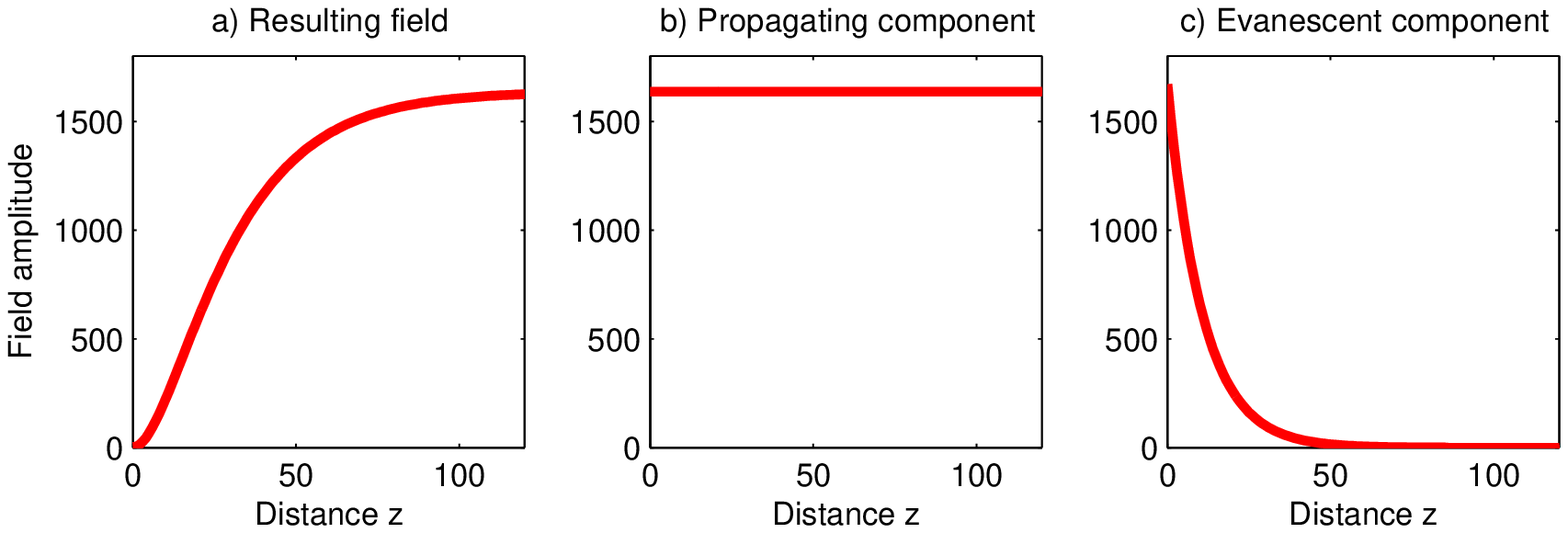}}
\end{center}
\caption{The transmitted electromagnetic field (\protect\ref{PsiT=pr+ev})
and its propagating and evanescent components inside semi-infinite slab in
close proximity of the frozen mode regime: (a) the amplitude $\left\vert
\Psi _{T}\left( z\right) \right\vert ^{2}$ of the resulting field, (b) the
amplitude $\left\vert \Psi_{pr}\left( z\right) \right\vert ^{2}$ of the
propagating contribution, (c) the amplitude $\left\vert \Psi_{ev}\left(
z\right) \right\vert ^{2}$ of the evanescent contribution. Due to
destructive interference of the propagating and evanescent components, the
resulting field amplitude at $z=0$ is small enough to satisfy the boundary
conditions (38). The amplitude $\left\vert \Psi_{I}\right\vert ^{2}$ of the
incident wave is unity. The distance $z$ from the slab boundary is expressed
in units of $L$.}
\label{AMz_PRE03}
\end{figure}

Let us consider the structure of the frozen mode. At $\omega \approx \omega
_{0}$, the transmitted wave $\Psi _{T}$ is a superposition (\ref{PsiT=pr+ev}%
) of one propagating and one evanescent Bloch component. In contrast to the
case of a photonic band edge, in the vicinity of $\omega _{0}$ both Bloch
components of $\Psi _{T}$ develop strong singularity. Specifically, as the
frequency $\omega $ approaches $\omega _{0}$, both contributions grow
sharply, while remaining nearly equal and opposite in sign at the slab
boundary%
\begin{equation}
\Psi _{pr}\left( 0\right) \approx -\Psi _{ev}\left( 0\right) \propto
\left\vert \omega -\omega _{0}\right\vert ^{-1/3},\ \ \text{as }\omega
\rightarrow \omega _{0}.  \label{DI 3}
\end{equation}%
Due to the destructive interference (\ref{DI 3}), the resulting field%
\begin{equation*}
\Psi _{T}\left( 0\right) =\Psi _{pr}\left( 0\right) +\Psi _{ev}\left(
0\right)
\end{equation*}%
at the slab boundary is small enough to satisfy the boundary condition (\ref%
{BC}), as illustrated in Fig. \ref{AM0_PRE05}. As the distance $z$ from the
slab boundary increases, the evanescent component $\Psi _{ev}$ decays
exponentially%
\begin{equation*}
\Psi _{ev}\left( z\right) \approx \Psi _{ev}\left( 0\right) \exp \left( -z%
\func{Im}k\right)
\end{equation*}%
while the amplitude of the propagating component $\Psi _{pr}$ remains
constant and very large. As a consequence, the amplitude of the resulting
transmitted wave $\Psi _{T}\left( z\right) $ sharply increases with the
distance $z$ from the slab boundary and, eventually, reaches its large
saturation value corresponding to the propagating component $\Psi _{pr}$, as
illustrated in Fig. \ref{AMz_PRE03}.

If the frequency $\omega$ of incident light is exactly equal to the frozen
mode frequency $\omega_{0}$, the transmitted wave $\Psi_{T}\left( z\right) $
does not reduce to the sum (\ref{PsiT=pr+ev}) of propagating and evanescent
contributions \cite{PRB03,PRE03}. Instead, it corresponds to a non-Bloch
Floquet eigenmode diverging linearly with $z$ 
\begin{equation*}
\Psi_{T}\left( z\right) -\Psi_{T}\left( 0\right) \propto z\sqrt{\frac{\tau
S_{I}}{\omega_{0}^{\prime\prime\prime}}}\Psi_{0},\;\ \text{at}\;\omega
=\omega_{0}.
\end{equation*}

Evidently, the frozen mode regime associated with stationary inflection
point (\ref{SIP}) provides an ideal and unique situation in terms of slow
mode excitation. Indeed, in this case virtually all the incident light
energy can be converted into the slow mode with greatly enhanced amplitude.

A consistent mathematical analysis of the frozen mode regime is rather
sophisticated and will take a great deal of our attention further in this
paper. Specifically, the fundamental fact that at $\omega=\omega_{0}$, the
energy flux of the frozen mode remains finite in spite of vanishing group
velocity, is rigorously proven in Section 9 (see Eqs. (\ref{tz1}) through (%
\ref{tz22}) and related explanations).

\subsection{Degenerate band edge}

\begin{figure}[tbph]
\begin{center}
\scalebox{0.8}{\includegraphics[viewport=0 0 500 200,clip]{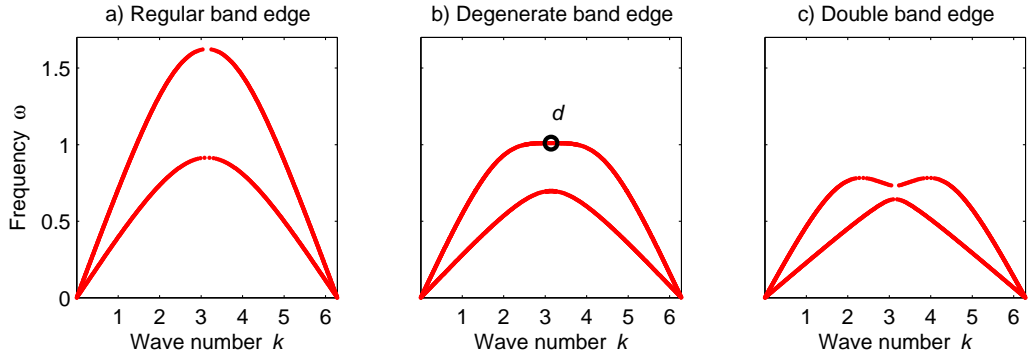}}
\end{center}
\caption{Dispersion relations $\protect\omega\left( k\right) $ of three
periodic stacks with slightly different layer thicknesses. The plot in the
middle displays a degenerate band edge (54).}
\label{DR3DBE}
\end{figure}

While the situation with the regular band edge appears quite obvious and not
particularly interesting from the perspective of slow light, the so-called
degenerate band edge proves to be quite different \cite{JMMM06}. An example
of an electromagnetic dispersion relation with degenerate band edge is shown
in Fig. \ref{DR3DBE}.

In the vicinity of the degenerate band edge $d$, the dispersion relation can
be approximated as%
\begin{equation}
\omega _{d}-\omega \approx \frac{\omega _{d}^{\prime \prime \prime \prime }}{%
24}\left( k-k_{d}\right) ^{4},\ \text{where }\ \omega _{d}^{\prime \prime
\prime \prime }=\left( \frac{\partial ^{4}\omega }{\partial k^{4}}\right)
_{k=k_{d}}.  \label{om_d}
\end{equation}%
Similarly to the regular band edge, below the degenerate band edge frequency 
$\omega _{d}$, the transmitted field $\Psi _{T}$ is a superposition (\ref%
{PsiT=pr+ev}) of one propagating and one evanescent components, while above $%
\omega _{d}$, the transmitted wave is a combination (\ref{PsiT=ev+ev}) of
two evanescent components. The critical difference though is that now both
the Bloch components display strong singularity. Specifically, as the
frequency $\omega $ approaches $\omega _{d}$, both Bloch contributions grow
sharply, while remaining nearly equal and opposite in sign at the slab
boundary%
\begin{equation}
\Psi _{pr}\left( 0\right) \approx -\Psi _{ev}\left( 0\right) \propto
\left\vert \omega _{d}-\omega \right\vert ^{-1/4},\ \ \text{as }\omega
\rightarrow \omega _{d}-0.  \label{DI 4}
\end{equation}%
The destructive interference (\ref{DI 4}) ensures that the boundary
condition (\ref{BC}) is satisfied, while both Bloch contributions to $\Psi
_{T}\left( z\right) $ diverge. As the distance $z$ from the slab boundary
increases, the evanescent component $\Psi _{ev}\left( z\right) $ dies out,
while the propagating component $\Psi _{pr}\left( z\right) $ remains huge.
Eventually, at $z\gg L$, the resulting transmitted wave $\Psi _{T}\left(
z\right) $ coincides with the propagating Bloch eigenmode $\Psi _{pr}\left(
z\right) $. If the frequency $\omega $ of incident light is exactly equal to 
$\omega _{d}$, the transmitted wave $\Psi _{T}\left( z\right) $ does not
reduce to the sum (\ref{PsiT=pr+ev}) of propagating and evanescent
contributions. Instead, it corresponds to a non-Bloch Floquet eigenmode
diverging linearly with $z$ 
\begin{equation*}
\Psi _{T}\left( z\right) -\Psi _{T}\left( 0\right) \propto z\Psi _{d},\;\ 
\text{at}\;\omega =\omega _{d}.
\end{equation*}

The above behavior appears to be very similar to that of the frozen mode
regime described in the previous subsection. In both cases, Figs. \ref%
{AM0_PRE05} and \ref{AMz_PRE03} provide a good graphical illustration of the
electromagnetic field distribution inside the slab in the vicinity of the
relevant stationary point. Yet, there is a crucial difference between the
two cases. In spite of its huge diverging amplitude (\ref{DI 4}), the
transmitted wave $\Psi _{T}$ does not provide any energy flux in the
immediate proximity of a degenerate band edge. Indeed, according to (\ref{DI
4}), as $\omega $ approaches $\omega _{d}$, the energy density $W$ of the
transmitted wave diverges as%
\begin{equation*}
W\propto \left\vert \Psi _{pr}\right\vert ^{2}\propto \left\vert \omega
_{d}-\omega \right\vert ^{-1/2},\text{ }\ \ \text{at }\omega \lessapprox
\omega _{d}.
\end{equation*}%
But, from Eq. (\ref{om_d}) one can derive that the respective slow mode
group velocity vanishes even faster%
\begin{equation*}
u\approx \frac{\omega _{d}^{\prime \prime \prime \prime }}{6}\left(
k-k_{d}\right) ^{3}\approx \frac{24^{3/4}}{6}\left( \omega _{d}^{\prime
\prime \prime \prime }\right) ^{1/4}\left\vert \omega _{d}-\omega
\right\vert ^{3/4},\text{ }\ \ \text{at }\omega \lessapprox \omega _{d}
\end{equation*}%
As the result, the energy flux of the transmitted wave vanishes, as one
approaches the degenerate band edge%
\begin{equation*}
S_{T}=Wu\propto \left\{ 
\begin{array}{c}
\left( \omega _{d}-\omega \right) ^{1/4},\text{ at }\omega \lessapprox
\omega _{d} \\ 
0,\text{ at }\omega \geq \omega _{d}%
\end{array}%
\right. ,
\end{equation*}%
and so does the slab transmittance $\tau $ 
\begin{equation*}
\text{ }\tau \propto \left\{ 
\begin{array}{c}
\left( \omega _{d}-\omega \right) ^{1/4},\text{ at }\omega \lessapprox
\omega _{d} \\ 
0,\text{ at }\omega \geq \omega _{d}%
\end{array}%
\right. .
\end{equation*}%
By contrast, in the case of the frozen mode regime the slab transmittance
remains finite and a significant fraction of the incident light energy goes
to the slow mode.

The situation at a degenerate band edge can be viewed as intermediate
between the frozen mode regime and the vicinity of a regular band edge\cite%
{JMMM06}. Indeed, on the one hand, the incident wave at $\omega =\omega _{d}$
is totally reflected back to space, as would be the case at a regular band
edge. On the other hand, the transmitted field amplitude inside the slab
becomes huge as $\omega \rightarrow \omega _{d}$, which is similar to what
occurs in the frozen mode regime. The large amplitude of the transmitted
wave at $\omega \approx \omega _{d}$ can be very attractive for a variety of
practical applications, although such a behavior cannot be qualified as a
slow light case. Detailed analysis of some peculiar electromagnetic
properties associated with degenerate frequency band edge (\ref{om_d}) can
be found in \cite{PRE 05,JMMM06}.

\section{Physical conditions for the frozen mode regime in layered media}

The frozen mode regime is associated with a stationary inflection point (\ref%
{SIP}) of the electromagnetic dispersion relation. Leaving the proof of this
statement to the following sections, here we establish the conditions under
which the dispersion relation of a periodic layered array can develop the
singularity (\ref{SIP}). We will see that only special layered structures
incorporating anisotropic layers can display this property. In the following
sections, based on the Maxwell equations, we will show that indeed the
stationary inflection point (\ref{SIP}) is uniquely associated with the
frozen mode regime.

\subsection{Axial dispersion relation: basic definitions}

We start with the generalization of the frozen mode concept to the case of
oblique light incidence.

Consider a monochromatic plane wave obliquely incident on a periodic
semi-infinite stack, as shown in Fig. \ref{SISn}. Let $\Psi _{I}$, $\Psi
_{R} $ and $\Psi _{T}$\ denote the incident, reflected and transmitted
waves, respectively. Due to the boundary conditions (\ref{BC}), all three
waves $\Psi _{I}$, $\Psi _{R}$ and $\Psi _{T}$ \ must be assigned the same
pair of tangential components $k_{x},k_{y}$ of the respective wave vector 
\cite{LLEM}%
\begin{equation}
\left( \vec{k}_{I}\right) _{x}=\left( \vec{k}_{R}\right) _{x}=\left( \vec{k}%
_{T}\right) _{x},\ \ \left( \vec{k}_{I}\right) _{y}=\left( \vec{k}%
_{R}\right) _{y}=\left( \vec{k}_{T}\right) _{y},  \label{k_x k_y}
\end{equation}%
while their axial (normal) components $k_{z}$ can be different. Hereinafter,
the normal component of the transmitted Bloch waves propagating inside the
periodic layered medium will be referred to as the \emph{wave number} and
denoted by the symbol $k$, rather than $k_{z}$, so that inside the periodic
stack (at $z>0$) 
\begin{equation}
\vec{k}=\left( k_{x},k_{y},k\right) .  \label{k=kx,ky,k}
\end{equation}%
Unlike $k_{x}$ and $k_{y}$, the $z$ component $k$ of the Bloch wave vector (%
\ref{k=kx,ky,k}) is defined up to a multiple of $2\pi /L$%
\begin{equation}
k\equiv k+2\pi N/L,  \label{k=k+b}
\end{equation}%
where $L$ is the period of the layered structure and $N$ is an integer. For
given $k_{x},k_{y}$ and $\omega $, the value $k$ is found by solving the
time-harmonic Maxwell equations (\ref{MEz}) in the periodic medium, as will
be done in the following sections. The result can be represented as the 
\emph{axial dispersion relation}, which gives the relation between $\omega $
and $k$ at fixed $k_{x},k_{y}$%
\begin{equation}
\omega =\omega \left( k\right) ,\text{\ \ at fixed }k_{x},k_{y}\text{.}
\label{ADR,k}
\end{equation}%
It can be more convenient to define the axial dispersion relation as the
relation between $\omega $ and $k$ at fixed direction $\vec{n}$ of incident
light propagation%
\begin{equation}
\omega =\omega \left( k\right) ,\text{\ \ at fixed }n_{x},n_{y},
\label{ADR,n}
\end{equation}%
where the unit vector $\vec{n}$ can be expressed in terms of the tangential
components (\ref{k_x k_y}) of the wave vector%
\begin{equation}
n_{x}=k_{x}c/\omega ,\ \ n_{y}=k_{y}c/\omega ,\ \ n_{z}=\sqrt{1-\left(
n_{x}^{2}+n_{y}^{2}\right) }.  \label{n(k)}
\end{equation}

\subsubsection{ Axial stationary inflection point and the frozen mode regime}

Suppose that at $\vec{k}=\vec{k}_{0}\;$and $\omega =\omega _{0}=\omega
\left( \vec{k}_{0}\right) $, one of the axial spectral branches (\ref{ADR,k}%
) develops a stationary inflection point for given $(k_{x},k_{y})$, namely 
\begin{equation}
\text{at }\omega =\omega _{0}\text{ \ and \ }\vec{k}=\vec{k}_{0}\text{: \ }%
\left( \frac{\partial \omega }{\partial k}\right) _{k_{x},k_{y}}=0,\;\left( 
\frac{\partial ^{2}\omega }{\partial k^{2}}\right) _{k_{x},k_{y}}=0,\;\left( 
\frac{\partial ^{3}\omega }{\partial k^{3}}\right) _{k_{x},k_{y}}\neq 0,
\label{kInfl}
\end{equation}%
The value%
\begin{equation}
u\equiv u_{z}=\left( \frac{\partial \omega }{\partial k}\right)
_{k_{x},k_{y}}  \label{u_z}
\end{equation}%
in Eq. (\ref{kInfl}) is the axial component of the group velocity, which
vanishes at $\vec{k}=\vec{k}_{0}$. Observe that%
\begin{equation}
u_{x}=\left( \frac{\partial \omega }{\partial k_{x}}\right) _{k,k_{y}}\text{
\ and \ }u_{y}=\left( \frac{\partial \omega }{\partial k_{y}}\right)
_{k,k_{x}},  \label{u_x, u_y}
\end{equation}%
representing the tangential components of the group velocity, may not be
zero at $\vec{k}=\vec{k}_{0}$. The spectral singularity (\ref{kInfl}) is
called the \emph{axial stationary inflection point}.

One can also use another definition of axial stationary inflection point (%
\ref{kInfl}), which is based on the axial dispersion relation (\ref{ADR,n})
rather than (\ref{ADR,k}), namely%
\begin{equation}
\text{at }\omega =\omega _{0}\text{ \ and \ }\vec{n}=\vec{n}_{0}\text{: \ }%
\left( \frac{\partial \omega }{\partial k}\right) _{n_{x},n_{y}}=0,\;\left( 
\frac{\partial ^{2}\omega }{\partial k^{2}}\right) _{n_{x},n_{y}}=0,\;\left( 
\frac{\partial ^{3}\omega }{\partial k^{3}}\right) _{n_{x},n_{y}}\neq 0.
\label{nInfl}
\end{equation}%
The partial derivatives in (\ref{nInfl}) are taken at constant $%
(n_{x},n_{y}) $,$\ $rather than at constant $(k_{x},k_{y})$. Both
definitions (\ref{kInfl}) and (\ref{nInfl}) are equivalent to one other. In
the particular case of normal incidence in which $\vec{n}\parallel \vec{k}%
\parallel z$, the axial stationary inflection point (\ref{kInfl}) or,
equivalently, (\ref{nInfl}) turns into a regular stationary inflection point
(\ref{SIP}).

The \emph{axial frozen mode regime} associated with the singularity (\ref%
{kInfl}) is very similar to its particular case, the regular frozen mode
regime, related to the regular stationary inflection point (\ref{SIP}).
Specifically, in the axial frozen mode regime, obliquely incident light can
enter the semi-infinite photonic crystal with little reflection, where it is
completely converted into a coherent mode with infinitesimal normal
component (\ref{u_z}) of the group velocity and drastically enhanced
amplitude. The energy density of the axial frozen mode displays the same
resonance-like behavior (\ref{W(SIP)}). The only difference between the
axial and the regular frozen mode regime is that in the former case, the
tangential component (\ref{u_x, u_y}) of the group velocity remains finite
at $\omega =\omega _{0}$. The specificity of the axial frozen mode regime as
compared to the regular one is discussed in \cite{PRB03}. Further in this
paper we will focus exclusively on the common features of these two cases.
Either of them will be referred to simply as the frozen mode regime.

\subsection{Spectral asymmetry in periodic stacks}

The (axial) stationary inflection point is indeed associated with the frozen
mode\ regime. But not every periodic layered media can display such a
spectral singularity. It turns out that a necessary condition for the
existence of an axial stationary inflection point and, therefore, a
necessary condition for the frozen mode regime is the following property of
the electromagnetic dispersion relation of the periodic stack%
\begin{equation}
\omega \left( k_{x},k_{y},k\right) \neq \omega \left( k_{x},k_{y},-k\right) 
\text{ \ or, equivalently,\ \ }\omega \left( n_{x},n_{y},k\right) \neq
\omega \left( n_{x},n_{y},-k\right) .  \label{A asym}
\end{equation}

The property (\ref{A asym}) is referred to as \emph{axial spectral asymmetry}%
. Further in this paper, we will use the simplified notation (\ref{ADR,k})
for the axial dispersion relation. In this notation, the requirement (\ref{A
asym}) of axial spectral asymmetry takes the following form%
\begin{equation}
\omega \left( k\right) \neq \omega \left( -k\right) ,  \label{asym}
\end{equation}%
where $k$ is the $z$ component (\ref{k=k+b}) of the Bloch wave vector $\vec{k%
}$. A\ robust frozen mode regime only occurs if the degree of spectral
asymmetry (\ref{asym}) is significant. For brevity, hereinafter, the
quantity $k$ will be referred to as the Bloch wave number, although in the
case of oblique propagation, $k$ is just the normal component of the Bloch
wave vector $\vec{k}$.

In the particular case of normal wave propagation, in which $\vec{n}%
\parallel \vec{k}\parallel z$, the requirement (\ref{A asym}) of axial
spectral asymmetry reduces to%
\begin{equation}
\omega \left( \vec{k}\right) \neq \omega \left( -\vec{k}\right) \text{, }%
\vec{k}\parallel z.  \label{N asym}
\end{equation}%
This kind of asymmetric dispersion relation can occur only in periodic
structures with some of the constitutive components being magnetic and
displaying nonreciprocal Faraday rotation \cite{PRE01,PRB03}. Significant
spectral asymmetry requires strong Faraday rotation. The simplest periodic
array supporting the spectral asymmetry (\ref{N asym}) is shown in Fig. \ref%
{StackAFA}.

\begin{figure}[tbph]
\begin{center}
\scalebox{0.8}{\includegraphics[viewport=0 0 400 300,clip]{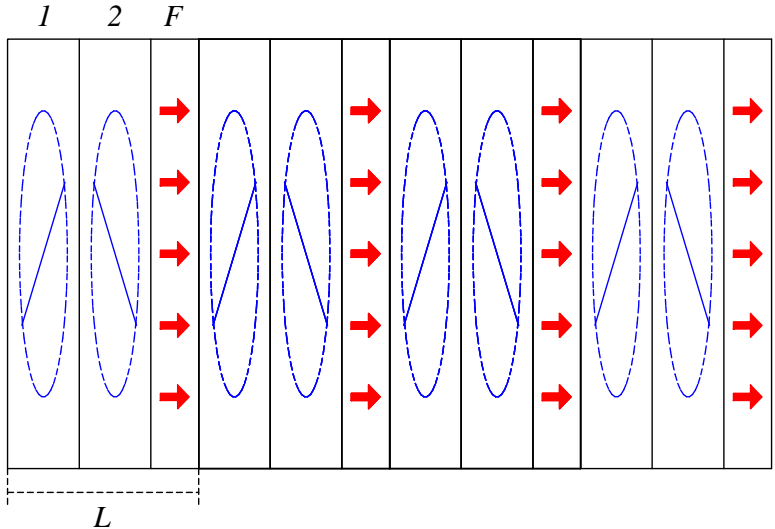}}
\end{center}
\caption{The simplest periodic stack supporting asymmetric dispersion
relation (68). A unit cell $L$ of this stack comprises three layers: two
anisotropic layers 1 and 2 with misaligned in-plane anisotropy (the $A$ -
layers), and one magnetic layer $F$ with magnetization shown by the arrows.}
\label{StackAFA}
\end{figure}

At microwave frequencies, there exist a number of magnetic materials
displaying low losses and strong Faraday rotation. But, at infrared and
optical frequencies, the magnetic materials with sufficiently strong Faraday
rotation are usually too lossy for our purposes. Therefore, if we are
interested in optical frequencies, we have to rely on non-magnetic stacks,
in which the regular spectral asymmetry (\ref{N asym}) is impossible. By
contrast, the axial spectral asymmetry (\ref{A asym}) or (\ref{asym}) does
not require magnetic layers and can occur even in perfectly reciprocal
non-magnetic stacks.

The physical conditions under which the electromagnetic dispersion relation
of a non-magnetic layered structure can develop the (axial) spectral
asymmetry (\ref{asym}) and, thereby, support the (axial) frozen mode regime
can be grouped in two categories. The first one comprises several symmetry
restrictions. The second category includes some basic qualitative
recommendations which would ensure the robustness of the frozen mode regime,
provided that the symmetry conditions for the regime are met. In what
follows we briefly describe those conditions and then show how they apply to
periodic stacks incorporating some real dielectric materials.

There are two fundamental necessary conditions for the frozen mode regime in
a non-magnetic (reciprocal) periodic layered structure. The first one is
that the Bloch dispersion function $\omega \left( \vec{k}\right) $ in the
periodic layered medium must display the axial spectral asymmetry (\ref{asym}%
). This condition is necessary for the existence of the axial stationary \
inflection point (\ref{kInfl}) in the electromagnetic dispersion relation of
an arbitrary periodic layered medium. The second necessary condition is that
for the given direction $\vec{k}$ of wave propagation, the Bloch eigenmodes $%
\Psi _{\vec{k}}$ with different polarizations must have the same symmetry.
In the case of oblique propagation in periodic layered media, the latter
condition implies that for the given $\vec{k}$, the Bloch eigenmodes\ are
neither TE nor TM:%
\begin{equation}
\Psi _{\vec{k}}\text{ \ is neither TE nor TM.}  \label{TE/TM}
\end{equation}

The condition (\ref{asym}) imposes certain restrictions on (i) the point
symmetry group $G$ of the periodic layered array and (ii) on the direction $%
\vec{k}$ of the transmitted wave propagation inside the layered medium,
while the condition (\ref{TE/TM}) may impose an additional restriction on
the direction of $\vec{k}$.

The restriction on the symmetry of the periodic stack stemming from the
requirement (\ref{asym}) of the axial spectral asymmetry is%
\begin{equation}
m_{z}\notin G\text{ \ and \ }2_{z}\notin G.  \label{asymC2}
\end{equation}%
where $m_{z}$ is the mirror plane parallel to the layers, $2_{z}$ is the
2-fold rotation about the $z$ axis. An immediate consequence of the
criterion (\ref{asymC2}) is that at least one of the alternating layers of
the periodic stack must be an anisotropic dielectric with%
\begin{equation}
\varepsilon _{xz}\neq 0\text{ \ \ and/or \ \ }\varepsilon _{yz}\neq 0
\label{eps<>0}
\end{equation}%
where the $z$ direction is normal to the layers. Otherwise, the operation $%
2_{z}$ will be present in the symmetry group $G$ of the periodic stack.

\begin{figure}[tbph]
\begin{center}
\scalebox{0.8}{\includegraphics[viewport=0 0 400 300,clip]{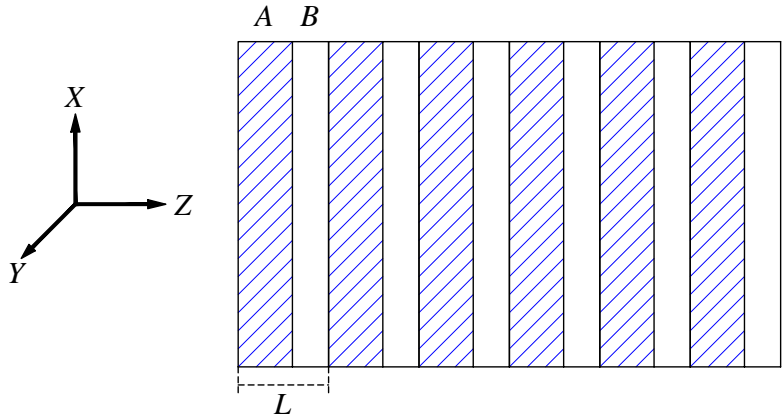}}
\end{center}
\caption{Periodic layered structure with two layers $A$ and $B$ in a unit
cell $L$. The $A$ layers (hatched) are anisotropic with one of the principle
axes of the dielectric permittivity tensor $\hat{\protect\varepsilon}$
making an oblique angle with the normal $z$ to the layers ($\protect%
\varepsilon_{xz}\neq0$). The $B$ layers can be isotropic. The $x-z$ plane
coincides with the mirror plane $m_{y}$ of the stack.}
\label{StackAB}
\end{figure}

The simplest and the most practical example of a non-magnetic periodic stack
satisfying the criterion (\ref{asymC2}) is shown in Fig. \ref{StackAB}. It
is made up of anisotropic $A$ layers alternating with isotropic $B$ layers.
The respective dielectric permittivity tensors are 
\begin{equation}
\hat{\varepsilon}_{A}=\left[ 
\begin{array}{ccc}
\varepsilon _{xx} & 0 & \varepsilon _{xz} \\ 
0 & \varepsilon _{yy} & 0 \\ 
\varepsilon _{xz} & 0 & \varepsilon _{zz}%
\end{array}%
\right] ,\;\hat{\varepsilon}_{B}=\left[ 
\begin{array}{ccc}
\varepsilon _{B} & 0 & 0 \\ 
0 & \varepsilon _{B} & 0 \\ 
0 & 0 & \varepsilon _{B}%
\end{array}%
\right] .  \label{epsilon}
\end{equation}%
For simplicity, we assume%
\begin{equation}
\hat{\mu}_{A}=\hat{\mu}_{B}=\hat{I}.  \label{mu}
\end{equation}%
The stack in Fig. \ref{StackAB} has the monoclinic symmetry%
\begin{equation}
2_{y}/m_{y}  \label{2/m}
\end{equation}%
with the mirror plane $m_{y}$ normal to the $y$ - axis. Such a symmetry is
compatible with the necessary condition (\ref{asymC2}) for the axial
spectral asymmetry (\ref{asym}). Therefore, the periodic array in Fig. \ref%
{StackAB} can support the frozen mode regime, provided that incident beam
lies neither in the $x-z$, nor in $y-x$ plane \cite{PRB03}%
\begin{equation*}
n_{x}\neq 0\text{ and }n_{y}\neq 0.
\end{equation*}

If all the above necessary conditions are met, then the (axial) frozen mode
regime is, at least, not forbidden by symmetry. More details on the symmetry
aspects of the frozen mode regime can be found in \cite{PRE03}, Section II
and \cite{PRB03}, Sections I and II.

In practice, as soon as the symmetry conditions are met, one can almost
certainly achieve the (axial) frozen mode regime at any desirable frequency $%
\omega $ within a certain frequency range. The frequency range is determined
by the layer thicknesses and the dielectric materials used, while a specific
value of $\omega $ within the range can be selected by the direction $\vec{n}
$ of the light incidence. The problem is that unless the physical parameters
of the stack layers lie within a certain range, the effects associated with
the frozen mode regime can be insignificant or even practically
undetectable. The basic guiding principle in choosing appropriate layer
materials are discussed in Ref. \cite{PRE03,PRE05}.

The biggest challenge at optical frequencies lies in the fact that most of
the commercially available optical anisotropic crystals have weak
anisotropy. According to \cite{PRE03}, this would push the axial stationary
inflection point (\ref{kInfl}) very close to the photonic band edge and make
the photonic crystal almost 100\% reflective. This indeed would be the case
if we tried to realize the frozen mode regime at the lowest frequency band.
But, in Ref. \cite{PRE05} it was shown that the above problem can be
successfully solved by moving to a higher frequency band. So, a robust
axially frozen mode regime with almost complete conversion of the incident
light into the frozen mode can be achieved with the commercially available
anisotropic dielectric materials such as $NbLiO_{3}$, $YVO_{4}$, etc.

\section{Electrodynamics of lossless stratified media}

This section starts with a description of some basic electrodynamic
properties of stratified media composed of lossless anisotropic layers. Then
we turn to the important particular case of unbounded periodic layered
arrays (periodic stacks), where the electromagnetic eigenmodes are Bloch
waves. Then we consider the problem of electromagnetic energy flux in
lossless stratified media. Finally, we outline the electromagnetic
scattering problem for a semi-infinite photonic slab. The material presented
in this section is sufficient for a numerical analysis of slow light
phenomena in periodic layered media. This approach was used in \cite%
{PRB03,PRE03,PRE05} to analyze the frozen mode regime in magnetic and
non-magnetic periodic stacks. Yet, to develop a consistent analytical
picture of the frozen mode regime we shall need a more sophisticated
mathematical framework based on a perturbation theory for non-diagonalizable
degenerate matrices. This problem will be addressed in the following
sections.

\subsection{Reduced time-harmonic Maxwell equations}

Our treatment is based on the time-harmonic Maxwell equations in
heterogeneous nonconducting media%
\begin{equation}
\nabla \times \mathbf{\vec{E}}\left( \vec{r}\right) =i\frac{\omega }{c}%
\mathbf{B}\left( \vec{r}\right) ,\;\nabla \times \mathbf{\vec{H}}\left( \vec{%
r}\right) =-i\frac{\omega }{c}\mathbf{D}\left( \vec{r}\right) .  \label{THME}
\end{equation}%
Electric and magnetic fields and inductions in Eq. (\ref{THME}) are related
through the linear constitutive equations%
\begin{equation}
\mathbf{\vec{D}}\left( \vec{r}\right) =\hat{\varepsilon}\left( \vec{r}%
\right) \mathbf{\vec{E}}\left( \vec{r}\right) ,\ \mathbf{\vec{B}}\left( \vec{%
r}\right) =\hat{\mu}\left( \vec{r}\right) \mathbf{\vec{H}}\left( \vec{r}%
\right) .  \label{MR}
\end{equation}%
All variables in Eqs. (\ref{THME}) and (\ref{MR}) are frequency dependent.
In lossless media, the material tensors $\hat{\varepsilon}\left( \vec{r}%
\right) $ and $\hat{\mu}\left( \vec{r}\right) $\ are Hermitian%
\begin{equation}
\text{In lossless media: \ \ }\hat{\varepsilon}\left( \vec{r}\right) =\hat{%
\varepsilon}^{\dagger }\left( \vec{r}\right) ,~\hat{\mu}\left( \vec{r}%
\right) =\hat{\mu}^{\dagger }\left( \vec{r}\right) ,  \label{Herm}
\end{equation}%
where the dagger $\dagger $ denotes the Hermitian conjugate. In lossless
non-magnetic media, both tensors $\hat{\varepsilon}$ and $\hat{\mu}$ are
also real and symmetric%
\begin{equation}
\text{In lossless non-magnetic media: \ \ }\hat{\varepsilon}\left( \vec{r}%
\right) =\hat{\varepsilon}^{\ast }\left( \vec{r}\right) =\hat{\varepsilon}%
^{T}\left( \vec{r}\right) ,~\hat{\mu}\left( \vec{r}\right) =\hat{\mu}^{\ast
}\left( \vec{r}\right) =\hat{\mu}^{T}\left( \vec{r}\right) ,  \label{Rcpr}
\end{equation}%
where the asterisk denotes the complex conjugate and the superscript $T$
denotes matrix transposition. In magnetically polarized lossless media, the
Hermitian material tensors (\ref{Herm}) may have a skew-symmetric imaginary
part which is responsible for the non-reciprocal effect of Faraday rotation 
\cite{LLEM}.

In a stratified medium, the second rank tensors $\hat{\varepsilon}\left( 
\vec{r}\right) $ and $\hat{\mu}\left( \vec{r}\right) $ depend on a single
Cartesian coordinate $z$, and$\;$the Maxwell equations (\ref{THME}) can be
recast as%
\begin{equation}
\nabla \times \mathbf{\vec{E}}\left( \vec{r}\right) =i\frac{\omega }{c}\hat{%
\mu}\left( z\right) \mathbf{\vec{H}}\left( \vec{r}\right) ,\;\nabla \times 
\mathbf{\vec{H}}\left( \vec{r}\right) =-i\frac{\omega }{c}\hat{\varepsilon}%
\left( z\right) \mathbf{\vec{E}}\left( \vec{r}\right) .  \label{MEz}
\end{equation}%
Solutions for Eq. (\ref{MEz}) can be sought in the following form%
\begin{equation}
\mathbf{\vec{E}}\left( \vec{r}\right) =e^{i\left( k_{x}x+k_{y}y\right) }\vec{%
E}\left( z\right) ,\ \mathbf{\vec{H}}\left( \vec{r}\right) =e^{i\left(
k_{x}x+k_{y}y\right) }\vec{H}\left( z\right) ,  \label{LEM}
\end{equation}%
$\allowbreak $which can be interpreted as the \textquotedblleft
tangential\textquotedblright\ Bloch representation. The substitution (\ref%
{LEM}) allows separation of the tangential components of the fields into a
closed system of four linear ordinary differential equations%
\begin{equation}
\partial _{z}\Psi \left( z\right) =i\frac{\omega }{c}M\left( z\right) \Psi
\left( z\right) ,\;\Psi \left( z\right) =\left[ 
\begin{array}{c}
E_{x}\left( z\right) \\ 
E_{y}\left( z\right) \\ 
H_{x}\left( z\right) \\ 
H_{y}\left( z\right)%
\end{array}%
\right] ,  \label{ME4}
\end{equation}%
where the $4\times 4$ matrix $M\left( z\right) $ is referred to as the
(reduced) Maxwell operator. The normal field components $E_{z}$ and $H_{z}$
do not enter the reduced Maxwell equations (\ref{ME4}) and can be expressed
in terms of the tangential field components from Eq. (\ref{ME4}) as 
\begin{equation}
\begin{array}{c}
E_{z}=\left( -n_{x}H_{y}+n_{y}H_{x}-\varepsilon _{xz}^{\ast
}E_{x}-\varepsilon _{yz}^{\ast }E_{y}\right) \varepsilon _{zz}^{-1}, \\ 
H_{z}=\left( n_{x}E_{y}-n_{y}E_{x}-\mu _{xz}^{\ast }H_{x}-\mu _{yz}^{\ast
}H_{y}\right) \mu _{zz}^{-1},%
\end{array}
\label{EzHz}
\end{equation}%
where $n_{x}=ck_{x}/\omega $, $n_{y}=ck_{y}/\omega $.

The explicit expression for the Maxwell operator $M\left( z\right) $ in Eq. (%
\ref{ME4}) is%
\begin{equation}
M\left( z\right) =\left[ 
\begin{array}{cc}
M_{11} & M_{12} \\ 
M_{21} & M_{22}%
\end{array}
\right]  \label{M}
\end{equation}
where%
\begin{align*}
M_{11} & =\left[ 
\begin{array}{cc}
-\frac{\varepsilon_{xz}^{\ast}}{\varepsilon_{zz}}n_{x}-\frac{\mu_{yz}}{%
\mu_{zz}}n_{y} & \left( -\frac{\varepsilon_{yz}^{\ast}}{\varepsilon_{zz}}+%
\frac{\mu_{yz}}{\mu_{zz}}\right) n_{x} \\ 
-\left( \frac{\varepsilon_{xz}^{\ast}}{\varepsilon_{zz}}-\frac{\mu_{xz}}{%
\mu_{zz}}\right) n_{y} & -\frac{\varepsilon_{yz}^{\ast}}{\varepsilon_{zz}}%
n_{y}-\frac{\mu_{xz}}{\mu_{zz}}n_{x}%
\end{array}
\right] , \\
M_{22} & =\left[ 
\begin{array}{cc}
-\frac{\varepsilon_{yz}}{\varepsilon_{zz}}n_{y}-\frac{\mu_{xz}^{\ast}}{%
\mu_{zz}}n_{x} & \left( \frac{\varepsilon_{yz}}{\varepsilon_{zz}}-\frac {%
\mu_{yz}^{\ast}}{\mu_{zz}}\right) n_{x} \\ 
\left( \frac{\varepsilon_{xz}}{\varepsilon_{zz}}-\frac{\mu_{xz}^{\ast}}{%
\mu_{zz}}\right) n_{y} & -\frac{\varepsilon_{xz}}{\varepsilon_{zz}}n_{x}-%
\frac{\mu_{yz}^{\ast}}{\mu_{zz}}n_{y}%
\end{array}
\right] , \\
M_{12} & =\left[ 
\begin{array}{cc}
\mu_{xy}^{\ast}-\frac{\mu_{xz}^{\ast}\mu_{yz}}{\mu_{zz}}+\frac{n_{x}n_{y}}{%
\varepsilon_{zz}} & \mu_{yy}-\frac{\mu_{yz}\mu_{yz}^{\ast}}{\mu_{zz}}-\frac{%
n_{x}^{2}}{\varepsilon_{zz}} \\ 
-\mu_{xx}+\frac{\mu_{xz}\mu_{xz}^{\ast}}{\mu_{zz}}+\frac{n_{y}^{2}}{%
\varepsilon_{zz}} & -\mu_{xy}+\frac{\mu_{xz}\mu_{yz}^{\ast}}{\mu_{zz}}-\frac{%
n_{x}n_{y}}{\varepsilon_{zz}}%
\end{array}
\right] , \\
M_{21} & =\left[ 
\begin{array}{cc}
-\varepsilon_{xy}^{\ast}+\frac{\varepsilon_{xz}^{\ast}\varepsilon_{yz}}{%
\varepsilon_{zz}}-\frac{n_{x}n_{y}}{\mu_{zz}} & -\varepsilon_{yy}+\frac{%
\varepsilon_{yz}\varepsilon_{yz}^{\ast}}{\varepsilon_{zz}}+\frac {n_{x}^{2}}{%
\mu_{zz}} \\ 
\varepsilon_{xx}-\frac{\varepsilon_{xz}\varepsilon_{xz}^{\ast}}{\varepsilon
_{zz}}-\frac{n_{y}^{2}}{\mu_{zz}} & \varepsilon_{xy}-\frac{\varepsilon
_{xz}\varepsilon_{yz}^{\ast}}{\varepsilon_{zz}}+\frac{n_{x}n_{y}}{\mu_{zz}}%
\end{array}
\right] .
\end{align*}

In the important particular case of $k_{x}=k_{y}=0$, the Maxwell operator (%
\ref{M}) has a simpler form%
\begin{equation}
M\left( z\right) =\left[ 
\begin{array}{cc}
0 & M_{12} \\ 
M_{21} & 0%
\end{array}%
\right] .  \label{Mn}
\end{equation}%
Only in this case of $\vec{k}\parallel z$, the fields $\vec{E}\left(
z\right) $ and $\vec{H}\left( z\right) $ coincide with the actual electric
and magnetic fields $\mathbf{\vec{E}}\left( \vec{r}\right) $ and $\mathbf{%
\vec{H}}\left( \vec{r}\right) $, as follows from the relation (\ref{LEM}).

The (reduced) Maxwell operator $M\left( z\right) $ is a function of:

\begin{itemize}
\item[-] the local values of material tensors $\hat{\varepsilon}\left(
z\right) $ and $\hat{\mu}\left( z\right) $,

\item[-] the space coordinate $z$, via the material tensors $\hat{%
\varepsilon }\left( z\right) $ and $\hat{\mu}\left( z\right) $,

\item[-] the tangential components $k_{x},k_{y}$ of the wave vector,

\item[-] the frequency $\omega$.
\end{itemize}

Different versions of the reduced Maxwell equation (\ref{ME4}) can be found
in the extensive literature on electrodynamics of stratified media (see, for
example, \cite{Tmatrix} and references therein).

The $4\times 4$ matrix $M\left( z\right) $ in Eq. (\ref{M}) has the
fundamental property of $J$-Hermitivity defined as%
\begin{equation}
M^{\dagger }=JMJ,  \label{JM}
\end{equation}%
where%
\begin{equation}
J=J^{\dagger }=J^{-1}=\left[ 
\begin{array}{cccc}
0 & 0 & 0 & 1 \\ 
0 & 0 & -1 & 0 \\ 
0 & -1 & 0 & 0 \\ 
1 & 0 & 0 & 0%
\end{array}%
\right] .  \label{J}
\end{equation}%
The property (\ref{JM}) relates to conservation of electromagnetic energy in
lossless media, it plays an important role in further consideration.

\subsection{The transfer matrix formalism}

The Cauchy problem%
\begin{equation}
\partial _{z}\Psi \left( z\right) =i\frac{\omega }{c}M\left( z\right) \Psi
\left( z\right) ,\;\Psi \left( z_{0}\right) =\Psi _{0}  \label{CauPsi}
\end{equation}%
for the reduced Maxwell equation (\ref{ME4}) has a unique solution%
\begin{equation}
\Psi \left( z\right) =T\left( z,z_{0}\right) \Psi \left( z_{0}\right)
\label{T(zz0)}
\end{equation}%
where the $4\times 4$ matrix $T\left( z,z_{0}\right) $ is referred to as the 
\emph{transfer matrix}. From the definition (\ref{T(zz0)}) it follows that%
\begin{equation}
T\left( z,z_{0}\right) =T\left( z,z^{\prime }\right) T\left( z^{\prime
},z_{0}\right) ,\;T\left( z,z_{0}\right) =T^{-1}\left( z_{0},z\right)
,\;T\left( z,z\right) =I,  \label{Tzz0}
\end{equation}%
where $I$ is the identity matrix. The transfer matrix $T\left(
z,z_{0}\right) $ allows determination of the time-harmonic electromagnetic
field $\Psi \left( z\right) $ at an arbitrary point $z$ of the stratified
medium once its value $\Psi \left( z_{0}\right) $ at any particular point $%
z_{0}$.

The matrix $T\left( z,z_{0}\right) $\ itself is uniquely defined by the
following Cauchy problem%
\begin{equation}
\partial _{z}T\left( z,z_{0}\right) =i\frac{\omega }{c}M\left( z\right)
T\left( z,z_{0}\right) ,\;T\left( z,z\right) =I.  \label{CauT}
\end{equation}%
The equation (\ref{CauT}), together with $J$-Hermitivity (\ref{JM}) of the
Maxwell operator $M\left( z\right) $, implies that the transfer matrix $%
T\left( z,z_{0}\right) $ is $J$-\emph{unitary}%
\begin{equation}
T^{\dagger }=JT^{-1}J,  \label{JU}
\end{equation}%
as shown in Ref. \cite{PRE03}. The $J$-unitarity (\ref{JU}) of the transfer
matrix $T=T\left( z,z_{0}\right) $ imposes the following constraint on its
set of four eigenvalues $\zeta _{i},~i=1,2,3,4$%
\begin{equation}
\left\{ \zeta _{1}^{\ast },\zeta _{2}^{\ast },\zeta _{3}^{\ast },\zeta
_{4}^{\ast }\right\} \equiv \left\{ \zeta _{1}^{-1},\zeta _{2}^{-1},\zeta
_{3}^{-1},\zeta _{4}^{-1}\right\} ,  \label{X*=1/X}
\end{equation}%
which also implies that%
\begin{equation}
\left\vert \det T\right\vert =1.  \label{|detT|=1}
\end{equation}

\subsubsection{The transfer matrix of a stack of uniform layers}

The greatest advantage of the transfer matrix approach stems from the fact
that the transfer matrix $T_{S}$ of an arbitrary stack of layers is a
sequential product of the transfer matrices $T_{m}$ of the constituent layers%
\begin{equation}
T_{S}=\prod_{m}T_{m}.  \label{TS}
\end{equation}%
According to Eq. (\ref{CauT}), if each individual layer $m$ is homogeneous,
the corresponding single-layer transfer matrices $T_{m}$ can be explicitly
expressed in terms of the respective Maxwell operators $M_{m}$%
\begin{equation}
T_{m}=\exp \left( i\frac{\omega }{c}z_{m}M_{m}\right) ,  \label{Tm}
\end{equation}%
where $z_{m}$ is the thickness of the $m$-th layer. The explicit expression
for the Maxwell operator $M_{m}$ of an arbitrary uniform layer of
anisotropic dielectric material is given by Eq. (\ref{M}). Thus, Eq. (\ref%
{TS}), together with (\ref{Tm}) and (\ref{M}), give an explicit expression
for the transfer matrix $T_{S}$ of an arbitrary stack of anisotropic
dielectric layers. The elements of the $J$-unitary matrix $T_{S}$ are
functions of:

\begin{itemize}
\item[-] the material tensors $\hat{\varepsilon}$ and $\hat{\mu}$ in each
layer of the stack,

\item[-] the layer thicknesses $d_{m}$,

\item[-] the frequency $\omega$,

\item[-] the tangential components $k_{x}=\frac{\omega}{c}n_{x},\ \ \ k_{y}=%
\frac{\omega}{c}n_{y}$ of the wave vector.
\end{itemize}

In the case of $k_{x}=k_{y}=0$, we have instead of Eq. (\ref{|detT|=1})%
\begin{equation}
\text{if }k_{x}=k_{y}=0\text{, }\det T_{S}=1,  \label{detT=1}
\end{equation}
that can be derived directly from Eqs. (\ref{Tm}) and (\ref{M}).

\subsection{Periodic layered arrays. Bloch eigenmodes.}

In a periodic layered medium, all material tensors are periodic functions of 
$z$%
\begin{equation*}
\hat{\varepsilon}\left( z\right) =\hat{\varepsilon}\left( z+L\right) ,~\hat{%
\mu}\left( z\right) =\hat{\mu}\left( z+L\right) ,
\end{equation*}%
and so is the Maxwell operator $M(z)$ in Eq. (\ref{ME4}),%
\begin{equation}
M\left( z+L\right) =M\left( z\right) ,  \label{A(z+L)}
\end{equation}%
where $L$ is the length of a unit cell of the periodic stack. A Bloch
solution $\Psi _{k}\left( z\right) $ of the reduced Maxwell equation (\ref%
{ME4}) with the periodic operator $M(z)$ should satisfy the following
relation%
\begin{equation}
\Psi _{k}\left( z+L\right) =e^{ikL}\Psi _{k}\left( z\right) ,  \label{Bloch}
\end{equation}%
where $k$ is the normal component of the Bloch wave vector%
\begin{equation}
k=k_{z}.  \label{k=kz}
\end{equation}%
Unlike $k_{x}$ and $k_{y}$, the $z$ component (\ref{k=kz}) of the Bloch wave
vector is defined up to a multiple of $2\pi /L$%
\begin{equation*}
k\equiv k+2\pi N/L,
\end{equation*}%
where $N$ is an integer. Hereinafter, the normal component $k_{z}$ of the
Bloch wave vector $\vec{k}$ will be referred to simply as the \emph{wave
number} and denoted with symbol $k$, rather than $k_{z}$.

The definition (\ref{T(zz0)}) of the $T$-matrix together with Eq. (\ref%
{Bloch}) give%
\begin{equation}
\Psi _{k}\left( z+L\right) =T\left( z+L,z\right) \Psi _{k}\left( z\right)
=e^{ikL}\Psi _{k}\left( z\right) .  \label{Psi=TPsi}
\end{equation}%
Introducing the transfer matrix of a primitive cell%
\begin{equation}
T_{L}=T\left( L,0\right)  \label{TL}
\end{equation}%
we have from Eq. (\ref{Psi=TPsi})%
\begin{equation}
T_{L}\Phi _{k}=e^{ikL}\Phi _{k},\text{ \ \ where }\;\Phi _{k}=\Psi
_{k}\left( 0\right) .  \label{T(L)=e(ikL)}
\end{equation}%
Thus, the eigenvectors of the transfer matrix $T_{L}$ of the unit cell are
uniquely related to the eigenmodes $\Psi _{k}\left( z\right) $ of the
reduced Maxwell equation (\ref{ME4}) through the relations%
\begin{equation}
\Phi _{i}=\Psi _{i}\left( 0\right) ,\;i=1,2,3,4.  \label{Psi 1234}
\end{equation}%
The respective four eigenvalues%
\begin{equation}
\zeta _{i}=e^{ik_{i}L},\;i=1,2,3,4  \label{X(k)}
\end{equation}%
of $T_{L}$ are the roots of the characteristic equation%
\begin{equation}
F\left( \zeta \right) =0,\text{ \ where \ }F\left( \zeta \right) =\det
\left( T_{L}-\zeta \hat{I}\right) .  \label{Char X}
\end{equation}%
For any given $\omega $ and $\left( k_{x},k_{y}\right) $, the characteristic
equation (\ref{Char X}) defines a set of four values $\left\{ \zeta
_{1},\zeta _{2},\zeta _{3},\zeta _{4}\right\} $, or equivalently, $\left\{
k_{1},k_{2},k_{3},k_{4}\right\} $. Real $k$ (or, equivalently, $\left\vert
\zeta \right\vert =1$) correspond to propagating Bloch waves (propagating
modes), while complex $k$ (or, equivalently, $\left\vert \zeta \right\vert
\neq 1$) correspond to evanescent modes. Evanescent modes are relevant near
photonic crystal boundaries and other structural irregularities.

The $J$-unitarity (\ref{JU}) of $T_{L}$ imposes the restriction (\ref{X*=1/X}%
) on the eigenvalues (\ref{X(k)}), which can be recast as%
\begin{equation}
\{k_{1},k_{2},k_{3},k_{4}\}\equiv \{k_{1}^{\ast },k_{2}^{\ast },k_{3}^{\ast
},k_{4}^{\ast }\},  \label{k=k*}
\end{equation}%
for any given $\omega $ and $k_{x},k_{y}$. In view of the relations (\ref%
{k=k*}) or (\ref{X*=1/X}), one can distinguish the following three different
situations.

\begin{enumerate}
\item[(i)] All four wave numbers are real%
\begin{equation}
k_{1}\equiv k_{1}^{\ast },\;k_{2}\equiv k_{2}^{\ast },\;k_{3}\equiv
k_{3}^{\ast },\;k_{4}\equiv k_{4}^{\ast }.  \label{4ex}
\end{equation}%
or, equivalently,%
\begin{equation}
\left\vert \zeta _{1}\right\vert =\left\vert \zeta _{2}\right\vert
=\left\vert \zeta _{3}\right\vert =\left\vert \zeta _{4}\right\vert =1.
\label{4 ex}
\end{equation}%
The respective four Bloch eigenmodes are propagating.

\item[(ii)] Two real and to complex wave numbers%
\begin{equation}
k_{1}=k_{1}^{\ast },\;k_{2}=k_{2}^{\ast },\;k_{4}=k_{3}^{\ast },\;\text{%
where }k_{3}\neq k_{3}^{\ast },\;k_{4}\neq k_{4}^{\ast }.  \label{2ex2ev}
\end{equation}%
or, equivalently,%
\begin{equation}
\left\vert \zeta _{1}\right\vert =\left\vert \zeta _{2}\right\vert =1,\
\zeta _{4}=1/\zeta _{3}^{\ast },\;\text{where }\left\vert \zeta
_{3}\right\vert ,\left\vert \zeta _{4}\right\vert \neq 1.  \label{2 ex 2 ev}
\end{equation}%
Two of the four Bloch eigenmodes are propagating and the remaining two are
evanescent with complex conjugate wave numbers.

\item[(iii)] All four wave numbers are complex%
\begin{equation}
k_{2}=k_{1}^{\ast },\;k_{4}=k_{3}^{\ast },\;\text{where }k_{1}\neq
k_{1}^{\ast },\;k_{2}\neq k_{2}^{\ast },\;k_{3}\neq k_{3}^{\ast
},\;k_{4}\neq k_{4}^{\ast }.  \label{4ev}
\end{equation}%
or, equivalently,%
\begin{equation}
\zeta _{2}=1/\zeta _{1}^{\ast },~\zeta _{4}=1/\zeta _{3}^{\ast },\;\text{%
where }\left\vert \zeta _{1}\right\vert ,\left\vert \zeta _{2}\right\vert
,\left\vert \zeta _{3}\right\vert ,\left\vert \zeta _{4}\right\vert \neq 1.
\label{4 ev}
\end{equation}%
This situation relates to a frequency gap, where for given $\omega $ and $%
k_{x},k_{y}$, all four Bloch eigenmodes are evanescent.
\end{enumerate}

\subsection{Symmetry of the dispersion relation}

Below we will see that the dispersion relation is symmetric under $%
k\rightarrow -k$ if and only if the transfer matrix $T_{L}$ is similar to
its inverse. Indeed, assume that%
\begin{equation}
T_{L}=U^{-1}T_{L}^{-1}U  \label{TL -- 1/TL}
\end{equation}%
where $U$\ is a nonsingular $4\times 4$ matrix. This assumption together
with the property (\ref{JU}) of $J$-unitarity, imply also the similarity of $%
T_{L}$ and $T_{L}^{\dagger }$%
\begin{equation}
T_{L}=V^{-1}T_{L}^{\dagger }V,  \label{TL -- TL*}
\end{equation}%
where \ \ $V=JU$. Either of the above two relations imposes the following
additional restriction on the eigenvalues (\ref{X(k)}) of $T_{L}$ for given $%
\omega $ and $k_{x},k_{y}$%
\begin{equation}
\{\zeta _{1},\zeta _{2},\zeta _{3},\zeta _{4}\}\equiv \{\zeta
_{1}^{-1},\zeta _{2}^{-1},\zeta _{3}^{-1},\zeta _{4}^{-1}\},  \label{X=1/X}
\end{equation}%
or, equivalently,%
\begin{equation}
\{k_{1},k_{2},k_{3},k_{4}\}\equiv \{-k_{1},-k_{2},-k_{3},-k_{4}\}.
\label{k=-k}
\end{equation}%
The relation (\ref{k=-k}) is referred to as \emph{axial spectral symmetry}.
It applies both to propagating and evanescent solutions.

Let us consider the symmetry relation (\ref{k=-k}) in more detail. Assume
that $k_{1}=k_{1}^{\ast }$ is a real wave number corresponding to a
propagating eigenmode. The relation (\ref{k=-k}) implies that for given $%
\omega $ and $k_{x},k_{y}$, there is another real wave number $%
k_{2}=k_{2}^{\ast }$ such that 
\begin{equation}
k_{2}=-k_{1}  \label{sym ex}
\end{equation}%
In terms of the (axial) dispersion relation $\omega \left( k\right) $, the
Eq. (\ref{sym ex}) boils down to a simple definition of axial spectral
symmetry%
\begin{equation*}
\omega \left( k_{x},k_{y},k\right) =\omega \left( k_{x},k_{y},-k\right) ,
\end{equation*}%
where $k_{1}=k$ and $k_{2}=-k$ constitute a pair of reciprocal real wave
numbers related to given $\omega $ and $k_{x},k_{y}$. In the case (\ref{4ex}%
) of four propagating eigenmodes, there will be an additional pair $k_{3}$
and $k_{4}=-k_{3}$ of reciprocal wave numbers.

Now assume that while $k_{1}$ and $k_{2}=-k_{1}$ are real, the remaining
wave numbers $k_{3}$ and $k_{4}$ from the set (\ref{k=-k}) are complex,
which constitutes the case (\ref{2ex2ev}) of two propagating and two
evanescent eigenmodes. In such a case, in addition to Eq. (\ref{sym ex}),
the relation (\ref{k=-k}) together with (\ref{k=k*}) yields%
\begin{equation}
k_{4}=-k_{3}=k_{3}^{\ast },  \label{3-4}
\end{equation}%
or equivalently%
\begin{equation}
\func{Re}k_{4}=\func{Re}k_{3}=0,\pi /L;\ \ \ \ \func{Im}k_{4}=-\func{Im}%
k_{3}.  \label{sym ev 34}
\end{equation}%
In Eq. (\ref{sym ev 34}) we took into account that $k\equiv k+2\pi /L$.

Consider now the case (\ref{4ev}) of a frequency gap, where all four
eigenmodes are evanescent. The relations (\ref{k=-k}) and (\ref{k=k*}) allow
for two different possibilities. The first one is similar to that of Eq. (%
\ref{3-4})%
\begin{equation}
\begin{array}{c}
k_{2}=-k_{1}=k_{1}^{\ast }, \\ 
k_{4}=-k_{3}=k_{3}^{\ast }.%
\end{array}
\label{1-2, 3-4}
\end{equation}%
or, equivalently,%
\begin{equation}
\begin{array}{c}
\func{Re}k_{2}=\func{Re}k_{1}=0,\pi /L;\ \ \func{Im}k_{2}=-\func{Im}k_{1},
\\ 
\func{Re}k_{4}=\func{Re}k_{3}=0,\pi /L;\ \ \func{Im}k_{4}=-\func{Im}k_{3}.%
\end{array}
\label{sym ev 12,34}
\end{equation}%
In the above situation, the four complex wave numbers split into two
reciprocal pairs $k_{1},k_{2}$ and $k_{3},k_{4}$ of the conjugate values.
The other possibility is%
\begin{equation}
k_{1}=-k_{2}^{\ast }=k_{3}^{\ast }=-k_{4},  \label{1-2-3-4}
\end{equation}%
or, equivalently,%
\begin{equation}
\func{Re}k_{1}=\func{Re}k_{3}=-\func{Re}k_{2}=-\func{Re}k_{4},\ \func{Im}%
k_{1}=-\func{Im}k_{3}=\func{Im}k_{2}=-\func{Im}k_{4}.  \label{sym ev 1234}
\end{equation}

\subsubsection{Spectral asymmetry}

If the sufficient condition (\ref{TL -- 1/TL}) for the axial spectral
symmetry is not in place, then we can have for given $\omega $ and $%
k_{x},k_{y}$%
\begin{equation}
\{k_{1},k_{2},k_{3},k_{4}\}\neq \{-k_{1},-k_{2},-k_{3},-k_{4}\},
\label{k<>-k}
\end{equation}%
which implies \emph{axial spectral asymmetry} (\ref{asym}). The relation (%
\ref{k=k*}), being a direct consequence of $J$-unitarity (\ref{JU}) of the
transfer matrix $T_{L}$, remains valid.

\subsection{Electromagnetic energy flux in stratified media}

\subsubsection{The $J$-scalar product}

For future reference, consider the following scalar product involving the $J$%
-matrix (\ref{J})%
\begin{equation*}
\left( \Psi _{1},J\Psi _{2}\right) =E_{1x}^{\ast }H_{2y}-E_{1y}^{\ast
}H_{2x}+H_{1,y}^{\ast }E_{2x}-H_{1x}^{\ast }E_{2y},
\end{equation*}%
which will be referred to as the $J$-scalar product. Given the importance of
the above quantity, hereinafter, we will use the following special notation
for it 
\begin{equation}
\left[ \Psi _{1},\Psi _{2}\right] \equiv \left( \Psi _{1},J\Psi _{2}\right) .
\label{Jbr}
\end{equation}%
The $J$-scalar product (\ref{Jbr}) is invariant under the following
transformation involving an arbitrary $J$-unitary matrix $T$%
\begin{equation}
\left[ T\Psi _{1},T\Psi _{2}\right] =\left[ \Psi _{1},\Psi _{2}\right] \text{
\ for any }\Psi _{1}\text{and }\Psi _{2}.  \label{JU DP}
\end{equation}%
The relation (\ref{JU DP}) can also be viewed as a criterion of $J$%
-unitarity of a matrix $T$. This relation is similar to that involving the
regular scalar product $\left( \Psi _{1},\Psi _{2}\right) $ and a unitary
matrix $U$%
\begin{equation*}
\left( U\Psi _{1},U\Psi _{2}\right) =\left( \Psi _{1},\Psi _{2}\right) \text{
\ for any }\Psi _{1}\text{and }\Psi _{2}.
\end{equation*}

Let $\Psi _{i}\left( z\right) $ and $\Psi _{j}\left( z\right) $ be two
arbitrary solutions of the time-harmonic Maxwell equation (\ref{ME4}). The
equality (\ref{JU DP}) together with the definition (\ref{T(zz0)}) of the
transfer matrix yields%
\begin{equation}
\left[ \Psi \left( z\right) _{i},\Psi _{j}\left( z\right) \right] =\left[
T\left( z,0\right) \Psi \left( 0\right) _{i},T\left( z,0\right) \Psi
_{j}\left( 0\right) \right] =\left[ \Psi \left( 0\right) _{i},\Psi
_{j}\left( 0\right) \right] ,  \label{JDP - inv}
\end{equation}%
which implies that the $J$-scalar product $\left[ \Psi \left( z\right)
_{i},\Psi _{j}\left( z\right) \right] $ does not depend on the coordinate $z$%
.

Consider now the $J$-scalar product%
\begin{equation}
\left[ \Phi _{i},\Phi _{j}\right]  \label{[ i , j ]}
\end{equation}%
of two eigenvectors $\Phi _{i}$ and $\Phi _{j}$ of the transfer matrix $%
T_{L} $%
\begin{equation}
T_{L}\Phi _{i}=\zeta _{i}\Phi _{i},\;i=1,2,3,4.  \label{TL -- X}
\end{equation}%
The $J$-unitarity (\ref{JU DP}) of $T_{L}$ implies that%
\begin{equation}
\left[ T_{L}\Phi _{i},T_{L}\Phi _{j}\right] =\zeta _{i}^{\ast }\zeta _{j}%
\left[ \Phi _{i},\Phi _{j}\right] =\left[ \Phi _{i},\Phi _{j}\right] ,
\label{Xi Xj}
\end{equation}%
which, in turn, yields the following important relation%
\begin{equation}
\left[ \Phi _{i},\Phi _{j}\right] =0,\text{ \ if \ }\zeta _{i}^{\ast }\zeta
_{j}\neq 1,  \label{[i,j] X}
\end{equation}%
or equivalently,%
\begin{equation}
\left[ \Phi _{i},\Phi _{j}\right] =0,\text{ \ if \ }k_{j}\neq k_{i}^{\ast }.
\label{[i,j] k}
\end{equation}%
In particular%
\begin{equation}
\left[ \Phi _{i},\Phi _{i}\right] \neq 0,\text{ \ only if \ }%
k_{i}=k_{i}^{\ast },  \label{[i,i] k}
\end{equation}%
which means that $\Phi _{i}$ in (\ref{[i,i] k}) should be a propagating
Bloch mode.

\subsubsection{Energy flux in stratified media}

The real-valued energy flux (the Poynting vector) associated with a
time-harmonic electromagnetic field is%
\begin{equation}
\mathbf{S}\left( \vec{r}\right) =\left[ \func{Re}\mathbf{E}\left( \vec{r}%
\right) \times \func{Re}\mathbf{H}\left( \vec{r}\right) \right] =\frac{1}{2}%
\func{Re}\left[ \mathbf{E}^{\ast }\left( \vec{r}\right) \times \mathbf{H}%
\left( \vec{r}\right) \right] .  \label{Pnt}
\end{equation}%
Substitution of the \textquotedblleft tangential\textquotedblright\ Bloch
representation (\ref{LEM}) for $\mathbf{E}\left( \vec{r}\right) $ and $%
\mathbf{H}\left( \vec{r}\right) $ in Eq. (\ref{Pnt}) yields%
\begin{equation}
\mathbf{S}\left( \vec{r}\right) =\mathbf{S}\left( z\right) =\frac{1}{2}\func{%
Re}\left[ \vec{E}^{\ast }\left( z\right) \times \vec{H}\left( z\right) %
\right] ,  \label{S=[EH]}
\end{equation}%
at fixed $\omega $ and $k_{x},k_{y}$. Eq. (\ref{S=[EH]}) implies that in a
stratified medium, at fixed $\omega $ and $k_{x},k_{y}$, all three Cartesian
components of the energy flux $\mathbf{S}\left( \vec{r}\right) $ are
independent of the tangential coordinates $x$\ and $y$. A simple energy
conservation argument shows that the\ normal component $S_{z}$ of the energy
flux does not depend on the coordinate $z$ either, while the tangential
components $S_{x}$\ and $S_{y}$ may depend on $z$. Indeed, in a lossless
stratified medium we have, with consideration for Eq. (\ref{S=[EH]})%
\begin{equation*}
\nabla \cdot \mathbf{S}\left( \vec{r}\right) =\partial _{z}S_{z}\left(
z\right) =0,
\end{equation*}%
which yields that at fixed $\omega $ and $k_{x},k_{y}$%
\begin{equation}
S_{z}\left( \vec{r}\right) =S_{z}=\text{const}.  \label{Sz - const}
\end{equation}%
By contrast, the tangential components of the steady-state energy flux are
dependent on the $z$ coordinate%
\begin{equation}
S_{x}\left( \vec{r}\right) =S_{x}\left( z\right) ,\;S_{y}\left( \vec{r}%
\right) =S_{y}\left( z\right) .  \label{Sx,Sy (z)}
\end{equation}

Hereinafter, the normal component of the energy flux will be referred to
simply as the \emph{energy flux}, unless otherwise specifically stated. It
also will be denoted as $S$, rather than $S_{z}$.

The explicit expression for the normal component of the energy flux (\ref%
{S=[EH]}) can be recast as%
\begin{equation}
S=\frac{1}{2}\left( E_{x}^{\ast }H_{y}-E_{y}^{\ast }H_{x}+E_{x}H_{y}^{\ast
}-E_{y}H_{x}^{\ast }\right) =\frac{1}{2}\left[ \Psi ,\Psi \right] ,
\label{Sz(Psi)}
\end{equation}%
where the $J$-scalar product $\left[ \Psi ,\Psi \right] \equiv \left( \Psi
,J\Psi \right) $ is defined in Eq. (\ref{Jbr}). The fact that $S$ in Eq. (%
\ref{Sz(Psi)}) is independent of $z$ implies that%
\begin{equation}
S=\frac{1}{2}\left[ \Psi \left( z\right) ,\Psi \left( z\right) \right] =%
\frac{1}{2}\left[ \Phi ,\Phi \right] ,\text{ \ where }\Phi =\Psi \left(
0\right) .  \label{Sz0}
\end{equation}

Eq. (\ref{Sz0}) can also be viewed as a direct consequence of $J$-unitarity (%
\ref{JU}) of the transfer matrix. Indeed, from the definition (\ref{T(zz0)})
of the transfer matrix we have 
\begin{equation}
\Psi \left( z\right) =T\left( z,0\right) \Psi \left( 0\right) =T\left(
z,0\right) \Phi .  \label{Psi(z) =TPhi}
\end{equation}%
Substituting (\ref{Psi(z) =TPhi}) into (\ref{Sz(Psi)}) yields%
\begin{equation*}
S=\frac{1}{2}\left[ \Psi \left( z\right) ,\Psi \left( z\right) \right] =%
\frac{1}{2}\left[ T\left( z,0\right) \Phi ,T\left( z,0\right) \Phi \right] .
\end{equation*}%
Taking into account the property (\ref{JU DP}) of a $J$-unitary matrix, we
again arrive at Eq. (\ref{Sz0}).

\subsubsection{Energy flux in periodic stratified media}

The direct relation (\ref{Sz0}) between the $J$-scalar product $\left[ \Psi
,\Psi \right] $ and the energy flux $S$ at fixed $\omega $ and $k_{x},k_{y}$
allows us to make some strong statements regarding electromagnetic energy
flux in periodic layered media.

Let us start with the simplest case of a single Bloch eigenmode. Eq. (\ref%
{Sz0}) together with (\ref{[i,i] k}) shows that only a propagating mode can
transfer electromagnetic energy%
\begin{equation}
S_{i}=\frac{1}{2}\left[ \Phi _{i},\Phi _{i}\right] \neq 0\text{ only if }%
k_{i}=k_{i}^{\ast }.  \label{Sz(i,i)}
\end{equation}%
A single evanescent eigenmode always has zero energy flux 
\begin{equation}
S_{i}=\left[ \Phi _{i},\Phi _{i}\right] =0,\text{ \ if \ }k_{i}\neq
k_{i}^{\ast }.  \label{Sz(i i)}
\end{equation}

Let us turn to the case of a superposition%
\begin{equation*}
\Phi=\dsum \limits_{i=1}^{4}a_{i}\Phi_{i}.
\end{equation*}
of different Bloch eigenmodes with fixed $\omega$ and $k_{x},k_{y}$. In such
a case, the energy flux is%
\begin{equation}
S=\frac{1}{2}\left[ \Phi,\Phi\right] =\frac{1}{2}\dsum
\limits_{i,j=1}^{4}a_{i}^{\ast}a_{j}\left[ \Phi_{i},\Phi_{j}\right] .
\label{Sz (Phi)}
\end{equation}
Taking into account Eqs. (\ref{[i,j] k}) we can draw the following
conclusions:

\begin{enumerate}
\item[1)] The contribution $S_{i}$ of each propagating eigenmode to the
total energy flux is independent of the presence or absence of other Bloch
eigenmodes with the same $\omega $ and $k_{x},k_{y}$%
\begin{equation}
S=\dsum\limits_{i=1}S_{i}=\frac{1}{2}\dsum\limits_{i=1}\left\vert
a_{i}\right\vert ^{2}\left[ \Phi _{i},\Phi _{i}\right] ,  \label{Sz(ex)}
\end{equation}%
where the summation runs over all propagating eigenmodes. The number of
propagating modes can be 4, 2, or 0, depending on which of the cases (\ref%
{4ev}), (\ref{2ex2ev}), or (\ref{4ev}) we are dealing with.

\item[2)] The contribution of evanescent Bloch eigenmodes to the energy flux
depends on their number.

\begin{enumerate}
\item In the case (\ref{2ex2ev}) of two evanescent modes $\Phi _{3}$ and $%
\Phi _{4}$ we have%
\begin{equation}
S=\func{Re}\left( a_{3}^{\ast }a_{4}\left[ \Phi _{3},\Phi _{4}\right]
\right) ,\text{ where }k_{4}=k_{3}^{\ast },  \label{Sz(2ev)}
\end{equation}%
which implies that only a pair of evanescent modes with conjugate wave
numbers can produce energy flux. The respective contribution (\ref{Sz(2ev)})
is independent of the presence of propagating modes $\Phi _{1}$ and $\Phi
_{2}$. In accordance with Eq. (\ref{Sz(i i)}), a single evanescent mode,
either $\Phi _{3}$ or $\Phi _{4}$, does not produce energy flux on its own.

\item In the case (\ref{4ev}) of four evanescent modes we have%
\begin{equation}
S=\func{Re}\left( a_{1}^{\ast }a_{2}\left[ \Phi _{1},\Phi _{2}\right]
\right) +\func{Re}\left( a_{3}^{\ast }a_{4}\left[ \Phi _{3},\Phi _{4}\right]
\right) ,\text{ where }k_{2}=k_{1}^{\ast },k_{4}=k_{3}^{\ast },
\label{Sz(4ev)}
\end{equation}%
which implies that either of the two pairs of evanescent modes with
conjugate wave numbers contribute to the energy flux independently of each
other.
\end{enumerate}
\end{enumerate}

\subsection{Scattering problem for a periodic semi-infinite stack}

In this final subsection we outline the standard procedure we use for
solving the scattering problem of a plane monochromatic wave incident on the
surface of a periodic semi-infinite stack.

In vacuum (to the left of the semi-infinite slab) the electromagnetic field $%
\Psi_{V}\left( z\right) $ is a superposition of the incident and reflected
waves%
\begin{equation}
\Psi_{V}\left( z\right) =\Psi_{I}\left( z\right) +\Psi_{R}\left( z\right) ,%
\text{ }z\leq0,  \label{L=I+R}
\end{equation}
where the indices $I$ and $R$ relate to the incident and reflected beams,
respectively. At the slab boundary\ we have%
\begin{equation}
\Psi_{V}\left( 0\right) =\Psi_{I}\left( 0\right) +\Psi_{R}\left( 0\right) .
\label{I,R}
\end{equation}

The transmitted wave $\Psi _{T}\left( z\right) $ inside the periodic
semi-infinite slab is a superposition of two forward Bloch eigenmodes%
\begin{equation}
\Psi _{T}\left( z\right) =\Psi _{1}\left( z\right) +\Psi _{2}\left( z\right)
,\text{ }z\geq 0.  \label{T=1+2}
\end{equation}%
The eigenmodes $\Psi _{1}\left( z\right) \ $and $\Psi _{2}\left( z\right) $
can be both propagating (with $u>0$), one propagating and one evanescent
(with $u>0$ and $\func{Im}k>0,$ respectively), or both evanescent (with $%
\func{Im}k>0$), depending on which of the three cases (\ref{PsiT=pr+pr}), (%
\ref{PsiT=pr+ev}), or (\ref{PsiT=ev+ev}) we are dealing with.

Assume now that for a given frequency $\omega $, the Bloch eigenmodes are
found, which can be readily done in the case of a periodic layered array.
Using the standard electromagnetic boundary conditions%
\begin{equation}
\Psi _{T}\left( 0\right) =\Psi _{I}\left( 0\right) +\Psi _{R}\left( 0\right)
,  \label{BC1}
\end{equation}%
one can express the reflected wave $\Psi _{R}$ and the eigenmode composition
of the transmitted wave $\Psi _{T}$, in terms of the amplitude and
polarization of the incident wave $\Psi _{I}$. This automatically gives the
electromagnetic field distribution $\Psi _{T}\left( z\right) $ inside the
slab, as a function of the incident wave frequency, polarization, and
direction of incidence.

The transmittance and reflectance coefficients of a lossless semi-infinite
slab are defined by the following expressions%
\begin{equation}
\tau=1-\rho=\frac{\left( \vec{S}_{T}\right) _{z}}{\left( \vec{S}_{I}\right)
_{z}},\;\;\rho=-\frac{\left( \vec{S}_{R}\right) _{z}}{\left( \vec{S}%
_{I}\right) _{z}}.  \label{te, re}
\end{equation}
where $\left( \vec{S}_{I}\right) _{z}$ , $\left( \vec{S}_{R}\right) _{z}$
and $\left( \vec{S}_{T}\right) _{z}$are the normal components of the energy
flux of the incident, reflected, and transmitted waves, respectively.
Knowing the value of the transmitted wave $\Psi_{T}$ or reflected wave $%
\Psi_{R}$ at the slab boundary, one can immediately find the respective
energy flux and, thereby, the transmittance/reflectance coefficients (\ref%
{te, re}).

The above-outlined standard procedure was used in all our numerical
simulations. It applies both to the case of normal and oblique incidence. In
the latter case, the explicit expressions for the column vectors $\Psi _{I}$
and $\Psi _{R}$ in (\ref{L=I+R}-\ref{BC1}) are%
\begin{equation}
\Psi _{I}=\left[ 
\begin{array}{c}
E_{I,x} \\ 
E_{I,y} \\ 
H_{I,x} \\ 
H_{I,y}%
\end{array}%
\right] ,\ \ \Psi _{R}=\left[ 
\begin{array}{c}
E_{R,x} \\ 
E_{R,y} \\ 
H_{R,x} \\ 
H_{R,y}%
\end{array}%
\right] ,  \label{Phi I, Phi R}
\end{equation}%
where the complex vectors $\vec{E}_{I},\vec{H}_{I}$ and $\vec{E}_{R},\vec{H}%
_{R}$ are related to the actual electromagnetic field components $\mathbf{E}%
_{I},\mathbf{H}_{I}$ and $\mathbf{E}_{R},\mathbf{H}_{R}$ as%
\begin{align}
\mathbf{\vec{E}}_{I}& =e^{i\frac{\omega }{c}\left( n_{x}x+n_{y}y\right) }%
\vec{E}_{I}\left( z\right) ,\ \mathbf{\vec{H}}_{I}=e^{i\frac{\omega }{c}%
\left( n_{x}x+n_{y}y\right) }\vec{H}_{I},  \label{LEM I} \\
\mathbf{\vec{E}}_{R}& =e^{i\frac{\omega }{c}\left( n_{x}x+n_{y}y\right) }%
\vec{E}_{R}\left( z\right) ,\ \mathbf{\vec{H}}_{R}=e^{i\frac{\omega }{c}%
\left( n_{x}x+n_{y}y\right) }\vec{H}_{R},  \label{LEM R}
\end{align}%
as prescribed by Eq. (\ref{LEM}). Here $\vec{n}$ is the unit vector in the
direction of light propagation%
\begin{equation}
\text{for incident beam: \ \ }\vec{n}=\vec{n}_{I}=\left( \frac{c}{\omega }%
k_{x},\frac{c}{\omega }k_{y},\frac{c}{\omega }k_{z}\right) ,  \label{n_I}
\end{equation}%
\begin{equation}
\text{for reflected beam: \ \ }\vec{n}=\vec{n}_{R}=\left( \frac{c}{\omega }%
k_{x},\frac{c}{\omega }k_{y},-\frac{c}{\omega }k_{z}\right) ,  \label{n_R}
\end{equation}%
where%
\begin{equation*}
k_{z}=\frac{c}{\omega }n_{z}=\frac{c}{\omega }\sqrt{1-\left(
n_{x}^{2}+n_{y}^{2}\right) }.
\end{equation*}%
Note that the tangential components of the unit vector $\vec{n}$ of the
incident wave are the same as those of the reflected wave. The electric and
magnetic fields of a plane monochromatic wave in a vacuum are uniquely
related to each other%
\begin{equation*}
\mathbf{\vec{H}}=\vec{n}\times \mathbf{\vec{E}}.
\end{equation*}%
The same relation holds for the complex vectors $\vec{E}_{I},\vec{H}_{I}$
and $\vec{E}_{R},\vec{H}_{R}$ defined in (\ref{LEM I}) and (\ref{LEM R}),
namely%
\begin{equation}
\vec{H}_{I}=\vec{n}_{I}\times \vec{E}_{I},\ \ \vec{H}_{R}=\vec{n}_{R}\times 
\vec{E}_{R}.  \label{H=n x E}
\end{equation}

\section{Matrix of reflection coefficients of a semi-infinite periodic stack}

This and the following sections are devoted to a rigorous mathematical
analysis of the scattering problem for a plane monochromatic wave incident
on a periodic semi-infinite stack. We focus on vicinities of stationary
points of the electromagnetic dispersion relation and our goal is to develop
an asymptotic analytical description of the frozen mode regime. Not only
that would allow to rigorously prove the physical results presented earlier
in this paper, it would also provide a better understanding of the very
essence of the frozen mode regime. The major part of the following analysis
is a perturbation theory of degenerate non-diagonalizable matrices.
Specifically, we refer to the transfer matrix $T_{L}$, which develops a
nontrivial Jordan block at any stationary point of the dispersion relations.
The latter circumstance implies the existence of diverging non-Bloch
eigenmodes, which usually do not contribute to the transmitted wave $\Psi
_{T}$ inside the semi-infinite photonic slab and, therefore, do not affect
the scattering problem at hand. Yet, there are two important exceptions. The
first one is the stationary inflection point (\ref{SIP}), where not only the
linearly diverging Floquet eigenmode dominates the transmitted wave, but it
also produces a finite energy flux inside the periodic medium. Another
exception is the degenerate band edge (\ref{om_d}), where the respective
linearly divergent non-Bloch eigenmode, although dominant, does not
contribute to the energy flux and, therefore, does not effectively transform
the incident radiation into the slow mode.

The rest of the paper is organized as follows. In this section we
re-formulate the scattering problem for a lossless periodic semi-infinite
stack, introducing basic notations and definitions. In the following
sections we develop a perturbation theory for degenerate non-diagonalizable $%
4\times 4$ matrices and apply this theory to the transfer matrix $T_{L}$
and, thereby, to the scattering problem. Special attention is given to the
comparative analysis of different stationary points of the electromagnetic
dispersion relation, such as a photonic band edge, a stationary inflection
point, and a degenerate band edge. To simplify the rather cumbersome
mathematical expressions of the following sections, we will use the
following new notations for the quantities already defined earlier%
\begin{align*}
x_{1}& \rightarrow x,\ x_{2}\rightarrow y,x_{3}\rightarrow z, \\
A& \rightarrow \frac{\omega }{c}JM,\;T\rightarrow T_{L}, \\
\mathbf{k}& =\left( k_{1},k_{2},k_{3}\right) \rightarrow \left(
ck_{x},ck_{y},ck_{z}\right) .
\end{align*}%
Observe that the $4\times 4$ matrix%
\begin{equation}
A=\frac{\omega }{c}JM  \label{A (M)}
\end{equation}%
is Hermitian, while the related Maxwell operator $M$ defined in Eqs. (\ref%
{ME4}-\ref{J}) is $J$-Hermitian.

\subsection{Basic definitions}

A periodic semi-infinite stack is defined in terms of the related matrix
function $A\left( x_{3}\right) $ satisfying%
\begin{equation}
A\left( x_{3}\right) =\limfunc{Const},\ -\infty <x_{3}<0;\ A\left(
x_{3}+L\right) =A\left( x_{3}\right) ,\ 0<x_{3}<\infty .  \label{si1}
\end{equation}%
In vacuum, the Hermitian matrix $A\left( x_{3}\right) $ defined in (\ref{A
(M)}) and has the form 
\begin{equation}
A\left( x_{3}\right) =A^{\left( 0\right) }=\left[ 
\begin{array}{cc}
\mathbf{a}^{\left( 0\right) } & \mathbf{0} \\ 
\mathbf{0} & \mathbf{a}^{\left( 0\right) }%
\end{array}%
\right] ,\ \mathbf{a}^{\left( 0\right) }=\frac{1}{\omega }\left[ 
\begin{array}{cc}
\omega ^{2}-k_{2}^{2} & k_{1}k_{2} \\ 
k_{1}k_{2} & \omega ^{2}-k_{1}^{2}%
\end{array}%
\right] ,  \label{A0kx1}
\end{equation}%
The above expressions immediately follow from Eqs. (\ref{M}) and (\ref{A (M)}%
). The tangential component $\mathbf{k}_{\tau }$ of the the wave vector $%
\mathbf{k}$ is related to its normal component $k_{3}$ by%
\begin{equation}
k_{3}=\sqrt{\omega ^{2}-\mathbf{k}_{\tau }^{2}}=\sqrt{\omega ^{2}-\left(
k_{1}^{2}+k_{2}^{2}\right) }.  \label{k3omx}
\end{equation}%
Let us introduce%
\begin{equation}
j_{2}=\left[ 
\begin{array}{cc}
0 & -1 \\ 
1 & 0%
\end{array}%
\right] ,\ J=\left[ 
\begin{array}{cc}
0 & -j_{2} \\ 
j_{2} & 0%
\end{array}%
\right] =j_{2}\otimes j_{2},  \label{jkk1}
\end{equation}%
and notice that%
\begin{equation}
JA^{\left( 0\right) }=\left[ 
\begin{array}{cc}
0 & -j_{2}\mathbf{a}^{\left( 0\right) } \\ 
j_{2}\mathbf{a}^{\left( 0\right) } & 0%
\end{array}%
\right] =\left[ 
\begin{array}{cc}
0 & -1 \\ 
1 & 0%
\end{array}%
\right] \otimes \left[ j_{2}\mathbf{a}^{\left( 0\right) }\right]
=j_{2}\otimes \left[ j_{2}\mathbf{a}^{\left( 0\right) }\right] .
\label{jkk2b}
\end{equation}%
Recall the basic properties of the tensor product operation: if $A$ and $B$
are square matrices and $u$ and $v$ are vectors of related dimensions then%
\begin{equation}
\left[ A\otimes B\right] \left( u\otimes v\right) =\left( Au\right) \otimes
\left( Bv\right) ,\ \left( u_{1}\otimes v_{1},u_{2}\otimes v_{2}\right)
=\left( u_{1},u_{2}\right) \left( v_{1},v_{2}\right) .  \label{tab1}
\end{equation}%
Suppose now that we know the set of eigenvectors and eigenvalues for two
square matrices $A$ and $B$, namely%
\begin{equation}
Au_{j}=\lambda _{j}u_{j},\ Bv_{m}=\mu _{m}v_{m}.  \label{tab2}
\end{equation}%
Then (\ref{tab1}) and (\ref{tab2}) imply%
\begin{equation}
\left[ A\otimes B\right] \left( u_{j}\otimes v_{m}\right) =\lambda _{j}\mu
_{m}u_{j}\otimes v_{m}.  \label{tab3}
\end{equation}%
Using (\ref{tab3}) and the tensor product representation (\ref{jkk2b}) for $%
JA^{\left( 0\right) }$ we can find \ its eigenvectors and eigenvalues as
follows. First, we find that%
\begin{align}
j_{2}u_{\pm }& =\pm \mathrm{i}u_{\pm },\ u_{\pm }=\frac{1}{\sqrt{2}}\left[ 
\begin{array}{c}
\pm \mathrm{i} \\ 
1%
\end{array}%
\right] ;\ \ \ j_{2}\mathbf{a}^{\left( 0\right) }v_{\pm }=\pm \mathrm{i}%
k_{3}v_{\pm },  \label{tab4} \\
v_{\pm }& =v_{\pm }\left( \omega ,\mathbf{k}_{\tau }\right) =\frac{1}{\gamma
_{\omega ,\mathbf{k}_{\tau }}}\left[ 
\begin{array}{c}
-\frac{k_{3}^{2}+k_{2}^{2}}{\pm \mathrm{i}\omega k_{3}+k_{1}k_{2}} \\ 
1%
\end{array}%
\right] ,\ \ \gamma _{\omega ,\mathbf{k}_{\tau }}=\sqrt{\frac{2\left(
k_{3}^{2}+k_{2}^{2}\right) k_{3}\omega }{\omega
^{2}k_{3}^{2}+k_{1}^{2}k_{2}^{2}}}.  \notag
\end{align}%
Notice also that%
\begin{align}
\left( v_{+},j_{2}v_{-}\right) & =\left( v_{-},j_{2}v_{+}\right) =0,\ \left(
v_{\pm },j_{2}v_{\pm }\right) =\pm \mathrm{i},  \label{tab4a} \\
\left( v_{\pm },v_{\pm }\right) & =\beta _{\omega ,\mathbf{k}_{\tau }}=\frac{%
\left( k_{3}^{2}+k_{2}^{2}\right) ^{2}+\left( \omega k_{3}\right)
^{2}+\left( k_{1}k_{2}\right) ^{2}}{2\left( k_{3}^{2}+k_{2}^{2}\right)
k_{3}\omega },  \notag \\
\left( v_{\mp },v_{\pm }\right) & =\frac{1}{\gamma _{\omega ,\mathbf{k}%
_{\tau }}^{2}}\left[ \left( \frac{k_{3}^{2}+k_{2}^{2}}{\pm \mathrm{i}\omega
k_{3}+k_{1}k_{2}}\right) ^{2}+1\right] ,  \notag
\end{align}%
\begin{equation}
\left( u_{\pm },u_{\pm }\right) =1,\ \left( u_{\mp },u_{\pm }\right) =0;\
\left( u_{\mp },j_{2}u_{\pm }\right) =0,\ \left( u_{\pm },j_{2}u_{\pm
}\right) =\pm \mathrm{i}.  \label{tab4b}
\end{equation}%
Using the tensor product representation (\ref{jkk2b}) for $JA^{\left(
0\right) }$ and (\ref{tab3}) and (\ref{tab4}), (\ref{tab4a}) we obtain%
\begin{equation}
\left[ JA^{\left( 0\right) }\right] Z_{1}^{\pm }=\pm k_{3}Z_{1}^{\pm },\ %
\left[ JA^{\left( 0\right) }\right] Z_{2}^{\pm }=\pm k_{3}Z_{2}^{\pm },\ \
Z_{j}^{\pm }=Z_{j}^{\pm }\left( \omega ,\mathbf{k}_{\tau }\right) .
\label{tab5}
\end{equation}%
where%
\begin{equation}
Z_{1}^{+}=u_{-}\otimes v_{+},\ Z_{2}^{+}=u_{+}\otimes v_{-},\
Z_{1}^{-}=u_{+}\otimes v_{+},\ Z_{2}^{-}=u_{-}\otimes v_{-}.  \label{tab6}
\end{equation}%
The component representations for $Z_{j}^{\pm }$ are as follows 
\begin{equation}
Z_{1}^{+}=\frac{1}{\sqrt{2}\gamma _{\omega ,\mathbf{k}_{\tau }}}\left[ 
\begin{array}{c}
\frac{\mathrm{i}\left( k_{3}^{2}+k_{2}^{2}\right) }{\mathrm{i}k_{3}\omega
+k_{1}k_{2}} \\ 
-\mathrm{i} \\ 
-\frac{k_{3}^{2}+k_{2}^{2}}{\mathrm{i}k_{3}\omega +k_{1}k_{2}} \\ 
1%
\end{array}%
\right] ,\ Z_{2}^{+}=\frac{1}{\sqrt{2}\gamma _{\omega ,\mathbf{k}_{\tau }}}%
\left[ 
\begin{array}{c}
-\frac{\mathrm{i}\left( k_{3}^{2}+k_{2}^{2}\right) }{-\mathrm{i}k_{3}\omega
+k_{1}k_{2}} \\ 
\mathrm{i} \\ 
-\frac{k_{3}^{2}+k_{2}^{2}}{-\mathrm{i}k_{3}\omega +k_{1}k_{2}} \\ 
1%
\end{array}%
\right] ,  \label{tab7}
\end{equation}%
\begin{equation}
Z_{1}^{-}=\frac{1}{\sqrt{2}\gamma _{\omega ,\mathbf{k}_{\tau }}}\left[ 
\begin{array}{c}
-\frac{\mathrm{i}\left( k_{3}^{2}+k_{2}^{2}\right) }{\mathrm{i}k_{3}\omega
+k_{1}k_{2}} \\ 
\mathrm{i} \\ 
-\frac{k_{3}^{2}+k_{2}^{2}}{\mathrm{i}k_{3}\omega +k_{1}k_{2}} \\ 
1%
\end{array}%
\right] ,\ Z_{2}^{-}=\frac{1}{\sqrt{2}\gamma _{\omega ,\mathbf{k}_{\tau }}}%
\left[ 
\begin{array}{c}
\frac{\mathrm{i}\left( k_{3}^{2}+k_{2}^{2}\right) }{-\mathrm{i}k_{3}\omega
+k_{1}k_{2}} \\ 
-\mathrm{i} \\ 
-\frac{k_{3}^{2}+k_{2}^{2}}{-\mathrm{i}k_{3}\omega +k_{1}k_{2}} \\ 
1%
\end{array}%
\right] .  \label{tab8}
\end{equation}%
Observe that (\ref{tab1}), (\ref{tab6}), (\ref{tab4b}) imply 
\begin{align}
\left[ Z_{1}^{+},Z_{2}^{+}\right] & =\left( Z_{1}^{+},JZ_{2}^{+}\right)
=\left( u_{-}\otimes v_{+},\left[ j_{2}\otimes j_{2}\right] u_{+}\otimes
v_{-}\right)  \label{tab9} \\
& =\left( u_{-}\otimes v_{+},j_{2}u_{+}\otimes j_{2}v_{-}\right) =\mathrm{i}%
\left( u_{-}\otimes v_{+},u_{+}\otimes j_{2}v_{-}\right)  \notag \\
& =\left( u_{-},u_{+}\right) \left( v_{+},\mathrm{i}j_{2}v_{-}\right) =0, 
\notag
\end{align}%
\begin{align}
\left[ Z_{1}^{+},Z_{1}^{+}\right] & =-\mathrm{i}\left( u_{-}\otimes
v_{+},u_{-}\otimes j_{2}v_{+}\right) =-\mathrm{i}\left( u_{-},u_{-}\right)
\left( v_{+},j_{2}v_{+}\right) =1,  \label{tab10} \\
\left[ Z_{1}^{+},Z_{1}^{-}\right] & =-\mathrm{i}\left( u_{-}\otimes
v_{+},u_{+}\otimes j_{2}v_{+}\right) =-\mathrm{i}\left( u_{-},u_{+}\right)
\left( v_{+},j_{2}v_{+}\right) =0.  \notag
\end{align}%
Carrying out more evaluations similar to (\ref{tab9}), (\ref{tab10}) we get%
\begin{equation}
\left[ Z_{j}^{\pm },Z_{m}^{\mp }\right] =0,\ \left[ Z_{j}^{\pm },Z_{m}^{\pm }%
\right] =\pm \delta _{jm},\ j,m=1,2,  \label{tab11}
\end{equation}%
where $\delta _{jm}$ is Kronecker symbol. The relations (\ref{tab11}) show
that the system of 4 vectors $Z_{j}^{\pm }$, $j=1,2$ is flux-orthonormal in
the sense that it is orthonormal with respect to the flux form $\left[ \Psi
_{1},\Psi _{2}\right] =\left( \Psi _{1},J\Psi _{2}\right) $.

Consider now the scalar products of 4 vectors $Z_{j}^{\pm}$, $j=1,2$: 
\begin{align}
\left( Z_{1}^{+},Z_{1}^{+}\right) & =\left( u_{-}\otimes v_{+},u_{-}\otimes
v_{+}\right) =\left( u_{-},u_{-}\right) \left( v_{+},v_{+}\right)
=\beta_{\omega,\mathbf{k}_{\tau}}\beta_{v},  \label{tab12} \\
\left( Z_{1}^{-},Z_{1}^{-}\right) & =\left( u_{+}\otimes v_{+},u_{+}\otimes
v_{+}\right) =\left( u_{+},u_{+}\right) \left( v_{+},v_{+}\right)
=\beta_{\omega,\mathbf{k}_{\tau}},  \notag
\end{align}%
\begin{align}
\left( Z_{1}^{+},Z_{1}^{-}\right) & =\left( u_{-}\otimes v_{+},u_{+}\otimes
v_{+}\right) =\left( u_{-},u_{+}\right) \left( v_{+},v_{+}\right) =0,
\label{tab13} \\
\left( Z_{1}^{+},Z_{2}^{+}\right) & =\left( u_{-}\otimes v_{+},u_{+}\otimes
v_{-}\right) =\left( u_{-},u_{+}\right) \left( v_{+},v_{-}\right) =0.  \notag
\end{align}
Carrying out evaluations similar to (\ref{tab12}), (\ref{tab13}) we get the
following complete set of equalities:%
\begin{align}
\left( Z_{j}^{\pm},Z_{m}^{\pm}\right) & =\beta_{\omega,\mathbf{k}%
_{\tau}}\delta_{jm},\ \left( Z_{j}^{\pm},Z_{m}^{\mp}\right) =0,\ \ j,m=1,2.
\label{tab14} \\
\beta_{\omega,\mathbf{k}_{\tau}} & =\frac{\left( k_{3}^{2}+k_{2}^{2}\right)
^{2}+\left( \omega k_{3}\right) ^{2}+\left( k_{1}k_{2}\right) ^{2}}{2\left(
k_{3}^{2}+k_{2}^{2}\right) k_{3}},  \notag
\end{align}
showing that the system $Z_{j}^{\pm}$, $j=1,2$ is orthogonal though,
evidently, it is not orthonormal.

The set of equalities (\ref{tab11}) and (\ref{tab14}) show the system of
vectors $Z_{j}^{\pm }$, $j=1,2$\ has a property that both the forms, namely,
the EM density form (the scalar product) and the flux form, become diagonal
if it is chosen\ to be a basis of the space $%
\mathbb{C}
^{4}$. Another advantage of choosing $Z_{j}^{\pm }$, $j=1,2$\ to be a basis
is that in this basis the flux balance equality for relevant modes takes its
simplest form as in the classical scattering theory (see (\ref{zpz6}), (\ref%
{zpz7}), (\ref{zpz13})). \emph{In fact, the latter is our primary motivation.%
}

For the periodic semi-infinite stack with $A\left( x_{3};\omega\right) $ we
have the following equation defining its eigenmodes $\Psi\left( x_{3}\right) 
$ at the frequency $\omega$%
\begin{equation}
\partial_{3}\Psi\left( x_{3}\right) =iJA\left( x_{3}\right) \Psi\left(
x_{3}\right) ,\ -\infty<x_{3}<\infty.  \label{si2}
\end{equation}

The eigenmodes of the periodic semi-infinite stack are the ones
corresponding to an incident wave which propagates from $-\infty $ to $%
\infty $, then it is partially reflected by the interface at $x_{3}=0$ and
partially transmitted into the dielectric substance in $0<x_{3}<\infty $. We
refer to such eigenmodes as relevant eigenmodes and denote the set of all
relevant eigenmodes by $\mathcal{S}_{\limfunc{T}}=\mathcal{S}_{\limfunc{T}%
}\left( \omega \right) =\mathcal{S}_{\limfunc{T}}\left( \omega ,\mathbf{k}%
_{\tau }\right) $.

The two extended eigenmodes $\Psi _{1}\left( x_{3}\right) $ and $\Psi
_{2}\left( x_{3}\right) $ describing the standard scattering problem satisfy
the following relations in the air, $x_{3}<0$, 
\begin{align}
\Psi _{1}\left( x_{3}\right) & =e^{\mathrm{i}k_{3}x_{3}}Z_{1}^{+}+e^{-%
\mathrm{i}k_{3}x_{3}}\left[ \rho _{11}Z_{1}^{-}+\rho _{21}Z_{2}^{-}\right] ,
\label{scatx1} \\
\Psi _{2}\left( x_{3}\right) & =e^{\mathrm{i}k_{3}x_{3}}Z_{2}^{+}+e^{-%
\mathrm{i}k_{3}x_{3}}\left[ \rho _{12}Z_{1}^{-}+\rho _{22}Z_{2}^{-}\right] ,
\notag
\end{align}%
where the matrix of reflection coefficients%
\begin{equation}
\rho =\rho _{\omega ,\mathbf{k}_{\tau }}=\left[ 
\begin{array}{cc}
\rho _{11}\left( \omega ,\mathbf{k}_{\tau }\right) & \rho _{12}\left( \omega
,\mathbf{k}_{\tau }\right) \\ 
\rho _{21}\left( \omega ,\mathbf{k}_{\tau }\right) & \rho _{22}\left( \omega
,\mathbf{k}_{\tau }\right)%
\end{array}%
\right]  \label{scatx1a}
\end{equation}%
carries the information about reflection properties of the slab. Its entries
can be called reflection coefficients.

The set $\mathcal{S}_{\limfunc{T}}$ of all relevant eigenmodes $\Psi \left(
x_{3}\right) $ happens to be a two-dimensional linear space. For every fixed
real $a$ it is uniquely determined by the two-dimensional space $\mathcal{S}%
_{\limfunc{T}}\left( a,\omega \right) =\mathcal{S}_{\limfunc{T}}\left(
a;\omega ,\mathbf{k}_{\tau }\right) $ of the values $\Psi \left( a\right) $
as $\Psi $ runs over $\mathcal{S}_{\limfunc{T}}\left( \mathbf{k}_{\tau
}\right) $, i.e.%
\begin{equation}
\mathcal{S}_{\limfunc{T}}\left( a;\omega ,\mathbf{k}_{\tau }\right) =\left\{
\Psi \left( a\right) :\Psi \in \mathcal{S}_{\limfunc{T}}\left( \mathbf{k}%
_{\tau }\right) \right\} .  \label{si2aa}
\end{equation}%
More precisely, all possible \emph{relevant eigenmodes} are described by
solutions to the following Cauchy problem%
\begin{equation}
\partial _{3}\Psi \left( x_{3}\right) =iJA\left( x_{3}\right) \Psi \left(
x_{3}\right) ,\ \Psi \left( a\right) =\Phi \in \mathcal{S}_{\limfunc{T}%
}\left( a,\omega \right) ,\ -\infty <x_{3}<\infty .  \label{si2a}
\end{equation}

The two-dimensional space $\mathcal{S}_{\limfunc{T}}\left( a,\omega ,\mathbf{%
k}_{\tau }\right) $ provides a convenient way to describe and parametrize
the relevant modes. For instance, assuming that we know $\mathcal{S}_{%
\limfunc{T}}\left( a,\omega ,\mathbf{k}_{\tau }\right) $ let us pick any $%
\Phi \in \mathcal{S}_{\limfunc{T}}\left( \omega ,\mathbf{k}_{\tau }\right) $
and find values of the eigenmode $\Psi \left( x_{3}\right) $ in the air. The
eigenmode $\Psi \left( x_{3}\right) $ can be represented as the following
linear combination for $-\infty <x_{3}<\infty :$%
\begin{align}
\Psi \left( x_{3}\right) & =e^{\mathrm{i}k_{3}x_{3}}\left[ \alpha
_{1}^{+}Z_{1}^{+}+\alpha _{2}^{+}Z_{2}^{+}\right] +e^{-\mathrm{i}k_{3}x_{3}}%
\left[ \alpha _{1}^{-}Z_{1}^{-}+\alpha _{2}^{-}Z_{2}^{-}\right] ,
\label{si2b} \\
\Psi \left( 0\right) & =\alpha _{1}^{+}Z_{1}^{+}+\alpha
_{2}^{+}Z_{2}^{+}+\alpha _{1}^{-}Z_{1}^{-}+\alpha _{2}^{-}Z_{2}^{-}\Phi \in 
\mathcal{S}_{\limfunc{T}}\left( 0\right) ,  \notag
\end{align}%
where evidently the two pairs of coefficients%
\begin{equation}
\alpha ^{+}=\left[ 
\begin{array}{c}
\alpha _{1}^{+} \\ 
\alpha _{2}^{+}%
\end{array}%
\right] \text{ and }\alpha ^{-}=\left[ 
\begin{array}{c}
\alpha _{1}^{-} \\ 
\alpha _{2}^{-}%
\end{array}%
\right]  \label{apa1}
\end{equation}%
are respectively related to the incident and the reflected waves. As is
commonly done, we choose arbitrarily the incident wave by picking the vector 
$\alpha ^{+}$ and then finding the the reflected wave as the vector $\alpha
^{-}$ using the relations (\ref{scatx1}), (\ref{scatx1a}) and (\ref{si2b})
by the following formula%
\begin{equation}
\alpha ^{-}=\rho \alpha ^{+},\ \alpha ^{\pm }=\left[ 
\begin{array}{c}
\alpha _{1}^{\pm } \\ 
\alpha _{2}^{\pm }%
\end{array}%
\right] ,\ \rho =\left[ 
\begin{array}{cc}
\rho _{11} & \rho _{12} \\ 
\rho _{21} & \rho _{22}%
\end{array}%
\right] .  \label{apa2}
\end{equation}%
Observe then that the matrix of reflection coefficients $\rho $ can be
viewed as the following mapping relating the incident wave $\alpha ^{+}$ to
the reflected wave $\alpha ^{-}$%
\begin{equation}
\rho :\alpha ^{+}\rightarrow \alpha ^{-}.  \label{apa2a}
\end{equation}%
Notice also that the reflection and the transmission coefficients $r\left(
\alpha ^{+}\right) $ and $t\left( \alpha ^{+}\right) $ corresponding to the
incident wave $\alpha ^{+}$ are defined by the formulae%
\begin{equation}
r^{2}\left( \alpha ^{+}\right) =\frac{\left\vert \rho \alpha ^{+}\left( \Phi
\right) \right\vert ^{2}}{\left\vert \alpha ^{+}\left( \Phi \right)
\right\vert ^{2}}=\frac{\left\vert \rho \alpha ^{+}\right\vert ^{2}}{%
\left\vert \alpha ^{+}\right\vert ^{2}},\ t^{2}\left( \alpha ^{+}\right)
=1-r^{2}\left( \alpha ^{+}\right) .  \label{apa3}
\end{equation}%
It follows from (\ref{scatx1}) that the space $\mathcal{S}_{\limfunc{T}%
}\left( 0;\omega ,\mathbf{k}_{\tau }\right) $ has the following
representation in terms of the vectors $Z_{j}^{\pm }$ and the reflection \
coefficients $\rho _{jm}$:%
\begin{equation}
\mathcal{S}_{\limfunc{T}}\left( 0;\omega ,\mathbf{k}_{\tau }\right) =%
\limfunc{Span}\left\{ \left( Z_{1}^{+}+\rho _{11}Z_{1}^{-}+\rho
_{21}Z_{2}^{-}\right) ,\left( Z_{2}^{+}+\rho _{12}Z_{1}^{-}+\rho
_{22}Z_{2}^{-}\right) \right\} .  \label{sst1}
\end{equation}%
The relation (\ref{sst1}) shows that the space $\mathcal{S}_{\limfunc{T}%
}\left( 0;\omega ,\mathbf{k}_{\tau }\right) $ is uniquely determined by the
matrix $\rho _{\omega ,\mathbf{k}_{\tau }}$. We show in the following
subsection that the matrix $\rho _{\omega ,\mathbf{k}_{\tau }}$ is uniquely
determined and can be constructed based on the space $\mathcal{S}_{\limfunc{T%
}}\left( 0;\omega ,\mathbf{k}_{\tau }\right) $.

\subsection{Basic properties of the space of relevant eigenmodes}

Let us consider now basic properties of the two-dimensional space $\mathcal{S%
}_{\limfunc{T}}\left( 0;\omega \right) =\mathcal{S}_{\limfunc{T}}\left(
0;\omega ,\mathbf{k}_{\tau }\right) $ suppressing in the notation its
dependence on $\mathbf{k}_{\tau }$. Notice first, that the space $\mathcal{S}%
_{\limfunc{T}}\left( 0;\omega \right) $ has the fundamental property that it
is always nonnegative with respect to flux form in the sense that%
\begin{equation}
\left[ \Phi ,\Phi \right] \geq 0\text{ for any }\Phi \in \mathcal{S}_{%
\limfunc{T}}\left( a;\omega \right) .  \label{si3}
\end{equation}%
The property (\ref{si3}) indicates that the modes related to $S_{\limfunc{T}%
}\left( a;\omega \right) $\ must transport energy in the chosen direction.

It is a well known result of spectral theory that no eigenmode $\Psi \left(
x_{3}\right) $ can grow at infinity faster then polynomially. In particular,
an eigenmode can not grow exponentially as $x_{3}\rightarrow \infty $. Since 
$\Psi \left( x_{3}\right) $ is a solution to (\ref{si2}) it must be a linear
combination of eigenmodes of the infinite periodic stack with the relevant
periodic $A\left( x_{3};\omega \right) $ on the interval $0<x_{3}<\infty $.
Consequently, such a linear combination can not include evanescent modes
growing exponentially as $x_{3}\rightarrow \infty $. Additionally, the
above-mentioned linear combination cannot include backward propagating
eigenmodes (those with negative group velocity). Notice that, the related
properties of $\mathcal{S}_{\limfunc{T}}\left( 0;\omega \right) $ can be
characterized by the spectral properties of the transfer matrix $\mathcal{T}%
\left( \omega \right) $. For instance, in the case when all eigenmodes are
propagating and have different wave numbers as described by (\ref{4 ex}),
the space $\mathcal{S}_{\limfunc{T}}\left( 0;\omega \right) $ is the span of
those two eigenvectors $\Phi _{\xi _{1}}=\Phi _{\xi _{1}}\left( \omega
\right) $ and $\Phi _{\xi _{2}}=\Phi _{\xi _{2}}\left( \omega \right) $ that
have positive fluxes, i.e.%
\begin{equation}
\mathcal{S}_{\limfunc{T}}\left( 0;\omega \right) =\limfunc{Span}\left\{ \Phi
_{\xi _{1}},\Phi _{\xi _{2}}\right\} ,\text{ where }\left[ \Phi _{\xi
_{1}},\Phi _{\xi _{1}}\right] ,\left[ \Phi _{\xi _{2}},\Phi _{\xi _{2}}%
\right] >0.  \label{si3a}
\end{equation}%
Hence, there are exactly two eigenvectors having positive fluxes.

In the case (\ref{2 ex 2 ev}) when there are two propagating and two
evanescent modes, $\mathcal{S}_{\limfunc{T}}\left( 0;\omega\right) $ is the
span of the two eigenvectors $\Phi_{\xi_{1}}\left( \omega\right) $, having a
positive flux, and $\Phi_{\zeta}\left( \omega\right) $ with $\left\vert
\zeta\right\vert <1$, i.e.%
\begin{equation}
\mathcal{S}_{\limfunc{T}}\left( 0;\omega\right) =\limfunc{Span}\left\{
\Phi_{\xi_{1}},\Phi_{\zeta}\right\} ,\text{ where }\left\vert
\xi_{1}\right\vert =1,\ \left[ \Phi_{\xi_{1}},\Phi_{\xi_{1}}\right] >0\text{
and }\left\vert \zeta\right\vert <1.  \label{si3b}
\end{equation}
Finally, in the case (\ref{4 ev}) when all modes are evanescent, $\mathcal{S}%
_{\limfunc{T}}\left( 0;\omega\right) $ is the span of the two eigenvectors $%
\Phi_{\zeta_{1}}\left( \omega\right) $ and $\Phi _{\zeta_{2}}\left(
\omega\right) $, i.e.%
\begin{equation}
\mathcal{S}_{\limfunc{T}}\left( 0;\omega\right) =\limfunc{Span}\left\{
\Phi_{\zeta_{1}},\Phi_{\zeta_{2}}\right\} ,\text{ where }\left\vert
\zeta_{1}\right\vert ,\left\vert \zeta_{2}\right\vert <1.  \label{si3c}
\end{equation}

If for a certain frequency$\ \omega _{0}$ the transfer matrix $\mathcal{T}%
\left( \omega _{0}\right) $ has a non-trivial Jordan block, then the space $%
\mathcal{S}_{\limfunc{T}}\left( 0;\omega _{0}\right) $ can be defined as the
following limit%
\begin{equation}
\mathcal{S}_{\limfunc{T}}\left( 0;\omega _{0}\right) =\lim_{\omega
\rightarrow \omega _{0}}\mathcal{S}_{\limfunc{T}}\left( 0;\omega \right) ,
\label{sts1}
\end{equation}%
where it is assumed that for $\omega \neq \omega _{0}$ the matrix $\mathcal{T%
}\left( \omega \right) $ is diagonalizable and $\mathcal{S}_{\limfunc{T}%
}\left( 0;\omega \right) $ is a well defined two-dimensional space. The
limit (\ref{sts1}) uses a distance $d$ between a two subspaces $S_{1}$ and $%
S_{2}$ defined by the formula, \cite{Kato}, Section IV, \S 2,%
\begin{equation}
d\left( S_{1},S_{2}\right) =\max \left\{ \delta \left( S_{1},S_{2}\right)
,\delta \left( S_{2},S_{1}\right) \right\} ,\ \delta \left(
S_{1},S_{2}\right) =\sup_{u\in S_{1}:\left\Vert u\right\Vert =1}\sup_{v\in
S_{2}}\left\Vert u-v\right\Vert .  \label{sts2}
\end{equation}%
Hence, the limit\ relation in (\ref{sts1}) is interpreted as%
\begin{equation}
\lim_{\omega \rightarrow \omega _{0}}d\left( \mathcal{S}_{\limfunc{T}}\left(
0;\omega \right) ,\mathcal{S}_{\limfunc{T}}\left( 0;\omega _{0}\right)
\right) =0.  \label{sts2a}
\end{equation}%
The distance $d\left( S_{1},S_{2}\right) $ defined by (\ref{sts2}) measures
the \textquotedblleft aperture\textquotedblright\ or \textquotedblleft
gap\textquotedblright\ between the subspaces $S_{1}$ and $S_{2}$. It has the
following important property, \cite{Kato}, Section IV, \S 2, Corollary 2.6, 
\begin{equation}
d\left( S_{1},S_{2}\right) <1\text{ implies }\dim S_{1}=\dim S_{2}.
\label{sts3}
\end{equation}%
The property (\ref{sts3}) implies that if the limit (\ref{sts1}) exists then
the dimension of the space $\mathcal{S}_{\limfunc{T}}\left( 0;\omega
_{0}\right) $ must be 2 since $\dim \mathcal{S}_{\limfunc{T}}\left( 0;\omega
\right) =2$ for $\omega \neq \omega _{0}$.

One can also verify that the limit relations (\ref{sts1}), (\ref{sts2a}) can
be conveniently recast as a limit relation between orthogonal projections
onto spaces $\mathcal{S}_{\limfunc{T}}\left( 0;\omega \right) $. \ Namely,
if we introduce%
\begin{align}
& P_{\mathcal{S}}\text{ to be the orthogonal projector on the space }%
\mathcal{S},  \label{sts2b} \\
\left\Vert P_{\mathcal{S}}\right\Vert & =\sup_{\Phi \in \mathcal{S}%
}\left\Vert P_{\mathcal{S}}\Phi \right\Vert ,\text{ where }\left\Vert \Phi
\right\Vert \text{ is length (norm) of }\left\Vert \Phi \right\Vert .
\label{sts2c}
\end{align}%
then (\ref{sts1}), (\ref{sts2a}) are equivalent to%
\begin{equation}
\lim_{\omega \rightarrow \omega _{0}}\left\Vert P_{\mathcal{S}_{\limfunc{T}%
}\left( 0;\omega \right) }-P_{\mathcal{S}_{\limfunc{T}}\left( 0;\omega
_{0}\right) }\right\Vert =0.  \label{sts2d}
\end{equation}%
Notice, that the relation (\ref{sts2d}) is equivalent, in turn, to the
relation%
\begin{equation}
\lim_{\omega \rightarrow \omega _{0}}P_{\mathcal{S}_{\limfunc{T}}\left(
0;\omega \right) }\Phi =P_{\mathcal{S}_{\limfunc{T}}\left( 0;\omega
_{0}\right) }\Phi \text{ for any }\Phi \in 
\mathbb{C}
^{4},  \label{sts2e}
\end{equation}%
and the relation (\ref{sts2e}) is equivalent to%
\begin{equation}
\lim_{\omega \rightarrow \omega _{0}}P_{\mathcal{S}_{\limfunc{T}}\left(
0;\omega \right) }\Phi =P_{\mathcal{S}_{\limfunc{T}}\left( 0;\omega
_{0}\right) }\Phi =\Phi \text{ for any }\Phi \in \mathcal{S}_{\limfunc{T}%
}\left( 0;\omega _{0}\right) .  \label{sts2f}
\end{equation}%
Notice now that for every vector $\Phi \in \mathcal{S}_{\limfunc{T}}\left(
0;\omega _{0}\right) $ we can define a family of vectors%
\begin{equation}
\Phi \left( \omega \right) =P_{\mathcal{S}_{\limfunc{T}}\left( 0;\omega
\right) }\Phi \in \mathcal{S}_{\limfunc{T}}\left( 0;\omega \right)
\label{sts2g}
\end{equation}%
converging, in view of (\ref{sts2f}), as $\omega \rightarrow \omega _{0}$ to
the vector $\Phi \left( \omega _{0}\right) $, i.e.%
\begin{equation}
\lim_{\omega \rightarrow \omega _{0}}\Phi \left( \omega \right) =\Phi \left(
\omega _{0}\right) =\Phi ,\ \text{for any }\Phi \in \mathcal{S}_{\limfunc{T}%
}\left( 0;\omega _{0}\right) .  \label{sts2h}
\end{equation}

\subsection{Matrix of reflection coefficients and the flux quadratic form}

In this section we look at the basic properties of the matrix of reflection
coefficients $\rho=\rho_{\omega,\mathbf{k}_{\tau}}$ as defined in (\ref%
{scatx1}), (\ref{scatx1a}), (\ref{apa2}), and its relation to the flux
quadratic form $\left[ \cdot,\cdot\right] $, and the space $\mathcal{S}_{%
\limfunc{T}}\left( 0;\omega,\mathbf{k}_{\tau}\right) $.

Observe that inserting $x_{3}=0$ into (\ref{si2b}) yields 
\begin{equation}
\Psi \left( 0\right) =\Phi =\Phi ^{+}+\Phi ^{-},\ \Phi _{\pm }=\alpha
_{1}^{\pm }\left( \Phi \right) Z_{1}^{\pm }+\alpha _{2}^{\pm }\left( \Phi
\right) Z_{2}^{\pm }.  \label{si2c}
\end{equation}%
The equality (\ref{si2c}) indicates that the numbers $\alpha _{j}^{\pm
}\left( \Phi \right) $ are the components of the vector $\Phi \in \mathcal{S}%
_{\limfunc{T}}\left( 0;\omega \right) $ with respect to the basis%
\begin{equation}
\left\{ Z_{1}^{+},Z_{2}^{+},Z_{1}^{-},Z_{2}^{-}\right\} ,  \label{si2d}
\end{equation}%
and they are determined by the following formulae. Let us introduce the
two-dimensional subspaces of $%
\mathbb{C}
^{4}$:%
\begin{equation}
\mathcal{Z}^{+}=\limfunc{Span}\left\{ Z_{1}^{+},Z_{2}^{+}\right\} ,\ 
\mathcal{Z}^{-}=\limfunc{Span}\left\{ Z_{1}^{-},Z_{2}^{-}\right\} ,
\label{zpz1}
\end{equation}%
and the respectively orthogonal projections:%
\begin{equation}
\pi ^{+}\text{ and }\pi ^{-}\text{ are respectively the orthogonal
projections on }\mathcal{Z}^{+}\text{ and }\mathcal{Z}^{-}.  \label{zpz2}
\end{equation}%
In view of (\ref{tab14}) we have the following representations for $\pi
_{\pm }$: 
\begin{equation}
\pi ^{\pm }\Phi =\frac{1}{\beta _{\omega ,\mathbf{k}_{\tau }}}\left[ \left(
Z_{1}^{\pm },\Phi \right) Z_{1}^{\pm }+\left( Z_{2}^{\pm },\Phi \right)
Z_{2}^{\pm }\right] ,\ \Phi \in 
\mathbb{C}
^{4},  \label{zpz3}
\end{equation}%
and, hence,%
\begin{equation}
\Phi =\Phi ^{+}+\Phi ^{-},\ \Phi ^{\pm }=\pi ^{\pm }\Phi =\alpha _{1}^{\pm
}\left( \Phi \right) Z_{1}^{\pm }+\alpha _{2}^{\pm }\left( \Phi \right)
Z_{2}^{\pm },  \label{zpz4}
\end{equation}%
\begin{align}
\alpha ^{\pm }\left( \Phi \right) & =\frac{1}{\beta _{\omega ,\mathbf{k}%
_{\tau }}}\left[ 
\begin{array}{c}
\left( Z_{1}^{+},\Phi \right) \\ 
\left( Z_{2}^{+},\Phi \right)%
\end{array}%
\right] ,  \label{zpz5} \\
\beta _{\omega ,\mathbf{k}_{\tau }}& =\frac{\left(
k_{3}^{2}+k_{2}^{2}\right) ^{2}+\left( \omega k_{3}\right) ^{2}+\left(
k_{1}k_{2}\right) ^{2}}{2\left( k_{3}^{2}+k_{2}^{2}\right) k_{3}},\ k_{3}=%
\sqrt{\omega ^{2}-\mathbf{k}_{\tau }^{2}}.  \notag
\end{align}%
\emph{Observe, in particular, that for }$\Phi \in S_{\limfunc{T}}\left(
0;\omega ,\mathbf{k}_{\tau }\right) $\emph{\ the equalities (\ref{zpz5})
provide relations between the value }$\Phi $\emph{\ of the mode at }$x_{3}=0$%
\emph{\ and the coefficients }$\alpha ^{\pm }\left( \Phi \right) $\emph{\
for the relevant incident and reflected waves.}

Another simple fundamental fact is that the two-dimensional vector $\alpha
^{+}\left( \Phi\right) $ can take any prescribed value from $%
\mathbb{C}
^{2}$, i.e.: 
\begin{equation}
\left\{ \alpha^{+}\left( \Phi\right) :\Phi\in S_{\limfunc{T}}\left( 0;\omega,%
\mathbf{k}_{\tau}\right) \right\} =%
\mathbb{C}
^{2},\ \alpha^{\pm}\left( \Phi\right) =\frac{1}{\beta_{\omega,\mathbf{k}%
_{\tau}}}\left[ 
\begin{array}{c}
\left( Z_{1}^{+},\Phi\right) \\ 
\left( Z_{2}^{+},\Phi\right)%
\end{array}
\right] ,  \label{zpz5a}
\end{equation}
The relation (\ref{zpz5a}) can be considered as another fundamental property
of the space $S_{\limfunc{T}}\left( 0;\omega,\mathbf{k}_{\tau}\right) $.

Using the coefficients $\alpha ^{\pm }\left( \Phi \right) $ and (\ref{tab11}%
) we get the following representation for the flux of the mode described by $%
\Phi $%
\begin{align}
\left[ \Phi ,\Phi \right] & =\left[ \Phi ^{+},\Phi ^{+}\right] -\left[ \Phi
^{-},\Phi ^{-}\right] =\left\vert \alpha ^{+}\left( \Phi \right) \right\vert
^{2}-\left\vert \alpha ^{-}\left( \Phi \right) \right\vert ^{2},
\label{zpz6} \\
\Phi & =\alpha _{1}^{+}Z_{1}^{+}+\alpha _{2}^{+}Z_{2}^{+}+\alpha
_{1}^{-}Z_{1}^{-}+\alpha _{2}^{-}Z_{2}^{-}\in \mathcal{S}_{\limfunc{T}%
}\left( 0;\omega ,\mathbf{k}_{\tau }\right) .  \notag
\end{align}%
The above equality reflects the fundamental energy flux balance of the
classical scattering theory in its the simplest form:%
\begin{align}
& \left\vert \alpha ^{+}\left( \Phi \right) \right\vert ^{2}\text{ (Incident
wave flux) }-\text{ }\left\vert \alpha ^{-}\left( \Phi \right) \right\vert
^{2}\text{ (Reflected wave flux)}  \label{zpz7} \\
& =\left\vert \alpha ^{+}\left( \Phi \right) \right\vert ^{2}-\left\vert
\alpha ^{-}\left( \Phi \right) \right\vert ^{2}\text{ (Transmitted wave
flux).}  \notag
\end{align}

The fundamental property (\ref{si3}) of the non-negativity of the flux on $%
\mathcal{S}_{\limfunc{T}}\left( 0;\omega ,\mathbf{k}_{\tau }\right) $ can be
recast as%
\begin{equation}
\left\vert \alpha ^{-}\left( \Phi \right) \right\vert ^{2}=\left[ \Phi
^{-},\Phi ^{-}\right] \leq \left\vert \alpha ^{+}\left( \Phi \right)
\right\vert ^{2}=\left[ \Phi ^{+},\Phi ^{+}\right] \text{ for any }\Phi \in
S_{\limfunc{T}}\left( 0;\omega ,\mathbf{k}_{\tau }\right) ,  \label{zpz8}
\end{equation}%
indicating the physically transparent fact that the flux of the reflected
wave cannot exceed the flux of the incident wave.

Combining now the relations (\ref{apa2}) and (\ref{zpz8}) and recalling that 
$\alpha ^{-}\left( \Phi \right) =\rho \alpha ^{+}\left( \Phi \right) $, gives%
$\ $ 
\begin{equation}
\left\vert \rho \alpha ^{+}\left( \Phi \right) \right\vert ^{2}\leq
\left\vert \alpha ^{+}\left( \Phi \right) \right\vert ^{2}\text{ for any }%
\alpha ^{+}\left( \Phi \right) \text{ (any }\Phi \in S_{\limfunc{T}}\left(
0;\omega ,\mathbf{k}_{\tau }\right) \text{),}  \label{zpz9}
\end{equation}%
which, in turn, together with (\ref{zpz5a}) implies%
\begin{equation}
\rho ^{\dag }\rho \leq I_{2}.  \label{zpz10}
\end{equation}%
The matrix inequality (\ref{zpz10}) signifies the fact that for any $\Phi $,
or any incident wave $\alpha ^{+}$, the reflection coefficient $r\left(
\alpha ^{+}\right) $ does not exceed $1$, i.e. 
\begin{equation}
r^{2}\left( \alpha ^{+}\right) =\frac{\left\vert \alpha ^{-}\left( \Phi
\right) \right\vert ^{2}}{\left\vert \alpha ^{+}\left( \Phi \right)
\right\vert ^{2}}=\frac{\left\vert \rho \alpha ^{+}\left( \Phi \right)
\right\vert ^{2}}{\left\vert \alpha ^{+}\left( \Phi \right) \right\vert ^{2}}%
=\frac{\left\vert \rho \alpha ^{+}\right\vert ^{2}}{\left\vert \alpha
^{+}\right\vert ^{2}}=\frac{\left( \alpha ^{+},\left[ \rho ^{\dag }\rho %
\right] \alpha ^{+}\right) }{\left( \alpha ^{+},\alpha ^{+}\right) }\leq 1.
\label{zpz11}
\end{equation}%
As to the EM energy density using (\ref{tab14}) we get%
\begin{align}
\left( \Phi ,\Phi \right) & =\beta _{v}\left( \left\vert \alpha
^{+}\right\vert ^{2}+\left\vert \alpha ^{-}\right\vert ^{2}\right) ,
\label{zpz12} \\
\beta _{\omega ,\mathbf{k}}& =\frac{\left( k_{3}^{2}+k_{2}^{2}\right)
^{2}+\left( \omega k_{3}\right) ^{2}+\left( k_{1}k_{2}\right) ^{2}}{2\left(
k_{3}^{2}+k_{2}^{2}\right) k_{3}\omega },\ k_{3}=\sqrt{\omega ^{2}-\mathbf{k}%
_{\tau }^{2}},  \notag \\
\Phi & =\alpha _{1}^{+}Z_{1}^{+}+\alpha _{2}^{+}Z_{2}^{+}+\alpha
_{1}^{-}Z_{1}^{-}+\alpha _{2}^{-}Z_{2}^{-}\in \mathcal{S}_{\limfunc{T}%
}\left( 0;\omega ,\mathbf{k}_{\tau }\right) .  \notag
\end{align}

Having the energy flux balance in the form (\ref{zpz6}) was our primary
motivation for choosing the vectors $Z_{j}^{\pm }$, $j=1,2$ as a basis in $%
\mathbb{C}
^{4}$. We also want to remind the reader that the vectors $Z_{j}^{\pm }$, $%
j=1,2$ reduce both the EM energy density (the scalar product) and the flux
quadratic forms to their diagonal form, as follows from the set of
equalities (\ref{tab11}) and (\ref{tab14}).

Let us look now at the limit case $\rho ^{\dag }\rho =I_{2}$ for which,
according to (\ref{zpz6}),%
\begin{align}
\left[ \Phi ,\Phi \right] & =\left\vert \alpha ^{+}\left( \Phi \right)
\right\vert ^{2}-\left\vert \alpha ^{-}\left( \Phi \right) \right\vert ^{2}=0%
\text{ for all }\alpha ^{+}\in 
\mathbb{C}
^{2},\ \alpha ^{-}=\rho \alpha _{+},  \label{zpz13} \\
\Phi & =\alpha _{1}^{+}Z_{1}^{+}+\alpha _{2}^{+}Z_{2}^{+}+\alpha
_{1}^{-}Z_{1}^{-}+\alpha _{2}^{-}Z_{2}^{-}\in \mathcal{S}_{\limfunc{T}%
}\left( 0;\omega ,\mathbf{k}_{\tau }\right) .  \notag
\end{align}%
If we denote by $\mathfrak{G}_{0}\left( J\right) $ the set of spaces on
which the flux is identically zero, then the relations (\ref{zpz13}) and (%
\ref{sst1}) imply that%
\begin{align}
& \mathcal{S}_{\limfunc{T}}\left( 0;\omega ,\mathbf{k}_{\tau }\right) =
\label{zpz13a} \\
& =\limfunc{Span}\left\{ \left( Z_{1}^{+}+\rho _{11}Z_{1}^{-}+\rho
_{21}Z_{2}^{-}\right) ,\left( Z_{2}^{+}+\rho _{12}Z_{1}^{-}+\rho
_{22}Z_{2}^{-}\right) \right\} \in \mathfrak{G}_{0}\left( J\right)  \notag
\end{align}%
or, in other words, all vectors of the space $\mathcal{S}_{\limfunc{T}%
}\left( 0;\omega ,\mathbf{k}_{\tau }\right) $ have zero flux. On the other
hand, if $\mathcal{S}_{\limfunc{T}}\left( 0;\omega ,\mathbf{k}_{\tau
}\right) \in \mathfrak{G}_{0}\left( J\right) $ then (\ref{zpz13}) holds
implying $\rho ^{\dag }\rho =I_{2}$. \emph{Consequently, the property that
the slab has complete reflection }$\rho ^{\dag }\rho =I_{2}$\emph{\ is
equivalent to the property of having zero flux for all relevant modes}, or,
symbolically, 
\begin{equation}
\rho ^{\dag }\rho =I_{2}\text{ is equivalent to }\left[ \Phi ,\Phi \right] =0%
\text{ for every }\Phi \in \mathcal{S}_{\limfunc{T}}\left( 0;\omega ,\mathbf{%
k}_{\tau }\right) ,  \label{zpz14}
\end{equation}%
\emph{or, in other words},%
\begin{align}
\text{the reflection coefficient }r\left( \alpha ^{+}\right) & =1\text{ for
every }\alpha ^{+}\in 
\mathbb{C}
^{2}\text{ is equivalent}  \label{zpz14aa} \\
\text{to }\left[ \Phi \left( \alpha ^{+}\right) ,\Phi \left( \alpha
^{+}\right) \right] & =0\text{ for every }\alpha ^{+}\in 
\mathbb{C}
^{2}.  \notag
\end{align}%
\emph{Therefore, to establish the state of complete reflectance it is
sufficient to verify that the fluxes of all relevant modes are zero.}

\emph{Observe also that as a consequence of (\ref{zpz14}), (\ref{zpz14aa})
we have}%
\begin{equation}
\text{if there exists }\Phi\in\mathcal{S}_{\limfunc{T}}\left( 0;\omega,%
\mathbf{k}_{\tau}\right) \text{ such that }\left[ \Phi,\Phi\right] \neq0%
\text{, \ then }\rho^{\dag}\rho\neq I_{2},  \label{zpz14a}
\end{equation}
\emph{\ or, in other words,}%
\begin{align}
\text{if there exist }\Phi & \in\mathcal{S}_{\limfunc{T}}\left( 0;\omega,%
\mathbf{k}_{\tau}\right) \text{ such that }\left[ \Phi,\Phi\right] \neq0
\label{zpz14b} \\
\text{then for almost all }\alpha^{+} & \in%
\mathbb{C}
^{2}:\text{ the reflection coefficient }r\left( \alpha^{+}\right) <1.  \notag
\end{align}

To establish a representation for the matrix $\rho $ in terms of the space $%
\mathcal{S}_{\limfunc{T}}\left( 0;\omega ,\mathbf{k}_{\tau }\right) $ let us
pick any two linearly independent vectors $\Phi _{1}$ and $\Phi _{2}$ in $%
\mathcal{S}_{\limfunc{T}}\left( 0;\omega ,\mathbf{k}_{\tau }\right) $. Then,
since $\mathcal{S}_{\limfunc{T}}\left( 0;\omega ,\mathbf{k}_{\tau }\right) $
is a two-dimensional space, we have%
\begin{equation}
\mathcal{S}_{\limfunc{T}}\left( 0;\omega ,\mathbf{k}_{\tau }\right) =%
\limfunc{Span}\left\{ \Phi _{1},\Phi _{2}\right\} .  \label{zpz15}
\end{equation}%
Having the basis $\left\{ \Phi _{1},\Phi _{2}\right\} $ of $\mathcal{S}_{%
\limfunc{T}}\left( 0;\omega ,\mathbf{k}_{\tau }\right) $ we introduce the
related component representation 
\begin{equation}
\check{\Phi}=\left[ 
\begin{array}{c}
\varphi _{1} \\ 
\varphi _{2}%
\end{array}%
\right] ,\ \Phi =\varphi _{1}\Phi _{1}+\varphi _{2}\Phi _{2}\in \mathcal{S}_{%
\limfunc{T}}\left( 0;\omega ,\mathbf{k}_{\tau }\right) ,  \label{phs1}
\end{equation}%
and%
\begin{align}
\pi ^{\pm }\Phi & =\alpha ^{\pm }\left( \Phi \right) =Q^{\pm }\check{\Phi},
\label{phs2} \\
Q^{\pm }& =\left[ Z_{1}^{+}Z_{2}^{+}\right] ^{\dag }\left[ \Phi _{1}\Phi _{2}%
\right] =\frac{1}{\beta _{\omega ,\mathbf{k}_{\tau }}}\left[ 
\begin{array}{cc}
\left( Z_{1}^{\pm },\Phi _{1}\right) & \left( Z_{1}^{\pm },\Phi _{2}\right)
\\ 
\left( Z_{2}^{\pm },\Phi _{1}\right) & \left( Z_{2}^{\pm },\Phi _{2}\right)%
\end{array}%
\right] .  \notag
\end{align}%
Observe now that the relation $\alpha ^{-}=\rho \alpha ^{+}$ together with (%
\ref{phs2}) implies 
\begin{equation}
Q^{-}\check{\Phi}=\rho Q^{+}\check{\Phi},  \label{phs3}
\end{equation}%
where%
\begin{equation}
\check{\Phi}=\check{\Phi}\left( \alpha ^{+}\right) =\left[ Q^{+}\right]
^{-1}\alpha ^{+},\ \Phi \left( \alpha ^{+}\right) =\left[ \Phi _{1}\Phi _{2}%
\right] \check{\Phi}\left( \alpha ^{+}\right) =\left[ \Phi _{1}\Phi _{2}%
\right] \left[ Q^{+}\right] ^{-1}\alpha ^{+}.  \label{phs3a}
\end{equation}%
The relations (\ref{phs3}) yields, in turn, the following representation for
the matrix $\rho $ 
\begin{equation}
\rho =Q^{-}\left[ Q^{+}\right] ^{-1}=\left[ 
\begin{array}{cc}
\left( Z_{1}^{-},\Phi _{1}\right) & \left( Z_{1}^{-},\Phi _{2}\right) \\ 
\left( Z_{2}^{-},\Phi _{1}\right) & \left( Z_{2}^{-},\Phi _{2}\right)%
\end{array}%
\right] \left[ 
\begin{array}{cc}
\left( Z_{1}^{+},\Phi _{1}\right) & \left( Z_{1}^{+},\Phi _{2}\right) \\ 
\left( Z_{2}^{+},\Phi _{1}\right) & \left( Z_{2}^{+},\Phi _{2}\right)%
\end{array}%
\right] ^{-1}.  \label{phs4}
\end{equation}%
Notice that the inequality (\ref{zpz8}) together with (\ref{zpz5a}) implies%
\begin{equation}
\left[ Q^{-}\right] ^{\dag }Q^{-}\leq \left[ Q^{+}\right] ^{\dag }Q^{+},
\label{phs5}
\end{equation}%
which is an alternative form of the inequalities (\ref{zpz8}), (\ref{zpz10})
and (\ref{zpz11}). Using (\ref{zpz11}), (\ref{phs2}) and (\ref{phs4}) we get
the following representation for the reflection coefficient%
\begin{equation}
r^{2}\left( \alpha ^{+}\right) =\frac{\left( Q^{-}\left[ Q^{+}\right]
^{-1}\alpha ^{+},Q^{-}\left[ Q^{+}\right] ^{-1}\alpha ^{+}\right) }{%
\left\vert \alpha ^{+}\right\vert ^{2}},  \label{phs6}
\end{equation}%
\begin{equation}
r^{2}\left( \alpha ^{+}\left( \Phi \right) \right) =\frac{\left\vert \alpha
^{-}\left( \Phi \right) \right\vert ^{2}}{\left\vert \alpha ^{+}\left( \Phi
\right) \right\vert ^{2}}=\frac{\left( Q^{-}\check{\Phi},Q^{-}\check{\Phi}%
\right) }{\left( Q^{+}\check{\Phi},Q^{+}\check{\Phi}\right) }.  \label{phs7}
\end{equation}%
Observe also that (\ref{zpz13}) and (\ref{phs6}) yield the following
expression for the flux associated with the incident wave described by $%
\alpha ^{+}$ 
\begin{align}
\left[ \Phi \left( \alpha ^{+}\right) ,\Phi \left( \alpha ^{+}\right) \right]
& =\left( 1-r^{2}\left( \alpha ^{+}\right) \right) \left\vert \alpha
^{+}\right\vert ^{2}  \label{phs8} \\
& =\left( 1-\frac{\left( Q^{-}\left[ Q^{+}\right] ^{-1}\alpha ^{+},Q^{-}%
\left[ Q^{+}\right] ^{-1}\alpha ^{+}\right) }{\left\vert \alpha
^{+}\right\vert ^{2}}\right) \left\vert \alpha ^{+}\right\vert ^{2}.  \notag
\end{align}%
The formula (\ref{phs8}) can be recast as the following representation for
the transmission coefficient $t=t\left( \alpha ^{+}\right) $ defined by (\ref%
{apa3})%
\begin{equation}
t^{2}\left( \alpha ^{+}\right) =1-r^{2}\left( \alpha ^{+}\right) =\frac{%
\left[ \Phi \left( \alpha ^{+}\right) ,\Phi \left( \alpha ^{+}\right) \right]
}{\left\vert \alpha ^{+}\right\vert ^{2}},\ \Phi \left( \alpha ^{+}\right) =%
\left[ \Phi _{1}\Phi _{2}\right] \left[ Q^{+}\right] ^{-1}\alpha ^{+}.
\label{phs9}
\end{equation}

\section{Transfer matrix at and near a point of degeneracy}

Let us recall first the definition of degenerate points including inflection
ones. A $n$ - degenerate point $k_{0}$ of a dispersion relation $\omega
\left( k\right) $ is defined as a point at which the following relations
holds%
\begin{equation}
\partial _{k}\omega \left( k_{0}\right) =\partial _{k}^{2}\omega \left(
k_{0}\right) =\cdots =\partial _{k}^{n-1}\omega \left( k_{0}\right) =0,\
\partial _{k}^{n}\omega \left( k_{0}\right) \neq 0.  \label{dgpx1}
\end{equation}%
In particular, an inflection point $k_{0}$ is a $3$-fold degenerate point if%
\begin{equation}
\omega ^{\prime }\left( k_{0}\right) =\omega ^{\prime \prime }\left(
k_{0}\right) =0,\ \omega ^{\prime \prime \prime }\left( k_{0}\right) \neq 0.
\label{dgpx2}
\end{equation}%
Hence, if $k_{0}$ is a $n$-degenerate point we have 
\begin{equation}
\omega \left( k\right) =\omega \left( k_{0}\right) +\frac{\partial
_{k}^{n}\omega \left( k_{0}\right) }{n!}\left( k-k_{0}\right) ^{n}+O\left(
\left( k-k_{0}\right) ^{n+1}\right) ,\ k\rightarrow k_{0}.  \label{dgpx3}
\end{equation}%
In particular, if $k_{0}$ is an inflection point then%
\begin{equation}
\omega \left( k\right) =\omega \left( k_{0}\right) +\frac{\omega ^{\prime
\prime \prime }\left( k_{0}\right) }{6}\left( k-k_{0}\right) ^{3}+O\left(
\left( k-k_{0}\right) ^{4}\right) ,\ k\rightarrow k_{0}.  \label{dgpx4}
\end{equation}

To study the behavior of the transfer matrix $T_{L}$ near $\omega _{0}$ we
introduce%
\begin{equation}
\mathcal{T}\left( \nu \right) =\mathbf{T}\left( \omega _{0}+\nu \right) ,\
\nu =\omega -\omega _{0}.  \label{tt1}
\end{equation}%
We assume the dependence of $\mathcal{T}\left( \nu \right) $ on $\nu $ to be
analytic in some vicinity of $\nu =0$. In our further analysis we use well
known statements from the analytic perturbation theory for matrices and
their spectra, \cite{Kato}.

To find the spectrum of $\mathcal{T}\left( \nu \right) $ we consider the
characteristic polynomial $\Delta _{\mathcal{T}\left( \nu \right) }\left(
\zeta \right) $ and the related characteristic equation%
\begin{equation}
\Delta _{\mathcal{T}\left( \nu \right) }\left( \zeta \right) =\det \left( 
\mathcal{T}\left( \nu \right) -\zeta I_{4}\right) =0,\ \zeta =e^{ik},
\label{tomx2}
\end{equation}%
where $I_{4}$ is the $4\times 4$ identity matrix, and $k$ is the
quasimomentum. Eq. (\ref{tomx2}) is the dispersion relation, namely it
relates to every frequency $\nu =\omega -\omega _{0}$ four values of $\
\zeta $ or, equivalently, four values of the wave number (see Eq. (\ref{X(k)}%
)).

Since $\mathcal{T}\left( \nu \right) $ is $4\times 4$ matrix the equation (%
\ref{tomx2}) can be written as%
\begin{equation}
\Delta _{\mathcal{T}\left( \nu \right) }\left( \zeta \right) =\zeta
^{4}+b_{3}\left( \nu \right) \zeta ^{3}+b_{2}\left( \nu \right) \zeta
^{2}+b_{1}\left( \nu \right) \zeta +b_{0}\left( \nu \right) =0,
\label{tomx3}
\end{equation}%
where the complex valued functions $b_{j}\left( \nu \right) $, \thinspace $%
j=0,1,2,3$ are analytic in $\nu $ in a vicinity of $\nu =0$.

For the frozen mode regime to occur at the frequency $\nu =0$, i.e., $\omega
=\omega _{0}$, the spectral decomposition of the transfer matrix $\mathcal{T}%
\left( 0\right) $ must have a Jordan block of rank $n\geq 2$ with an
algebraic eigenvalue $\zeta _{0}$. In this situation the characteristic
polynomial $\Delta _{\mathcal{T}\left( 0\right) }\left( \zeta \right) $
takes the following special form 
\begin{equation}
\Delta _{\mathcal{T}\left( 0\right) }\left( \zeta \right) =\left( \zeta
-\zeta _{0}\right) ^{n}Q_{n}\left( \zeta \right) ,  \label{tomx4}
\end{equation}%
where 
\begin{equation}
Q_{n}\left( \zeta \right) =\zeta ^{4-n}+\ldots \text{ is a polynomial of the
degree }4-n\text{ such that }Q_{n}\left( \zeta _{0}\right) \neq 0.
\label{tomx4xa}
\end{equation}%
It is an additional property of the transfer matrix $\mathcal{T}\left( \nu
\right) $ that 
\begin{equation}
\left\vert \zeta _{0}\right\vert =1,  \label{tomx4xb}
\end{equation}%
where the eigenvalues $\zeta _{0}=e^{ik_{0}}$ is $n$-degenerate and it
corresponds to a Floquet mode. Because of this degeneracy at $\nu =0$, the
perturbation theory, \cite{Kato}, Section II, classifies the point $\nu =0$
as an exceptional one, and the dependence $\zeta _{0}\left( \nu \right) $ is
described by the Puiseux series of the form%
\begin{equation}
\zeta _{0}\left( \nu \right) =\zeta _{0}\left( 1+\alpha _{1}\nu
^{1/n}+\alpha _{2}\nu ^{2/n}+\cdots \right) .  \label{tomx4a}
\end{equation}%
The corresponding eigenprojectors can be singular. In fact, in our case they
are singular.

If the characteristic equation (\ref{tomx2}) takes the special form (\ref%
{tomx4}) near $\nu =0$ then $\mathcal{T}\left( \nu \right) $ can be reduced
and represented as follows%
\begin{equation}
\mathcal{T}\left( \nu \right) =\mathcal{G}\left( \nu \right) \left[ 
\begin{array}{cc}
T\left( \nu \right) & 0 \\ 
0 & W\left( \nu \right)%
\end{array}%
\right] \mathcal{G}^{-1}\left( \nu \right) ,  \label{tomx5}
\end{equation}%
where $\mathcal{G}\left( \nu \right) $ is an invertible $4\times 4$ matrix
depending analytically on $\nu $, $T\left( \nu \right) $ and $W\left( \nu
\right) $ are respectively $n\times n$ and $\left( 4-n\right) \times \left(
4-n\right) $ matrices depending analytically on $\nu $. Additionally%
\begin{equation}
T\left( \nu \right) =T_{0}+T_{1}\nu +\cdots ,  \label{tomx7}
\end{equation}%
where $T_{0}$ has the following Jordan form%
\begin{equation}
T_{0}=\zeta _{0}\left( I_{n}+D_{0}\right) ,  \label{tomx7a}
\end{equation}%
with $I_{n}$ being the $n\times n$ identity matrix, and $D_{0}$ being a
nilpotent matrix, \cite{LanTi}, Section 6, such that%
\begin{equation}
D_{0}^{n}=0.  \label{tomx8}
\end{equation}%
We would like to show $D_{0}\neq 0$ and, even more, that,%
\begin{equation}
D_{0}^{n-1}\neq 0.  \label{tomx8a}
\end{equation}%
Notice that the characteristic equation for $T\left( \nu \right) $ is%
\begin{equation}
\det \left( T\left( \nu \right) -\zeta I_{n}\right) =0,\ \zeta =e^{ik},
\label{tomx9}
\end{equation}%
which, in view of (\ref{tomx7a}), takes the following form 
\begin{equation}
\det \left( T\left( \nu \right) -\zeta I_{3}\right) =\left( \zeta -\zeta
_{0}\right) ^{n}+\sum_{s=1}^{n-1}a_{n-s}\left( \nu \right) \left( \zeta
-\zeta _{0}\right) ^{n-s}+a_{0}\left( \nu \right) ,  \label{tomx9a}
\end{equation}%
where the functions $a_{s}\left( \nu \right) $ for small $\nu $ have the
following expansions $a_{s}\left( \nu \right) =a_{s}\nu +O\left( \nu
^{2}\right) $ for some number $a_{s}$, $0\leq s\leq n$. Hence, for small $%
\nu $ the characteristic equation (\ref{tomx9}) can be recast as%
\begin{equation}
\left( \zeta -\zeta _{0}\right) ^{n}+\sum_{s=1}^{n-1}\left[ a_{n-s}\nu
+O\left( \nu ^{2}\right) \right] \left( \zeta -\zeta _{0}\right)
^{n-s}+a_{0}\nu +O\left( \nu ^{2}\right) =0,  \label{tomx10}
\end{equation}%
where, importantly, we assume that%
\begin{equation}
a_{0}\neq 0.  \label{tomx10a}
\end{equation}%
It turns out, that the assumption $a_{0}\neq 0$ is equivalent to the
following assumption on the dispersion relation%
\begin{equation}
\omega ^{\left( n\right) }\left( k_{0}\right) \text{ is finite and nonzero,
i.e. }0<\left\vert \omega ^{\left( n\right) }\left( k_{0}\right) \right\vert
<\infty ,  \label{tomx10aa}
\end{equation}%
and the following representation holds%
\begin{equation}
a_{0}=\frac{n!\left( \mathrm{i}\zeta _{0}\right) ^{n}}{\omega ^{\left(
n\right) }\left( k_{0}\right) }.  \label{tomx10aaa}
\end{equation}%
To establish this representation we recall that $\zeta =e^{\mathrm{i}k}$,
where $k$ is the wave number, and notice that equation (\ref{tomx9}) or (\ref%
{tomx10}) relate to every $\nu =\omega -\omega _{0}$ certain $\zeta
_{j}\left( \omega \right) $ and, consequently, wave vectors $k_{j}\left(
\omega \right) $ determining dispersion relations. We can also add that the
algebraic equation (\ref{tomx10}) for $\zeta =e^{\mathrm{i}k}$ is just
another form of the dispersion relation (\ref{dgpx3}) for $\omega \left(
k\right) $. Using this observation we can derive (\ref{tomx10aaa}) from (\ref%
{tomx10}) by inserting in it $\zeta =e^{\mathrm{i}k}$ and $\zeta _{0}=e^{%
\mathrm{i}k_{0}}$ and, assuming $k-k_{0}$ to be small, we get%
\begin{equation}
\zeta _{0}^{n}\left[ \mathrm{i}\left( k-k_{0}\right) \right] ^{n}+a_{0}\nu
+O\left( \nu ^{2}\right) +O\left( \left( k-k_{0}\right) \nu \right) =0,
\label{tomx10a1}
\end{equation}%
which implies%
\begin{equation}
a_{0}\left( \omega \left( k\right) -\omega _{0}\right) =\left( \mathrm{i}%
\zeta _{0}\right) ^{n}\left( k-k_{0}\right) ^{n}+O\left[ \left(
k-k_{0}\right) ^{n+1}\right] .  \label{tomx10a2}
\end{equation}%
Differentiating (\ref{tomx10a2}) with respect to $k$ at $k=k_{0}$ we get%
\begin{equation}
a_{0}\omega ^{\left( n\right) }\left( k_{0}\right) =n!\left( \mathrm{i}\zeta
_{0}\right) ^{n},  \label{tomx11}
\end{equation}%
implying (\ref{tomx10aaa}). Notice also that the substitution $\zeta =\zeta
_{0}$ in (\ref{tomx9a}) yields%
\begin{equation}
\det \left( T\left( \nu \right) -\zeta _{0}I_{3}\right) =a_{0}\nu +O\left(
\nu ^{2}\right) .  \label{tomx12}
\end{equation}

Recall now that by the Cayley-Hamilton theorem, \cite{LanTi}, Section 6.2,
any matrix $T$ is annulled by its characteristic polynomial, i.e. $\Delta
_{T}\left( T\right) =0$. Hence, (\ref{tomx10}) holds if we substitute $\zeta
=T\left( \nu \right) $ treating all other complex numbers as scalar
matrices, i.e.%
\begin{equation}
\left( T\left( \nu \right) -\zeta _{0}I_{n}\right) ^{n}+\sum_{s=1}^{n-1} 
\left[ a_{n-s}\nu +O\left( \nu ^{2}\right) \right] \left( T\left( \nu
\right) -\zeta _{0}I_{n}\right) ^{n-s}+a_{0}\nu I_{n}+O\left( \nu
^{2}\right) =0.  \label{tom11}
\end{equation}%
Now substituting $T\left( \nu \right) =T_{0}+T_{1}\nu +O\left( \nu
^{2}\right) $ into (\ref{tom11}) and taking in account (\ref{tomx7a}) we
single out the terms linear with respect to $\nu $ getting the following
matrix equation 
\begin{equation}
\zeta _{0}^{n-1}\sum_{s=1}^{n}\zeta
_{0}D_{0}^{n-s}T_{1}D_{0}^{s-1}+\sum_{s=1}^{n-1}a_{n-s}\zeta
_{0}^{n-s}D_{0}^{n-s}=-a_{0}I_{n}.  \label{tom12}
\end{equation}%
Suppose now for the sake of argument that (\ref{tomx8a}) does not hold, and,
hence, $D_{0}^{n-1}=0$. Then in the case of $n=2$ we would have $%
D_{0}^{n-1}=D_{0}=0$ and\ the right-hand side of the equation (\ref{tom12})
becomes $0$ implying $a_{0}=0$ that contradicts the assumption (\ref{tomx10a}%
). Hence, for $n=2$, (\ref{tomx8a}) holds. In the case of $n\geq 3$ the
equation (\ref{tom12}) turns into%
\begin{equation}
\zeta
_{0}^{n-1}\sum_{s=2}^{n-1}D_{0}^{n-s}T_{1}D_{0}^{s-1}+%
\sum_{s=2}^{n-1}a_{n-s}\zeta _{0}^{n-s}D_{0}^{n-s}=-a_{0}I_{n},
\label{tom13}
\end{equation}%
so, taking the determinant of the both sides of (\ref{tom13}) implies,%
\begin{equation}
\det D_{0}\left( \zeta
_{0}^{n-1}\sum_{s=2}^{n-1}D_{0}^{n-s}T_{1}D_{0}^{s-2}+%
\sum_{s=2}^{n-1}a_{n-s}\zeta _{0}^{n-s}D_{0}^{n-s-1}\right) =\left(
-a_{0}\right) ^{n}.  \label{tom13a}
\end{equation}%
But, in view of (\ref{tomx8}), evidently $\det D_{0}=0$ implying together
with (\ref{tom13a}) $a_{0}=0$ that contradicts (\ref{tomx10a}). Therefore, (%
\ref{tomx8a}) is correct and the matrix $T_{0}=\zeta _{0}\left(
I_{n}+D_{0}\right) $ has nontrivial Jordan structure. In fact, in view of (%
\ref{tomx8}) $T_{0}=\zeta _{0}\left( I_{n}+D_{0}\right) $ is similar to the
Jordan block of rank $n$, i.e.%
\begin{equation}
T_{0}=\zeta _{0}S_{0}\left[ 
\begin{array}{ccccc}
1 & 1 & 0 & \cdots & 0 \\ 
0 & \ddots & \ddots & \ddots & \vdots \\ 
0 & 0 & \ddots & \ddots & 0 \\ 
\vdots & \ddots & \ddots & \ddots & 1 \\ 
0 & \cdots & 0 & 0 & 1%
\end{array}%
\right] S_{0}^{-1}  \label{tom14}
\end{equation}%
for an invertible $n\times n$ matrix $S_{0}$. In other words, there exists a
basis $f_{0},f_{1},\ldots ,f_{n}$ such that%
\begin{equation}
\zeta _{0}^{-1}T_{0}f_{0}=f_{0},\ \zeta
_{0}^{-1}T_{0}f_{1}=f_{1}+f_{0},\cdots ,\ \zeta
_{0}^{-1}T_{0}f_{n}=f_{n}+f_{n-1}.  \label{tom15}
\end{equation}%
The basis $f_{0},f_{1},\ldots ,f_{n}$ reducing $T_{0}$ to its canonical form
is not unique. What is unique is the following set of spans%
\begin{equation}
\limfunc{Span}\left\{ f_{0}\right\} ,\ \limfunc{Span}\left\{
f_{0},f_{1}\right\} ,\ \limfunc{Span}\left\{ f_{0},f_{1},\ldots
,f_{n-1}\right\} .  \label{tom16}
\end{equation}

Possible bases preserving the canonical matrix to the right of $S_{0}$ in (%
\ref{tom14}) and (\ref{tom15}) are described by the following transformations%
\begin{equation}
S\left[ 
\begin{array}{ccccc}
1 & 1 & 0 & \cdots & 0 \\ 
0 & 1 & \ddots & \ddots & \vdots \\ 
0 & 0 & \ddots & \ddots & 0 \\ 
\vdots & \ddots & \ddots & 1 & 1 \\ 
0 & \cdots & 0 & 0 & 1%
\end{array}%
\right] S^{-1}=\left[ 
\begin{array}{ccccc}
1 & 1 & 0 & \cdots & 0 \\ 
0 & 1 & \ddots & \ddots & \vdots \\ 
0 & 0 & \ddots & \ddots & 0 \\ 
\vdots & \ddots & \ddots & 1 & 1 \\ 
0 & \cdots & 0 & 0 & 1%
\end{array}%
\right]  \label{sg1}
\end{equation}%
where%
\begin{equation}
S=\left[ 
\begin{array}{ccccc}
\gamma _{1} & \gamma _{2} & \cdots & \gamma _{n-1} & \gamma _{n} \\ 
0 & \gamma _{1} & \ddots & \ddots & \gamma _{n-1} \\ 
0 & 0 & \ddots & \ddots & \vdots \\ 
\vdots & \ddots & \ddots & \gamma _{1} & \gamma _{2} \\ 
0 & \cdots & 0 & 0 & \gamma _{1}%
\end{array}%
\right] ,\ \gamma _{j}\in 
\mathbb{C}
,\ j=1,\cdots n.  \label{sg1a}
\end{equation}

Let us introduce now a $4\times 4$ matrix $Q$ reducing the matrix $\mathcal{G%
}^{-1}\mathcal{T}\left( 0\right) \mathcal{G}\left( 0\right) $ to its
canonical Jordan form $Q^{-1}\mathcal{G}^{-1}\mathcal{T}\left( 0\right) 
\mathcal{G}\left( 0\right) Q$. In other words, if we denote%
\begin{equation}
\mathcal{G}_{0}\left( \nu \right) =\mathcal{G}\left( \nu \right) Q
\label{sg1b}
\end{equation}%
then we have%
\begin{equation}
\mathcal{G}_{0}^{-1}\left( 0\right) \mathcal{T}\left( 0\right) \mathcal{G}%
_{0}\left( 0\right) =\left[ 
\begin{array}{cc}
T\left( 0\right) & 0 \\ 
0 & W\left( 0\right)%
\end{array}%
\right] ,  \label{sg2}
\end{equation}%
where the both matrices $\zeta _{0}^{-1}T\left( 0\right) $ and $W\left(
0\right) $ are of the canonical Jordan form. Namely, $T\left( 0\right)
=T_{0} $ takes the following form as in (\ref{tom14}) 
\begin{equation}
\zeta _{0}^{-1}T\left( 0\right) =\left[ 
\begin{array}{ccccc}
1 & 1 & 0 & \cdots & 0 \\ 
0 & \ddots & \ddots & \ddots & \vdots \\ 
0 & 0 & \ddots & \ddots & 0 \\ 
\vdots & \ddots & \ddots & \ddots & 1 \\ 
0 & \cdots & 0 & 0 & 1%
\end{array}%
\right] ,\ \text{and }W\left( 0\right) \text{ has canonical Jordan form.}
\label{sg3}
\end{equation}%
In the most interesting case of the inflection point for $n=3$ the matrix $%
W\left( 0\right) $ reduces to a scalar. In the case $n=4$ there is no $%
W\left( 0\right) $, and in the case $n=2$ in a generic situation $W\left(
0\right) $ will be just a diagonal matrix.

Consequently, the basis $\mathfrak{f}_{j}$, $j=0,1,2,3$ reducing $\mathcal{T}%
\left( 0\right) $ to the above mentioned Jordan form (\ref{sg2}), (\ref{sg3}%
) can represented as follows%
\begin{align}
\mathfrak{f}_{j}& =\mathcal{G}_{0}\left( 0\right) \mathfrak{b}_{j},\
j=0,1,2,3\text{ where}  \label{sg4} \\
\mathfrak{b}_{0}& =\left[ 
\begin{array}{c}
1 \\ 
0 \\ 
0 \\ 
0%
\end{array}%
\right] ,\ \mathfrak{b}_{1}=\left[ 
\begin{array}{c}
0 \\ 
1 \\ 
0 \\ 
0%
\end{array}%
\right] ,\ \mathfrak{b}_{2}=\left[ 
\begin{array}{c}
0 \\ 
0 \\ 
1 \\ 
0%
\end{array}%
\right] ,\ \mathfrak{b}_{3}=\left[ 
\begin{array}{c}
0 \\ 
0 \\ 
0 \\ 
1%
\end{array}%
\right] .  \notag
\end{align}

\section{Spectral perturbation theory of the transfer matrix at a point of
degeneracy}

In this section we develop the spectral perturbation theory for the transfer
matrix $\mathbf{T}\left( \omega \right) $ defined by (\ref{TL}). This
problem has been considered in \cite{PRE03} for a stationary inflection
point. For an inflection the essential part of the perturbation theory is
related to perturbational spectral analysis of the Jordan block of the rank
3, i.e.%
\begin{equation}
D_{0}=D_{0}^{\left( 3\right) }=\left[ 
\begin{array}{ccc}
0 & 1 & 0 \\ 
0 & 0 & 1 \\ 
0 & 0 & 0%
\end{array}%
\right] .  \label{dd1}
\end{equation}%
Below we extend the spectral constructions from \cite{PRE03} to the case of
degenerate points of the ranks 4 and 2. It turns out that as in the case of
an inflection point, which is a degenerate point of the rank 3, the
essential part of perturbational spectral analysis is reduced analysis of
Jordan blocks of the ranks 4 and 2, i.e.%
\begin{equation}
D_{0}=D_{0}^{\left( 4\right) }=\left[ 
\begin{array}{cccc}
0 & 1 & 0 & 0 \\ 
0 & 0 & 1 & 0 \\ 
0 & 0 & 0 & 1 \\ 
0 & 0 & 0 & 0%
\end{array}%
\right] ,\ D_{0}=D_{0}^{\left( 2\right) }=\left[ 
\begin{array}{cc}
0 & 1 \\ 
0 & 0%
\end{array}%
\right] .  \label{dd2}
\end{equation}%
We use the notations 
\begin{equation}
\mathbf{T}\left( \omega \right) =\mathbf{T}\left( L;\omega \right) ;\ 
\mathcal{T}\left( \nu \right) =\mathbf{T}\left( \omega _{0}+\nu \right) ,\
\nu =\omega -\omega _{0}.  \label{dd3}
\end{equation}%
The transfer matrix $\mathcal{T}\left( \nu \right) $ depends analytically on 
$\nu $ in a vicinity of $\nu =0$ and it can be be reduced to its canonical
Jordan form (see (\ref{tomx5})-(\ref{tomx8a}), (\ref{tom14}) and (\ref{sg2}%
)-(\ref{sg4}))%
\begin{equation}
\mathcal{T}\left( \nu \right) =\mathcal{G}_{0}\left( \nu \right) \left[ 
\begin{array}{cc}
T\left( \nu \right) & 0 \\ 
0 & W\left( \nu \right)%
\end{array}%
\right] \mathcal{G}_{0}^{-1}\left( \nu \right)  \label{dd4}
\end{equation}%
with help of the $4\times 4$ invertible matrix $\mathcal{G}_{0}\left( \nu
\right) $ depending analytically on $\nu $ in a vicinity of $\nu =0$. The
matrix $T\left( \nu \right) $ in (\ref{dd4}) is also analytic at $\nu =0$
and has the following representation%
\begin{equation}
T\left( \nu \right) =T_{0}+T_{1}\nu +\cdots ,\ T_{0}=\zeta _{0}\left(
I_{n}+D_{0}\right) ,  \label{dd5}
\end{equation}%
where in our case $D_{0}=D_{0}^{\left( n\right) }$ is a Jordan block (\ref%
{dd1}), (\ref{dd2}) of the order $n=2,3,4$ correspondingly to the rank of
the degenerate point, i.e. 
\begin{equation}
T\left( 0\right) =T_{0}=\zeta _{0}\left( I_{n}+D_{0}^{\left( n\right)
}\right) =\zeta _{0}\left[ 
\begin{array}{ccccc}
1 & 1 & 0 & \cdots & 0 \\ 
0 & \ddots & \ddots & \ddots & \vdots \\ 
0 & 0 & \ddots & \ddots & 0 \\ 
\vdots & \ddots & \ddots & \ddots & 1 \\ 
0 & \cdots & 0 & 0 & 1%
\end{array}%
\right] .  \label{dd5a}
\end{equation}%
It is convenient to recast (\ref{dd5}) as%
\begin{align}
T\left( \nu \right) & =\zeta _{0}\left[ I_{n}+\mathfrak{T}\left( \nu \right) %
\right] ,\ \mathfrak{T}\left( \nu \right) =D_{0}^{\left( n\right) }+%
\mathfrak{T}_{1}\nu +\mathfrak{T}_{2}\nu ^{2}+\ldots .,  \label{dd6} \\
\mathfrak{T}_{s}& =\zeta _{0}^{-1}T_{s},\ s=1,2,\ldots .  \notag
\end{align}%
Let us introduce also matrices $K_{0}^{\left( n\right) }$ by%
\begin{equation}
K_{0}^{\left( 2\right) }=\left[ 
\begin{array}{cc}
0 & 0 \\ 
1 & 0%
\end{array}%
\right] ,\ K_{0}^{\left( 3\right) }=\left[ 
\begin{array}{ccc}
0 & 0 & 0 \\ 
0 & 0 & 0 \\ 
1 & 0 & 0%
\end{array}%
\right] ,\ K_{0}^{\left( 4\right) }=\left[ 
\begin{array}{cccc}
0 & 0 & 0 & 0 \\ 
0 & 0 & 0 & 0 \\ 
0 & 0 & 0 & 0 \\ 
1 & 0 & 0 & 0%
\end{array}%
\right] ,  \label{dd7}
\end{equation}%
and for every $n\times n$ matrix $\mathfrak{T}$ define the matrix%
\begin{equation}
\mathfrak{T}^{\natural }=\mathfrak{T}-\left[ \mathfrak{T}\right]
_{n1}K_{0}^{\left( n\right) },\text{ where }\left[ \mathfrak{T}\right] _{n1}%
\text{ is the named entry of }\mathfrak{T}.  \label{dd8}
\end{equation}%
It turns out that the following very special case of $\mathfrak{T}\left( \nu
\right) $%
\begin{equation}
\mathfrak{T}_{0}\left( \nu \right) =\mathfrak{T}_{0}^{\left( n\right)
}\left( \nu \right) =D_{0}^{\left( n\right) }+\nu K_{0}^{\left( n\right) }=%
\left[ 
\begin{array}{ccccc}
0 & 1 & 0 & \ldots & 0 \\ 
0 & 0 & \ddots & \ddots & \vdots \\ 
\vdots & \ddots & \ddots & \ddots & 0 \\ 
0 & \ddots & \ddots & 0 & 1 \\ 
\nu & 0 & \ldots & 0 & 0%
\end{array}%
\right]  \label{dd9}
\end{equation}%
being an exact solution to the equation 
\begin{equation}
\mathfrak{T}_{0}^{n}\left( \nu \right) =\nu I_{n}  \label{dd10}
\end{equation}%
plays the key role in the spectral analysis of $\mathfrak{T}\left( \nu
\right) $. For that reason we study first spectral properties of $\mathfrak{T%
}_{0}\left( \nu \right) $.

Notice that the characteristic equation $\det \left( \mathfrak{T}_{0}\left(
\nu \right) -\zeta I_{n}\right) =0$ for the eigenvalues of $\mathfrak{T}%
_{0}\left( \nu \right) $ is%
\begin{equation}
\zeta ^{n}-\nu =0,  \label{dd11}
\end{equation}%
and that the matrix $\mathfrak{T}_{0}\left( \nu \right) $ is a companion
matrix of the polynomial $\zeta ^{n}-\nu $, \cite{LanTi}, Sections 2.2, 2.3.
Hence, if we introduce $n$-th roots of $1$ 
\begin{equation}
\varsigma _{0}=1,\ \varsigma _{1}=e^{\mathrm{i}\frac{2\pi }{n}},\ \varsigma
_{2}=\varsigma _{1}^{2},\ldots ,  \label{dd12}
\end{equation}%
then the $n$ eigenvalues of $\mathfrak{T}_{0}\left( \nu \right) $ are%
\begin{equation}
\nu ^{\frac{1}{n}},\ \varsigma _{1}\nu ^{\frac{1}{n}},\ \varsigma _{2}\nu ^{%
\frac{1}{n}},\ldots .  \label{dd13}
\end{equation}%
For the most interesting case of an inflection point $n=3$ we use another
natural notation for the roots%
\begin{equation}
\varsigma _{0}=1,\ \varsigma _{1}=\varsigma _{+}=e^{\mathrm{i}\frac{2\pi }{3}%
}=-\frac{1}{2}+\frac{1}{2}\mathrm{i}\sqrt{3},\ \varsigma _{2}=\varsigma
_{1}^{2}=\varsigma _{-}=-\frac{1}{2}-\frac{1}{2}\mathrm{i}\sqrt{3}.
\label{dd12a}
\end{equation}%
The corresponding eigenvectors of the companion matrix $\mathfrak{T}%
_{0}\left( \nu \right) $ can be also found, and, if one puts them as columns
in a $n\times n$ matrix $\mathfrak{S}_{0}\left( \nu \right) $, it takes the
form, \cite{Wilk}, Section I.10-I.13, \cite{LanTi}, Section 2.11(Problem 21),%
\begin{equation}
\mathfrak{S}_{0}\left( \nu \right) =\left[ 
\begin{array}{cccc}
1 & 1 & 1 & \cdots \\ 
\nu ^{\frac{1}{n}} & \varsigma _{1}\nu ^{\frac{1}{n}} & \varsigma _{2}\nu ^{%
\frac{1}{n}} & \cdots \\ 
\nu ^{\frac{2}{n}} & \varsigma _{1}^{2}\nu ^{\frac{2}{n}} & \varsigma
_{2}^{2}\nu ^{\frac{2}{n}} & \cdots \\ 
\vdots & \vdots & \vdots & \ddots%
\end{array}%
\right] .  \label{dd14}
\end{equation}%
Hence,%
\begin{equation}
\mathfrak{T}_{0}\left( \nu \right) =\nu ^{\frac{1}{n}}\mathfrak{S}_{0}\left(
\nu \right) \Lambda _{0}\mathfrak{S}_{0}^{-1}\left( \nu \right) ,\ \Lambda
_{0}=\left[ 
\begin{array}{cccc}
1 & 0 & 0 & \cdots \\ 
0 & \varsigma _{1} & 0 & \cdots \\ 
0 & 0 & \varsigma _{2} & \cdots \\ 
\vdots & \vdots & \vdots & \ddots%
\end{array}%
\right] .  \label{dd15}
\end{equation}

Observe that $\mathfrak{S}_{0}\left( \nu \right) $ is a Vandermonde matrix, 
\cite{LanTi}, of order $n$ corresponding to $n$ numbers $1,\varsigma
_{1},\varsigma _{2},\cdots $, and%
\begin{align}
\det \mathfrak{S}_{0}\left( \nu \right) & =\det \left[ 
\begin{array}{cccc}
1 & 1 & 1 & \cdots \\ 
1 & \varsigma _{1} & \varsigma _{2} & \cdots \\ 
1 & \varsigma _{1}^{2} & \varsigma _{2}^{2} & \cdots \\ 
\vdots & \vdots & \vdots & \ddots%
\end{array}%
\right] =\dprod\limits_{1\leq j<s\leq n}\left( \varsigma _{s}-\varsigma
_{j}\right)  \label{dd16} \\
& =\left( -1\right) ^{\frac{\left( n+2\right) \left( n-1\right) }{4}}n^{%
\frac{n}{2}}\nu ^{\frac{n-1}{2}}=e^{-\mathrm{i}\frac{\left( n+2\right)
\left( n-1\right) \pi }{4}}n^{\frac{n}{2}}\nu ^{\frac{n-1}{2}}.  \notag
\end{align}%
Notice also that%
\begin{align}
\mathfrak{S}_{0}^{-1}\left( \nu \right) & =\frac{1}{n}\mathfrak{S}%
_{0}^{\dagger }\left( \left( \bar{\nu}\right) ^{-1}\right) =\frac{1}{n}\left[
\begin{array}{cccc}
1 & \nu ^{-\frac{1}{n}} & \nu ^{-\frac{2}{n}} & \cdots \\ 
1 & \varsigma _{1}^{-1}\nu ^{-\frac{1}{n}} & \varsigma _{1}^{-2}\nu ^{-\frac{%
2}{n}} & \cdots \\ 
1 & \varsigma _{2}^{-1}\nu ^{-\frac{1}{n}} & \varsigma _{2}^{-2}\nu ^{-\frac{%
2}{n}} & \cdots \\ 
\vdots & \vdots & \vdots & \ddots%
\end{array}%
\right]  \label{dd17} \\
& =\nu ^{-\frac{n-1}{n}}\mathfrak{S}_{0}^{\flat }\left( \nu \right) ,\ 
\mathfrak{S}_{0}^{\flat }\left( \nu \right) =\frac{1}{n}\left[ 
\begin{array}{cccc}
\nu ^{\frac{n-1}{n}} & \nu ^{\frac{n-2}{n}} & \nu ^{\frac{n-3}{n}} & \cdots
\\ 
\nu ^{\frac{n-1}{n}} & \varsigma _{1}^{-1}\nu ^{\frac{n-2}{n}} & \varsigma
_{1}^{-2}\nu ^{\frac{n-3}{n}} & \cdots \\ 
\nu ^{\frac{n-1}{n}} & \varsigma _{2}^{-1}\nu ^{\frac{n-2}{n}} & \varsigma
_{2}^{-2}\nu ^{\frac{n-3}{n}} & \cdots \\ 
\vdots & \vdots & \vdots & \ddots%
\end{array}%
\right]  \notag
\end{align}%
where a $\mathfrak{S}^{\dagger }$ is conjugate transpose to a matrix $%
\mathfrak{S}$, and $\bar{\zeta}$ is the conjugate to a complex number $\zeta 
$. Evidently the matrix $\mathfrak{S}_{0}^{\flat }\left( \nu \right) $ is
analytic in $\nu ^{1/n}$ and%
\begin{equation}
\mathfrak{S}_{0}\left( \nu \right) \mathfrak{S}_{0}^{\flat }\left( \nu
\right) =\nu ^{-\frac{n-1}{n}}I_{n}.  \label{dd18}
\end{equation}

Let us consider now the $\left[ n,1\right] $ entry of the perturbed matrix $%
\mathfrak{T}\left( \nu \right) $ as a new variable $\tilde{\nu}$, namely 
\begin{equation}
\tilde{\nu}=\left[ \mathfrak{T}\left( \nu \right) \right] _{n1}=\sum_{s\geq
1}\mathfrak{t}_{s}\nu ^{s},\ \mathfrak{t}_{s}=\left[ \mathfrak{T}_{s}\right]
_{31},\ s\geq 1.  \label{dd19}
\end{equation}%
We will consider the generic case when%
\begin{equation}
\mathfrak{t}_{1}=\left[ \mathfrak{T}_{1}\right] _{n1}\neq 0.  \label{dd20}
\end{equation}%
The above assumption (\ref{dd20}), as we will show, is equivalent to the
fundamental assumption (\ref{tomx10aa}) on the dispersion relation at the
point $k_{0}$. Under the condition (\ref{dd20}) the relation (\ref{dd19})
can be inverted as%
\begin{equation}
\nu =\sum_{s\geq 1}\mathfrak{r}_{s}\tilde{\nu}^{s},  \label{dd21}
\end{equation}%
where the coefficients $\mathfrak{r}_{s}$ can be expressed recurrently in
terms of $\mathfrak{t}_{q},\ q\leq s$. In particular%
\begin{equation}
\mathfrak{r}_{1}=\frac{1}{\mathfrak{t}_{1}},\ \mathfrak{r}_{2}=-\frac{%
\mathfrak{t}_{2}}{\mathfrak{t}_{1}^{3}},\ \mathfrak{r}_{3}=\frac{2\mathfrak{t%
}_{2}^{2}-\mathfrak{t}_{1}\mathfrak{t}_{3}}{\mathfrak{t}_{1}^{5}}.
\label{dd22}
\end{equation}%
Hence, from (\ref{dd21}) and (\ref{dd22}) we have%
\begin{equation}
\nu =\frac{1}{\mathfrak{t}_{1}}\tilde{\nu}-\frac{\mathfrak{t}_{2}}{\mathfrak{%
t}_{1}^{3}}\tilde{\nu}^{2}+\frac{2\mathfrak{t}_{2}^{2}-\mathfrak{t}_{1}%
\mathfrak{t}_{3}}{\mathfrak{t}_{1}^{5}}\tilde{\nu}^{3}+\cdots .  \label{dd23}
\end{equation}%
Using the new variable $\tilde{\nu}$ and (\ref{dd9}) we recast the perturbed
matrix (\ref{dd6}) as a series in $\tilde{\nu}$:%
\begin{align}
\mathfrak{T}\left( \nu \right) & =\mathfrak{T}_{0}\left( \tilde{\nu}\right)
+\sum_{s\geq 1}\tilde{\nu}^{s}\widetilde{\mathfrak{T}}_{s},\ \left[ 
\widetilde{\mathfrak{T}}_{s}\right] _{n1}=0,\ s\geq 1,  \label{dd24} \\
\mathfrak{T}_{0}\left( \nu \right) & =\mathfrak{T}_{0}^{\left( n\right)
}\left( \nu \right) =D_{0}^{\left( n\right) }+\nu K_{0}^{\left( n\right) }=%
\left[ 
\begin{array}{ccccc}
0 & 1 & 0 & \ldots & 0 \\ 
0 & 0 & \ddots & \ddots & \vdots \\ 
\vdots & \ddots & \ddots & \ddots & 0 \\ 
0 & \ddots & \ddots & 0 & 1 \\ 
\nu & 0 & \ldots & 0 & 0%
\end{array}%
\right] ,  \notag
\end{align}%
where the matrix $\mathfrak{T}_{0}\left( \tilde{\nu}\right) $ satisfies also
(\ref{dd14}), (\ref{dd15}), and the matrices $\widetilde{\mathfrak{T}}_{s}$
can be expressed recurrently in terms of $\mathfrak{T}_{q}^{\sharp },\ q\leq
s$. In particular,%
\begin{equation}
\widetilde{\mathfrak{T}}_{1}=\frac{\mathfrak{T}_{1}^{\sharp }}{\mathfrak{t}%
_{1}},\ \widetilde{\mathfrak{T}}_{2}=-\frac{\mathfrak{t}_{2}\mathfrak{T}%
_{1}^{\sharp }}{\mathfrak{t}_{1}^{3}}+\frac{\mathfrak{T}_{2}^{\sharp }}{%
\mathfrak{t}_{1}^{2}},\ \widetilde{\mathfrak{T}}_{3}=\frac{\left( 2\mathfrak{%
t}_{2}^{2}-\mathfrak{t}_{1}\mathfrak{t}_{3}\right) \mathfrak{T}_{1}^{\sharp }%
}{\mathfrak{t}_{1}^{5}}-\frac{2\mathfrak{T}_{2}^{\sharp }}{\mathfrak{t}%
_{1}^{4}}+\frac{\mathfrak{T}_{3}^{\sharp }}{\mathfrak{t}_{1}^{3}}.
\label{dd25}
\end{equation}%
In particular, in view of (\ref{dd6}), the equalities (\ref{dd25}) yield%
\begin{equation}
\widetilde{\mathfrak{T}}_{1}=\frac{T_{1}^{\sharp }}{\zeta _{0}\mathfrak{t}%
_{1}},\ \widetilde{\mathfrak{T}}_{2}=-\frac{\mathfrak{t}_{2}T_{1}^{\sharp }}{%
\zeta _{0}\mathfrak{t}_{1}^{3}}+\frac{T_{2}^{\sharp }}{\zeta _{0}\mathfrak{t}%
_{1}^{2}},\ \widetilde{\mathfrak{T}}_{3}=\frac{\left( 2\mathfrak{t}_{2}^{2}-%
\mathfrak{t}_{1}\mathfrak{t}_{3}\right) T_{1}^{\sharp }}{\zeta _{0}\mathfrak{%
t}_{1}^{5}}-\frac{2T_{2}^{\sharp }}{\zeta _{0}\mathfrak{t}_{1}^{4}}+\frac{%
T_{3}^{\sharp }}{\zeta _{0}\mathfrak{t}_{1}^{3}}.  \label{dd25a}
\end{equation}%
Based on (\ref{dd14}) and (\ref{dd17}) we get the following representation
for an arbitrary $n\times n$ matrix $\mathfrak{A}$ 
\begin{equation}
\nu ^{\frac{n-1}{n}}\mathfrak{S}_{0}^{-1}\left( \nu \right) \mathfrak{AS}%
_{0}\left( \nu \right) =\sum_{q=0}^{\left( n-1\right) ^{2}}\left\langle 
\mathfrak{A}\right\rangle _{q}\nu ^{\frac{q}{n}},  \label{dd26}
\end{equation}%
where, evidently, $\left\langle \mathfrak{A}\right\rangle _{q}$ are $n\times
n$ matrices can be found based on the matrix $\mathfrak{A}$ from the very
relation (\ref{dd26}). In particular, one can find that%
\begin{equation}
\left\langle \mathfrak{A}\right\rangle _{0}=\frac{1}{n}\left[ \mathfrak{A}%
\right] _{n1}\left[ 
\begin{array}{cccc}
1 & 1 & \cdots & 1 \\ 
\varsigma _{1} & \varsigma _{1} & \cdots & \varsigma _{1} \\ 
\vdots & \vdots & \ddots & \cdots \\ 
\varsigma _{n-1} & \varsigma _{n-1} & \cdots & \varsigma _{n-1}%
\end{array}%
\right] ,  \label{dd27}
\end{equation}%
\begin{equation}
\left\langle \mathfrak{A}\right\rangle _{1}=\frac{1}{n}\left[ \mathfrak{A}%
\right] _{n-1,1}\left[ 
\begin{array}{cccc}
1 & 1 & \cdots & 1 \\ 
\varsigma _{1}^{2} & \varsigma _{1}^{2} & \cdots & \varsigma _{1}^{2} \\ 
\vdots & \vdots & \ddots & \cdots \\ 
\varsigma _{n-1}^{2} & \varsigma _{n-1}^{2} & \cdots & \varsigma _{n-1}^{2}%
\end{array}%
\right] +\frac{1}{n}\left[ \mathfrak{A}\right] _{n,2}\left[ 
\begin{array}{cccc}
1 & \varsigma _{1} & \cdots & \varsigma _{n-1} \\ 
\varsigma _{1} & \varsigma _{1}\varsigma _{1} & \cdots & \varsigma
_{1}\varsigma _{n-1} \\ 
\vdots & \vdots & \ddots & \cdots \\ 
\varsigma _{n-1} & \varsigma _{n-1}\varsigma _{1} & \cdots & \varsigma
_{n-1}\varsigma _{n-1}%
\end{array}%
\right]  \label{dd28}
\end{equation}%
showing as $\nu \rightarrow 0$ the most significant zero term in the
representation (\ref{dd26}) depends only on the entry $\left[ \mathfrak{A}%
\right] _{n1}$ of the entire matrix $\mathfrak{A}$. This elucidates the
special role played by the matrix entry $\left[ \mathfrak{A}\right] _{n1}$.
Then from (\ref{dd15}), (\ref{dd24})-(\ref{dd28}) we get the following
important representation%
\begin{align}
& \mathfrak{S}_{0}^{-1}\left( \tilde{\nu}\right) \mathfrak{T}\left( \nu
\right) \mathfrak{S}_{0}\left( \tilde{\nu}\right) =  \label{dd29} \\
& =\tilde{\nu}^{\frac{1}{n}}\Lambda _{0}+\sum_{s\geq 1}\tilde{\nu}^{s}%
\mathfrak{S}_{0}^{-1}\left( \tilde{\nu}\right) \widetilde{\mathfrak{T}}_{s}%
\mathfrak{S}_{0}\left( \tilde{\nu}\right) =\tilde{\nu}^{\frac{1}{n}}\Lambda
_{0}+\sum_{s\geq 1}\tilde{\nu}^{s-\frac{n-1}{n}}\sum_{q=0}^{\left(
n-1\right) ^{2}}\left\langle \widetilde{\mathfrak{T}}_{s}\right\rangle _{q}%
\tilde{\nu}^{\frac{q}{n}}  \notag \\
& =\tilde{\nu}^{\frac{1}{n}}\Lambda _{0}+\sum_{s\geq 1}\tilde{\nu}^{s-\frac{%
n-1}{n}}\sum_{q=1}^{\left( n-1\right) ^{2}}\left\langle \widetilde{\mathfrak{%
T}}_{s}\right\rangle _{q}\tilde{\nu}^{\frac{q-1}{n}}  \notag \\
& =\tilde{\nu}^{\frac{1}{n}}\left[ \Lambda _{0}+\tilde{\nu}^{\frac{1}{n}%
}\left\langle \widetilde{\mathfrak{T}}_{1}\right\rangle _{1}+\tilde{\nu}^{%
\frac{2}{n}}\left\langle \widetilde{\mathfrak{T}}_{1}\right\rangle _{2}+%
\tilde{\nu}^{\frac{3}{n}}\left\langle \widetilde{\mathfrak{T}}%
_{1}\right\rangle _{3}+\cdots +\tilde{\nu}^{\frac{n-1}{n}}\left\langle 
\widetilde{\mathfrak{T}}_{1}\right\rangle _{n-1}+\cdots \right] ,  \notag
\end{align}%
where the matrix $\Lambda _{0}$ is a diagonal matrix defined by (\ref{dd15})
and, evidently its entries $1,\varsigma _{1},\ldots ,\varsigma _{n-1}$
defined by (\ref{dd12}) are all distinct. Notice that the representation (%
\ref{dd29}) reduces the perturbation analysis of the initial series to the
last series in (\ref{dd29}). The perturbation theory of that series
involving a diagonal matrix $\Lambda _{0}$ with different elements is much
simpler and elementary. The relevant perturbational statements needed for
the analysis are collected in the following section.

To analyze perturbations of the matrix $\Lambda _{0}$ we introduce first the
following auxiliary variable%
\begin{equation}
\tilde{\nu}^{\frac{1}{n}}=\mathrm{i}\acute{\nu}.  \label{nv1}
\end{equation}%
Then based on the described results and general facts on the perturbation
theory for diagonal matrices \cite{BPov} (the sketch of the theory is
presented in the Appendix 2) we get 
\begin{align}
\mathfrak{T}\left( \nu \right) & =\mathrm{i}\acute{\nu}\mathfrak{S}%
_{0}\left( \tilde{\nu}\right) e^{-S\left( \mathrm{i}\acute{\nu}\right)
}\left( \Lambda _{0}+\left( \mathrm{i}\acute{\nu}\right) \Lambda _{1}+\left( 
\mathrm{i}\acute{\nu}\right) ^{2}\Lambda _{2}+\cdots \right) e^{S\left( 
\mathrm{i}\acute{\nu}\right) }\mathfrak{S}_{0}^{-1}\left( \tilde{\nu}\right)
,  \label{nv2} \\
S\left( \mathrm{i}\acute{\nu}\right) & =\left( \mathrm{i}\acute{\nu}\right)
S_{1}+\left( \mathrm{i}\acute{\nu}\right) ^{2}S_{2}+\cdots ,  \notag
\end{align}%
where $\Lambda _{s},s\geq 1$ are diagonal matrices. The above formula can be
also written in the form 
\begin{align}
\mathfrak{T}\left( \nu \right) & =\left( \mathrm{i}\acute{\nu}\right) ^{-1}%
\mathfrak{S}_{0}\left( \tilde{\nu}\right) e^{-S\left( \mathrm{i}\acute{\nu}%
\right) }\left( \Lambda _{0}+\Lambda _{1}\left( \mathrm{i}\acute{\nu}\right)
+\Lambda _{2}\left( \mathrm{i}\acute{\nu}\right) ^{2}+\cdots \right)
e^{S\left( \mathrm{i}\acute{\nu}\right) }\mathfrak{S}_{0}^{\flat }\left( \nu
\right) ,  \label{nv3} \\
S\left( \mathrm{i}\acute{\nu}\right) & =\left( \mathrm{i}\acute{\nu}\right)
S_{1}+\left( \mathrm{i}\acute{\nu}\right) ^{2}S_{2}+\cdots .  \notag
\end{align}

\emph{We would like to point out that in the representation (\ref{nv2}), (%
\ref{nv3}) the eigenvectors collected in the Vandermonde matrix }$\mathfrak{S%
}_{0}\left( \tilde{\nu}\right) $\emph{\ defined by (\ref{dd14})\ are
invariant under any change of variables described by (\ref{sg1}), (\ref{sg1a}%
). The dependence of the eigenvectors on the parameters }$\gamma_{j}$\emph{, 
}$j=1,\ldots,n$\emph{\ comes through terms of proper higher powers of }$%
\acute{\nu}$\emph{.}

The representation (\ref{nv2}) and (\ref{nv1}) imply that \emph{the
eigenvectors of the matrix }$T\left( \nu \right) $\emph{\ (and, hence, in
view of (\ref{dd6}), of the matrix }$T\left( \nu \right) $\emph{) are the
columns of the following matrix}%
\begin{equation}
\mathfrak{S}_{0}\left( \tilde{\nu}\right) e^{-S\left( \mathrm{i}\acute{\nu}%
\right) }=\left[ 
\begin{array}{cccc}
1 & 1 & 1 & \cdots \\ 
\left( \mathrm{i}\acute{\nu}\right) & \varsigma _{1}\left( \mathrm{i}\acute{%
\nu}\right) & \varsigma _{2}\left( \mathrm{i}\acute{\nu}\right) & \cdots \\ 
\left( \mathrm{i}\acute{\nu}\right) ^{2} & \varsigma _{1}^{2}\left( \mathrm{i%
}\acute{\nu}\right) ^{2} & \varsigma _{2}^{2}\left( \mathrm{i}\acute{\nu}%
\right) ^{2} & \cdots \\ 
\vdots & \vdots & \vdots & \ddots%
\end{array}%
\right] \left\{ I_{n}-\left( \mathrm{i}\acute{\nu}\right) S_{1}+O\left( 
\acute{\nu}^{2}\right) \right\} .  \label{nv2a}
\end{equation}%
Observe also that 
\begin{equation}
\det \mathfrak{T}\left( \nu \right) =\left( \mathrm{i}\acute{\nu}\right)
^{n}\det \Lambda _{0}\left( 1+O\left( \acute{\nu}\right) \right) =\left( 
\mathrm{i}\acute{\nu}\right) ^{n}\left( 1+O\left( \acute{\nu}\right) \right)
=\tilde{\nu}\left( 1+O\left( \tilde{\nu}^{1/n}\right) \right) .
\label{nv3aa}
\end{equation}%
>From (\ref{dd6}) and (\ref{nv3aa}) we get%
\begin{equation}
\det \left( T\left( \nu \right) -\zeta _{0}I_{n}\right) =\zeta _{0}^{n}\det 
\mathfrak{T}\left( \nu \right) =\zeta _{0}^{n}\tilde{\nu}\left( 1+O\left( 
\tilde{\nu}^{1/n}\right) \right) .  \label{nv3aab}
\end{equation}%
Comparing (\ref{nv3aab}) with (\ref{tomx12}) and taking into account (\ref%
{tomx10aaa}) we get%
\begin{equation}
\zeta _{0}^{n}\tilde{\nu}=a_{0}\nu ,\ a_{0}=\frac{n!\left( \mathrm{i}\zeta
_{0}\right) ^{n}}{\omega ^{\left( n\right) }\left( k_{0}\right) }.
\label{zaa1}
\end{equation}%
The relation (\ref{zaa1}) combined with (\ref{dd23}) yields the following
representation for the important quantity $\mathfrak{t}_{1}=\left[ \mathfrak{%
T}_{1}\right] _{n1}$%
\begin{equation}
\mathfrak{t}_{1}=\frac{n!\mathrm{i}^{n}}{\omega ^{\left( n\right) }\left(
k_{0}\right) },\ \mathfrak{t}_{1}^{\frac{1}{n}}=\alpha _{0}\mathrm{i},\
\alpha _{0}=\left[ \frac{n!}{\omega ^{\left( n\right) }\left( k_{0}\right) }%
\right] ^{\frac{1}{n}}.  \label{zaa2}
\end{equation}%
\emph{The representation (\ref{zaa2}), in turn, implies the equivalence of
the assumption (\ref{dd20}) (}$t_{1}\neq 0$\emph{) to the fundamental
assumption (\ref{tomx10aa}) on the dispersion relation at the point }$k_{0}$%
\emph{.} Combining (\ref{zaa2}) with (\ref{dd23}) and (\ref{nv1}) we get%
\begin{equation}
\tilde{\nu}^{\frac{1}{n}}=\mathrm{i}\acute{\nu}=\alpha _{0}\mathrm{i}\nu ^{%
\frac{1}{n}}+O\left( \nu ^{\frac{2}{n}}\right) ,\ \alpha _{0}=\left[ \frac{n!%
}{\omega ^{\left( n\right) }\left( k_{0}\right) }\right] ^{\frac{1}{n}}.
\label{zaa3}
\end{equation}%
\emph{\ }

The diagonal matrices $\Lambda_{s},s\geq1$ as well the terms of the Taylor
series for $S\left( \mathrm{i}\acute{\nu}\right) $ can be found recursively
(see the Appendix 2 for the details).

Notice that the eigenvectors of the transfer matrix $\mathcal{T}\left( \nu
\right) $ in view of (\ref{nv2}) and (\ref{sg2})-(\ref{sg4}) take the form 
\begin{equation}
\mathfrak{e}_{j}\left( \nu \right) =\mathcal{G}_{0}\left( \nu \right) \left[ 
\begin{array}{cc}
\mathfrak{S}_{0}\left( \tilde{\nu}\right) e^{-S\left( \mathrm{i}\acute{\nu}%
\right) } & 0 \\ 
0 & I_{4-n}%
\end{array}%
\right] \mathfrak{b}_{j},\ j=0,1,2,3,  \label{zaa4}
\end{equation}%
where, according to (\ref{nv1}),%
\begin{equation}
\mathcal{G}_{0}\left( \nu \right) =\mathcal{G}_{0}\left( 0\right) +O\left(
\nu \right) =\mathcal{G}_{0}\left( 0\right) +O\left( \acute{\nu}^{n}\right) ,%
\text{ where }n\text{ is the degeneracy order.}  \label{zaa5}
\end{equation}

\subsection{Spectrum of the transfer matrix at an inflection point}

In this section we derive the asymptotic formulae for the eigenvalues and
eigenvectors of the transfer matrix $T\left( \nu \right) $ as $\nu =\omega
-\omega _{0}\rightarrow 0$ in the case when the frequency $\omega _{0}$ is
an inflection point, i.e. a degeneracy point of the order 3. We remind the
reader that in this case according to (\ref{dd4}), (\ref{dd5}), (\ref{dd5a})
and (\ref{sg2})-(\ref{sg4}) we have%
\begin{equation}
\mathcal{T}\left( \nu \right) =\mathcal{G}_{0}\left( \nu \right) \left[ 
\begin{array}{cc}
T\left( \nu \right) & 0 \\ 
0 & W\left( \nu \right)%
\end{array}%
\right] \mathcal{G}_{0}^{-1}\left( \nu \right) ,  \label{gtg1}
\end{equation}%
where $\mathcal{G}_{0}\left( \nu \right) $ is a $4\times 4$ invertible
matrix $\mathcal{G}_{0}\left( \nu \right) $ depending analytically on $\nu $
in a vicinity of $\nu =0$, $T\left( \nu \right) $ is a $3\times 3$ matrix
depending analytically on $\nu $ in a vicinity of $\nu =0$, $W\left( \nu
\right) $ is a complex valued function analytic in $\nu $\ in a vicinity of $%
\nu =0$. In addition to that, (see (\ref{tz3b})), we have 
\begin{equation}
T\left( 0\right) =\zeta _{0}\left[ 
\begin{array}{ccc}
1 & 1 & 0 \\ 
0 & 1 & 1 \\ 
0 & 0 & 1%
\end{array}%
\right] ,\ \left\vert W\left( 0\right) \right\vert =1.  \label{gtg2}
\end{equation}%
In other words, the basis $\mathfrak{f}_{j}$, $j=0,1,2,3$ defined by (\ref%
{sg4}) reduces $\mathcal{T}\left( 0\right) $ to its canonical form%
\begin{equation}
\mathcal{T}\left( 0\right) =\mathcal{G}_{0}\left( 0\right) \left[ 
\begin{array}{cccc}
\zeta _{0} & \zeta _{0} & 0 & 0 \\ 
0 & \zeta _{0} & \zeta _{0} & 0 \\ 
0 & 0 & \zeta _{0} & 0 \\ 
0 & 0 & 0 & W\left( 0\right)%
\end{array}%
\right] \mathcal{G}_{0}^{-1}\left( 0\right) .  \label{gtg4}
\end{equation}

\subsubsection{Eigenvalues of the transfer matrix}

Observe that it follows from (\ref{dd15}), (\ref{nv2}) the eigenvalues$\
\eta _{0}\left( \nu \right) $, $\eta _{+}\left( \nu \right) $, $\eta
_{-}\left( \nu \right) $ of the matrix $I_{3}+\mathfrak{T}\left( \nu \right) 
$ from (\ref{dd6}) are 
\begin{align}
\eta _{0}\left( \nu \right) & =1+\tilde{\nu}^{1/3}+O\left( \tilde{\nu}%
^{2/3}\right) ,\ \eta _{+}\left( \nu \right) =1+\varsigma _{+}\tilde{\nu}%
^{1/3}+O\left( \tilde{\nu}^{2/3}\right) ,  \label{etv1} \\
\eta _{-}\left( \nu \right) & =1+\varsigma _{-}\tilde{\nu}^{1/3}+O\left( 
\tilde{\nu}^{2/3}\right) ,\ \tilde{\nu}=\mathfrak{t}_{1}^{1/3}\nu
^{1/3}+O\left( \nu ^{2/3}\right) ,  \notag
\end{align}%
or, in view of (\ref{dd19}),%
\begin{align}
\eta _{0}\left( \nu \right) & =1+\mathfrak{t}_{1}^{1/3}\nu ^{1/3}+O\left(
\nu ^{2/3}\right) ,\ \eta _{+}\left( \nu \right) =1+\mathfrak{t}%
_{1}^{1/3}\varsigma _{+}\nu ^{1/3}+O\left( \nu ^{2/3}\right) ,  \label{etv2}
\\
\eta _{-}\left( \nu \right) & =1+\mathfrak{t}_{1}^{1/3}\varsigma _{-}\nu
^{1/3}+O\left( \nu ^{2/3}\right) ,  \notag
\end{align}%
where according to (\ref{dd12a})%
\begin{equation}
\varsigma _{0}=1,\ \varsigma _{1}=\varsigma _{+}=e^{\mathrm{i}\frac{2\pi }{3}%
}=-\frac{1}{2}+\frac{1}{2}\mathrm{i}\sqrt{3},\ \varsigma _{2}=\varsigma
_{1}^{2}=\varsigma _{-}=-\frac{1}{2}-\frac{1}{2}\mathrm{i}\sqrt{3}.
\label{etv2b}
\end{equation}%
Notice that we can recast (\ref{etv2}) as%
\begin{align}
\eta _{0}\left( \nu \right) & =\exp \left\{ \mathfrak{t}_{1}^{1/3}\nu
^{1/3}+O\left( \nu ^{2/3}\right) \right\} ,\ \eta _{+}\left( \nu \right)
=\exp \left\{ \mathfrak{t}_{1}^{1/3}\varsigma _{+}\nu ^{1/3}+O\left( \nu
^{2/3}\right) \right\} ,  \label{etv2a} \\
\eta _{-}\left( \nu \right) & =\exp \left\{ \mathfrak{t}_{1}^{1/3}\varsigma
_{-}\nu ^{1/3}+O\left( \nu ^{2/3}\right) \right\} .  \notag
\end{align}%
As follows from the statement (\ref{trip3}) at least one of $\eta _{j}\left(
\nu \right) $ must satisfy $\left\vert \eta _{j}\left( \nu \right)
\right\vert =1$. Without lost of generality we can choose that one to be $%
\eta _{0}\left( \nu \right) $, and, hence for sufficiently small $\delta >0$
we have%
\begin{equation}
\left\vert \eta _{0}\left( \nu \right) \right\vert =1\text{ for }\left\vert
\nu \right\vert \leq \delta .  \label{etv3}
\end{equation}%
The representation (\ref{etv2}) together with (\ref{etv3}) (see also (\ref%
{zaa2}) and (\ref{zaa3})) yields%
\begin{equation}
\mathfrak{t}_{1}^{1/3}=\alpha _{0}\mathrm{i}\text{ with a real }\alpha _{0}=%
\left[ \frac{6}{\omega ^{\prime \prime \prime }\left( k_{0}\right) }\right]
^{\frac{1}{3}},  \label{etv4}
\end{equation}%
\begin{align}
\mathfrak{t}_{1}& =\left[ \mathfrak{T}_{1}\right] _{31}=-\alpha _{0}^{3}%
\mathrm{i}\text{ with a real }\alpha _{0}=\left[ \frac{6}{\omega ^{\prime
\prime \prime }\left( k_{0}\right) }\right] ^{\frac{1}{3}}.  \label{etv5} \\
\tilde{\nu}^{1/3}& =\mathrm{i}\alpha _{0}\nu ^{1/3}+O\left( \nu ^{2/3}\right)
\label{etv5a}
\end{align}%
Hence, (\ref{etv2a}) takes the form%
\begin{align}
\eta _{0}\left( \nu \right) & =\exp \left\{ \mathrm{i}\alpha _{0}\nu
^{1/3}+O\left( \nu ^{2/3}\right) \right\} ,\ \eta _{+}\left( \nu \right)
=\exp \left\{ \mathrm{i}\alpha _{0}\varsigma _{+}\nu ^{1/3}+O\left( \nu
^{2/3}\right) \right\} ,  \label{etv6} \\
\eta _{-}\left( \nu \right) & =\exp \left\{ \mathrm{i}\alpha _{0}\varsigma
_{-}\nu ^{1/3}+O\left( \nu ^{2/3}\right) \right\} .  \notag
\end{align}%
Observe that (\ref{etv2b}) and (\ref{etv6}) imply that%
\begin{align}
\text{if }\alpha _{0}\nu & >0\text{ then }\left\vert \eta _{0}\left( \nu
\right) \right\vert =1,\ \left\vert \eta _{+}\left( \nu \right) \right\vert
<1,\ \left\vert \eta _{-}\left( \nu \right) \right\vert >1,  \label{etv8} \\
\text{if }\alpha _{0}\nu & <0\text{ then }\left\vert \eta _{0}\left( \nu
\right) \right\vert =1,\ \left\vert \eta _{-}\left( \nu \right) \right\vert
<1,\ \left\vert \eta _{+}\left( \nu \right) \right\vert >1.  \notag
\end{align}%
Now, as follows from (\ref{dd6}), the eigenvalues $\theta _{j}\left( \nu
\right) $ of the $3\times 3$ transfer matrix $T\left( \nu \right) $ take the
form 
\begin{equation}
\theta _{j}\left( \nu \right) =\zeta _{0}\eta _{j}\left( \nu \right) ,\
j=0,\pm .  \label{etv9}
\end{equation}%
If now we denote%
\begin{equation}
\zeta _{0}=e^{\mathrm{i}k_{0}}\text{ where }k_{0}\text{ is real,}
\label{etv10}
\end{equation}%
then (\ref{etv6})-(\ref{etv10}) imply%
\begin{align}
\theta _{0}\left( \nu \right) & =\exp \left\{ \mathrm{i}k_{0}+\mathrm{i}%
\alpha _{0}\nu ^{1/3}+O\left( \nu ^{2/3}\right) \right\} =e^{\mathrm{i}k_{0}+%
\mathrm{i}\alpha _{0}\nu ^{1/3}}\left( 1+O\left( \nu ^{2/3}\right) \right) ,
\label{etv11} \\
\theta _{+}\left( \nu \right) & =\exp \left\{ \mathrm{i}k_{0}+\mathrm{i}%
\alpha _{0}\varsigma _{+}\nu ^{1/3}+O\left( \nu ^{2/3}\right) \right\} =e^{%
\mathrm{i}k_{0}+\mathrm{i}\alpha _{0}\varsigma _{+}\nu ^{1/3}}\left(
1+O\left( \nu ^{2/3}\right) \right) ,  \notag \\
\theta _{-}\left( \nu \right) & =\exp \left\{ \mathrm{i}k_{0}+\mathrm{i}%
\alpha _{0}\varsigma _{-}\nu ^{1/3}+O\left( \nu ^{2/3}\right) \right\} =e^{%
\mathrm{i}k_{0}+\mathrm{i}\alpha _{0}\varsigma _{-}\nu ^{1/3}}\left(
1+O\left( \nu ^{2/3}\right) \right) .  \notag
\end{align}%
Observe that%
\begin{align}
\text{if }\alpha _{0}\nu & >0\text{ then }\left\vert \theta _{0}\left( \nu
\right) \right\vert =1,\ \left\vert \theta _{+}\left( \nu \right)
\right\vert <1,\ \left\vert \theta _{-}\left( \nu \right) \right\vert >1,
\label{etv12} \\
\text{if }\alpha _{0}\nu & <0\text{ then }\left\vert \theta _{0}\left( \nu
\right) \right\vert =1,\ \left\vert \theta _{-}\left( \nu \right)
\right\vert <1,\ \left\vert \theta _{+}\left( \nu \right) \right\vert >1, 
\notag
\end{align}%
which can be recast as%
\begin{equation}
\left\vert \theta _{0}\left( \nu \right) \right\vert =1,\ \left\vert \theta
_{\limfunc{sign}\left( \alpha _{0}\nu \right) }\left( \nu \right)
\right\vert <1,\ \left\vert \theta _{-\limfunc{sign}\left( \alpha _{0}\nu
\right) }\left( \nu \right) \right\vert >1.  \label{etv12a}
\end{equation}%
Notice that if $\Delta k=k-k_{0}$ and $\Delta \omega =\omega -\omega
_{0}=\nu $ (\ref{etv4}), (\ref{etv5}) and (\ref{etv11}) yield 
\begin{equation}
\mathrm{i}\Delta k=\mathfrak{t}_{1}^{1/3}\nu ^{1/3}+O\left( \nu
^{2/3}\right) \text{ or }\Delta \omega =\nu =\frac{-\mathrm{i}}{\mathfrak{t}%
_{1}}\Delta k^{3}+O\left( \Delta k^{4}\right) ,  \label{etv12b}
\end{equation}%
implying%
\begin{equation}
\omega ^{\prime \prime \prime }\left( k_{0}\right) =\frac{6}{\mathrm{i}%
\mathfrak{t}_{1}}\text{ or }\mathrm{i}\mathfrak{t}_{1}=\alpha _{0}^{3}=\frac{%
6}{\omega ^{\prime \prime \prime }\left( k_{0}\right) }.  \label{etv13}
\end{equation}%
Notice also that (\ref{tomx10aaa}) implies%
\begin{equation}
a_{0}=-\mathrm{i}\alpha _{0}^{3}\zeta _{0}^{3}=-\mathrm{i}\frac{6}{\omega
^{\prime \prime \prime }\left( k_{0}\right) }\zeta _{0}^{3}.  \label{etv13a}
\end{equation}%
It is convenient to introduce%
\begin{equation}
\acute{\nu}=\alpha _{0}\nu ^{1/3}  \label{etv14}
\end{equation}%
and to rewrite (\ref{etv11}) and (\ref{etv12a}) as%
\begin{align}
\theta _{0}\left( \nu \right) & =\exp \left\{ \mathrm{i}k_{0}+\mathrm{i}%
\acute{\nu}+O\left( \acute{\nu}^{2/3}\right) \right\} =e^{\mathrm{i}k_{0}+%
\mathrm{i}\acute{\nu}}\left( 1+O\left( \acute{\nu}^{2/3}\right) \right) ,
\label{etv15} \\
\theta _{+}\left( \nu \right) & =\exp \left\{ \mathrm{i}k_{0}+\mathrm{i}%
\varsigma _{+}\acute{\nu}+O\left( \acute{\nu}^{2/3}\right) \right\} =e^{%
\mathrm{i}k_{0}+\mathrm{i}\varsigma _{+}\acute{\nu}}\left( 1+O\left( \acute{%
\nu}^{2/3}\right) \right) ,  \notag \\
\theta _{-}\left( \nu \right) & =\exp \left\{ \mathrm{i}k_{0}+\mathrm{i}%
\varsigma _{-}\acute{\nu}+O\left( \acute{\nu}^{2/3}\right) \right\} =e^{%
\mathrm{i}k_{0}+\mathrm{i}\varsigma _{-}\acute{\nu}}\left( 1+O\left( \acute{%
\nu}^{2/3}\right) \right) .  \notag
\end{align}%
The relations (\ref{etv15}), in turn, together with (\ref{etv2a}) imply%
\begin{equation}
\left\vert \theta _{0}\left( \nu \right) \right\vert =1,\ \left\vert \theta
_{\limfunc{sign}\left( \acute{\nu}\right) }\left( \nu \right) \right\vert
<1,\ \left\vert \theta _{-\limfunc{sign}\left( \acute{\nu}\right) }\left(
\nu \right) \right\vert >1.  \label{etv16}
\end{equation}

\subsubsection{Eigenvectors of the transfer matrix}

We recall that we work with the basis in which $T\left( 0\right) $\emph{\
has its canonical Jordan form as in (\ref{dd5})}, namely%
\begin{equation*}
T\left( 0\right) =\zeta _{0}\left[ 
\begin{array}{ccc}
1 & 1 & 0 \\ 
0 & 1 & 1 \\ 
0 & 0 & 1%
\end{array}%
\right] .
\end{equation*}%
Recall also that the eigenvectors $e_{j}\left( \nu \right) $, $j=0,1,2$ of
the matrix $T\left( \nu \right) $ are the respective columns of the matrix $%
\mathfrak{S}_{0}\left( \tilde{\nu}\right) e^{-S\left( \mathrm{i}\acute{\nu}%
\right) }$ represented by the asymptotic equality (\ref{nv2a}). To use (\ref%
{nv2a}) we need to find the matrix $S_{1}$. Following the Appendix 2, we
first introduce a decomposition of a square matrix $W$ into its the diagonal
component $\limfunc{diag}\left( W\right) $ and the remaining part $\mathring{%
W}=W-\limfunc{diag}\left( W\right) $ with zero diagonal elements, i.e. 
\begin{align}
W& =\left[ W_{mj}\right] =\limfunc{diag}\left( W\right) +\mathring{W},\text{
where }\limfunc{diag}\left( W\right) =\left[ W_{mj}\delta _{mj}\right]
\label{nv3a} \\
\mathring{W}& =W-\limfunc{diag}\left( W\right) ,  \notag
\end{align}%
where $\delta _{mj}$ is the Kronecker symbol. Then we get the following
expressions for the matrices $\Lambda _{1}$ and $S_{1}$as follows: 
\begin{align}
\Lambda _{1}& =\limfunc{diag}\left( W_{1}\right) ,\ \left[ S_{1}\right]
_{nm}=\frac{1}{w_{m}-w_{n}}\left[ \mathring{W}_{1}\right] _{nm},\ n\neq m;\ %
\left[ S_{1}\right] _{nn}=0,  \label{nv4} \\
\text{where }W_{1}& =\left\langle \widetilde{\mathfrak{T}}_{1}\right\rangle
_{1}\text{ and }w_{1}=\varsigma _{0}=1,\text{ }w_{2}=\varsigma
_{1}=\varsigma _{+},\text{ }w_{3}=\varsigma _{2}=\varsigma _{-}.  \notag
\end{align}%
Carrying out the operations described in (\ref{nv4}), and using (\ref{dd6}),
(\ref{dd8}), (\ref{dd25}), (\ref{dd28}) we obtain%
\begin{align}
\left\langle \widetilde{\mathfrak{T}}_{1}\right\rangle _{1}& =\frac{1}{3%
\mathfrak{t}_{1}}\left[ 
\begin{array}{rrr}
\left[ \mathfrak{T}_{1}\right] _{21}+\left[ \mathfrak{T}_{1}\right] _{32} & 
\left[ \mathfrak{T}_{1}\right] _{21}+\varsigma _{+}\left[ \mathfrak{T}_{1}%
\right] _{32} & \left[ \mathfrak{T}_{1}\right] _{21}+\varsigma _{-}\left[ 
\mathfrak{T}_{1}\right] _{32} \\ 
\varsigma _{-}\left[ \mathfrak{T}_{1}\right] _{21}+\varsigma _{+}\left[ 
\mathfrak{T}_{1}\right] _{32} & \varsigma _{-}\left( \left[ \mathfrak{T}_{1}%
\right] _{21}+\left[ \mathfrak{T}_{1}\right] _{32}\right) & \varsigma _{-} 
\left[ \mathfrak{T}_{1}\right] _{21}+\left[ \mathfrak{T}_{1}\right] _{32} \\ 
\varsigma _{+}\left[ \mathfrak{T}_{1}\right] _{21}+\varsigma _{-}\left[ 
\mathfrak{T}_{1}\right] _{32} & \varsigma _{+}\left[ \mathfrak{T}_{1}\right]
_{21}+\left[ \mathfrak{T}_{1}\right] _{32} & \varsigma _{+}\left( \left[ 
\mathfrak{T}_{1}\right] _{21}+\left[ \mathfrak{T}_{1}\right] _{32}\right)%
\end{array}%
\right]  \label{nv6} \\
& =\frac{1}{3\zeta _{0}\mathfrak{t}_{1}}\left[ 
\begin{array}{rrr}
\left[ T_{1}\right] _{21}+\left[ T_{1}\right] _{32} & \left[ T_{1}\right]
_{21}+\varsigma _{+}\left[ T_{1}\right] _{32} & \left[ T_{1}\right]
_{21}+\varsigma _{-}\left[ T_{1}\right] _{32} \\ 
\varsigma _{-}\left[ T_{1}\right] _{21}+\varsigma _{+}\left[ T_{1}\right]
_{32} & \varsigma _{-}\left( \left[ T_{1}\right] _{21}+\left[ \mathfrak{T}%
_{1}\right] _{32}\right) & \varsigma _{-}\left[ T_{1}\right] _{21}+\left[
T_{1}\right] _{32} \\ 
\varsigma _{+}\left[ T_{1}\right] _{21}+\varsigma _{-}\left[ T_{1}\right]
_{32} & \varsigma _{+}\left[ T_{1}\right] _{21}+\left[ \mathfrak{T}_{1}%
\right] _{32} & \varsigma _{+}\left( \left[ T_{1}\right] _{21}+\left[ T_{1}%
\right] _{32}\right)%
\end{array}%
\right] ,  \notag
\end{align}%
\begin{align}
\Lambda _{1}& =\frac{\left[ T_{1}\right] _{21}+\left[ T_{1}\right] _{32}}{%
3\zeta _{0}\mathfrak{t}_{1}}\left[ 
\begin{array}{ccc}
1 & 0 & 0 \\ 
0 & \varsigma _{-} & 0 \\ 
0 & 0 & \varsigma _{+}%
\end{array}%
\right]  \label{nv7} \\
S_{1}& =\frac{1}{3\mathfrak{t}_{1}}\left[ 
\begin{array}{ccc}
0 & \frac{\left[ \mathfrak{T}_{1}\right] _{21}+\varsigma _{1}\left[ 
\mathfrak{T}_{1}\right] _{32}}{\varsigma _{1}-1} & \frac{\left[ \mathfrak{T}%
_{1}\right] _{21}+\varsigma _{2}\left[ \mathfrak{T}_{1}\right] _{32}}{%
\varsigma _{2}-1} \\ 
\frac{\varsigma _{2}\left[ \mathfrak{T}_{1}\right] _{21}+\varsigma _{1}\left[
\mathfrak{T}_{1}\right] _{32}}{1-\varsigma _{1}} & 0 & \frac{\varsigma _{2}%
\left[ \mathfrak{T}_{1}\right] _{21}+\left[ \mathfrak{T}_{1}\right] _{32}}{%
\varsigma _{2}-\varsigma _{1}} \\ 
\frac{\varsigma _{1}\left[ \mathfrak{T}_{1}\right] _{21}+\varsigma _{2}\left[
\mathfrak{T}_{1}\right] _{32}}{1-\varsigma _{2}} & \frac{\varsigma _{1}\left[
\mathfrak{T}_{1}\right] _{21}+\left[ \mathfrak{T}_{1}\right] _{32}}{%
\varsigma _{1}-\varsigma _{2}} & 0%
\end{array}%
\right]  \notag \\
& =\left[ 
\begin{array}{ccc}
0 & \frac{\tau _{1}+\varsigma _{+}\tau _{2}}{\varsigma _{+}-1} & \frac{\tau
_{1}+\varsigma _{-}\tau _{2}}{\varsigma _{-}-1} \\ 
\frac{\varsigma _{-}\tau _{1}+\varsigma _{+}\tau _{2}}{1-\varsigma _{+}} & 0
& \frac{\varsigma _{-}\tau _{1}+\tau _{2}}{\varsigma _{-}-\varsigma _{+}} \\ 
\frac{\varsigma _{+}\tau _{1}+\varsigma _{-}\tau _{2}}{1-\varsigma _{-}} & 
\frac{\varsigma _{+}\tau _{1}+\tau _{2}}{\varsigma _{+}-\varsigma _{-}} & 0%
\end{array}%
\right] ,\tau _{1}=\frac{\left[ T_{1}\right] _{21}}{3\zeta _{0}\mathfrak{t}%
_{1}},\tau _{2}=\frac{\left[ T_{1}\right] _{32}}{3\zeta _{0}\mathfrak{t}_{1}}%
.  \notag
\end{align}

The eigenvectors $e_{0}\left( \nu \right) $, $e_{1}\left( \nu \right) $ and $%
e_{2}\left( \nu \right) $ of the matrix $\mathfrak{T}\left( \nu \right) $
(and, hence, the matrix $T\left( \nu \right) $), corresponding respectively
to the eigenvalues $\varsigma _{0}=1$, and $\varsigma _{1}=\varsigma _{+}=%
\frac{1}{2}\left( -1+\mathrm{i}\sqrt{3}\right) $ and $\varsigma
_{2}=\varsigma _{-}=-\frac{1}{2}\left( 1+\mathrm{i}\sqrt{3}\right) $ in view
of (\ref{nv2a}) and (\ref{nv7}) take the following form 
\begin{align}
e_{0}\left( \nu \right) & =\left[ 
\begin{array}{c}
1+\mathrm{i}\tau _{2}\acute{\nu}+O\left( \acute{\nu}^{2}\right) \\ 
\mathrm{i}\acute{\nu}+\mathrm{i}\tau _{1}\acute{\nu}^{2}+O\left( \acute{\nu}%
^{3}\right) \\ 
-\acute{\nu}^{2}+\mathrm{i}\left( \tau _{2}-\tau _{1}\right) \acute{\nu}%
^{3}+O\left( \acute{\nu}^{4}\right)%
\end{array}%
\right] ,\ e_{1}\left( \nu \right) =\left[ 
\begin{array}{c}
1-\frac{\mathrm{i}+\sqrt{3}}{2}\tau _{2}\acute{\nu}+O\left( \acute{\nu}%
^{2}\right) \\ 
-\frac{\mathrm{i}+\sqrt{3}}{2}\acute{\nu}-\frac{\mathrm{i}\sqrt{3}+1}{2}\tau
_{1}\acute{\nu}^{2}+O\left( \acute{\nu}^{3}\right) \\ 
\frac{\mathrm{i}\sqrt{3}+1}{2}\acute{\nu}^{2}+\mathrm{i}\left( \tau
_{2}-\tau _{1}\right) \acute{\nu}^{3}+O\left( \acute{\nu}^{4}\right)%
\end{array}%
\right] ,  \label{nv8} \\
e_{2}\left( \nu \right) & =\left[ 
\begin{array}{c}
1-\frac{\mathrm{i}-\sqrt{3}}{2}\tau _{2}\acute{\nu}+O\left( \acute{\nu}%
^{2}\right) \\ 
\frac{\sqrt{3}-\mathrm{i}}{2}\acute{\nu}+\frac{\mathrm{i}\sqrt{3}-1}{2}\tau
_{1}\acute{\nu}^{2}+O\left( \acute{\nu}^{3}\right) \\ 
\frac{1-\mathrm{i}\sqrt{3}}{2}\acute{\nu}^{2}+\mathrm{i}\left( \tau
_{2}-\tau _{1}\right) \acute{\nu}^{3}+O\left( \acute{\nu}^{4}\right)%
\end{array}%
\right] ,\ \tau _{1}=\frac{\left[ T_{1}\right] _{21}}{3\zeta _{0}\mathfrak{t}%
_{1}},\ \tau _{2}=\frac{\left[ T_{1}\right] _{32}}{3\zeta _{0}\mathfrak{t}%
_{1}}.  \notag
\end{align}%
The equality (\ref{nv8}), in turn, implies%
\begin{equation}
\frac{e_{1}\left( \nu \right) -e_{0}\left( \nu \right) }{\left( \frac{%
\mathrm{i}}{2}-\frac{\sqrt{3}}{6}\right) \acute{\nu}}=\left[ 
\begin{array}{c}
\tau _{2}+O\left( \acute{\nu}\right) \\ 
1+\frac{\sqrt{3}-\mathrm{i}}{2}\tau _{1}\acute{\nu}+O\left( \acute{\nu}%
^{2}\right) \\ 
\frac{\mathrm{i}-\sqrt{3}}{2}\acute{\nu}+O\left( \acute{\nu}^{3}\right)%
\end{array}%
\right] ,\ \tau _{1}=\frac{\left[ T_{1}\right] _{21}}{3\zeta _{0}\mathfrak{t}%
_{1}},\ \tau _{2}=\frac{\left[ T_{1}\right] _{32}}{3\zeta _{0}\mathfrak{t}%
_{1}}.  \label{nv9}
\end{equation}%
Consequently%
\begin{equation}
\lim_{\nu \rightarrow 0}e_{0}\left( \nu \right) =\left[ 
\begin{array}{c}
1 \\ 
0 \\ 
0%
\end{array}%
\right] ,\ \lim_{\nu \rightarrow 0}\frac{e_{1}\left( \nu \right)
-e_{0}\left( \nu \right) }{\left( \frac{\mathrm{i}}{2}-\frac{\sqrt{3}}{6}%
\right) \acute{\nu}}=\left[ 
\begin{array}{c}
\tau _{2} \\ 
1 \\ 
0%
\end{array}%
\right] ,\ \tau _{2}=\frac{\left[ T_{1}\right] _{32}}{3\zeta _{0}\mathfrak{t}%
_{1}}.  \label{nv10}
\end{equation}%
Thus we have the following set of eigenvalues and corresponding eigenvectors%
\begin{align}
\theta _{0}\left( \nu \right) & =e^{\mathrm{i}k_{0}+\mathrm{i}\acute{\nu}%
}\left( 1+O\left( \acute{\nu}^{2}\right) \right) ,\ e_{0}\left( \nu \right) =%
\left[ 
\begin{array}{c}
1+\mathrm{i}\tau _{2}\acute{\nu}+O\left( \acute{\nu}^{2}\right) \\ 
\mathrm{i}\acute{\nu}+\mathrm{i}\tau _{1}\acute{\nu}^{2}+O\left( \acute{\nu}%
^{3}\right) \\ 
-\acute{\nu}^{2}+\mathrm{i}\left( \tau _{2}-\tau _{1}\right) \acute{\nu}%
^{3}+O\left( \acute{\nu}^{4}\right)%
\end{array}%
\right] ,  \label{Fj4} \\
\theta _{+}\left( \nu \right) & =e^{\mathrm{i}k_{0}+\mathrm{i}\acute{\nu}%
\varsigma _{+}}\left( 1+O\left( \acute{\nu}^{2}\right) \right) ,\
e_{+}\left( \nu \right) =\left[ 
\begin{array}{c}
1-\frac{\mathrm{i}+\sqrt{3}}{2}\tau _{2}\acute{\nu}+O\left( \acute{\nu}%
^{2}\right) \\ 
-\frac{\mathrm{i}+\sqrt{3}}{2}\acute{\nu}-\frac{\mathrm{i}\sqrt{3}+1}{2}\tau
_{1}\acute{\nu}^{2}+O\left( \acute{\nu}^{3}\right) \\ 
\frac{\mathrm{i}\sqrt{3}+1}{2}\acute{\nu}^{2}+\mathrm{i}\left( \tau
_{2}-\tau _{1}\right) \acute{\nu}^{3}+O\left( \acute{\nu}^{4}\right)%
\end{array}%
\right] ,  \notag \\
\theta _{-}\left( \nu \right) & =e^{\mathrm{i}k_{0}+\mathrm{i}\acute{\nu}%
\varsigma _{-}}\left( 1+O\left( \acute{\nu}^{2}\right) \right) ,\
e_{\_}\left( \nu \right) =\left[ 
\begin{array}{c}
1-\frac{\mathrm{i}-\sqrt{3}}{2}\tau _{2}\acute{\nu}+O\left( \acute{\nu}%
^{2}\right) \\ 
\frac{\sqrt{3}-\mathrm{i}}{2}\acute{\nu}+\frac{\mathrm{i}\sqrt{3}-1}{2}\tau
_{1}\acute{\nu}^{2}+O\left( \acute{\nu}^{3}\right) \\ 
\frac{1-\mathrm{i}\sqrt{3}}{2}\acute{\nu}^{2}+\mathrm{i}\left( \tau
_{2}-\tau _{1}\right) \acute{\nu}^{3}+O\left( \acute{\nu}^{4}\right)%
\end{array}%
\right] ,  \notag \\
\tau _{1}& =\frac{\left[ T_{1}\right] _{21}}{3\zeta _{0}\mathfrak{t}_{1}},\
\tau _{2}=\frac{\left[ T_{1}\right] _{32}}{3\zeta _{0}\mathfrak{t}_{1}};\
e_{+}\left( \nu \right) =e_{1}\left( \nu \right) ,\ e_{-}\left( \nu \right)
=e_{2}\left( \nu \right) .  \notag
\end{align}%
Notice that in view of (\ref{etv16}) we always have%
\begin{equation}
\left\vert \theta _{0}\left( \nu \right) \right\vert =1,\ \left\vert \theta
_{\limfunc{sign}\left( \acute{\nu}\right) }\left( \nu \right) \right\vert <1,
\label{Fj5}
\end{equation}%
implying that the vector $e_{0}\left( \nu \right) $ always corresponds to
the frozen mode and the vector $e_{\limfunc{sign}\left( \acute{\nu}\right)
}\left( \nu \right) $ always corresponds to the evanescent mode, i.e. the
one decaying exponentially away from the surface of the photonic crystal. In
particular, the two-dimensional space $\limfunc{Span}\left\{ e_{0}\left( \nu
\right) ,e_{\limfunc{sign}\left( \acute{\nu}\right) }\left( \nu \right)
\right\} $ describes all possible values of the EM field of the ST
(scattering theory) eigenmodes on the surface of the photonic crystal.

It readily follows from (\ref{nv10}) that%
\begin{equation}
\lim_{\nu \rightarrow 0}\limfunc{Span}\left\{ e_{0}\left( \nu \right)
,e_{_{\pm }}\left( \nu \right) \right\} =\limfunc{Span}\left\{
f_{0},f_{1}\right\} ,\ f_{0}=\left[ 
\begin{array}{c}
1 \\ 
0 \\ 
0%
\end{array}%
\right] ,\ f_{1}=\left[ 
\begin{array}{c}
0 \\ 
1 \\ 
0%
\end{array}%
\right] .  \label{Fj6}
\end{equation}%
Hence, in particular, the two-dimensional space $\limfunc{Span}\left\{
e_{0}\left( \nu \right) ,e_{\limfunc{sign}\left( \acute{\nu}\right) }\left(
\nu \right) \right\} $, which describes all possible values of the EM field
of ST eigenmodes on the surface of the photonic crystal, converges as $\nu
\rightarrow 0$ to the space $\limfunc{Span}\left\{ f_{0},f_{1}\right\} $,
which describes the two-dimensional space of all possible values on EM field
of ST eigenmodes on the surface of the photonic crystal for $\nu =0$, i.e.
at the frequency $\omega _{0}$ of the frozen eigenmode.

Hence, in view of (\ref{zaa4}) and (\ref{zaa5}), we have the following
representation for the eigenvectors $\mathfrak{e}_{j}\left( \nu \right) =%
\mathcal{G}_{0}\left( \nu \right) e_{j}\left( \nu \right) $, $j=0,1,2,4$ of
the transfer matrix $\mathcal{T}\left( \nu \right) $%
\begin{align}
\mathfrak{e}_{0}\left( \nu \right) & =\mathcal{G}_{0}\left( 0\right) \left[ 
\begin{array}{c}
1+\mathrm{i}\tau _{2}\acute{\nu}+O\left( \acute{\nu}^{2}\right) \\ 
\mathrm{i}\acute{\nu}+\mathrm{i}\tau _{1}\acute{\nu}^{2}+O\left( \acute{\nu}%
^{3}\right) \\ 
-\acute{\nu}^{2}+O\left( \acute{\nu}^{3}\right) \\ 
O\left( \acute{\nu}^{3}\right)%
\end{array}%
\right] ,\ \tau _{1}=\frac{\left[ T_{1}\right] _{21}}{3\zeta _{0}\mathfrak{t}%
_{1}},\ \tau _{2}=\frac{\left[ T_{1}\right] _{32}}{3\zeta _{0}\mathfrak{t}%
_{1}}.  \label{e3nu1} \\
\mathfrak{e}_{1}\left( \nu \right) & =\mathfrak{e}_{+}\left( \nu \right) =%
\mathcal{G}_{0}\left( 0\right) \left[ 
\begin{array}{c}
1-\frac{\mathrm{i}+\sqrt{3}}{2}\tau _{2}\acute{\nu}+O\left( \acute{\nu}%
^{2}\right) \\ 
-\frac{\mathrm{i}+\sqrt{3}}{2}\acute{\nu}-\frac{\mathrm{i}\sqrt{3}+1}{2}\tau
_{1}\acute{\nu}^{2}+O\left( \acute{\nu}^{3}\right) \\ 
\frac{\mathrm{i}\sqrt{3}+1}{2}\acute{\nu}^{2}+O\left( \acute{\nu}^{3}\right)
\\ 
O\left( \acute{\nu}^{3}\right)%
\end{array}%
\right] ,  \notag
\end{align}%
\begin{equation}
\mathfrak{e}_{2}\left( \nu \right) =\mathcal{G}_{0}\left( 0\right) \left[ 
\begin{array}{c}
1-\frac{\mathrm{i}-\sqrt{3}}{2}\tau _{2}\acute{\nu}+O\left( \acute{\nu}%
^{2}\right) \\ 
\frac{\sqrt{3}-\mathrm{i}}{2}\acute{\nu}+\frac{\mathrm{i}\sqrt{3}-1}{2}\tau
_{1}\acute{\nu}^{2}+O\left( \acute{\nu}^{3}\right) \\ 
\frac{1-\mathrm{i}\sqrt{3}}{2}\acute{\nu}^{2}+O\left( \acute{\nu}^{3}\right)
\\ 
O\left( \acute{\nu}^{3}\right)%
\end{array}%
\right] ,\ \mathfrak{e}_{3}\left( \nu \right) =\mathcal{G}_{0}\left(
0\right) \left[ 
\begin{array}{c}
0 \\ 
0 \\ 
0 \\ 
1+O\left( \acute{\nu}^{3}\right)%
\end{array}%
\right] .  \label{e3nu2}
\end{equation}%
Combining now (\ref{e3nu1}), (\ref{e3nu2}) with (\ref{zaa5}) and (\ref{sg4})
we get the following representations for the eigenvectors $\mathfrak{e}%
_{j}\left( \nu \right) $%
\begin{align}
\mathfrak{e}_{0}\left( \nu \right) & =\left( 1+\mathrm{i}\tau _{2}\acute{\nu}%
+O\left( \acute{\nu}^{2}\right) \right) \mathfrak{f}_{0}+\left( \mathrm{i}%
\acute{\nu}+\mathrm{i}\tau _{1}\acute{\nu}^{2}\right) \mathfrak{f}_{1}-%
\acute{\nu}^{2}\mathfrak{f}_{2}+O\left( \acute{\nu}^{3}\right) ,
\label{e3nu3} \\
\mathfrak{e}_{+}\left( \nu \right) & =\mathfrak{e}_{1}\left( \nu \right)
=\left( 1-\frac{\mathrm{i}+\sqrt{3}}{2}\tau _{2}\acute{\nu}+O\left( \acute{%
\nu}^{2}\right) \right) \mathfrak{f}_{0}+  \notag \\
& \left( -\frac{\mathrm{i}+\sqrt{3}}{2}\acute{\nu}-\frac{\mathrm{i}\sqrt{3}+1%
}{2}\tau _{1}\acute{\nu}^{2}\right) \mathfrak{f}_{1}+\frac{\mathrm{i}\sqrt{3}%
+1}{2}\acute{\nu}^{2}\mathfrak{f}_{2}+O\left( \acute{\nu}^{3}\right) , 
\notag \\
\mathfrak{e}_{-}\left( \nu \right) & =\mathfrak{e}_{2}\left( \nu \right)
=\left( 1+\frac{\sqrt{3}-\mathrm{i}}{2}\tau _{2}\acute{\nu}+O\left( \acute{%
\nu}^{2}\right) \right) \mathfrak{f}_{0}+  \notag \\
& \left( \frac{\sqrt{3}-\mathrm{i}}{2}\acute{\nu}+\frac{\mathrm{i}\sqrt{3}-1%
}{2}\tau _{1}\acute{\nu}^{2}\right) \mathfrak{f}_{1}+\frac{1-\mathrm{i}\sqrt{%
3}}{2}\acute{\nu}^{2}\mathfrak{f}_{2}+O\left( \acute{\nu}^{3}\right) , 
\notag
\end{align}%
or, using the symbols $\varsigma _{\pm }=-\frac{1}{2}\pm \frac{\sqrt{3%
\mathrm{i}}}{2}$ from (\ref{dd12a}) we have%
\begin{align}
\mathfrak{e}_{0}\left( \nu \right) & =\left( 1+\mathrm{i}\tau _{2}\acute{\nu}%
+O\left( \acute{\nu}^{2}\right) \right) \mathfrak{f}_{0}+\left( \mathrm{i}%
\acute{\nu}+\mathrm{i}\tau _{1}\acute{\nu}^{2}\right) \mathfrak{f}_{1}-%
\acute{\nu}^{2}\mathfrak{f}_{2}+O\left( \acute{\nu}^{3}\right) ,
\label{e3nu4} \\
\mathfrak{e}_{+}\left( \nu \right) & =\mathfrak{e}_{1}\left( \nu \right)
=\left( 1+\mathrm{i}\varsigma _{+}\tau _{2}\acute{\nu}+O\left( \acute{\nu}%
^{2}\right) \right) \mathfrak{f}_{0}+\left( \mathrm{i}\varsigma _{+}\acute{%
\nu}+\varsigma _{\_}\tau _{1}\acute{\nu}^{2}\right) \mathfrak{f}%
_{1}-\varsigma _{-}\acute{\nu}^{2}\mathfrak{f}_{2}+O\left( \acute{\nu}%
^{3}\right) ,  \notag \\
\mathfrak{e}_{-}\left( \nu \right) & =\mathfrak{e}_{2}\left( \nu \right)
=\left( 1+\mathrm{i}\varsigma _{-}\tau _{2}\acute{\nu}+O\left( \acute{\nu}%
^{2}\right) \right) \mathfrak{f}_{0}+\left( \mathrm{i}\varsigma _{-}\acute{%
\nu}+\varsigma _{+}\tau _{1}\acute{\nu}^{2}\right) \mathfrak{f}%
_{1}-\varsigma _{+}\acute{\nu}^{2}\mathfrak{f}_{2}+O\left( \acute{\nu}%
^{3}\right) .  \notag
\end{align}%
Notice that the equalities (\ref{e3nu4}) imply%
\begin{equation}
\lim_{\nu \rightarrow 0}\mathfrak{e}_{0}\left( \nu \right) =\lim_{\nu
\rightarrow 0}\mathfrak{e}_{\pm }\left( \nu \right) =\mathfrak{f}_{0},
\label{e3nu5}
\end{equation}%
indicating, in particular, that the three vectors $\mathfrak{e}_{0}\left(
\nu \right) $ and $\mathfrak{e}_{\pm }\left( \nu \right) $ become almost
parallel as $\nu \rightarrow 0$. To have a nicer way to trace the
two-dimensional spaces $\limfunc{Span}\left\{ \mathfrak{e}_{0}\left( \nu
\right) ,\mathfrak{e}_{\pm }\left( \nu \right) \right\} $ we introduce the
following two vectors 
\begin{align}
\mathfrak{h}_{+}\left( \nu \right) & =\frac{\mathfrak{e}_{+}\left( \nu
\right) -\mathfrak{e}_{0}\left( \nu \right) }{\mathrm{i}\acute{\nu}\left(
\varsigma _{+}-1\right) }=\left[ \tau _{2}+O\left( \acute{\nu}\right) \right]
\mathfrak{f}_{0}+\left[ 1+\frac{\varsigma _{\_}-\mathrm{i}}{\mathrm{i}\left(
\varsigma _{+}-1\right) }\acute{\nu}\right] \mathfrak{f}_{1}+\frac{%
1-\varsigma _{\_}}{\mathrm{i}\left( \varsigma _{+}-1\right) }\acute{\nu}%
\mathfrak{f}_{2}+O\left( \acute{\nu}^{2}\right) ;  \label{e3nu7} \\
\mathfrak{h}_{-}\left( \nu \right) & =\frac{\mathfrak{e}_{-}\left( \nu
\right) -\mathfrak{e}_{0}\left( \nu \right) }{\mathrm{i}\acute{\nu}\left(
\varsigma _{-}-1\right) }=\left[ \tau _{2}+O\left( \acute{\nu}\right) \right]
\mathfrak{f}_{0}+\left[ 1+\frac{\varsigma _{+}-\mathrm{i}}{\mathrm{i}\left(
\varsigma _{-}-1\right) }\acute{\nu}\right] \mathfrak{f}_{1}+\frac{%
1-\varsigma _{+}}{\mathrm{i}\left( \varsigma _{-}-1\right) }\acute{\nu}%
\mathfrak{f}_{2}+O\left( \acute{\nu}^{2}\right) .  \notag
\end{align}%
Notice that the equalities (\ref{e3nu1}), (\ref{e3nu7}) imply%
\begin{equation}
\lim_{\nu \rightarrow 0}\mathfrak{h}_{\pm }\left( \nu \right) =\tau _{2}%
\mathfrak{f}_{0}+\mathfrak{f}_{1},\ \tau _{2}=\frac{\left[ T_{1}\right] _{32}%
}{3\zeta _{0}\mathfrak{t}_{1}}.  \label{e3nu8}
\end{equation}%
Then the relations (\ref{e3nu5}), (\ref{e3nu7}) and (\ref{e3nu8}) yield 
\begin{align}
\limfunc{Span}\left\{ \mathfrak{e}_{0}\left( \nu \right) ,\mathfrak{e}_{\pm
}\left( \nu \right) \right\} & =\limfunc{Span}\left\{ \mathfrak{e}_{0}\left(
\nu \right) ,\mathfrak{h}_{\pm }\left( \nu \right) \right\} ,  \label{e3nu9}
\\
\lim_{\nu \rightarrow 0}\limfunc{Span}\left\{ \mathfrak{e}_{0}\left( \nu
\right) ,\mathfrak{e}_{\pm }\left( \nu \right) \right\} & =\limfunc{Span}%
\left\{ \mathfrak{f}_{0},\mathfrak{f}_{1}\right\} .  \notag
\end{align}

\subsection{Spectrum of the transfer matrix at a degeneracy point of the
order 4}

In this section we derive the asymptotic formulae for the eigenvalues and
eigenvectors of the transfer matrix $T\left( \nu \right) $ as $\nu =\omega
-\omega _{0}\rightarrow 0$ in the case when the frequency $\omega _{0}$ is a
degenerate point of order 4. In this case since $n=4$ the transfer matrix $%
\mathcal{T}\left( \nu \right) $ defined by (\ref{tt1}) is such that $%
\mathcal{T}\left( 0\right) $ is a Jordan block of order 4, as it follows
from the analysis carried out in the previous Section. The mentioned
analysis implies also that the matrix $T\left( \nu \right) $ defined by (\ref%
{tomx5}) is a $4\times 4$ matrix and that $\mathcal{T}\left( \nu \right)
=T\left( \nu \right) $.

Without loss of generality we assume that $\omega _{0}=\omega \left(
k_{0}\right) $ is a point of local minimum of the dispersion relation $%
\omega \left( k\right) $ in a vicinity of $k_{0}$. In the later case for $%
\nu =\omega -\omega _{0}>0$ and $\nu $ small there must be two propagating
Bloch modes and two evanescent modes. Consequently, there will be two
eigenvalues of the matrix $T\left( \nu \right) $ with absolute value 1, one
eigenvalue with absolute values lesser than 1 and one eigenvalue with
absolute value larger than 1.

\subsubsection{Eigenvalues of the transfer matrix}

Observe that it follows from (\ref{dd15}), (\ref{nv2}) that the eigenvalues$%
\ \eta _{j}\left( \nu \right) $, $j=0,1,2,3$ of the matrix $I_{4}+\mathfrak{T%
}\left( \nu \right) $ from (\ref{dd6}) are 
\begin{equation}
\eta _{j}\left( \nu \right) =1+\varsigma _{j}\tilde{\nu}^{1/4}+O\left( 
\tilde{\nu}^{1/2}\right) ,\ j=0,1,2,3,  \label{etv41}
\end{equation}%
or, in view of (\ref{dd19}),%
\begin{equation}
\eta _{j}\left( \nu \right) =1+\mathfrak{t}_{1}^{1/4}\varsigma _{j}\nu
^{1/4}+O\left( \tilde{\nu}^{1/2}\right) ,\ j=0,1,2,3,  \label{etv42}
\end{equation}%
where%
\begin{equation}
\varsigma _{j}=e^{\mathrm{i}\frac{\pi }{2}j},\ j=0,1,2,3.  \label{etv42b}
\end{equation}%
Notice, we can recast (\ref{etv42}) as%
\begin{equation}
\eta _{j}\left( \nu \right) =\exp \left\{ \mathfrak{t}_{1}^{1/4}\varsigma
_{j}\nu ^{1/4}+O\left( \tilde{\nu}^{1/2}\right) \right\} ,\ j=0,1,2,3.
\label{etv42a}
\end{equation}%
As we have found at the beginning of the section, $\left\vert \eta
_{j}\left( \nu \right) \right\vert =1$ for exactly two values of $j=0,1,2,3$%
. Using (\ref{zaa2}) and (\ref{zaa3}) we get%
\begin{equation}
\mathfrak{t}_{1}^{1/4}=\alpha _{0}\mathrm{i}\text{ with a real }\alpha _{0}=%
\left[ \frac{24}{\omega ^{\left( 4\right) }\left( k_{0}\right) }\right] ^{%
\frac{1}{4}}>0,  \label{etv44}
\end{equation}%
and%
\begin{align}
\mathfrak{t}_{1}& =\left[ \mathfrak{T}_{1}\right] _{41}=\alpha _{0}^{4}\text{
with a real }\alpha _{0}=\left[ \frac{24}{\omega ^{\left( 4\right) }\left(
k_{0}\right) }\right] ^{\frac{1}{4}}>0.  \label{etv45} \\
\tilde{\nu}^{1/4}& =\mathrm{i}\alpha _{0}\nu ^{1/4}+O\left( \nu
^{1/2}\right) .  \label{etv45a}
\end{align}%
Hence, (\ref{etv42a}) takes the form%
\begin{equation}
\eta _{j}\left( \nu \right) =\exp \left\{ \mathrm{i}\alpha _{0}\varsigma
_{j}\nu ^{1/4}+O\left( \tilde{\nu}^{1/2}\right) \right\} ,\ j=0,1,2,3.
\label{etv46}
\end{equation}%
Observe that for $\alpha _{0}>0$, (\ref{etv46}) implies that 
\begin{equation}
\left\vert \eta _{0}\left( \nu \right) \right\vert =\left\vert \eta
_{2}\left( \nu \right) \right\vert =1,\ \left\vert \eta _{1}\left( \nu
\right) \right\vert <1,\ \left\vert \eta _{3}\left( \nu \right) \right\vert
>1.  \label{etv48}
\end{equation}%
Now as it follows from (\ref{dd6}) the eigenvalues $\theta _{j}\left( \nu
\right) $ of the $4\times 4$ transfer matrix $T\left( \nu \right) $ take the
form 
\begin{equation}
\theta _{j}\left( \nu \right) =\zeta _{0}\eta _{j}\left( \nu \right) ,\
j=0,1,2,3.  \label{etv49}
\end{equation}%
If now we denote%
\begin{equation}
\zeta _{0}=e^{\mathrm{i}k_{0}}\text{ where }k_{0}\text{ is real,}
\label{etv410}
\end{equation}%
then (\ref{etv46})-(\ref{etv410}) imply%
\begin{align}
\theta _{j}\left( \nu \right) & =\exp \left\{ \mathrm{i}k_{0}+\mathrm{i}%
\alpha _{0}\varsigma _{j}\nu ^{1/4}+O\left( \nu ^{1/2}\right) \right\}
\label{etv411} \\
& =e^{\mathrm{i}k_{0}+\mathrm{i}\alpha _{0}\nu ^{1/4}}\left( 1+O\left( \nu
^{1/2}\right) \right) ,\ j=0,1,2,3.  \notag
\end{align}%
Observe that%
\begin{equation}
\left\vert \theta _{0}\left( \nu \right) \right\vert =\left\vert \theta
_{2}\left( \nu \right) \right\vert =1,\ \left\vert \theta _{1}\left( \nu
\right) \right\vert <1,\ \left\vert \theta _{3}\left( \nu \right)
\right\vert >1.  \label{etv412}
\end{equation}%
Notice also that (\ref{tomx10aaa}) implies%
\begin{equation}
a_{0}=\alpha _{0}^{4}\zeta _{0}^{4}=\frac{24}{\omega ^{\left( 4\right)
}\left( k_{0}\right) }\zeta _{0}^{4}.  \label{etv413a}
\end{equation}%
It is convenient to introduce%
\begin{equation}
\acute{\nu}=\alpha _{0}\nu ^{1/4},  \label{etv414}
\end{equation}%
and to rewrite (\ref{etv411}) as%
\begin{equation}
\theta _{j}\left( \nu \right) =\exp \left\{ \mathrm{i}k_{0}+\mathrm{i}%
\varsigma _{j}\acute{\nu}+O\left( \acute{\nu}^{2}\right) \right\} =e^{%
\mathrm{i}k_{0}+\mathrm{i}\varsigma _{j}\acute{\nu}}\left( 1+O\left( \acute{%
\nu}^{2}\right) \right) ,\ j=0,1,2,3.  \label{etv415}
\end{equation}

\subsubsection{Eigenvectors of the transfer matrix}

We begin this section with the reminder that we work with the basis in which 
$T\left( 0\right) $\emph{\ has its canonical Jordan form as in (\ref{dd5})},
namely%
\begin{equation*}
T\left( 0\right) =\zeta _{0}\left[ 
\begin{array}{cccc}
1 & 1 & 0 & 0 \\ 
0 & 1 & 1 & 0 \\ 
0 & 0 & 1 & 1 \\ 
0 & 0 & 0 & 1%
\end{array}%
\right] .
\end{equation*}%
Based on (\ref{dd6}), (\ref{nv2}), (\ref{nv3}), (\ref{zaa3}) and (\ref{zaa4}%
) we can find the eigenvectors $e_{j}\left( \nu \right) $ of $T\left( \nu
\right) $ corresponding to its eigenvalues $\theta _{j}\left( \nu \right) $.

Recall also that the eigenvectors $e_{j}\left( \nu \right) $, $j=0,1,2,4$ of
the matrix $T\left( \nu \right) $ are the respective columns of the matrix $%
\mathfrak{S}_{0}\left( \tilde{\nu}\right) e^{-S\left( \mathrm{i}\acute{\nu}%
\right) }$ represented by the asymptotic equality (\ref{nv2a}). To use (\ref%
{nv2a}) we need to find the matrix $S_{1}$. Following the Appendix 2, first
we introduce a decomposition of a square matrix $W$ into its diagonal part $%
\limfunc{diag}\left( W\right) $ and the remaining part $\mathring{W}=W-%
\limfunc{diag}\left( W\right) $ with zero diagonal elements (as in (\ref%
{nv3a})) 
\begin{equation}
W=\left[ W_{mj}\right] =\limfunc{diag}\left( W\right) +\mathring{W},\text{
where }\limfunc{diag}\left( W\right) =\left[ W_{mj}\delta _{mj}\right] ,\ 
\mathring{W}=W-\limfunc{diag}\left( W\right) ,  \notag
\end{equation}%
where $\delta _{mj}$ is the Kronecker symbol. Then we get the following
expressions for the matrices $\Lambda _{1}$ and $S_{1}$ 
\begin{align}
\Lambda _{1}& =\limfunc{diag}\left( W_{1}\right) ,\ \left[ S_{1}\right]
_{nm}=\frac{1}{\varsigma _{m-1}-\varsigma _{n-1}}\left[ \mathring{W}_{1}%
\right] _{nm},\ n\neq m;\ \left[ S_{1}\right] _{nn}=0,  \label{s1w1} \\
\text{where }W_{1}& =\left\langle \widetilde{\mathfrak{T}}_{1}\right\rangle
_{1},\varsigma _{0}=1,\ \varsigma _{1}=\mathrm{i},\ \varsigma _{2}=-1,\
\varsigma _{3}=-\mathrm{i}.  \notag
\end{align}%
Carrying out the operations described in (\ref{s1w1}), and using (\ref{dd6}%
), (\ref{dd8}), (\ref{dd25}), (\ref{dd28}) we obtain%
\begin{align}
\left\langle \widetilde{\mathfrak{T}}_{1}\right\rangle _{1}& =\left[ 
\begin{array}{cccc}
\tau _{1}+\tau _{2} & \tau _{1}+\tau _{2}\varsigma _{1} & \tau _{1}+\tau
_{2}\varsigma _{2} & \tau _{1}+\tau _{2}\varsigma _{3} \\ 
\varsigma _{1}\left( \tau _{1}\varsigma _{1}+\tau _{2}\right) & \varsigma
_{1}^{2}\left( \tau _{1}+\tau _{2}\right) & \varsigma _{1}\left( \tau
_{1}\varsigma _{1}+\tau _{2}\varsigma _{2}\right) & \varsigma _{1}\left(
\tau _{1}\varsigma _{1}+\tau _{2}\varsigma _{3}\right) \\ 
\varsigma _{2}\left( \tau _{1}\varsigma _{2}+\tau _{2}\right) & \varsigma
_{2}\left( \tau _{1}\varsigma _{2}+\tau _{2}\varsigma _{1}\right) & 
\varsigma _{2}^{2}\left( \tau _{1}+\tau _{2}\right) & \varsigma _{2}\left(
\tau _{1}\varsigma _{2}+\tau _{2}\varsigma _{3}\right) \\ 
\varsigma _{3}\left( \tau _{1}\varsigma _{3}+\tau _{2}\right) & \varsigma
_{3}\left( \tau _{1}\varsigma _{3}+\tau _{2}\varsigma _{1}\right) & 
\varsigma _{3}\left( \tau _{1}\varsigma _{3}+\tau _{2}\varsigma _{2}\right)
& \varsigma _{3}^{2}\left( \tau _{1}+\tau _{2}\right)%
\end{array}%
\right] ,  \label{s1w2} \\
\tau _{1}& =\frac{\left[ T_{1}\right] _{31}}{4\zeta _{0}\mathfrak{t}_{1}}%
,\tau _{2}=\frac{\left[ T_{1}\right] _{42}}{4\zeta _{0}\mathfrak{t}_{1}}, 
\notag
\end{align}%
\begin{equation}
S_{1}=\left[ 
\begin{array}{cccc}
0 & \frac{\tau _{1}+\tau _{2}\varsigma _{1}}{\varsigma _{1}-1} & \frac{\tau
_{1}+\tau _{2}\varsigma _{2}}{\varsigma _{2}-1} & \frac{\tau _{1}+\tau
_{2}\varsigma _{3}}{\varsigma _{3}-1} \\ 
\frac{\varsigma _{1}\left( \tau _{1}\varsigma _{1}+\tau _{2}\right) }{%
1-\varsigma _{1}} & 0 & \frac{\varsigma _{1}\left( \tau _{1}\varsigma
_{1}+\tau _{2}\varsigma _{2}\right) }{\varsigma _{2}-\varsigma _{1}} & \frac{%
\varsigma _{1}\left( \tau _{1}\varsigma _{1}+\tau _{2}\varsigma _{3}\right) 
}{\varsigma _{3}-1} \\ 
\frac{\varsigma _{2}\left( \tau _{1}\varsigma _{2}+\tau _{2}\right) }{%
1-\varsigma _{2}} & \frac{\varsigma _{2}\left( \tau _{1}\varsigma _{2}+\tau
_{2}\varsigma _{1}\right) }{\varsigma _{1}-\varsigma _{2}} & 0 & \frac{%
\varsigma _{2}\left( \tau _{1}\varsigma _{2}+\tau _{2}\varsigma _{3}\right) 
}{\varsigma _{3}-1} \\ 
\frac{\varsigma _{3}\left( \tau _{1}\varsigma _{3}+\tau _{2}\right) }{%
1-\varsigma _{3}} & \frac{\varsigma _{3}\left( \tau _{1}\varsigma _{3}+\tau
_{2}\varsigma _{1}\right) }{\varsigma _{1}-\varsigma _{3}} & \frac{\varsigma
_{3}\left( \tau _{1}\varsigma _{3}+\tau _{2}\varsigma _{2}\right) }{%
\varsigma _{2}-\varsigma _{3}} & 0%
\end{array}%
\right] .  \label{s1w3}
\end{equation}

The eigenvectors $e_{j}\left( \nu \right) $, $j=0,1,2,3$ of the matrix $%
\mathfrak{T}\left( \nu \right) $ (and, hence, the matrix $T\left( \nu
\right) $), corresponding to respectively the eigenvalues $\varsigma
_{j}=\exp \left( \frac{2\pi \mathrm{i}j}{4}\right) $, $j=0,1,2,3$ in view of
(\ref{nv2a}) and (\ref{nv7}) take the following form 
\begin{align}
e_{0}\left( \nu \right) & =\left[ 
\begin{array}{c}
1+\mathrm{i}\left( \tau _{1}+3\tau _{2}\right) \acute{\nu}+O\left( \acute{\nu%
}^{2}\right) \\ 
\mathrm{i}\acute{\nu}+\left( \tau _{1}-\tau _{2}\right) \acute{\nu}%
^{2}+O\left( \acute{\nu}^{3}\right) \\ 
-\acute{\nu}^{2}+\mathrm{i}\left( 3\tau _{1}+\tau _{2}\right) \acute{\nu}%
^{3}+O\left( \acute{\nu}^{4}\right) \\ 
-\mathrm{i}\acute{\nu}^{3}+3\left( \tau _{1}-\tau _{2}\right) \acute{\nu}%
^{4}+O\left( \acute{\nu}^{5}\right)%
\end{array}%
\right] ,\ e_{1}\left( \nu \right) =\left[ 
\begin{array}{c}
1-\left( \tau _{1}+3\tau _{2}\right) \acute{\nu}+O\left( \acute{\nu}%
^{2}\right) \\ 
-\acute{\nu}+\left( \tau _{2}-\tau _{1}\right) \acute{\nu}^{2}+O\left( 
\acute{\nu}^{3}\right) \\ 
\acute{\nu}^{2}+\left( 3\tau _{1}+\tau _{2}\right) \acute{\nu}^{3}+O\left( 
\acute{\nu}^{4}\right) \\ 
-\acute{\nu}^{3}+3\left( \tau _{1}-\tau _{2}\right) \acute{\nu}^{4}+O\left( 
\acute{\nu}^{5}\right)%
\end{array}%
\right] ,  \label{s1w3a} \\
e_{2}\left( \nu \right) & =\left[ 
\begin{array}{c}
1-\mathrm{i}\left( \tau _{1}+3\tau _{2}\right) \acute{\nu}+O\left( \acute{\nu%
}^{2}\right) \\ 
-\mathrm{i}\acute{\nu}+\left( \tau _{1}-\tau _{2}\right) \acute{\nu}%
^{2}+O\left( \acute{\nu}^{3}\right) \\ 
-\acute{\nu}^{2}-\mathrm{i}\left( 3\tau _{1}+\tau _{2}\right) \acute{\nu}%
^{3}+O\left( \acute{\nu}^{4}\right) \\ 
\mathrm{i}\acute{\nu}^{3}+3\left( \tau _{1}-\tau _{2}\right) \acute{\nu}%
^{4}+O\left( \acute{\nu}^{5}\right)%
\end{array}%
\right] ,\ e_{3}\left( \nu \right) =\left[ 
\begin{array}{c}
1+\left( \tau _{1}+3\tau _{2}\right) \acute{\nu}+O\left( \acute{\nu}%
^{2}\right) \\ 
\acute{\nu}+\left( \tau _{2}-\tau _{1}\right) \acute{\nu}^{2}+O\left( \acute{%
\nu}^{3}\right) \\ 
\acute{\nu}^{2}-\left( 3\tau _{1}+\tau _{2}\right) \acute{\nu}^{3}+O\left( 
\acute{\nu}^{4}\right) \\ 
\acute{\nu}^{3}+3\left( \tau _{1}-\tau _{2}\right) \acute{\nu}^{4}+O\left( 
\acute{\nu}^{5}\right)%
\end{array}%
\right] ,  \notag \\
\tau _{1}& =\frac{\left[ T_{1}\right] _{31}}{8\zeta _{0}\mathfrak{t}_{1}},\
\tau _{2}=\frac{\left[ T_{1}\right] _{42}}{8\zeta _{0}\mathfrak{t}_{1}}. 
\notag
\end{align}%
The equality (\ref{s1w3a}), in turn, implies%
\begin{equation}
\frac{e_{0}\left( \nu \right) -e_{1}\left( \nu \right) }{\left( 1+\mathrm{i}%
\right) \acute{\nu}}=\left[ 
\begin{array}{c}
\tau _{1}+3\tau _{2}+O\left( \acute{\nu}\right) \\ 
1+\left( 1-\mathrm{i}\right) \left( \tau _{1}-\tau _{2}\right) \acute{\nu}%
+O\left( \acute{\nu}^{2}\right) \\ 
\left( \mathrm{i}-1\right) \acute{\nu}+\left( 3\tau _{1}+\tau _{2}\right) 
\acute{\nu}^{2}+O\left( \acute{\nu}^{3}\right) \\ 
-\mathrm{i}\acute{\nu}^{2}+O\left( \acute{\nu}^{3}\right)%
\end{array}%
\right] .  \label{s1w3b}
\end{equation}%
Consequently, 
\begin{align}
\lim_{\nu \rightarrow 0}e_{0}\left( \nu \right) & =\left[ 
\begin{array}{c}
1 \\ 
0 \\ 
0 \\ 
0%
\end{array}%
\right] ,\ \lim_{\nu \rightarrow 0}\frac{e_{j}\left( \nu \right)
-e_{0}\left( \nu \right) }{\mathrm{i}\left( \varsigma _{j}-1\right) \acute{%
\nu}}=\left[ 
\begin{array}{c}
\tau _{1}+3\tau _{2} \\ 
1 \\ 
0 \\ 
0%
\end{array}%
\right] ,\ j=1,2,3,  \label{s1w3c} \\
\tau _{1}& =\frac{\left[ T_{1}\right] _{31}}{8\zeta _{0}\mathfrak{t}_{1}},\
\tau _{2}=\frac{\left[ T_{1}\right] _{42}}{8\zeta _{0}\mathfrak{t}_{1}}. 
\notag
\end{align}

In view of the above and (\ref{etv415}), we have the following complete set
of eigenvalues and corresponding eigenvectors of $T\left( \nu \right) $%
\begin{align}
\theta _{0}\left( \nu \right) & =e^{\mathrm{i}k_{0}+\mathrm{i}\acute{\nu}%
}\left( 1+O\left( \acute{\nu}^{2}\right) \right) ,\ e_{0}\left( \nu \right) =%
\left[ 
\begin{array}{c}
1+\mathrm{i}\left( \tau _{1}+3\tau _{2}\right) \acute{\nu}+O\left( \acute{\nu%
}^{2}\right) \\ 
\mathrm{i}\acute{\nu}+\left( \tau _{1}-\tau _{2}\right) \acute{\nu}%
^{2}+O\left( \acute{\nu}^{3}\right) \\ 
-\acute{\nu}^{2}+\mathrm{i}\left( 3\tau _{1}+\tau _{2}\right) \acute{\nu}%
^{3}+O\left( \acute{\nu}^{4}\right) \\ 
-\mathrm{i}\acute{\nu}^{3}+3\left( \tau _{1}-\tau _{2}\right) \acute{\nu}%
^{4}+O\left( \acute{\nu}^{5}\right)%
\end{array}%
\right] ,  \label{4Fj4} \\
\theta _{1}\left( \nu \right) & =e^{\mathrm{i}k_{0}+\mathrm{i}\acute{\nu}%
\varsigma _{1}}\left( 1+O\left( \acute{\nu}^{2}\right) \right) ,\
e_{1}\left( \nu \right) =\left[ 
\begin{array}{c}
1-\left( \tau _{1}+3\tau _{2}\right) \acute{\nu}+O\left( \acute{\nu}%
^{2}\right) \\ 
-\acute{\nu}+\left( \tau _{2}-\tau _{1}\right) \acute{\nu}^{2}+O\left( 
\acute{\nu}^{3}\right) \\ 
\acute{\nu}^{2}+\left( 3\tau _{1}+\tau _{2}\right) \acute{\nu}^{3}+O\left( 
\acute{\nu}^{4}\right) \\ 
-\acute{\nu}^{3}+3\left( \tau _{1}-\tau _{2}\right) \acute{\nu}^{4}+O\left( 
\acute{\nu}^{5}\right)%
\end{array}%
\right] ,  \notag \\
\theta _{2}\left( \nu \right) & =e^{\mathrm{i}k_{0}+\mathrm{i}\acute{\nu}%
\varsigma _{2}}\left( 1+O\left( \acute{\nu}^{2}\right) \right) ,\
e_{2}\left( \nu \right) =\left[ 
\begin{array}{c}
1-\mathrm{i}\left( \tau _{1}+3\tau _{2}\right) \acute{\nu}+O\left( \acute{\nu%
}^{2}\right) \\ 
-\mathrm{i}\acute{\nu}+\left( \tau _{1}-\tau _{2}\right) \acute{\nu}%
^{2}+O\left( \acute{\nu}^{3}\right) \\ 
-\acute{\nu}^{2}-\mathrm{i}\left( 3\tau _{1}+\tau _{2}\right) \acute{\nu}%
^{3}+O\left( \acute{\nu}^{4}\right) \\ 
\mathrm{i}\acute{\nu}^{3}+3\left( \tau _{1}-\tau _{2}\right) \acute{\nu}%
^{4}+O\left( \acute{\nu}^{5}\right)%
\end{array}%
\right] ,  \notag \\
\theta _{3}\left( \nu \right) & =e^{\mathrm{i}k_{0}+\mathrm{i}\acute{\nu}%
\varsigma _{2}}\left( 1+O\left( \acute{\nu}^{2}\right) \right) ,\
e_{3}\left( \nu \right) =\left[ 
\begin{array}{c}
1+\left( \tau _{1}+3\tau _{2}\right) \acute{\nu}+O\left( \acute{\nu}%
^{2}\right) \\ 
\acute{\nu}+\left( \tau _{2}-\tau _{1}\right) \acute{\nu}^{2}+O\left( \acute{%
\nu}^{3}\right) \\ 
\acute{\nu}^{2}-\left( 3\tau _{1}+\tau _{2}\right) \acute{\nu}^{3}+O\left( 
\acute{\nu}^{4}\right) \\ 
\acute{\nu}^{3}+3\left( \tau _{1}-\tau _{2}\right) \acute{\nu}^{4}+O\left( 
\acute{\nu}^{5}\right)%
\end{array}%
\right] ,  \notag \\
\tau _{1}& =\frac{\left[ T_{1}\right] _{31}}{8\zeta _{0}\mathfrak{t}_{1}},\
\tau _{2}=\frac{\left[ T_{1}\right] _{42}}{8\zeta _{0}\mathfrak{t}_{1}}. 
\notag
\end{align}%
Notice that in view of (\ref{etv412}) we always have%
\begin{equation}
\left\vert \theta _{0}\left( \nu \right) \right\vert =1,\ \left\vert \theta
_{1}\left( \nu \right) \right\vert <1,  \label{4Fj5}
\end{equation}%
implying that the vector $e_{0}\left( \nu \right) $ always corresponds to
the frozen mode and the vector $e_{1}\left( \nu \right) $ always corresponds
to the evanescent mode, i.e. the one decaying exponentially away from the
surface of the photonic crystal.

Hence, in view of (\ref{zaa4}) and (\ref{zaa5}), we have the following
representation for the eigenvectors $\mathfrak{e}_{j}\left( \nu \right) =%
\mathcal{G}_{0}\left( \nu \right) e_{j}\left( \nu \right) $, $j=0,1,2,4$ of
the transfer matrix $\mathcal{T}\left( \nu \right) $%
\begin{align}
\mathfrak{e}_{0}\left( \nu \right) & =\mathcal{G}_{0}\left( 0\right) \left[ 
\begin{array}{c}
1+\mathrm{i}\left( \tau _{1}+3\tau _{2}\right) \acute{\nu}+O\left( \acute{\nu%
}^{2}\right) \\ 
\mathrm{i}\acute{\nu}+\left( \tau _{1}-\tau _{2}\right) \acute{\nu}%
^{2}+O\left( \acute{\nu}^{3}\right) \\ 
-\acute{\nu}^{2}+\mathrm{i}\left( 3\tau _{1}+\tau _{2}\right) \acute{\nu}%
^{3}+O\left( \acute{\nu}^{4}\right) \\ 
-\mathrm{i}\acute{\nu}^{3}+O\left( \acute{\nu}^{4}\right)%
\end{array}%
\right] ,  \label{4enu1} \\
\mathfrak{e}_{1}\left( \nu \right) & =\mathcal{G}_{0}\left( 0\right) \left[ 
\begin{array}{c}
1-\left( \tau _{1}+3\tau _{2}\right) \acute{\nu}+O\left( \acute{\nu}%
^{2}\right) \\ 
-\acute{\nu}+\left( \tau _{2}-\tau _{1}\right) \acute{\nu}^{2}+O\left( 
\acute{\nu}^{3}\right) \\ 
\acute{\nu}^{2}+\left( 3\tau _{1}+\tau _{2}\right) \acute{\nu}^{3}+O\left( 
\acute{\nu}^{4}\right) \\ 
-\acute{\nu}^{3}+O\left( \acute{\nu}^{4}\right)%
\end{array}%
\right] ,  \notag
\end{align}%
\begin{align}
\mathfrak{e}_{2}\left( \nu \right) & =\mathcal{G}_{0}\left( 0\right) \left[ 
\begin{array}{c}
1-\mathrm{i}\left( \tau _{1}+3\tau _{2}\right) \acute{\nu}+O\left( \acute{\nu%
}^{2}\right) \\ 
-\mathrm{i}\acute{\nu}+\left( \tau _{1}-\tau _{2}\right) \acute{\nu}%
^{2}+O\left( \acute{\nu}^{3}\right) \\ 
-\acute{\nu}^{2}-\mathrm{i}\left( 3\tau _{1}+\tau _{2}\right) \acute{\nu}%
^{3}+O\left( \acute{\nu}^{4}\right) \\ 
\mathrm{i}\acute{\nu}^{3}+O\left( \acute{\nu}^{4}\right)%
\end{array}%
\right] ,  \label{4enu2} \\
\mathfrak{e}_{3}\left( \nu \right) & =\mathcal{G}_{0}\left( 0\right) \left[ 
\begin{array}{c}
1+\left( \tau _{1}+3\tau _{2}\right) \acute{\nu}+O\left( \acute{\nu}%
^{2}\right) \\ 
\acute{\nu}+\left( \tau _{2}-\tau _{1}\right) \acute{\nu}^{2}+O\left( \acute{%
\nu}^{3}\right) \\ 
\acute{\nu}^{2}-\left( 3\tau _{1}+\tau _{2}\right) \acute{\nu}^{3}+O\left( 
\acute{\nu}^{4}\right) \\ 
\acute{\nu}^{3}+O\left( \acute{\nu}^{4}\right)%
\end{array}%
\right] .  \notag
\end{align}

Combining now (\ref{4enu1}), (\ref{4enu2}) with (\ref{zaa5}) and (\ref{sg4})
we get the following representations for the eigenvectors $\mathfrak{e}%
_{j}\left( \nu \right) $%
\begin{align}
\mathfrak{e}_{0}\left( \nu \right) & =\left[ 1+\mathrm{i}\left( \tau
_{1}+3\tau _{2}\right) \acute{\nu}+O\left( \acute{\nu}^{2}\right) \right] 
\mathfrak{f}_{0}+\left[ \mathrm{i}\acute{\nu}+\left( \tau _{1}-\tau
_{2}\right) \acute{\nu}^{2}+O\left( \acute{\nu}^{3}\right) \right] \mathfrak{%
f}_{1}+  \label{4enu3} \\
& \left[ -\acute{\nu}^{2}+\mathrm{i}\left( 3\tau _{1}+\tau _{2}\right) 
\acute{\nu}^{3}\right] \mathfrak{f}_{2}-\mathrm{i}\acute{\nu}^{3}\mathfrak{f}%
_{3}+O\left( \acute{\nu}^{4}\right) ,  \notag \\
\mathfrak{e}_{1}\left( \nu \right) & =\left[ 1-\left( \tau _{1}+3\tau
_{2}\right) \acute{\nu}+O\left( \acute{\nu}^{2}\right) \right] \mathfrak{f}%
_{0}+\left[ -\acute{\nu}+\left( \tau _{2}-\tau _{1}\right) \acute{\nu}%
^{2}+O\left( \acute{\nu}^{3}\right) \right] \mathfrak{f}_{1}+  \notag \\
& \left[ \acute{\nu}^{2}+\left( 3\tau _{1}+\tau _{2}\right) \acute{\nu}^{3}%
\right] \mathfrak{f}_{2}-\acute{\nu}^{3}\mathfrak{f}_{3}+O\left( \acute{\nu}%
^{4}\right) ,  \notag \\
\mathfrak{e}_{2}\left( \nu \right) & =\left[ 1-\mathrm{i}\left( \tau
_{1}+3\tau _{2}\right) \acute{\nu}+O\left( \acute{\nu}^{2}\right) \right] 
\mathfrak{f}_{0}+\left[ -\mathrm{i}\acute{\nu}+\left( \tau _{1}-\tau
_{2}\right) \acute{\nu}^{2}+O\left( \acute{\nu}^{3}\right) \right] \mathfrak{%
f}_{1}+  \notag \\
& \left[ -\acute{\nu}^{2}-\mathrm{i}\left( 3\tau _{1}+\tau _{2}\right) 
\acute{\nu}^{3}\right] \mathfrak{f}_{2}+\mathrm{i}\acute{\nu}^{3}\mathfrak{f}%
_{3}+O\left( \acute{\nu}^{4}\right) ,  \notag \\
\mathfrak{e}_{3}\left( \nu \right) & =\left[ 1+\left( \tau _{1}+3\tau
_{2}\right) \acute{\nu}+O\left( \acute{\nu}^{2}\right) \right] \mathfrak{f}%
_{0}+\left[ \acute{\nu}+\left( \tau _{2}-\tau _{1}\right) \acute{\nu}%
^{2}+O\left( \acute{\nu}^{3}\right) \right] \mathfrak{f}_{1}+  \notag \\
& \left[ \acute{\nu}^{2}-\left( 3\tau _{1}+\tau _{2}\right) \acute{\nu}^{3}%
\right] \mathfrak{f}_{2}+\acute{\nu}^{3}\mathfrak{f}_{3}+O\left( \acute{\nu}%
^{4}\right) .  \notag
\end{align}%
\bigskip It readily follows from (\ref{4Fj4}) and (\ref{4enu3})that%
\begin{align}
\lim_{\nu \rightarrow 0}e_{j}\left( \nu \right) & =\left[ 
\begin{array}{c}
1 \\ 
0 \\ 
0 \\ 
0%
\end{array}%
\right] ,\ j=0,1,2,3,  \label{4enu4} \\
\lim_{\nu \rightarrow 0}\frac{e_{j}\left( \nu \right) -e_{0}\left( \nu
\right) }{\mathrm{i}\acute{\nu}\left( \varsigma _{j}-1\right) }& =\left[ 
\begin{array}{c}
\tau _{1}+3\tau _{2} \\ 
1 \\ 
0 \\ 
0%
\end{array}%
\right] ,\ j=1,2,3.  \notag
\end{align}%
\begin{align}
\lim_{\nu \rightarrow 0}\mathfrak{e}_{j}\left( \nu \right) & =\mathfrak{f}%
_{0},\ j=0,1,2,3,  \label{4enu5} \\
\lim_{\nu \rightarrow 0}\frac{\mathfrak{e}_{j}\left( \nu \right) -\mathfrak{e%
}_{0}\left( \nu \right) }{\mathrm{i}\acute{\nu}\left( \varsigma
_{j}-1\right) }& =\left( \tau _{1}+3\tau _{2}\right) \mathfrak{f}_{0}+%
\mathfrak{f}_{1},\ j=1,2,3,  \notag
\end{align}%
indicating, in particular, that all four vectors $\mathfrak{e}_{j}\left( \nu
\right) $, $j=0,1,2,3$ become almost parallel as $\nu \rightarrow 0$. To
have a nicer way to trace the two-dimensional spaces $\limfunc{Span}\left\{ 
\mathfrak{e}_{0}\left( \nu \right) ,\mathfrak{e}_{1}\left( \nu \right)
\right\} $ we introduce the following vector 
\begin{align}
\mathfrak{h}_{1}\left( \nu \right) & =\frac{\mathfrak{e}_{1}\left( \nu
\right) -\mathfrak{e}_{0}\left( \nu \right) }{\mathrm{i}\acute{\nu}\left(
\varsigma _{1}-1\right) }=\frac{\mathfrak{e}_{1}\left( \nu \right) -%
\mathfrak{e}_{0}\left( \nu \right) }{\mathrm{i}\acute{\nu}\left( \mathrm{i}%
-1\right) }=  \label{e3nu6} \\
\frac{\mathfrak{e}_{0}\left( \nu \right) -\mathfrak{e}_{1}\left( \nu \right) 
}{\acute{\nu}\left( \mathrm{i}+1\right) }& =\left( \tau _{1}+3\tau
_{2}\right) \mathfrak{f}_{0}+\mathfrak{f}_{1}+O\left( \acute{\nu}\right) . 
\notag
\end{align}%
Notice that the equalities (\ref{4enu5}) imply%
\begin{equation}
\lim_{\nu \rightarrow 0}\mathfrak{h}_{1}\left( \nu \right) =\left( \tau
_{1}+3\tau _{2}\right) \mathfrak{f}_{0}+\mathfrak{f}_{1}.  \label{4enu7}
\end{equation}%
Then the relations (\ref{4enu5})-(\ref{4enu7}) yield 
\begin{align}
\limfunc{Span}\left\{ \mathfrak{e}_{0}\left( \nu \right) ,\mathfrak{e}%
_{1}\left( \nu \right) \right\} & =\limfunc{Span}\left\{ \mathfrak{e}%
_{0}\left( \nu \right) ,\mathfrak{h}_{1}\left( \nu \right) \right\} ,
\label{4enu8} \\
\lim_{\nu \rightarrow 0}\limfunc{Span}\left\{ \mathfrak{e}_{0}\left( \nu
\right) ,\mathfrak{e}_{1}\left( \nu \right) \right\} & =\lim_{\nu
\rightarrow 0}\limfunc{Span}\left\{ \mathfrak{e}_{0}\left( \nu \right) ,%
\mathfrak{h}_{1}\left( \nu \right) \right\} =\limfunc{Span}\left\{ \mathfrak{%
f}_{0},\mathfrak{f}_{1}\right\} .  \notag
\end{align}

\subsection{Spectrum of the transfer matrix at a degenerate point of order 2}

In this section we derive the asymptotic formulae for the eigenvalues and
eigenvectors of the transfer matrix $T\left( \nu \right) $ as $\nu =\omega
-\omega _{0}\rightarrow 0$ in the case when the frequency $\omega _{0}$ is a
degenerate point of order 2. In this case since $n=2$ the transfer matrix $%
\mathcal{T}\left( \nu \right) $ defined by (\ref{tt1}) is such that $%
\mathcal{T}\left( 0\right) $ is a Jordan block of $2$, that follows from the
analysis carried out in the previous Section.

Without loss of generality we assume that $\omega _{0}=\omega \left(
k_{0}\right) $ is a point, say, of local minimum of the dispersion relation $%
\omega \left( k\right) $ in a vicinity of $k_{0}$. In the later case for $%
\nu =\omega -\omega _{0}>0$ and $\nu $ small there must be two propagating
Bloch modes associated with the chosen branch $\omega \left( k\right) $ and
two modes which either both are propagating or both are evanescent modes.

\subsubsection{Eigenvalues of the transfer matrix}

Observe that it follows from (\ref{dd15}), (\ref{nv2}) the eigenvalues$\
\eta _{0}\left( \nu \right) $, $\eta _{1}\left( \nu \right) $ of the matrix $%
I_{2}+\mathfrak{T}\left( \nu \right) $ from (\ref{dd6}) are 
\begin{equation}
\eta _{0}\left( \nu \right) =1+\sqrt{\tilde{\nu}}+O\left( \tilde{\nu}\right)
,\ \eta _{1}\left( \nu \right) =1-\sqrt{\tilde{\nu}}+O\left( \tilde{\nu}%
\right) ,  \label{etv21}
\end{equation}%
or, in view of (\ref{dd19}),%
\begin{equation}
\eta _{0}\left( \nu \right) =1+\sqrt{\mathfrak{t}_{1}}\sqrt{\nu }+O\left(
\nu \right) ,\ \eta _{1}\left( \nu \right) =1-\sqrt{\mathfrak{t}_{1}}\sqrt{%
\nu }+O\left( \nu \right) ,  \label{etv22}
\end{equation}%
where, we recall, according to (\ref{dd12a})%
\begin{equation}
\varsigma _{0}=1,\ \varsigma _{1}=-1.  \label{etv22b}
\end{equation}%
Notice, that we can recast (\ref{etv22}) as%
\begin{equation}
\eta _{0}\left( \nu \right) =\exp \left\{ \sqrt{\mathfrak{t}_{1}}\sqrt{\nu }%
+O\left( \nu \right) \right\} ,\ \eta _{1}\left( \nu \right) =\exp \left\{ -%
\sqrt{\mathfrak{t}_{1}}\sqrt{\nu }+O\left( \nu \right) \right\} .
\label{etv22a}
\end{equation}%
Since we have two propagating modes for sufficiently small positive $\nu $
there exists a sufficiently small $\delta >0$ such that%
\begin{equation}
\left\vert \eta _{0}\left( \nu \right) \right\vert =\eta _{1}\left( \nu
\right) =1\text{ for }\left\vert \nu \right\vert \leq \delta .  \label{etv23}
\end{equation}%
Using (\ref{zaa2}) and (\ref{zaa3})we get%
\begin{equation}
\sqrt{\mathfrak{t}_{1}}=\alpha _{0}\mathrm{i}\text{ with a real }\alpha _{0}=%
\sqrt{\frac{2}{\omega ^{\prime \prime }\left( k_{0}\right) }}>0,
\label{etv24}
\end{equation}%
\begin{align}
\mathfrak{t}_{1}& =\left[ \mathfrak{T}_{1}\right] _{31}=-\alpha _{0}^{2}%
\text{ with a real }\alpha _{0}=\sqrt{\frac{2}{\omega ^{\prime \prime
}\left( k_{0}\right) }}>0,  \label{etv25} \\
\sqrt{\tilde{\nu}}& =\mathrm{i}\alpha _{0}\sqrt{\nu }+O\left( \nu \right)
\label{etv25a}
\end{align}%
Hence, (\ref{etv22a}) takes the form%
\begin{equation}
\eta _{0}\left( \nu \right) =\exp \left\{ \mathrm{i}\alpha _{0}\sqrt{\nu }%
+O\left( \nu \right) \right\} ,\ \eta _{1}\left( \nu \right) =\exp \left\{ -%
\mathrm{i}\alpha _{0}\sqrt{\nu }+O\left( \nu \right) \right\} .
\label{etv26}
\end{equation}%
Now as follows from (\ref{dd6}), the eigenvalues $\theta _{j}\left( \nu
\right) $ of the $2\times 2$ transfer matrix $T\left( \nu \right) $ take the
form 
\begin{equation}
\theta _{j}\left( \nu \right) =\zeta _{0}\eta _{j}\left( \nu \right) ,\
j=0,1.  \label{etv29}
\end{equation}%
Recalling that%
\begin{equation}
\zeta _{0}=e^{\mathrm{i}k_{0}},  \label{etv210}
\end{equation}%
where $k_{0}$ is real, we get from (\ref{etv26})-(\ref{etv210})%
\begin{align}
\theta _{0}\left( \nu \right) & =\exp \left\{ \mathrm{i}k_{0}+\mathrm{i}%
\alpha _{0}\sqrt{\nu }+O\left( \nu \right) \right\} =e^{\mathrm{i}k_{0}+%
\mathrm{i}\alpha _{0}\sqrt{\nu }}\left( 1+O\left( \nu \right) \right) ,
\label{etv211} \\
\theta _{1}\left( \nu \right) & =\exp \left\{ \mathrm{i}k_{0}-\mathrm{i}%
\alpha _{0}\sqrt{\nu }+O\left( \nu \right) \right\} =e^{\mathrm{i}k_{0}-%
\mathrm{i}\alpha _{0}\sqrt{\nu }}\left( 1+O\left( \nu \right) \right) . 
\notag
\end{align}%
Notice also that (\ref{tomx10aaa}) implies%
\begin{equation}
a_{0}=-\alpha _{0}^{2}\zeta _{0}^{2}=\frac{2}{\omega ^{\prime \prime }\left(
k_{0}\right) }\zeta _{0}^{2}.  \label{etv213a}
\end{equation}%
It is convenient to introduce%
\begin{equation}
\acute{\nu}=\alpha _{0}\sqrt{\nu },  \label{etv214}
\end{equation}%
and to rewrite (\ref{etv11}) and (\ref{etv12a}) as%
\begin{align}
\theta _{0}\left( \nu \right) & =\exp \left\{ \mathrm{i}k_{0}+\mathrm{i}%
\acute{\nu}+O\left( \acute{\nu}^{2}\right) \right\} =e^{\mathrm{i}k_{0}+%
\mathrm{i}\acute{\nu}}\left( 1+O\left( \acute{\nu}^{2}\right) \right) ,
\label{etv215} \\
\theta _{1}\left( \nu \right) & =\exp \left\{ \mathrm{i}k_{0}-\mathrm{i}%
\acute{\nu}+O\left( \acute{\nu}^{2}\right) \right\} =e^{\mathrm{i}k_{0}-%
\mathrm{i}\acute{\nu}}\left( 1+O\left( \acute{\nu}^{2}\right) \right) . 
\notag
\end{align}

\subsubsection{Eigenvectors of the transfer matrix.}

We recall that we work with the basis in which $T\left( 0\right) $\emph{\
has its canonical Jordan form as in (\ref{dd5})}, namely%
\begin{equation*}
T\left( 0\right) =\zeta _{0}\left[ 
\begin{array}{cc}
1 & 1 \\ 
0 & 1%
\end{array}%
\right] .
\end{equation*}%
Based on (\ref{dd6}), (\ref{nv2}), (\ref{nv3}), (\ref{zaa3}) and (\ref{zaa4}%
) we can find the eigenvectors $e_{j}\left( \nu \right) $ of $T\left( \nu
\right) $ corresponding to its eigenvalues $\theta _{j}\left( \nu \right) $.

Recall also that the eigenvectors $e_{j}\left( \nu \right) $, $j=0,1$ of the
matrix $T\left( \nu \right) $ are the respective columns of the matrix $%
\mathfrak{S}_{0}\left( \tilde{\nu}\right) e^{-S\left( \mathrm{i}\acute{\nu}%
\right) }$ represented by the asymptotic equality (\ref{nv2a}). To use (\ref%
{nv2a}) we need to find the matrix $S_{1}$ following the section
\textquotedblleft Perturbation\ theory for diagonal
matrix\textquotedblright\ in the Appendix. Namely, first we introduce a
decomposition of a square matrix $W$ into its the diagonal component $%
\limfunc{diag}\left( W\right) $ and the remaining part $\mathring{W}=W-%
\limfunc{diag}\left( W\right) $ with zero diagonal elements, i.e.%
\begin{align}
\Lambda _{1}& =\limfunc{diag}\left( W_{1}\right) ,\ \left[ S_{1}\right]
_{nm}=\frac{1}{\varsigma _{m-1}-\varsigma _{n-1}}\left[ \mathring{W}_{1}%
\right] _{nm},\ n\neq m;\ \left[ S_{1}\right] _{nn}=0,  \label{s1w4} \\
\text{where }W_{1}& =\left\langle \widetilde{\mathfrak{T}}_{1}\right\rangle
_{1},\ \varsigma _{0}=1,\ \varsigma _{1}=-1,  \notag
\end{align}%
Carrying out the operations described in (\ref{s1w4}), and using (\ref{dd6}%
), (\ref{dd8}), (\ref{dd25}), (\ref{dd28}) we obtain%
\begin{equation}
\left\langle \widetilde{\mathfrak{T}}_{1}\right\rangle _{1}=\frac{\left[
T_{1}\right] _{22}}{2\zeta _{0}\mathfrak{t}_{1}}\left[ 
\begin{array}{cc}
1 & -1 \\ 
-1 & 1%
\end{array}%
\right] ,\ S_{1}=\frac{\left[ T_{1}\right] _{22}}{4\zeta _{0}\mathfrak{t}_{1}%
}\left[ 
\begin{array}{cc}
0 & 1 \\ 
-1 & 0%
\end{array}%
\right] .  \label{s1w5}
\end{equation}%
The first two eigenvectors $e_{0}\left( \nu \right) $ and $e_{1}\left( \nu
\right) $ of the matrix $\mathfrak{T}\left( \nu \right) $ (and, hence, the
matrix $T\left( \nu \right) $), corresponding to eigenvalues $\varsigma
_{0}=1$ and $\varsigma _{1}=\mathrm{i}$ in view of (\ref{nv2a}) and (\ref%
{nv7}) take the following form 
\begin{align}
e_{0}\left( \nu \right) & =\left[ 
\begin{array}{c}
1+\mathrm{i}\tau \acute{\nu}+O\left( \acute{\nu}^{2}\right) \\ 
\mathrm{i}\acute{\nu}+\mathrm{i}\tau \acute{\nu}^{2}+O\left( \acute{\nu}%
^{3}\right)%
\end{array}%
\right] ,\ e_{1}\left( \nu \right) =\left[ 
\begin{array}{c}
1-\mathrm{i}\tau \acute{\nu}+O\left( \acute{\nu}^{2}\right) \\ 
-\mathrm{i}\acute{\nu}+\mathrm{i}\tau \acute{\nu}^{2}+O\left( \acute{\nu}%
^{3}\right)%
\end{array}%
\right] ,  \label{s1w5a} \\
\tau & =\frac{\left[ T_{1}\right] _{22}}{4\zeta _{0}\mathfrak{t}_{1}}. 
\notag
\end{align}%
Thus using (\ref{s1w5a}) and (\ref{etv211}) we get the following complete
set of the eigenvalues and corresponding eigenvectors of $T\left( \nu
\right) $%
\begin{align}
\theta _{0}\left( \nu \right) & =e^{\mathrm{i}k_{0}+\mathrm{i}\acute{\nu}%
}\left( 1+O\left( \acute{\nu}^{2}\right) \right) ,\ e_{0}\left( \nu \right) =%
\left[ 
\begin{array}{c}
1+\mathrm{i}\tau \acute{\nu}+O\left( \acute{\nu}^{2}\right) \\ 
\mathrm{i}\acute{\nu}+\mathrm{i}\tau \acute{\nu}^{2}+O\left( \acute{\nu}%
^{3}\right)%
\end{array}%
\right] ,  \label{Fj24} \\
\theta _{1}\left( \nu \right) & =e^{\mathrm{i}k_{0}-\mathrm{i}\acute{\nu}%
}\left( 1+O\left( \acute{\nu}^{2}\right) \right) ,\ e_{1}\left( \nu \right) =%
\left[ 
\begin{array}{c}
1-\mathrm{i}\tau \acute{\nu}+O\left( \acute{\nu}^{2}\right) \\ 
-\mathrm{i}\acute{\nu}+\mathrm{i}\tau \acute{\nu}^{2}+O\left( \acute{\nu}%
^{3}\right)%
\end{array}%
\right] .  \notag
\end{align}

Hence, in view of (\ref{zaa4}) and (\ref{zaa5}), we have the following
representation for the eigenvectors $\mathfrak{e}_{j}\left( \nu \right) =%
\mathcal{G}_{0}\left( \nu \right) e_{j}\left( \nu \right) $, $j=0,1,2,3$ of
the transfer matrix $\mathcal{T}\left( \nu \right) $%
\begin{equation}
\mathfrak{e}_{0}\left( \nu \right) =\mathcal{G}_{0}\left( 0\right) \left[ 
\begin{array}{c}
1+\mathrm{i}\tau \acute{\nu}+O\left( \acute{\nu}^{2}\right) \\ 
\mathrm{i}\acute{\nu}+O\left( \acute{\nu}^{2}\right) \\ 
O\left( \acute{\nu}^{2}\right) \\ 
O\left( \acute{\nu}^{2}\right)%
\end{array}%
\right] ,\ \mathfrak{e}_{1}\left( \nu \right) =\mathcal{G}_{0}\left(
0\right) \left[ 
\begin{array}{c}
1-\mathrm{i}\tau \acute{\nu}+O\left( \acute{\nu}^{2}\right) \\ 
-\mathrm{i}\acute{\nu}+O\left( \acute{\nu}^{2}\right) \\ 
O\left( \acute{\nu}^{2}\right) \\ 
O\left( \acute{\nu}^{2}\right)%
\end{array}%
\right] ,  \label{Fj25}
\end{equation}%
\begin{equation}
\mathfrak{e}_{2}\left( \nu \right) =\mathcal{G}_{0}\left( 0\right) \left[ 
\begin{array}{c}
O\left( \acute{\nu}^{2}\right) \\ 
O\left( \acute{\nu}^{2}\right) \\ 
1+O\left( \acute{\nu}^{2}\right) \\ 
O\left( \acute{\nu}^{2}\right)%
\end{array}%
\right] ,\mathfrak{e}_{3}\left( \nu \right) =\mathcal{G}_{0}\left( 0\right) %
\left[ 
\begin{array}{c}
O\left( \acute{\nu}^{2}\right) \\ 
O\left( \acute{\nu}^{2}\right) \\ 
O\left( \acute{\nu}^{2}\right) \\ 
1+O\left( \acute{\nu}^{2}\right)%
\end{array}%
\right] .  \label{Fj25a}
\end{equation}%
Combining now (\ref{Fj25}) with (\ref{zaa5}) and (\ref{sg4}) we get the
following representations for the eigenvectors $\mathfrak{e}_{j}\left( \nu
\right) $%
\begin{align}
\mathfrak{e}_{0}\left( \nu \right) & =\left[ 1+\mathrm{i}\tau \acute{\nu}%
\right] \mathfrak{f}_{0}+\mathrm{i}\acute{\nu}\mathfrak{f}_{1}+O\left( 
\acute{\nu}^{2}\right) ,\ \mathfrak{e}_{1}\left( \nu \right) =\left[ 1-%
\mathrm{i}\tau \acute{\nu}\right] \mathfrak{f}_{0}-\mathrm{i}\acute{\nu}%
\mathfrak{f}_{1}+O\left( \acute{\nu}^{2}\right) ,  \label{Fj26} \\
\mathfrak{e}_{2}\left( \nu \right) & =\mathfrak{f}_{2}+O\left( \acute{\nu}%
^{2}\right) ,\ \mathfrak{e}_{3}\left( \nu \right) =\mathfrak{f}_{3}+O\left( 
\acute{\nu}^{2}\right) .  \notag
\end{align}%
It readily follows from (\ref{Fj25}) and (\ref{Fj26}) that,%
\begin{equation}
\lim_{\nu \rightarrow 0}e_{j}\left( \nu \right) =\left[ 
\begin{array}{c}
1 \\ 
0%
\end{array}%
\right] ,\ j=0,1;\ \lim_{\nu \rightarrow 0}\frac{e_{1}\left( \nu \right)
-e_{0}\left( \nu \right) }{-2\mathrm{i}\acute{\nu}}=\left[ 
\begin{array}{c}
\tau \\ 
1%
\end{array}%
\right] ,\ \tau =\frac{\left[ T_{1}\right] _{22}}{4\zeta _{0}\mathfrak{t}_{1}%
}.  \label{Fj27}
\end{equation}%
\begin{equation}
\lim_{\nu \rightarrow 0}\mathfrak{e}_{j}\left( \nu \right) =\mathfrak{f}%
_{0},\ j=0,1;\lim_{\nu \rightarrow 0}\frac{\mathfrak{e}_{1}\left( \nu
\right) -\mathfrak{e}_{0}\left( \nu \right) }{-2\mathrm{i}\acute{\nu}}=\tau 
\mathfrak{f}_{0}+\mathfrak{f}_{1},\ \tau =\frac{\left[ T_{1}\right] _{22}}{%
4\zeta _{0}\mathfrak{t}_{1}}.  \label{Fj28}
\end{equation}%
indicating, in particular, that the two vectors $\mathfrak{e}_{0}\left( \nu
\right) $ and $\mathfrak{e}_{1}\left( \nu \right) $ become almost parallel
as $\nu \rightarrow 0$. To have a nicer way to trace the two-dimensional
spaces $\limfunc{Span}\left\{ \mathfrak{e}_{0}\left( \nu \right) ,\mathfrak{e%
}_{1}\left( \nu \right) \right\} $ we introduce the following vector 
\begin{equation}
\mathfrak{h}_{1}\left( \nu \right) =\frac{\mathfrak{e}_{1}\left( \nu \right)
-\mathfrak{e}_{0}\left( \nu \right) }{-2\mathrm{i}\acute{\nu}}.  \label{Fj29}
\end{equation}%
Notice that the equalities (\ref{Fj28}) and (\ref{Fj29}) imply%
\begin{equation}
\lim_{\nu \rightarrow 0}\mathfrak{h}_{1}\left( \nu \right) =\tau \mathfrak{f}%
_{0}+\mathfrak{f}_{1},\ \tau =\frac{\left[ T_{1}\right] _{22}}{4\zeta _{0}%
\mathfrak{t}_{1}}.  \label{Fj30}
\end{equation}%
Then the relations (\ref{Fj29})-(\ref{Fj30}) yield 
\begin{align}
\limfunc{Span}\left\{ \mathfrak{e}_{0}\left( \nu \right) ,\mathfrak{e}%
_{1}\left( \nu \right) \right\} & =\limfunc{Span}\left\{ \mathfrak{e}%
_{0}\left( \nu \right) ,\mathfrak{h}_{1}\left( \nu \right) \right\} ,
\label{Fj31} \\
\lim_{\nu \rightarrow 0}\limfunc{Span}\left\{ \mathfrak{e}_{0}\left( \nu
\right) ,\mathfrak{e}_{1}\left( \nu \right) \right\} & =\lim_{\nu
\rightarrow 0}\limfunc{Span}\left\{ \mathfrak{e}_{0}\left( \nu \right) ,%
\mathfrak{h}_{1}\left( \nu \right) \right\} =\limfunc{Span}\left\{ \mathfrak{%
f}_{0},\mathfrak{f}_{1}\right\} .  \notag
\end{align}

\section{Transfer matrix and the flux}

\subsection{Transfer matrix and the flux for an inflection point}

The case of an inflection point is the case with degeneracy index $n=3$. In
this case according to (\ref{tomx5}), the matrix $T\left( \nu \right) $ is a 
$3\times 3$ matrix and $W\left( \nu \right) $ is a $1\times 1$ matrix, i.e.
just a complex number $W\left( \nu \right) $. Let the matrix $T\left( \nu
\right) $ be defined by (\ref{tomx5}). Consider first the matrix at the
frequency of the frozen mode $\omega _{0}$, i.e. for $\nu =0$:%
\begin{equation}
T\left( 0\right) =T_{0}=\zeta _{0}\left( I_{3}+D_{0}\right) ,\text{ }%
\left\vert \zeta _{0}\right\vert =1,\ D_{0}^{3}=0\text{ and }D_{0}^{2}\neq 0.
\label{tz1}
\end{equation}%
Then there exists a canonical basis $f_{0}$, $f_{1}$, $f_{2}$ related to the
matrix $D$ such that%
\begin{equation}
D_{0}^{3}f_{2}=0,\ f_{0}=D_{0}^{2}f_{2},\ f_{1}=D_{0}f_{2}.  \label{tz2}
\end{equation}%
In fact, the basis $f_{0}$, $f_{1}$, $f_{2}$ is not unique and is defined up
to some transformations. The equalities (\ref{tz1}) and (\ref{tz2}) imply
the following representation for $T_{0}$ 
\begin{equation}
\zeta _{0}^{-1}T_{0}f_{0}=f_{0},\ \zeta _{0}^{-1}T_{0}f_{1}=f_{1}+f_{0},\
\zeta _{0}^{-1}T_{0}f_{2}=f_{2}+f_{1}.  \label{tz3}
\end{equation}%
or, in the basis $f_{0}$, $f_{1}$, $f_{2}$ we have%
\begin{equation}
T_{0}=\zeta _{0}\left[ 
\begin{array}{ccc}
1 & 1 & 0 \\ 
0 & 1 & 1 \\ 
0 & 0 & 1%
\end{array}%
\right] ,\ \det T_{0}=\zeta _{0}^{3}.  \label{tz3a}
\end{equation}%
In addition to that, (\ref{tomx5}) and (\ref{tz3a}) imply $\det \mathcal{T}%
\left( \nu \right) =\det T\left( 0\right) W\left( 0\right) =\zeta
_{0}^{3}W\left( 0\right) =1$, and, since $\left\vert \zeta _{0}\right\vert
=1 $ we get%
\begin{equation}
W\left( 0\right) =\zeta _{0}^{-3},\ \left\vert W\left( 0\right) \right\vert
=1.  \label{tz3b}
\end{equation}%
We reiterate that our fundamental assumption is that $\zeta _{0}$ is triply
degenerate and $W\left( 0\right) $ is an eigenvalue of $\mathcal{T}\left(
0\right) $ that differs from $\zeta _{0}$, i.e.%
\begin{equation}
W\left( 0\right) \neq \zeta _{0}.  \label{tz3c}
\end{equation}%
Observe now that since according to (\ref{tz3b}) $\left\vert W\left(
0\right) \right\vert =1$ then $\overline{W\left( 0\right) }=\left[ W\left(
0\right) \right] ^{-1}$ and, hence, the relation (\ref{tz3c}) can be
rewritten as%
\begin{equation}
\zeta _{0}\overline{W\left( 0\right) }\neq 1.  \label{tz3d}
\end{equation}

Recalling again the relation (\ref{tomx5}) between the $3\times 3$ matrix $%
T\left( 0\right) $ and the original $4\times 4$ matrix $\mathcal{T}\left(
0\right) $ we introduce a basis $\mathfrak{f}_{0}$, $\mathfrak{f}_{1}$, $%
\mathfrak{f}_{2}$, $\mathfrak{f}_{3}$ in the four-dimensional space such
that 
\begin{equation}
f_{j}=P_{3}\mathcal{S}^{-1}\left( 0\right) \mathfrak{f}_{j},\ j=0,1,2,\text{
where }P_{3}\left[ 
\begin{array}{c}
X_{1} \\ 
X_{2} \\ 
X_{3} \\ 
X_{4}%
\end{array}%
\right] =\left[ 
\begin{array}{c}
X_{1} \\ 
X_{2} \\ 
X_{3}%
\end{array}%
\right] .  \label{tz4}
\end{equation}%
and the fourth vector $\mathfrak{f}_{3}$ is the eigenvector $\mathcal{T}%
\left( 0\right) $ related to the eigenvalue $W\left( 0\right) $. Evidently
the vectors $\mathfrak{f}_{0}$, $\mathfrak{f}_{1}$, $\mathfrak{f}_{2}$ are
4-dimensional representation of respective vectors $f_{0}$, $f_{1}$, $f_{2}$%
. Based on the above and the relations (\ref{tz3}) we get%
\begin{equation}
\mathcal{T}_{0}\mathfrak{f}_{0}=\zeta _{0}\mathfrak{f}_{0},\ \mathcal{T}_{0}%
\mathfrak{f}_{1}=\zeta _{0}\mathfrak{f}_{1}+\zeta _{0}\mathfrak{f}_{0},\ 
\mathcal{T}_{0}\mathfrak{f}_{2}=\zeta _{0}\mathfrak{f}_{2}+\zeta _{0}%
\mathfrak{f}_{1},  \label{tz5}
\end{equation}%
\begin{equation}
\mathcal{T}_{0}\mathfrak{f}_{3}=W\left( 0\right) \mathfrak{f}_{3}.
\label{tz5a}
\end{equation}%
In particular, the matrix $\mathcal{T}_{0}$ has just two genuine
eigenvectors $\mathfrak{f}_{0}$ and $\mathfrak{f}_{3}$ with corresponding
distinct eigenvalues $\zeta _{0}$ and $W\left( 0\right) $. The vectors $%
\mathfrak{f}_{0}$ and $\mathfrak{f}_{3}$ correspond respectively to the
frozen mode and the only propagating mode at the origin. The vectors $%
\mathfrak{f}_{1}$ and $\mathfrak{f}_{2}$ correspond respectively to linearly
and quadratically growing Floquet modes at the origin.

Recall now that $\mathcal{T}_{0}$ is a $J$-unitary matrix, i.e.%
\begin{equation}
\mathcal{T}_{0}^{\dag }J\mathcal{T}_{0}=J,\ \left[ \mathcal{T}_{0}\Phi _{1},%
\mathcal{T}_{0}\Phi _{2}\right] =\left[ \Phi _{1},\Phi _{2}\right] \text{
for any }\Phi _{1},\Phi _{2}.  \label{tz6}
\end{equation}%
Using (\ref{tz5}) and (\ref{tz6}) we obtain the following identities 
\begin{equation}
\left[ \mathfrak{f}_{0},\mathfrak{f}_{1}\right] =\left[ \mathcal{T}_{0}%
\mathfrak{f}_{0},\mathcal{T}_{0}\mathfrak{f}_{1}\right] =\left[ \mathfrak{f}%
_{0},\mathfrak{f}_{1}+\mathfrak{f}_{0}\right] =\left[ \mathfrak{f}_{0},%
\mathfrak{f}_{1}\right] +\left[ \mathfrak{f}_{0},\mathfrak{f}_{0}\right] ,
\label{tz7}
\end{equation}%
\begin{equation}
\left[ \mathfrak{f}_{0},\mathfrak{f}_{2}\right] =\left[ \mathcal{T}_{0}%
\mathfrak{f}_{0},\mathcal{T}_{0}\mathfrak{f}_{2}\right] =\left[ \mathfrak{f}%
_{0},\mathfrak{f}_{2}+\mathfrak{f}_{1}\right] =\left[ \mathfrak{f}_{0},%
\mathfrak{f}_{2}\right] +\left[ \mathfrak{f}_{0},\mathfrak{f}_{1}\right] ,
\label{tz8}
\end{equation}%
readily implying that%
\begin{equation}
\left[ \mathfrak{f}_{0},\mathfrak{f}_{0}\right] =0,\ \left[ \mathfrak{f}_{0},%
\mathfrak{f}_{1}\right] =0.  \label{tz9}
\end{equation}%
Then using (\ref{tz5}), (\ref{tz6}) again we get 
\begin{align}
\left[ \mathfrak{f}_{1},\mathfrak{f}_{2}\right] & =\left[ \mathcal{T}_{0}%
\mathfrak{f}_{1},\mathcal{T}_{0}\mathfrak{f}_{2}\right] =\left[ \mathfrak{f}%
_{1}+\mathfrak{f}_{0},\mathfrak{f}_{2}+\mathfrak{f}_{1}\right]  \label{tz10}
\\
& =\left[ \mathfrak{f}_{1},\mathfrak{f}_{2}\right] +\left[ \mathfrak{f}_{0},%
\mathfrak{f}_{2}\right] +\left[ \mathfrak{f}_{1},\mathfrak{f}_{1}\right] +%
\left[ \mathfrak{f}_{0},\mathfrak{f}_{1}\right] .  \notag
\end{align}%
The equalities (\ref{tz9}) together with (\ref{tz10}) yield%
\begin{equation}
\left[ \mathfrak{f}_{0},\mathfrak{f}_{2}\right] +\left[ \mathfrak{f}_{1},%
\mathfrak{f}_{1}\right] =0.  \label{tz11}
\end{equation}%
and, since $\left[ \mathfrak{f}_{1},\mathfrak{f}_{1}\right] $ is a real
number, we consequently have%
\begin{equation}
\left[ \mathfrak{f}_{0},\mathfrak{f}_{2}\right] =\left[ \mathfrak{f}_{2},%
\mathfrak{f}_{0}\right] =-\left[ \mathfrak{f}_{1},\mathfrak{f}_{1}\right] ,\ 
\func{Im}\left\{ \left[ \mathfrak{f}_{0},\mathfrak{f}_{2}\right] \right\} =0.
\label{tz11a}
\end{equation}%
We also have the relation 
\begin{align}
\left[ \mathfrak{f}_{1},\mathfrak{f}_{1}\right] & =\left[ \mathcal{T}_{0}%
\mathfrak{f}_{1},\mathcal{T}_{0}\mathfrak{f}_{1}\right] =\left[ \mathfrak{f}%
_{1}+\mathfrak{f}_{0},\mathfrak{f}_{1}+\mathfrak{f}_{0}\right]  \label{tz12}
\\
& =\left[ \mathfrak{f}_{1},\mathfrak{f}_{1}\right] +\left[ \mathfrak{f}_{0},%
\mathfrak{f}_{1}\right] +\left[ \mathfrak{f}_{1},\mathfrak{f}_{0}\right] +%
\left[ \mathfrak{f}_{0},\mathfrak{f}_{0}\right] ,  \notag
\end{align}%
but, in view of (\ref{tz9}), it is already satisfied and does not produce a
new relation. The remaining relation is%
\begin{align}
\left[ \mathfrak{f}_{2},\mathfrak{f}_{2}\right] & =\left[ \mathcal{T}_{0}%
\mathfrak{f}_{2},\mathcal{T}_{0}\mathfrak{f}_{2}\right] =\left[ \mathfrak{f}%
_{2}+\mathfrak{f}_{1},\mathfrak{f}_{2}+\mathfrak{f}_{1}\right]  \label{tz13}
\\
& =\left[ \mathfrak{f}_{2},\mathfrak{f}_{2}\right] +\left[ \mathfrak{f}_{1},%
\mathfrak{f}_{2}\right] +\left[ \mathfrak{f}_{2},\mathfrak{f}_{1}\right] +%
\left[ \mathfrak{f}_{1},\mathfrak{f}_{1}\right] ,  \notag
\end{align}%
yielding%
\begin{equation}
\left[ \mathfrak{f}_{1},\mathfrak{f}_{2}\right] +\left[ \mathfrak{f}_{2},%
\mathfrak{f}_{1}\right] +\left[ \mathfrak{f}_{1},\mathfrak{f}_{1}\right] =0.
\label{tz14}
\end{equation}%
Notice that for a natural number $m\geq 2$ we have%
\begin{equation}
\left[ 
\begin{array}{ccc}
1 & 1 & 0 \\ 
0 & 1 & 1 \\ 
0 & 0 & 1%
\end{array}%
\right] ^{m}=\left[ 
\begin{array}{ccc}
1 & m & \frac{\left( m-1\right) m}{2} \\ 
0 & 1 & m \\ 
0 & 0 & 1%
\end{array}%
\right] .  \label{tz14a}
\end{equation}%
This identity together (\ref{tz5}) and (\ref{tz6}) imply%
\begin{align}
\left[ \zeta _{0}^{-1}\mathcal{T}_{0}\right] ^{m}\mathfrak{f}_{0}& =%
\mathfrak{f}_{0},\ \mathcal{T}_{0}^{m}\mathfrak{f}_{1}=\mathfrak{f}_{1}+m%
\mathfrak{f}_{0},  \label{tz15} \\
\left[ \zeta _{0}^{-1}\mathcal{T}_{0}\right] ^{m}\mathfrak{f}_{2}& =%
\mathfrak{f}_{2}+m\mathfrak{f}_{1}+\frac{\left( m-1\right) m}{2}\mathfrak{f}%
_{0},  \notag
\end{align}%
\begin{equation}
\left( \left[ \zeta _{0}^{-1}\mathcal{T}_{0}\right] ^{m}\right) ^{\dag }J%
\left[ \zeta _{0}^{-1}\mathcal{T}_{0}\right] ^{m}=J.  \label{tz16}
\end{equation}%
Notice that the vectors $\mathcal{T}_{0}^{m}\mathfrak{f}_{1}$ and $\mathcal{T%
}_{0}^{m}\mathfrak{f}_{2}$ representing the EM field at points $mL$ grow
respectively linearly and quadratically as $m\rightarrow \infty $.

Using (\ref{tz15}) and (\ref{tz16}) we obtain%
\begin{align}
\left[ \mathfrak{f}_{2},\mathfrak{f}_{2}\right] & =\left[ \mathfrak{f}_{2}+m%
\mathfrak{f}_{1}+\frac{\left( m-1\right) m}{2}\mathfrak{f}_{0},\mathfrak{f}%
_{2}+m\mathfrak{f}_{1}+\frac{\left( m-1\right) m}{2}\mathfrak{f}_{0}\right]
\label{tz17} \\
& =\left[ \mathfrak{f}_{2},\mathfrak{f}_{2}\right] +m^{2}\left[ \mathfrak{f}%
_{1},\mathfrak{f}_{1}\right] +m\left[ \mathfrak{f}_{1},\mathfrak{f}_{2}%
\right] +m\left[ \mathfrak{f}_{2},\mathfrak{f}_{1}\right]  \notag \\
& +\frac{\left( m-1\right) m}{2}\left( \left[ \mathfrak{f}_{0},\mathfrak{f}%
_{2}\right] +\left[ \mathfrak{f}_{2},\mathfrak{f}_{0}\right] \right) , 
\notag
\end{align}%
implying%
\begin{equation}
m\left[ \mathfrak{f}_{1},\mathfrak{f}_{1}\right] +\left[ \mathfrak{f}_{1},%
\mathfrak{f}_{2}\right] +\left[ \mathfrak{f}_{2},\mathfrak{f}_{1}\right] +%
\frac{\left( m-1\right) }{2}\left( \left[ \mathfrak{f}_{0},\mathfrak{f}_{2}%
\right] +\left[ \mathfrak{f}_{2},\mathfrak{f}_{0}\right] \right) =0.
\label{tz18}
\end{equation}%
Combining (\ref{tz18}) with (\ref{tz11a}) we get%
\begin{equation}
\left[ \mathfrak{f}_{1},\mathfrak{f}_{1}\right] +\left[ \mathfrak{f}_{1},%
\mathfrak{f}_{2}\right] +\left[ \mathfrak{f}_{2},\mathfrak{f}_{1}\right] =0.
\label{tz19}
\end{equation}%
which is identical to (\ref{tz14}). Consider now%
\begin{align}
\left[ \mathfrak{f}_{2},\mathfrak{f}_{1}\right] & =\left[ \mathfrak{f}_{2}+m%
\mathfrak{f}_{1}+\frac{\left( m-1\right) m}{2}\mathfrak{f}_{0},\mathfrak{f}%
_{1}+m\mathfrak{f}_{0}\right]  \label{tz20} \\
& =\left[ \mathfrak{f}_{2},\mathfrak{f}_{1}\right] +m\left[ \mathfrak{f}_{1},%
\mathfrak{f}_{1}\right] +m\left[ \mathfrak{f}_{2},\mathfrak{f}_{0}\right] , 
\notag
\end{align}%
implying%
\begin{equation}
\left[ \mathfrak{f}_{1},\mathfrak{f}_{1}\right] +\left[ \mathfrak{f}_{2},%
\mathfrak{f}_{0}\right] =0,  \label{tz21}
\end{equation}%
which is equivalent to (\ref{tz11}). So, consideration of powers $\mathcal{T}%
_{0}^{m}$ of the transfer matrix have not produced new identities. Observe
now that (\ref{tz3d}), (\ref{tz5a}) and (\ref{[i,j] X}) imply 
\begin{equation}
\left[ \mathfrak{f}_{0},\mathfrak{f}_{3}\right] =0.  \label{tz22}
\end{equation}%
Collecting (\ref{tz9}), (\ref{tz11}), (\ref{tz14}) and (\ref{tz22}) we get
the following system%
\begin{align}
\left[ \mathfrak{f}_{0},\mathfrak{f}_{0}\right] & =0,  \label{sJe1} \\
\left[ \mathfrak{f}_{0},\mathfrak{f}_{1}\right] & =0,  \label{sJe2} \\
\left[ \mathfrak{f}_{0},\mathfrak{f}_{2}\right] +\left[ \mathfrak{f}_{1},%
\mathfrak{f}_{1}\right] & =0,  \label{sJe3} \\
\left[ \mathfrak{f}_{1},\mathfrak{f}_{2}\right] +\left[ \mathfrak{f}_{2},%
\mathfrak{f}_{1}\right] +\left[ \mathfrak{f}_{1},\mathfrak{f}_{1}\right] &
=0,  \label{sJe4} \\
\left[ \mathfrak{f}_{0},\mathfrak{f}_{3}\right] & =0.  \label{sJe5}
\end{align}%
Notice that since $J$ is Hermitian (\ref{sJe3}) implies%
\begin{equation}
\func{Im}\left\{ \left[ \mathfrak{f}_{0},\mathfrak{f}_{2}\right] \right\} =%
\func{Im}\left\{ \left[ \mathfrak{f}_{1},\mathfrak{f}_{1}\right] \right\}
=0,\ \left[ \mathfrak{f}_{0},\mathfrak{f}_{2}\right] =\func{Re}\left\{ \left[
\mathfrak{f}_{0},\mathfrak{f}_{2}\right] \right\} .  \label{sJe6}
\end{equation}%
In addition to that, (\ref{sJe3}) and (\ref{sJe4}) yield%
\begin{equation}
\left[ \mathfrak{f}_{0},\mathfrak{f}_{2}\right] =\func{Re}\left\{ \left[ 
\mathfrak{f}_{0},\mathfrak{f}_{2}\right] \right\} =-\left[ \mathfrak{f}_{1},%
\mathfrak{f}_{1}\right] ,  \label{sJe7}
\end{equation}%
\begin{equation}
2\func{Re}\left\{ \left[ \mathfrak{f}_{1},\mathfrak{f}_{2}\right] \right\} =-%
\left[ \mathfrak{f}_{1},\mathfrak{f}_{1}\right] .  \label{sJe8}
\end{equation}

Let us show now that%
\begin{equation}
\left[ \mathfrak{f}_{1},\mathfrak{f}_{1}\right] \neq0.  \label{ee1}
\end{equation}
Indeed, assume for the sake of the argument that $\left[ \mathfrak{f}_{1},%
\mathfrak{f}_{1}\right] =0$. Then, in view of (\ref{sJe1})-(\ref{sJe3}) and (%
\ref{sJe5}) we have%
\begin{equation}
\left[ \mathfrak{f}_{0},\mathfrak{f}_{j}\right] =0,\ j=0,1,2,3,  \label{ee2}
\end{equation}
or, in other words,%
\begin{equation}
\left( J\mathfrak{f}_{0},\mathfrak{f}_{j}\right) ,\ j=0,1,2,3.  \label{ee3}
\end{equation}
Since $\mathfrak{f}_{0}$, $\mathfrak{f}_{1}$, $\mathfrak{f}_{2}$, $\mathfrak{%
f}_{3}$ is a basis in the $4$-dimensional space the relations (\ref{ee3})
imply that $J\mathfrak{f}_{0}=0$, and, consequently, that $\mathfrak{f}%
_{0}=0 $ since evidently $J$ is an invertible matrix. $\mathfrak{f}_{0}=0$
is impossible, and we must conclude that the relation (\ref{ee1}) holds.

\emph{Since }$\left[ \mathfrak{f}_{1},\mathfrak{f}_{1}\right] $\emph{\ is
the flux corresponding to the Floquet mode described by }$f_{1}$\emph{\ the
relation (\ref{ee1}) signifies a fundamental fact that the Floquet mode
described by }$f_{1}$\emph{\ has nonzero flux.}

Observe that (\ref{sJe7}), (\ref{sJe8}) and (\ref{ee1}) imply 
\begin{equation}
2\func{Re}\left\{ \left[ \mathfrak{f}_{1},\mathfrak{f}_{2}\right] \right\} =%
\func{Re}\left\{ \left[ \mathfrak{f}_{0},\mathfrak{f}_{2}\right] \right\} =-%
\left[ \mathfrak{f}_{1},\mathfrak{f}_{1}\right] \neq 0.  \label{ee4}
\end{equation}%
Notice also that in view of (\ref{sJe1}), (\ref{sJe2}) we have%
\begin{equation}
\left[ \alpha f_{0}+\beta f_{1},\alpha f_{0}+\beta f_{1}\right] =\left\vert
\beta \right\vert ^{2}\left[ f_{1},f_{1}\right] ,  \label{uff1}
\end{equation}%
\begin{align}
& \left[ f_{2}+\alpha f_{0}+\beta f_{1},f_{2}+\alpha f_{0}+\beta f_{1}\right]
\label{uff1a} \\
& =\left[ f_{2},f_{2}\right] -2\func{Re}\left\{ \alpha \right\} \left[
f_{1},f_{1}\right] +2\func{Re}\left\{ \beta \left[ f_{2},f_{1}\right]
\right\} +\left\vert \beta \right\vert ^{2}\left[ f_{1},f_{1}\right] , 
\notag
\end{align}%
and, hence, we have:%
\begin{equation}
\text{as }u\text{ runs over }\limfunc{Span}\left( f_{0},f_{1}\right) \text{ }%
\frac{\left[ u,u\right] }{\left[ f_{1},f_{1}\right] }\text{, then runs over }%
\left[ 0,+\infty \right) ,  \label{uff2}
\end{equation}%
\begin{equation}
\text{as }u\text{ runs over }\limfunc{Span}\left( f_{0},f_{1},f_{2}\right) 
\text{ }\frac{\left[ u,u\right] }{\left[ f_{1},f_{1}\right] }\text{, then
runs over }\left( -\infty ,+\infty \right) .  \label{uff2a}
\end{equation}%
The relation (\ref{uff2a}) follows from (\ref{uff1a}) if we set $\beta =0$
and let $\alpha $ run over all real values $\left( -\infty ,+\infty \right) $%
. In other words, for all vectors $u$ from the $\limfunc{Span}\left(
f_{0},f_{1}\right) $ the corresponding fluxes have the same sign, whereas in
the case of $\limfunc{Span}\left( f_{0},f_{1},f_{2}\right) $ the flux can be
any real number.

\subsection{Transfer matrix and the fluxes for a degenerate point of order 4}

In the case of a degenerate point of order 4 the transfer matrix becomes a
Jordan block of rank 4 and according to (\ref{tom15}) there exists a basis $%
\mathfrak{f}_{j}$, $j=0,1,2,3$ in $%
\mathbb{C}
^{4}$ for which we have 
\begin{equation}
\mathcal{T}_{0}\mathfrak{f}_{0}=\zeta _{0}\mathfrak{f}_{0},\ \mathcal{T}_{0}%
\mathfrak{f}_{1}=\zeta _{0}\mathfrak{f}_{1}+\zeta _{0}\mathfrak{f}_{0},\ 
\mathcal{T}_{0}\mathfrak{f}_{2}=\zeta _{0}\mathfrak{f}_{2}+\zeta _{0}%
\mathfrak{f}_{1},  \label{tzf1}
\end{equation}%
\begin{equation}
\mathcal{T}_{0}\mathfrak{f}_{3}=\zeta _{0}\mathfrak{f}_{3}+\zeta _{0}%
\mathfrak{f}_{2}.  \label{tzf2}
\end{equation}%
Notice that the three equations (\ref{tzf1}) are exactly the same as the
three equations (\ref{tz5}) for the inflection point. Hence, the identities (%
\ref{sJe1})-(\ref{sJe4}) in this case are%
\begin{align}
\left[ \mathfrak{f}_{0},\mathfrak{f}_{0}\right] & =0,  \label{tzf3} \\
\left[ \mathfrak{f}_{0},\mathfrak{f}_{1}\right] & =0,  \label{tzf4} \\
\left[ \mathfrak{f}_{0},\mathfrak{f}_{2}\right] +\left[ \mathfrak{f}_{1},%
\mathfrak{f}_{1}\right] & =0,  \label{tzf5} \\
\left[ \mathfrak{f}_{1},\mathfrak{f}_{2}\right] +\left[ \mathfrak{f}_{2},%
\mathfrak{f}_{1}\right] +\left[ \mathfrak{f}_{1},\mathfrak{f}_{1}\right] &
=0.  \label{tzf6}
\end{align}%
Using now (\ref{tz6}), (\ref{tzf1}) and (\ref{tzf2}) we obtain%
\begin{equation}
\left[ \mathfrak{f}_{0},\mathfrak{f}_{3}\right] =\left[ \mathcal{T}_{0}%
\mathfrak{f}_{0},\mathcal{T}_{0}\mathfrak{f}_{3}\right] =\left[ \mathfrak{f}%
_{0},\mathfrak{f}_{3}+\mathfrak{f}_{2}\right] ,  \label{tzf7}
\end{equation}%
implying%
\begin{equation}
\left[ \mathfrak{f}_{0},\mathfrak{f}_{2}\right] =0,  \label{tzf8}
\end{equation}%
which together with (\ref{tzf5}) and (\ref{tzf6}) yields%
\begin{equation}
\left[ \mathfrak{f}_{1},\mathfrak{f}_{1}\right] =0,\ \left[ \mathfrak{f}_{1},%
\mathfrak{f}_{2}\right] +\left[ \mathfrak{f}_{2},\mathfrak{f}_{1}\right] =0.
\label{tzf9}
\end{equation}

Observe an important difference of the case of a degenerate band edge
compared to the case of an inflection point. Namely, as follows from (\ref%
{tzf9}) in the case of a degenerate point of order 4 we have $\left[ 
\mathfrak{f}_{1},\mathfrak{f}_{1}\right] =0$ where in the case of an
inflection point, according to (\ref{ee1}), $\left[ \mathfrak{f}_{1},%
\mathfrak{f}_{1}\right] \neq 0$.

Using again (\ref{tz6}) together with (\ref{tzf1}), (\ref{tzf2}) and (\ref%
{tzf8}) we get%
\begin{equation}
\left[ \mathfrak{f}_{1},\mathfrak{f}_{3}\right] =\left[ \mathcal{T}_{0}%
\mathfrak{f}_{1},\mathcal{T}_{0}\mathfrak{f}_{3}\right] =\left[ \mathfrak{f}%
_{1}+\mathfrak{f}_{0},\mathfrak{f}_{3}+\mathfrak{f}_{2}\right] =\left[ 
\mathfrak{f}_{1},\mathfrak{f}_{3}\right] +\left[ \mathfrak{f}_{0},\mathfrak{f%
}_{3}\right] +\left[ \mathfrak{f}_{1},\mathfrak{f}_{2}\right] +\left[ 
\mathfrak{f}_{0},\mathfrak{f}_{2}\right] ,  \label{tzf10}
\end{equation}%
readily implying%
\begin{equation}
\left[ \mathfrak{f}_{0},\mathfrak{f}_{3}\right] +\left[ \mathfrak{f}_{1},%
\mathfrak{f}_{2}\right] =0.  \label{tzf11}
\end{equation}%
Summarizing (\ref{tzf3})-(\ref{tzf6}), (\ref{tzf8}), (\ref{tzf9}), (\ref%
{tzf11}) 
\begin{equation}
\left[ \mathfrak{f}_{0},\mathfrak{f}_{0}\right] =0,\ \left[ \mathfrak{f}_{0},%
\mathfrak{f}_{1}\right] =0,\ \left[ \mathfrak{f}_{1},\mathfrak{f}_{1}\right]
=0,\ \left[ \mathfrak{f}_{0},\mathfrak{f}_{2}\right] =0,  \label{tzf12}
\end{equation}%
and%
\begin{equation}
\left[ \mathfrak{f}_{1},\mathfrak{f}_{2}\right] +\left[ \mathfrak{f}_{2},%
\mathfrak{f}_{1}\right] =0,\ \left[ \mathfrak{f}_{0},\mathfrak{f}_{3}\right]
+\left[ \mathfrak{f}_{1},\mathfrak{f}_{2}\right] =0.  \label{tzf13}
\end{equation}%
Notice that the first identity in (\ref{tzf13}) implies that $\left[ 
\mathfrak{f}_{1},\mathfrak{f}_{2}\right] $ is pure imaginary, i.e.$\ $ 
\begin{equation}
\func{Re}\left\{ \left[ \mathfrak{f}_{1},\mathfrak{f}_{2}\right] \right\}
=0,\ \left[ \mathfrak{f}_{1},\mathfrak{f}_{2}\right] =\mathrm{i}\func{Im}%
\left[ \mathfrak{f}_{1},\mathfrak{f}_{2}\right] .  \label{tzf13a}
\end{equation}%
In particular, the first three identities in (\ref{tzf12}) imply that%
\begin{equation}
\text{for any }\mathfrak{f}\in \limfunc{Span}\left\{ \mathfrak{f}_{0},%
\mathfrak{f}_{1}\right\} :\left[ \mathfrak{f},\mathfrak{f}\right] =0.
\label{tzf14}
\end{equation}%
As we have already pointed out this behavior of fluxes reflected by (\ref%
{tzf14}) is very different from the case of an inflection point for which
always $\left[ \mathfrak{f}_{1},\mathfrak{f}_{1}\right] \neq 0$.

\subsection{Transfer matrix and the fluxes for a degenerate point of order 2}

In the case of a degenerate point of order 2 the transfer matrix has a
Jordan block of rank 2 and according to (\ref{tom15}) there exists a basis $%
\mathfrak{f}_{j}$, $j=0,1,2,3$ in $%
\mathbb{C}
^{4}$ for which we have 
\begin{equation}
\mathcal{T}_{0}\mathfrak{f}_{0}=\zeta _{0}\mathfrak{f}_{0},\ \mathcal{T}_{0}%
\mathfrak{f}_{1}=\zeta _{0}\mathfrak{f}_{1}+\zeta _{0}\mathfrak{f}_{0},\
\left\vert \zeta _{0}\right\vert =1,  \label{fzz1}
\end{equation}%
\begin{equation}
\mathcal{T}_{0}\mathfrak{f}_{2}=\zeta \mathfrak{f}_{2},\ \zeta \neq \zeta
_{0},  \label{fzz2}
\end{equation}%
where $\left\vert \zeta \right\vert =1$ or $\left\vert \zeta \right\vert
\neq 1$. There are some additional relations not given here. Notice that for
(\ref{fzz1}) the relation (\ref{tz7}) applies yielding%
\begin{equation}
\left[ \mathfrak{f}_{0},\mathfrak{f}_{0}\right] =0.  \label{fzz3}
\end{equation}%
Observe that since $\zeta \neq \zeta _{0}$, in both cases we have $%
\left\vert \zeta \right\vert =1$ or $\left\vert \zeta \right\vert \neq 1$,
in view of (\ref{[i,j] X}) and (\ref{[i,i] k}), then 
\begin{equation}
\left[ \mathfrak{f}_{0},\mathfrak{f}_{2}\right] =0.  \label{fzz4}
\end{equation}%
Notice that $\zeta \neq \zeta _{0}$ implies $\overline{\zeta _{0}}\zeta \neq
1$. Using (\ref{tz6}) together with (\ref{fzz1}) and (\ref{fzz2}) we obtain 
\begin{equation}
\left[ \mathfrak{f}_{1},\mathfrak{f}_{2}\right] =\left[ \mathcal{T}_{0}%
\mathfrak{f}_{1},\mathcal{T}_{0}\mathfrak{f}_{2}\right] =\overline{\zeta _{0}%
}\zeta \left[ \mathfrak{f}_{1}+\mathfrak{f}_{0},\mathfrak{f}_{2}\right] =%
\overline{\zeta _{0}}\zeta \left[ \mathfrak{f}_{1},\mathfrak{f}_{2}\right] ,
\label{fzz5}
\end{equation}%
implying, in view of $\overline{\zeta _{0}}\zeta \neq 1$,%
\begin{equation}
\left[ \mathfrak{f}_{1},\mathfrak{f}_{2}\right] =0.  \label{fzz6}
\end{equation}%
Notice that (\ref{tz12}) applied in this case yielding%
\begin{equation}
\left[ \mathfrak{f}_{0},\mathfrak{f}_{1}\right] +\left[ \mathfrak{f}_{1},%
\mathfrak{f}_{0}\right] =0.  \label{fzz7}
\end{equation}%
Evidently,%
\begin{equation}
\func{Re}\left[ \mathfrak{f}_{0},\mathfrak{f}_{1}\right] =0,\ \left[ 
\mathfrak{f}_{0},\mathfrak{f}_{1}\right] =\mathrm{i}\func{Im}\left[ 
\mathfrak{f}_{0},\mathfrak{f}_{1}\right] .  \label{fzz7a}
\end{equation}

\section{Perturbation theory for the matrix of reflection coefficients.}

In Section 6 we have introduce and studied the matrix $\rho$ of reflection
coefficients and its relation to the space $S_{\limfunc{T}}\left( 0;\omega,%
\mathbf{k}_{\tau}\right) $. In this section we study the behavior of the
matrix $\rho$ at frequencies $\omega$ close to the frequency of a degenerate
point $\omega_{0}$, i.e. as $\nu=\omega-\omega_{0}\rightarrow0$. To do that
we first describe the space $S_{\limfunc{T}}\left( 0;\omega,\mathbf{k}%
_{\tau}\right) $ by the formula (\ref{zpz15}) where the vectors $\Phi_{1}$
and $\Phi_{2}$ depend on the frequency $\nu$. As we know by now that this
dependence has the form (see (\ref{nv1}) and the section on the perturbation
theory, and also (\ref{zaa3}))%
\begin{align}
\Phi_{j}\left( \nu\right) & =\Phi_{j0}+\Phi_{j0}\acute{\nu}++O\left( \acute{%
\nu}^{2}\right) ,\ j=1,2,  \label{php1} \\
\acute{\nu} & =-\mathrm{i}\tilde{\nu}^{\frac{1}{n}}=\alpha_{0}\nu^{\frac {1}{%
n}}+O\left( \nu^{\frac{2}{n}}\right) ,\ \alpha_{0}=\left[ \frac {n!}{%
\omega^{\left( n\right) }\left( k_{0}\right) }\right] ^{\frac{1}{n}},  \notag
\end{align}
where $n=2,3,4$ is the degeneracy index. To get an expansion for\ $%
\rho\left( \nu\right) $ we use the relations (\ref{phs1})-(\ref{phs7}).
First we need obtain an expansion for the matrices $Q^{\pm}\left( \nu\right) 
$

\begin{equation}
Q^{\pm }\left( \nu \right) =Q_{0}^{\pm }+Q_{1}^{\pm }\acute{\nu}+O\left( 
\acute{\nu}^{2}\right) ,\ \text{where }Q_{0}^{\pm }=Q^{\pm }\left( 0\right) ,
\label{php2}
\end{equation}%
based on (\ref{phs2}) and (\ref{phs4}). Notice that according to (\ref{phs2}%
) we have 
\begin{equation}
Q_{0}^{\pm }=Q^{\pm }\left( 0\right) =\left[ Z_{1}^{+}Z_{2}^{+}\right]
^{\dag }\left[ \Phi _{1}\left( 0\right) \Phi _{2}\left( 0\right) \right] =%
\frac{1}{\beta _{\omega ,\mathbf{k}_{\tau }}}\left[ 
\begin{array}{cc}
\left( Z_{1}^{\pm },\Phi _{1}\left( 0\right) \right) & \left( Z_{1}^{\pm
},\Phi _{2}\left( 0\right) \right) \\ 
\left( Z_{2}^{\pm },\Phi _{1}\left( 0\right) \right) & \left( Z_{2}^{\pm
},\Phi _{2}\left( 0\right) \right)%
\end{array}%
\right] .  \label{php2a}
\end{equation}%
We assume that the vectors $\Phi _{1}\left( \nu \right) $ and $\Phi
_{2}\left( \nu \right) $ are chosen so that for $\nu =0$ they are linearly
independent, i.e.%
\begin{equation}
\left\{ \Phi _{1}\left( 0\right) ,\Phi _{2}\left( 0\right) \right\} =\left\{
\Phi _{10},\Phi _{20}\right\} \text{ are linearly independent.}  \label{php3}
\end{equation}%
The fulfillment of the condition (\ref{php3}) allows the limit space $%
\limfunc{Span}\left\{ \Phi _{1}\left( \nu \right) ,\Phi _{2}\left( \nu
\right) \right\} $ as $\nu \rightarrow 0$ to be described as the
two-dimensional space $\limfunc{Span}\left\{ \Phi _{1}\left( 0\right) ,\Phi
_{2}\left( 0\right) \right\} $. It is also necessary for the invertibility
of the matrix $Q^{\pm }\left( 0\right) $ defined by (\ref{php2a}), i.e. for 
\begin{equation}
\det Q^{\pm }\left( 0\right) \neq 0.  \label{php3a}
\end{equation}%
In fact, we should always have%
\begin{equation}
\det Q^{\pm }\left( \nu \right) \neq 0\text{ for any }\nu ,  \label{php3b}
\end{equation}%
for any semi-infinite slab problem.

Based on the above, we get the following asymptotic expansions for $\left[
Q^{+}\left( \nu \right) \right] ^{-1}$ and $\rho \left( \nu \right) $ 
\begin{align}
\left[ Q^{+}\left( \nu \right) \right] ^{-1}& =\left[ Q_{0}^{+}\right]
^{-1}-Q_{0}^{+}Q_{1}^{+}\left[ Q_{0}^{+}\right] ^{-1}\acute{\nu}+O\left( 
\acute{\nu}^{2}\right) ,  \label{php4} \\
\rho \left( \nu \right) & =Q^{-}\left( \nu \right) \left[ Q^{+}\left( \nu
\right) \right] ^{-1}=\left\{ Q_{0}^{-}+Q_{1}^{-}\acute{\nu}+O\left( \acute{%
\nu}^{2}\right) \right\} \left\{ Q_{0}^{+}+Q_{1}^{+}\acute{\nu}+O\left( 
\acute{\nu}^{2}\right) \right\} ^{-1}  \notag \\
& =\left\{ Q_{0}^{-}+Q_{1}^{-}\acute{\nu}+O\left( \acute{\nu}^{2}\right)
\right\} \left\{ \left[ Q_{0}^{+}\right] ^{-1}-Q_{0}^{+}Q_{1}^{+}\left[
Q_{0}^{+}\right] ^{-1}\acute{\nu}+O\left( \acute{\nu}^{2}\right) \right\} 
\notag \\
& =Q_{0}^{-}\left[ Q_{0}^{+}\right] ^{-1}-\left[ Q_{0}^{-}\right]
^{-1}\left\{ Q_{0}^{-}Q_{1}^{-}-Q_{0}^{+}Q_{1}^{+}\right\} \left[ Q_{0}^{+}%
\right] ^{-1}\acute{\nu}+O\left( \acute{\nu}^{2}\right) ,  \notag
\end{align}%
or%
\begin{align}
\rho \left( \nu \right) & =\rho _{0}+\rho _{1}\acute{\nu}+O\left( \acute{\nu}%
^{2}\right) ,\text{ where}  \label{php5} \\
\rho _{0}& =Q_{0}^{-}\left[ Q_{0}^{+}\right] ^{-1},\text{ }\rho _{1}=\left[
Q_{0}^{-}\right] ^{-1}\left\{ Q_{0}^{-}Q_{1}^{-}-Q_{0}^{+}Q_{1}^{+}\right\} %
\left[ Q_{0}^{+}\right] ^{-1}.  \notag
\end{align}%
The relation (\ref{php5}) readily implies%
\begin{equation}
\rho ^{\dag }\left( \nu \right) \rho \left( \nu \right) =\rho _{0}^{\dag
}\rho _{0}+\rho _{0}^{\dag }\rho _{1}\acute{\nu}+\rho _{1}^{\dag }\rho _{0}%
\overline{\acute{\nu}}+O\left( \acute{\nu}^{2}\right)  \label{php6}
\end{equation}%
\begin{align}
r^{2}\left( \alpha ^{+};\nu \right) & =\frac{\left\vert \rho _{0}\alpha
^{+}\right\vert ^{2}}{\left\vert \alpha ^{+}\right\vert ^{2}}+\frac{2\func{Re%
}\left\{ \left( \rho _{1}\alpha ^{+},\rho _{0}\alpha ^{+}\right) \right\} }{%
\left\vert \alpha ^{+}\right\vert ^{2}}\acute{\nu}+O\left( \acute{\nu}%
^{2}\right)  \label{php7} \\
& =r^{2}\left( \alpha ^{+};0\right) +\frac{2\func{Re}\left\{ \left( \rho
_{1}\alpha ^{+},\rho _{0}\alpha ^{+}\right) \right\} }{\left\vert \alpha
^{+}\right\vert ^{2}}\acute{\nu}+O\left( \acute{\nu}^{2}\right)  \notag
\end{align}%
The relation (\ref{php7}) together with (\ref{phs8}) yield the following
expression for the flux associated with incident wave described by $\alpha
^{+}$ 
\begin{align}
& \left[ \Phi \left( \alpha ^{+};\nu \right) ,\Phi \left( \alpha ^{+};\nu
\right) \right]  \label{php8} \\
& =\left( 1-r^{2}\left( \alpha ^{+};\nu \right) \right) \left\vert \alpha
^{+}\right\vert ^{2}=\left( 1-\frac{\left\vert \rho _{0}\alpha
^{+}\right\vert ^{2}}{\left\vert \alpha ^{+}\right\vert ^{2}}-\frac{2\func{Re%
}\left\{ \left( \rho _{1}\alpha ^{+},\rho _{0}\alpha ^{+}\right) \right\} }{%
\left\vert \alpha ^{+}\right\vert ^{2}}\acute{\nu}+O\left( \acute{\nu}%
^{2}\right) \right) .  \notag
\end{align}

\section{Relevant modes near a degenerate point.}

In this section we describe in detail the properties of the space of
relevant modes $\mathcal{S}_{\limfunc{T}}\left( 0;\omega \right) =\mathcal{S}%
_{\limfunc{T}}\left( 0;\omega ,\mathbf{k}_{\tau }\right) $ (suppressing in
the notation its dependence on $\mathbf{k}_{\tau }$) in a vicinity of a
degenerate point for all the three cases, namely, an inflection point, $n=3$%
, and band edges of orders $n=2,4$.

The general framework determining the basic properties of the space $%
\mathcal{S}_{\limfunc{T}}\left( 0;\omega \right) $ has been considered in
the subsections \textquotedblleft Basic properties of the space of relevant
eigenmodes\textquotedblright\ and \textquotedblleft Matrix of reflection
coefficients and the flux quadratic form\textquotedblright . At this point
having investigated the spectral properties of the transfer matrix $\mathcal{%
T}\left( \nu \right) $, $\nu =\omega -\omega _{0}$ at a degenerate point $%
\omega _{0}$ and as $\nu \rightarrow 0$ (see Section \textquotedblleft
Spectral perturbation theory of the transfer matrix a point of
degeneracy\textquotedblright ), we can provide more details of the
properties of $\mathcal{S}_{\limfunc{T}}\left( 0;\omega \right) $ including
the asymptotic behavior of the flux and the reflection coefficients of the
relevant eigenmodes for a semi-infinite slab as $\nu \rightarrow 0$.

\subsection{Relevant modes near an inflection point.}

In this section we study the basic properties of the relevant eigenmodes
and, in particular, the space $\mathcal{S}_{\limfunc{T}}\left(
0;\omega\right) $ as functions of the frequency $\omega$ in a vicinity of an
inflection point $\omega_{0}$, i.e. for $\omega=\omega_{0}+\nu$ when $\nu$
is small.

Using the equalities (\ref{e3nu4}) and (\ref{sJe1})-(\ref{sJe5}) we get the
following asymptotic formulae as $\nu \rightarrow 0$ for the fluxes%
\begin{align}
\left[ \mathfrak{e}_{0}\left( \nu \right) ,\mathfrak{e}_{0}\left( \nu
\right) \right] & =\acute{\nu}^{2}\left[ \mathfrak{f}_{1},\mathfrak{f}_{1}%
\right] -2\acute{\nu}^{2}\func{Re}\left[ \mathfrak{f}_{0},\mathfrak{f}_{2}%
\right] +O\left( \acute{\nu}^{3}\right) =  \label{etaux5} \\
& =3\acute{\nu}^{2}\left[ \mathfrak{f}_{1},\mathfrak{f}_{1}\right] +O\left( 
\acute{\nu}^{3}\right) =3\alpha _{0}^{2}\nu ^{2/3}\left[ \mathfrak{f}_{1},%
\mathfrak{f}_{1}\right] +O\left( \nu \right) .  \notag
\end{align}%
where, in view of (\ref{etv13})-(\ref{etv14}),%
\begin{equation}
\acute{\nu}=\alpha _{0}\nu ^{1/3},\mathrm{i}\mathfrak{t}_{1}=\alpha _{0}^{3}=%
\frac{6}{\omega ^{\prime \prime \prime }\left( k_{0}\right) }.
\label{etaux5a}
\end{equation}

Notice now that the relations (\ref{[i,j] X}-\ref{[i,i] k}), together with (%
\ref{e3nu4}) and (\ref{tz3c}) yield the following formulae for the fluxes%
\begin{align}
\left[ \mathfrak{e}_{1}\left( \nu \right) ,\mathfrak{e}_{1}\left( \nu
\right) \right] & =\left[ \mathfrak{e}_{2}\left( \nu \right) ,\mathfrak{e}%
_{2}\left( \nu \right) \right] =0,\ \left[ \mathfrak{e}_{3}\left( \nu
\right) ,\mathfrak{e}_{3}\left( \nu \right) \right] =\left[ \mathfrak{f}_{3},%
\mathfrak{f}_{3}\right] +O\left( \nu \right)  \label{etaux6} \\
\left[ \mathfrak{e}_{0}\left( \nu \right) ,\mathfrak{e}_{1}\left( \nu
\right) \right] & =\left[ \mathfrak{e}_{0}\left( \nu \right) ,\mathfrak{e}%
_{2}\left( \nu \right) \right] =\left[ \mathfrak{e}_{0}\left( \nu \right) ,%
\mathfrak{e}_{3}\left( \nu \right) \right] =0,  \notag \\
\left[ \mathfrak{e}_{3}\left( \nu \right) ,\mathfrak{e}_{0}\left( \nu
\right) \right] & =\left[ \mathfrak{e}_{3}\left( \nu \right) ,\mathfrak{e}%
_{1}\left( \nu \right) \right] =\left[ \mathfrak{e}_{3}\left( \nu \right) ,%
\mathfrak{e}_{2}\left( \nu \right) \right] =0.  \notag
\end{align}

To handle in a uniform fashion both positive and negative $\nu $ we introduce%
\begin{align}
\mathfrak{e}_{1}^{\sharp }\left( \nu \right) & =\mathfrak{e}_{\limfunc{sign}%
\nu }\left( \nu \right) =\left\{ 
\begin{array}{ccc}
\mathfrak{e}_{+}\left( \nu \right) & \text{if} & \nu \geq 0 \\ 
\mathfrak{e}_{-}\left( \nu \right) & \text{if} & \nu <0%
\end{array}%
\right. ,\ \theta _{1}^{\sharp }\left( \nu \right) =\left\{ 
\begin{array}{ccc}
\theta _{1}\left( \nu \right) & \text{if} & \nu \geq 0 \\ 
\theta _{2}\left( \nu \right) & \text{if} & \nu <0%
\end{array}%
\right. .  \label{etaux7} \\
\mathfrak{e}_{+}\left( \nu \right) & =\mathfrak{e}_{1}\left( \nu \right) ,\ 
\mathfrak{e}_{-}\left( \nu \right) =\mathfrak{e}_{2}\left( \nu \right) . 
\notag
\end{align}%
Then it follows from (\ref{etv15}), (\ref{Fj4}) that%
\begin{equation}
\left\vert \theta _{1}^{\sharp }\left( \nu \right) \right\vert =e^{-\frac{%
\sqrt{3}}{2}\left\vert \acute{\nu}\right\vert }\left( 1+O\left( \acute{\nu}%
^{2/3}\right) \right) .  \label{etaux7a}
\end{equation}%
Notice that,\ as follows from (\ref{etaux5}), the vector $\mathfrak{e}%
_{0}\left( \nu \right) $ has a positive flux and, hence, corresponds to a
propagating mode. As to $\mathfrak{e}_{1}^{\sharp }\left( \nu \right) $, in
view of (\ref{etaux7a}), it corresponds to an evanescent mode decaying as $%
x_{3}\rightarrow \infty $. So, based on (\ref{si3b}), we obtain%
\begin{equation}
\mathcal{S}_{\limfunc{T}}\left( 0;\omega _{0}+\nu \right) =\limfunc{Span}%
\left\{ \mathfrak{e}_{0}\left( \nu \right) ,\mathfrak{e}_{1}^{\sharp }\left(
\nu \right) \right\} .  \label{etaux7b}
\end{equation}%
Then using (\ref{e3nu9}) one verifies that the following limit exists%
\begin{equation}
\mathcal{S}_{\limfunc{T}}\left( 0;\omega _{0}\right) =\lim_{\nu \rightarrow
0}\mathcal{S}_{\limfunc{T}}\left( 0;\omega _{0}+\nu \right) =\limfunc{Span}%
\left\{ \mathfrak{f}_{0},\mathfrak{f}_{1}\right\} .  \label{etaux7c}
\end{equation}%
Observe that the representation (\ref{etaux7c}) for the space $\mathcal{S}_{%
\limfunc{T}}\left( 0;\omega _{0}\right) $ together with (\ref{ee1}) and (\ref%
{uff1}) yield%
\begin{equation}
\left[ \beta _{0}\mathfrak{f}_{0}+\beta _{1}\mathfrak{f}_{1},\beta _{0}%
\mathfrak{f}_{0}+\beta _{1}\mathfrak{f}_{1}\right] =\left\vert \beta
_{1}\right\vert ^{2}\left[ \mathfrak{f}_{1},\mathfrak{f}_{1}\right] \neq 0%
\text{ if }\beta _{1}\neq 0.  \label{etaux7d}
\end{equation}%
The relation (\ref{etaux7d}) combined with (\ref{zpz14b}) imply the
following very important property of the reflection coefficient $r\left(
\alpha ^{+};0\right) $ at the inflection point $\omega _{0}$, i.e. $\nu =0$, 
\begin{equation}
\text{for almost all }\alpha ^{+}\in 
\mathbb{C}
^{2}:\text{ the reflection coefficient }r\left( \alpha ^{+};0\right) <1.
\label{etaux7e}
\end{equation}%
\emph{The relation (\ref{etaux7e}) clearly indicates that the reflection
coefficients }$r\left( \alpha ^{+};0\right) $\emph{\ are always strictly
less than }$1$\emph{\ for all the relevant eigenmodes of the semi-infinite
periodic stack, with the only exception when EM field value of the eigenmode
at the surface of the slab is }$f_{0}$\emph{. In other words, at an
inflection point there always will be a positive fraction of the incident
energy transmitted through the infinite slab. In fact, by proper design of
the slab one can achieve almost 100\% transmission of the incident energy.
In contrast, at any band edges the transmission is always exactly zero and
the reflection is always 100\%, as we will see from the analysis in the
following sections.}

More elaborate analysis yields asymptotic expressions for the matrix of
reflection coefficients $\rho $, as determined by (\ref{phs4}) and (\ref%
{php5}), and other related quantities for nonzero by small $\nu =\omega
-\omega _{0}$. Indeed, let us use in the relations (\ref{phs2})-(\ref{phs4})
the vectors $\Phi _{1}$ and $\Phi _{2}$ defined by%
\begin{equation}
\Phi _{1}\left( \nu \right) =\mathfrak{e}_{0}\left( \nu \right) ,\ \Phi
_{2}\left( \nu \right) =\mathfrak{h}^{\sharp }\left( \nu \right) =\frac{%
\mathfrak{e}_{\limfunc{sign}\left( \acute{\nu}\right) }\left( \nu \right) -%
\mathfrak{e}_{0}\left( \nu \right) }{\mathrm{i}\acute{\nu}\left( \varsigma _{%
\limfunc{sign}\left( \acute{\nu}\right) }-1\right) }.  \label{ff31}
\end{equation}%
Notice that (\ref{e3nu3}) and (\ref{e3nu7}) yield the following
representation%
\begin{align}
\mathfrak{e}_{0}\left( \nu \right) & =\left( 1+\mathrm{i}\tau _{2}\acute{\nu}%
+O\left( \acute{\nu}^{2}\right) \right) \mathfrak{f}_{0}+\left( \mathrm{i}%
\acute{\nu}+\mathrm{i}\tau _{1}\acute{\nu}^{2}\right) \mathfrak{f}_{1}-%
\acute{\nu}^{2}\mathfrak{f}_{2}+O\left( \acute{\nu}^{3}\right) ,
\label{ff32} \\
\mathfrak{h}^{\sharp }\left( \nu \right) & =\left[ \tau _{2}+O\left( \acute{%
\nu}\right) \right] \mathfrak{f}_{0}+\left[ 1+\frac{\varsigma _{-\limfunc{%
sign}\left( \acute{\nu}\right) }-\mathrm{i}}{\mathrm{i}\left( \varsigma _{%
\limfunc{sign}\left( \acute{\nu}\right) }-1\right) }\acute{\nu}\right] 
\mathfrak{f}_{1}+\frac{1-\varsigma _{-\limfunc{sign}\left( \acute{\nu}%
\right) }}{\mathrm{i}\left( \varsigma _{\limfunc{sign}\left( \acute{\nu}%
\right) }-1\right) }\acute{\nu}\mathfrak{f}_{2}+O\left( \acute{\nu}%
^{2}\right) ,  \notag \\
\tau _{2}& =\frac{\left[ T_{1}\right] _{32}}{3\zeta _{0}\mathfrak{t}_{1}},\ 
\mathrm{i}\mathfrak{t}_{1}=\alpha _{0}^{3}=\frac{6}{\omega ^{\prime \prime
\prime }\left( k_{0}\right) }.  \notag
\end{align}%
In particular, (\ref{ff31}) and (\ref{ff32}) yield for $\nu =0$%
\begin{equation}
\Phi _{1}\left( 0\right) =\mathfrak{f}_{0},\ \Phi _{1}\left( 0\right) =\tau
_{2}\mathfrak{f}_{0}+\mathfrak{f}_{1},  \label{ff33}
\end{equation}%
implying that%
\begin{equation}
\left\{ \Phi _{1}\left( 0\right) ,\Phi _{2}\left( 0\right) \right\} \text{
are linearly independent.}  \label{ff34}
\end{equation}%
The relation (\ref{ff34}) implies that the condition (\ref{php3}) is
satisfied.

Now, let find the value $\Phi \left( \alpha ^{+};\nu \right) $ of the
eigenmode corresponding to the incident wave $\alpha ^{+}$. Using (\ref{phs1}%
), (\ref{phs3a}), (\ref{php4}) we consequently obtain 
\begin{equation}
\check{\Phi}\left( \alpha ^{+};\nu \right) =\left[ 
\begin{array}{c}
\varphi _{1}\left( \nu \right) \\ 
\varphi _{2}\left( \nu \right)%
\end{array}%
\right] =\left[ 
\begin{array}{c}
\varphi _{1}\left( 0\right) \\ 
\varphi _{2}\left( 0\right)%
\end{array}%
\right] +O\left( \acute{\nu}\right) ,\ \left[ 
\begin{array}{c}
\varphi _{1}\left( 0\right) \\ 
\varphi _{2}\left( 0\right)%
\end{array}%
\right] =\left[ Q^{+}\left( 0\right) \right] ^{-1}\alpha ^{+}.  \label{ff35}
\end{equation}%
\begin{align}
& \Phi \left( \alpha ^{+};\nu \right)  \label{ff36} \\
& =\varphi _{1}\left( \nu \right) \Phi _{1}\left( \nu \right) +\varphi
_{2}\left( \nu \right) \Phi _{2}\left( \nu \right) =\varphi _{1}\left( \nu
\right) \mathfrak{e}_{0}\left( \nu \right) +\varphi _{2}\left( \nu \right) 
\frac{\mathfrak{e}_{\limfunc{sign}\left( \acute{\nu}\right) }\left( \nu
\right) -\mathfrak{e}_{0}\left( \nu \right) }{\mathrm{i}\acute{\nu}\left(
\varsigma _{\limfunc{sign}\left( \acute{\nu}\right) }-1\right) }  \notag \\
& =\frac{\varphi _{2}\left( 0\right) }{\mathrm{i}\acute{\nu}\left( \varsigma
_{\limfunc{sign}\left( \acute{\nu}\right) }-1\right) }\left[ \mathfrak{e}_{%
\limfunc{sign}\left( \acute{\nu}\right) }\left( \nu \right) -\mathfrak{e}%
_{0}\left( \nu \right) \right] +O\left( 1\right) .  \notag
\end{align}%
Observe that the decomposition (\ref{ff36}) of the vector $\Phi \left(
\alpha ^{+};\nu \right) $ into a linear combination of eigenvectors $%
\mathfrak{e}_{0}\left( \nu \right) $ and $\mathfrak{e}_{\limfunc{sign}\left( 
\acute{\nu}\right) }\left( \nu \right) $ of the transfer matrix $\mathcal{T}%
\left( \nu \right) $ signifies that%
\begin{equation}
\text{the amplitude of the eigenmode inside the slab is }\frac{\varphi
_{2}\left( 0\right) }{\mathrm{i}\acute{\nu}\left( \varsigma _{\limfunc{sign}%
\left( \acute{\nu}\right) }-1\right) }+O\left( 1\right) .  \label{ff37}
\end{equation}%
Combining (\ref{etaux5}), (\ref{etaux6}) with (\ref{ff32}), (\ref{ff35}) we
get the following formula for the flux%
\begin{align}
& \left[ \Phi \left( \alpha ^{+};\nu \right) ,\Phi \left( \alpha ^{+};\nu
\right) \right]  \label{ff38} \\
& =\left\vert \frac{\varphi _{2}\left( 0\right) }{\varsigma _{+}-1}%
\right\vert ^{2}\left[ \mathfrak{f}_{1},\mathfrak{f}_{1}\right] +O\left( 
\acute{\nu}\right) =\frac{\left\{ \left[ Q^{+}\left( 0\right) \right]
^{-1}\alpha ^{+}\right\} _{2}^{2}}{3}\left[ \mathfrak{f}_{1},\mathfrak{f}_{1}%
\right] +O\left( \acute{\nu}\right) .  \notag
\end{align}%
The formula (\ref{ff38}) together with (\ref{phs9}) yield%
\begin{equation}
t^{2}\left( \alpha ^{+};\nu \right) =1-r^{2}\left( \alpha ^{+};\nu \right) =%
\frac{\left\{ \left[ Q^{+}\left( 0\right) \right] ^{-1}\alpha ^{+}\right\}
_{2}^{2}}{3\left\vert \alpha ^{+}\right\vert ^{2}}\left[ \mathfrak{f}_{1},%
\mathfrak{f}_{1}\right] +O\left( \acute{\nu}\right) .  \label{ff39}
\end{equation}%
More accurate computation based on (\ref{php8}), (\ref{php5}) and (\ref%
{etaux7e}) implies the following asymptotic formulae for the transmission
and reflection coefficients%
\begin{align}
t^{2}\left( \alpha ^{+};\nu \right) & =1-r^{2}\left( \alpha ^{+};\nu \right)
=1-\frac{\left\vert \rho _{0}\alpha ^{+}\right\vert ^{2}}{\left\vert \alpha
^{+}\right\vert ^{2}}-\frac{2\func{Re}\left\{ \left( \rho _{1}\alpha
^{+},\rho _{0}\alpha ^{+}\right) \right\} }{\left\vert \alpha
^{+}\right\vert ^{2}}\acute{\nu}+O\left( \acute{\nu}^{2}\right) ,
\label{ff310} \\
r^{2}\left( \alpha ^{+};\nu \right) & =\frac{\left\vert \rho _{0}\alpha
^{+}\right\vert ^{2}}{\left\vert \alpha ^{+}\right\vert ^{2}}+\frac{2\func{Re%
}\left\{ \left( \rho _{1}\alpha ^{+},\rho _{0}\alpha ^{+}\right) \right\} }{%
\left\vert \alpha ^{+}\right\vert ^{2}}\acute{\nu}+O\left( \acute{\nu}%
^{2}\right) ,\ \frac{\left\vert \rho _{0}\alpha ^{+}\right\vert ^{2}}{%
\left\vert \alpha ^{+}\right\vert ^{2}}<1,  \notag \\
\rho _{0}& =Q_{0}^{-}\left[ Q_{0}^{+}\right] ^{-1},\ \rho _{1}=\left[
Q_{0}^{-}\right] ^{-1}\left\{ Q_{0}^{-}Q_{1}^{-}-Q_{0}^{+}Q_{1}^{+}\right\} %
\left[ Q_{0}^{+}\right] ^{-1}.  \notag
\end{align}%
Observe that the formulae (\ref{ff310}) involve the matrix $\rho _{1}$
requiring more terms in the expressions for the eigenvectors $\mathfrak{e}%
_{0}\left( \nu \right) $ and $\mathfrak{e}_{\pm }\left( \nu \right) $
(namely we need to compute the matrix $\Lambda _{2}$ as defined in the
Appendix 2). When the exact value of the matrix $\rho _{1}$ is found we can
find the exact value of the coefficient $\frac{2\func{Re}\left\{ \left( \rho
_{1}\alpha ^{+},\rho _{0}\alpha ^{+}\right) \right\} }{\left\vert \alpha
^{+}\right\vert ^{2}}$ in (\ref{ff310}). At this point we are interested in
the concrete value of $\frac{2\func{Re}\left\{ \left( \rho _{1}\alpha
^{+},\rho _{0}\alpha ^{+}\right) \right\} }{\left\vert \alpha
^{+}\right\vert ^{2}}$ and for that reason we have not carried out the
computation of the matrix $\Lambda _{2}$.

\subsection{Relevant modes near a degeneracy point of order 4.}

In this section we study the basic properties of the relevant eigenmodes
and, in particular, the space $\mathcal{S}_{\limfunc{T}}\left(
0;\omega\right) $ as functions of the frequency $\omega$ in a vicinity of a
degenerate point $\omega_{0}$ of order $n=4$, i.e. for $\omega=\omega_{0}+%
\nu $ when $\nu$ is small. Without loss of generality we assume $\nu\geq0$.

Notice that the eigenvector $\mathfrak{e}_{0}\left( \nu\right) $ corresponds
to the eigenvalue $\theta_{1}\left( \nu\right) $ for which%
\begin{equation}
\left\vert \theta_{0}\left( \nu\right) \right\vert =1,  \label{ff40}
\end{equation}
and, hence, the corresponding eigenmode is a propagating one.

Using the equalities (\ref{4enu3}) and (\ref{tzf12})-(\ref{tzf13a}) we get
the following asymptotic formulae as $\nu \rightarrow 0$ for the fluxes%
\begin{align}
\left[ \mathfrak{e}_{0}\left( \nu \right) ,\mathfrak{e}_{0}\left( \nu
\right) \right] & =2\acute{\nu}^{3}\func{Im}\left[ \mathfrak{f}_{1},%
\mathfrak{f}_{2}\right] +O\left( \acute{\nu}^{4}\right) ,  \label{ff41} \\
\tau _{1}& =\frac{\left[ T_{1}\right] _{31}}{8\zeta _{0}\mathfrak{t}_{1}},\
\tau _{2}=\frac{\left[ T_{1}\right] _{42}}{8\zeta _{0}\mathfrak{t}_{1}},\ 
\acute{\nu}=\alpha _{0}\nu ^{1/4},\ \alpha _{0}^{4}=\frac{4!}{\omega
^{\left( 4\right) }\left( k_{0}\right) }.  \notag
\end{align}%
Notice now in view of (\ref{4Fj4}) the eigenvector $\mathfrak{e}_{1}\left(
\nu \right) $ corresponds to the eigenvalue $\theta _{1}\left( \nu \right) $
for which evidently%
\begin{equation}
\left\vert \theta _{1}\left( \nu \right) \right\vert =\left\vert e^{\mathrm{i%
}k_{0}+\mathrm{i}\acute{\nu}\varsigma _{1}}\right\vert \left( 1+O\left( 
\acute{\nu}^{2}\right) \right) =e^{-\acute{\nu}},\ \acute{\nu}\geq 0,
\label{ff41a}
\end{equation}%
and, hence, the corresponding eigenmode is an evanescent one.

In view of (\ref{ff40}) and (\ref{ff41a}), we can use the relations (\ref%
{[i,j] X}) and (\ref{[i,j] k}) yielding%
\begin{equation}
\left[ \mathfrak{e}_{1}\left( \nu \right) ,\mathfrak{e}_{1}\left( \nu
\right) \right] =\left[ \mathfrak{e}_{1}\left( \nu \right) ,\mathfrak{e}%
_{0}\left( \nu \right) \right] =0.  \label{ff42}
\end{equation}%
Hence, as follows from (\ref{ff40}) and (\ref{ff41}) the vectors $\mathfrak{e%
}_{0}\left( \nu \right) $ and $\mathfrak{e}_{1}\left( \nu \right) $
correspond respectively to a propagating and evanescent modes. So based on (%
\ref{si3b}) we get%
\begin{equation}
\mathcal{S}_{\limfunc{T}}\left( 0;\omega _{0}+\nu \right) =\limfunc{Span}%
\left\{ \mathfrak{e}_{0}\left( \nu \right) ,\mathfrak{e}_{1}\left( \nu
\right) \right\} .  \label{ff43}
\end{equation}%
Then using (\ref{4enu8}) one verifies that the following limit exists%
\begin{equation}
\mathcal{S}_{\limfunc{T}}\left( 0;\omega _{0}\right) =\lim_{\nu \rightarrow
0}\mathcal{S}_{\limfunc{T}}\left( 0;\omega _{0}+\nu \right) =\limfunc{Span}%
\left\{ \mathfrak{f}_{0},\mathfrak{f}_{1}\right\} .  \label{ff44}
\end{equation}%
Notice the representation (\ref{ff44}) for the space $\mathcal{S}_{\limfunc{T%
}}\left( 0;\omega _{0}\right) $ together with (\ref{tzf14}) yield%
\begin{equation}
\left[ \mathfrak{f},\mathfrak{f}\right] =0\text{ for any }\mathfrak{f}\in 
\mathcal{S}_{\limfunc{T}}\left( 0;\omega _{0}\right) .  \label{ff45}
\end{equation}%
The relation (\ref{ff45}) combined with (\ref{zpz14}) and (\ref{zpz14aa})
implies the following very important property of the reflection coefficient $%
r\left( \alpha ^{+};0\right) $ at the inflection point $\omega _{0}$, i.e. $%
\nu =0$,%
\begin{equation}
\text{for every }\alpha ^{+}\in 
\mathbb{C}
^{2}\text{ the reflection coefficient }r\left( \alpha ^{+}\right) =1\text{
and }\rho ^{\dag }\rho =I_{2}\text{.}  \label{ff46}
\end{equation}%
\emph{The relation (\ref{ff46}) clearly indicates that the reflection
coefficient }$r\left( \alpha ^{+};0\right) $\emph{\ is always exactly }$1$%
\emph{\ for all the relevant eigenmodes of semi-infinite periodic stack In
other words, at any degenerate point of order }$n=4$\emph{,} \emph{100\% of
the incident energy is reflected and, hence, no energy is transmitted. In
contrast, at any inflection point a positive fraction of the incident energy
is always transmitted.}

More elaborate analysis provides asymptotic expressions for the matrix of
reflection coefficients $\rho $, as determined by (\ref{phs4}) and (\ref%
{php5}), and other related quantities for nonzero, but small, $\nu =\omega
-\omega _{0}$. Indeed, let us use in the relations (\ref{phs2})-(\ref{phs4})
the vectors $\Phi _{1}$ and $\Phi _{2}$ defined by%
\begin{equation}
\Phi _{1}\left( \nu \right) =\mathfrak{e}_{0}\left( \nu \right) ,\ \Phi
_{2}\left( \nu \right) =\mathfrak{h}_{1}\left( \nu \right) =\frac{\mathfrak{e%
}_{0}\left( \nu \right) -\mathfrak{e}_{1}\left( \nu \right) }{\acute{\nu}%
\left( \mathrm{i}+1\right) }.  \label{ff47}
\end{equation}%
Notice that (\ref{e3nu3})-(\ref{e3nu7}) yields the following representation%
\begin{align}
\mathfrak{e}_{0}\left( \nu \right) & =\left( 1+\mathrm{i}\tau _{2}\acute{\nu}%
+O\left( \acute{\nu}^{2}\right) \right) \mathfrak{f}_{0}+\left( \mathrm{i}%
\acute{\nu}+\mathrm{i}\tau _{1}\acute{\nu}^{2}\right) \mathfrak{f}_{1}-%
\acute{\nu}^{2}\mathfrak{f}_{2}+O\left( \acute{\nu}^{3}\right) ,
\label{ff48} \\
\mathfrak{h}_{1}\left( \nu \right) & =\left( \tau _{1}+3\tau _{2}\right) 
\mathfrak{f}_{0}+\mathfrak{f}_{1}+O\left( \acute{\nu}\right) .  \notag
\end{align}%
In particular, (\ref{ff47}) and (\ref{ff48}) yield for $\nu =0$%
\begin{equation*}
\Phi _{1}\left( 0\right) =\mathfrak{f}_{0},\ \Phi _{1}\left( 0\right)
=\left( \tau _{1}+3\tau _{2}\right) \mathfrak{f}_{0}+\mathfrak{f}_{1},
\end{equation*}%
implying that%
\begin{equation}
\left\{ \Phi _{1}\left( 0\right) ,\Phi _{2}\left( 0\right) \right\} \text{
are linearly independent.}  \label{ff49}
\end{equation}%
The relation (\ref{ff49}) shows that condition (\ref{php3}) is satisfied.

Now let find the value $\Phi \left( \alpha ^{+};\nu \right) $ of the
eigenmode corresponding to the incident wave $\alpha ^{+}$. Using (\ref{phs1}%
), (\ref{phs3a}) and (\ref{php4}) we consequently obtain 
\begin{equation}
\check{\Phi}\left( \alpha ^{+};\nu \right) =\left[ 
\begin{array}{c}
\varphi _{1}\left( \nu \right) \\ 
\varphi _{2}\left( \nu \right)%
\end{array}%
\right] =\left[ 
\begin{array}{c}
\varphi _{1}\left( 0\right) \\ 
\varphi _{2}\left( 0\right)%
\end{array}%
\right] +O\left( \acute{\nu}\right) ,\ \left[ 
\begin{array}{c}
\varphi _{1}\left( 0\right) \\ 
\varphi _{2}\left( 0\right)%
\end{array}%
\right] =\left[ Q^{+}\left( 0\right) \right] ^{-1}\alpha ^{+}.  \label{ff410}
\end{equation}%
\begin{align}
& \Phi \left( \alpha ^{+};\nu \right)  \label{ff411} \\
& =\varphi _{1}\left( \nu \right) \Phi _{1}\left( \nu \right) +\varphi
_{2}\left( \nu \right) \Phi _{2}\left( \nu \right) =\varphi _{1}\left( \nu
\right) \mathfrak{e}_{0}\left( \nu \right) +\varphi _{2}\left( \nu \right) 
\frac{\mathfrak{e}_{0}\left( \nu \right) -\mathfrak{e}_{1}\left( \nu \right) 
}{\acute{\nu}\left( \mathrm{i}+1\right) }  \notag \\
& =\frac{\varphi _{2}\left( 0\right) }{\acute{\nu}\left( \mathrm{i}+1\right) 
}\left[ \mathfrak{e}_{0}\left( \nu \right) -\mathfrak{e}_{1}\left( \nu
\right) \right] +O\left( 1\right) .  \notag
\end{align}%
Observe that the decomposition (\ref{ff411}) of the vector $\Phi \left(
\alpha ^{+};\nu \right) $ into a linear combination of eigenvectors $%
\mathfrak{e}_{0}\left( \nu \right) $ and $\mathfrak{e}_{1}\left( \nu \right) 
$ of the transfer matrix $\mathcal{T}\left( \nu \right) $ signifies that%
\begin{equation}
\text{the amplitude of the eigenmode inside the slab is }\frac{\varphi
_{2}\left( 0\right) }{\acute{\nu}\left( \mathrm{i}+1\right) }+O\left(
1\right) .  \label{ff412}
\end{equation}%
Combining (\ref{ff41}) and (\ref{ff42}) with (\ref{ff410}) and (\ref{ff41})
we get the following formula for the flux%
\begin{align}
& \left[ \Phi \left( \alpha ^{+};\nu \right) ,\Phi \left( \alpha ^{+};\nu
\right) \right] =\left\vert \frac{\varphi _{2}\left( 0\right) }{\mathrm{i}+1}%
\right\vert ^{2}2\acute{\nu}\func{Im}\left[ \mathfrak{f}_{1},\mathfrak{f}_{2}%
\right] +O\left( \acute{\nu}^{2}\right)  \label{ff413} \\
& =\left\{ \left[ Q^{+}\left( 0\right) \right] ^{-1}\alpha ^{+}\right\}
_{2}^{2}\func{Im}\left[ \mathfrak{f}_{1},\mathfrak{f}_{2}\right] \acute{\nu}%
+O\left( \acute{\nu}^{2}\right)  \notag
\end{align}%
The formula (\ref{ff413}) together with (\ref{phs9}) yields%
\begin{align}
t^{2}\left( \alpha ^{+};\nu \right) & =1-r^{2}\left( \alpha ^{+};\nu \right)
\label{ff414} \\
& =\frac{\left\{ \left[ Q^{+}\left( 0\right) \right] ^{-1}\alpha
^{+}\right\} _{2}^{2}\func{Im}\left[ \mathfrak{f}_{1},\mathfrak{f}_{2}\right]
\acute{\nu}}{\left\vert \alpha ^{+}\right\vert ^{2}}\acute{\nu}+O\left( 
\acute{\nu}^{2}\right) .  \notag
\end{align}

An alternative computation based on (\ref{php8}), (\ref{php5}) and (\ref%
{ff46}) (which turns into $\left\vert \rho _{0}\alpha ^{+}\right\vert
^{2}=\left\vert \alpha ^{+}\right\vert ^{2}$ for all $\alpha ^{+}$) implies
the following asymptotic formulae for the transmission and reflection
coefficients%
\begin{align}
t^{2}\left( \alpha ^{+};\nu \right) & =1-r^{2}\left( \alpha ^{+};\nu \right)
=-\frac{2\func{Re}\left\{ \left( \rho _{1}\alpha ^{+},\rho _{0}\alpha
^{+}\right) \right\} }{\left\vert \alpha ^{+}\right\vert ^{2}}\acute{\nu}%
+O\left( \acute{\nu}^{2}\right) ,  \label{ff415} \\
r^{2}\left( \alpha ^{+};\nu \right) & =1+\frac{2\func{Re}\left\{ \left( \rho
_{1}\alpha ^{+},\rho _{0}\alpha ^{+}\right) \right\} }{\left\vert \alpha
^{+}\right\vert ^{2}}\acute{\nu}+O\left( \acute{\nu}^{2}\right) ,  \notag \\
\rho _{0}& =Q_{0}^{-}\left[ Q_{0}^{+}\right] ^{-1},\ \rho _{1}=\left[
Q_{0}^{-}\right] ^{-1}\left\{ Q_{0}^{-}Q_{1}^{-}-Q_{0}^{+}Q_{1}^{+}\right\} %
\left[ Q_{0}^{+}\right] ^{-1}.  \notag
\end{align}

\subsection{Relevant modes near a degenerate point of order 2}

In this section we study the basic properties of the relevant eigenmodes
and, in particular, the space $\mathcal{S}_{\limfunc{T}}\left(
0;\omega\right) $ as functions of the frequency $\omega$ in a vicinity of a
degenerate point $\omega_{0}$ of order $n=2$, i.e. for $\omega=\omega_{0}+%
\nu $ when $\nu$ is small. Without loss of generality we assume $\nu\geq0$.

Notice that the eigenvector $\mathfrak{e}_{0}\left( \nu \right) $
corresponds to the eigenvalue $\theta _{1}\left( \nu \right) $ for which%
\begin{equation}
\theta _{0}\left( \nu \right) =e^{\mathrm{i}k_{0}+\mathrm{i}\acute{\nu}%
}\left( 1+O\left( \acute{\nu}^{2}\right) \right) :\left\vert \theta
_{0}\left( \nu \right) \right\vert =1,  \label{ff20}
\end{equation}%
with the corresponding eigenmode propagating in the positive direction.

Using the equalities (\ref{Fj26}), (\ref{fzz3}) and (\ref{fzz6})-(\ref{fzz7a}%
) we get the following asymptotic formulae as $\nu \rightarrow 0$ for the
fluxes%
\begin{equation}
\left[ \mathfrak{e}_{0}\left( \nu \right) ,\mathfrak{e}_{0}\left( \nu
\right) \right] =-2\func{Im}\left[ \mathfrak{f}_{0},\mathfrak{f}_{1}\right] 
\acute{\nu}+O\left( \acute{\nu}^{2}\right) ,\ \acute{\nu}=\alpha _{0}\sqrt{%
\nu },\ \alpha _{0}=\sqrt{\frac{2}{\omega ^{\prime \prime }\left(
k_{0}\right) }}.  \label{ff21}
\end{equation}%
In particular, for $\nu =0$ the equality (\ref{ff21}) implies%
\begin{equation}
\left[ \mathfrak{e}_{0}\left( 0\right) ,\mathfrak{e}_{0}\left( 0\right) %
\right] =0.  \label{ff21b}
\end{equation}%
Notice that in view of (\ref{4Fj4}) the eigenvector $\mathfrak{e}_{1}\left(
\nu \right) $ corresponds to the eigenvalue $\theta _{1}\left( \nu \right) $
for which%
\begin{equation}
\theta _{0}\left( \nu \right) =e^{\mathrm{i}k_{0}-\mathrm{i}\acute{\nu}%
}\left( 1+O\left( \acute{\nu}^{2}\right) \right) :\left\vert \theta
_{1}\left( \nu \right) \right\vert =1,\ \acute{\nu}\geq 0,  \label{ff21a}
\end{equation}%
with the corresponding eigenmode propagating in the negative direction.

In view of (\ref{ff20}) and (\ref{ff21a}), we can use the relations (\ref%
{[i,j] X}) and (\ref{[i,i] k}) yielding%
\begin{equation}
\left[ \mathfrak{e}_{1}\left( \nu \right) ,\mathfrak{e}_{0}\left( \nu
\right) \right] =0.  \label{ff22}
\end{equation}%
So, unlike in situations for $n=3,4$ in the case $n=2\,$only the vector $%
\mathfrak{e}_{0}\left( \nu \right) $ belongs to the space $\mathcal{S}_{%
\limfunc{T}}\left( 0;\omega _{0}+\nu \right) $, where another one, namely $%
\mathfrak{e}_{1}\left( \nu \right) $, does not belong to $\mathcal{S}_{%
\limfunc{T}}\left( 0;\omega _{0}+\nu \right) $ since it corresponds to an
eigenmode propagating in the negative direction. Hence, the second vector in 
$\mathcal{S}_{\limfunc{T}}\left( 0;\omega _{0}+\nu \right) $ must be one of $%
\mathfrak{e}_{2}\left( \nu \right) $ and $\mathfrak{e}_{3}\left( \nu \right) 
$. Without loss of generality we assume that is $\mathfrak{e}_{2}\left( \nu
\right) $, and, hence 
\begin{equation}
\mathcal{S}_{\limfunc{T}}\left( 0;\omega _{0}+\nu \right) =\limfunc{Span}%
\left\{ \mathfrak{e}_{0}\left( \nu \right) ,\mathfrak{e}_{2}\left( \nu
\right) \right\} .  \label{ff23}
\end{equation}%
Now there can be two possibilities: $\left\vert \theta _{2}\left( \nu
\right) \right\vert <1$ or $\left\vert \theta _{2}\left( \nu \right)
\right\vert =1$. The most interesting case is%
\begin{equation}
\left\vert \theta _{2}\left( \nu \right) \right\vert <1,\ \nu \geq 0,
\label{ff23a}
\end{equation}%
when the corresponding mode is an evanescent one. Since we are interested to
know if there can be any transmission of energy by a mode related to a
regular band edge under assumption (\ref{ff23a}), the only possibility of
the transmission will the mode related to the band edge.

In the case of $\left\vert \theta _{2}\left( \nu \right) \right\vert =1$ the
corresponding mode will be a common one propagating in the positive
direction with non-zero velocity. In this case the calculation is similar to
the case (\ref{ff23a}) with the only difference that we have to pick the
single eigenmode related to the band edge and find the corresponding flux
and the reflection coefficient. For that mode the result is the same as in
the case (\ref{ff23a}).

So, we assume now that the condition (\ref{ff23a}) is satisfied. Notice that
under the condition (\ref{ff23a}) in view of (\ref{[i,j] X}) and (\ref{[i,j]
k}) we have%
\begin{equation}
\left[ \mathfrak{e}_{0}\left( \nu \right) ,\mathfrak{e}_{2}\left( \nu
\right) \right] =0,\ \left[ \mathfrak{e}_{2}\left( \nu \right) ,\mathfrak{e}%
_{2}\left( \nu \right) \right] =0,\ \nu \geq 0  \label{ff23b}
\end{equation}%
Then it follows from (\ref{ff23}) that%
\begin{equation}
\mathcal{S}_{\limfunc{T}}\left( 0;\omega _{0}\right) =\limfunc{Span}\left\{ 
\mathfrak{e}_{0}\left( 0\right) ,\mathfrak{e}_{2}\left( 0\right) \right\} .
\label{ff24}
\end{equation}%
Notice the representation (\ref{ff24}) for the space $\mathcal{S}_{\limfunc{T%
}}\left( 0;\omega _{0}\right) $ together with (\ref{ff21b}) and (\ref{ff23b}%
) yields%
\begin{equation}
\left[ \mathfrak{f},\mathfrak{f}\right] =0\text{ for any }\mathfrak{f}\in 
\mathcal{S}_{\limfunc{T}}\left( 0;\omega _{0}\right) .  \label{ff25}
\end{equation}%
The relation (\ref{ff25}) combined with (\ref{zpz14}) and (\ref{zpz14aa})
implies the following very important property of the reflection coefficient $%
r\left( \alpha ^{+};0\right) $ at the inflection point $\omega _{0}$, i.e. $%
\nu =0$,%
\begin{equation}
\text{for every }\alpha ^{+}\in 
\mathbb{C}
^{2}\text{ the reflection coefficient }r\left( \alpha ^{+}\right) =1\text{
and }\rho ^{\dag }\rho =I_{2}\text{.}  \label{ff26}
\end{equation}%
\emph{The relation (\ref{ff46}) clearly indicates that the reflection
coefficient }$r\left( \alpha ^{+};0\right) $\emph{\ is always exactly }$1$%
\emph{\ for all the relevant eigenmodes of semi-infinite periodic stack
related to the band edge.}

More elaborate analysis yields asymptotic expressions for the matrix of
reflection coefficients $\rho $, as determined by (\ref{phs4}) and (\ref%
{php5}), and other related quantities for nonzero, but small, $\nu =\omega
-\omega _{0}$. Indeed, let us use in the relations (\ref{phs2})-(\ref{phs4})
the vectors $\Phi _{1}$ and $\Phi _{2}$ defined by%
\begin{equation}
\Phi _{1}\left( \nu \right) =\mathfrak{e}_{0}\left( \nu \right) ,\ \Phi
_{2}\left( \nu \right) =\mathfrak{e}_{2}\left( \nu \right) .  \label{ff27}
\end{equation}%
Notice that generically $\mathfrak{e}_{0}\left( 0\right) $ and $\mathfrak{e}%
_{2}\left( 0\right) $ are always linearly independent and, hence,%
\begin{equation}
\left\{ \Phi _{1}\left( 0\right) ,\Phi _{2}\left( 0\right) \right\} \text{
are linearly independent.}  \label{ff29}
\end{equation}%
The relation (\ref{ff29}) shows that condition (\ref{php3}) is satisfied.

Now let us find the value $\Phi \left( \alpha ^{+};\nu \right) $ of the
eigenmode corresponding to the incident wave $\alpha ^{+}$. Using (\ref{phs1}%
), (\ref{phs3a}), (\ref{php4}) we consequently obtain 
\begin{equation}
\check{\Phi}\left( \alpha ^{+};\nu \right) =\left[ 
\begin{array}{c}
\varphi _{1}\left( \nu \right) \\ 
\varphi _{2}\left( \nu \right)%
\end{array}%
\right] =\left[ 
\begin{array}{c}
\varphi _{1}\left( 0\right) \\ 
\varphi _{2}\left( 0\right)%
\end{array}%
\right] +O\left( \acute{\nu}\right) ,\ \left[ 
\begin{array}{c}
\varphi _{1}\left( 0\right) \\ 
\varphi _{2}\left( 0\right)%
\end{array}%
\right] =\left[ Q^{+}\left( 0\right) \right] ^{-1}\alpha ^{+}.  \label{ff210}
\end{equation}%
\begin{equation}
\Phi \left( \alpha ^{+};\nu \right) =\varphi _{1}\left( \nu \right) \Phi
_{1}\left( \nu \right) +\varphi _{2}\left( \nu \right) \Phi _{2}\left( \nu
\right) =\varphi _{1}\left( 0\right) \mathfrak{e}_{0}\left( 0\right)
+\varphi _{2}\left( 0\right) \mathfrak{e}_{2}\left( 0\right) +O\left( \acute{%
\nu}\right)  \label{ff211}
\end{equation}%
Observe that the decomposition (\ref{ff211}) of the vector $\Phi \left(
\alpha ^{+};\nu \right) $ into a linear combination of eigenvectors $%
\mathfrak{e}_{0}\left( \nu \right) $ and $\mathfrak{e}_{1}\left( \nu \right) 
$ of the transfer matrix $\mathcal{T}\left( \nu \right) $ signifies that%
\begin{equation}
\text{the amplitude of the eigenmode inside the slab is }\varphi _{1}\left(
0\right) +O\left( \acute{\nu}\right) .  \label{ff212}
\end{equation}%
Combining (\ref{ff21}), (\ref{ff23b}) with (\ref{ff210}), (\ref{ff211}) we
get the following formula for the flux%
\begin{align}
& \left[ \Phi \left( \alpha ^{+};\nu \right) ,\Phi \left( \alpha ^{+};\nu
\right) \right] =-2\left\vert \varphi _{1}\left( 0\right) \right\vert ^{2}%
\func{Im}\left[ \mathfrak{f}_{0},\mathfrak{f}_{1}\right] \acute{\nu}+O\left( 
\acute{\nu}^{2}\right)  \label{ff213} \\
& =-2\left\{ \left[ Q^{+}\left( 0\right) \right] ^{-1}\alpha ^{+}\right\}
_{1}^{2}\func{Im}\left[ \mathfrak{f}_{0},\mathfrak{f}_{1}\right] \acute{\nu}%
+O\left( \acute{\nu}^{2}\right) .  \notag
\end{align}%
The formula (\ref{ff213}) together with (\ref{phs9}) yields%
\begin{equation}
t^{2}\left( \alpha ^{+};\nu \right) =1-r^{2}\left( \alpha ^{+};\nu \right) =-%
\frac{2\left\{ \left[ Q^{+}\left( 0\right) \right] ^{-1}\alpha ^{+}\right\}
_{1}^{2}\func{Im}\left[ \mathfrak{f}_{0},\mathfrak{f}_{1}\right] }{%
\left\vert \alpha ^{+}\right\vert ^{2}}\acute{\nu}+O\left( \acute{\nu}%
^{2}\right) .  \label{ff214}
\end{equation}

An alternative computation based on (\ref{php8}), (\ref{php5}) and (\ref%
{ff26}) (which turns into $\left\vert \rho _{0}\alpha ^{+}\right\vert
^{2}=\left\vert \alpha ^{+}\right\vert ^{2}$ for all $\alpha ^{+}$) implies
the following asymptotic formulae for the transmission and reflection
coefficients%
\begin{align}
t^{2}\left( \alpha ^{+};\nu \right) & =1-r^{2}\left( \alpha ^{+};\nu \right)
=-\frac{2\func{Re}\left\{ \left( \rho _{1}\alpha ^{+},\rho _{0}\alpha
^{+}\right) \right\} }{\left\vert \alpha ^{+}\right\vert ^{2}}\acute{\nu}%
+O\left( \acute{\nu}^{2}\right) ,  \label{ff215} \\
r^{2}\left( \alpha ^{+};\nu \right) & =1+\frac{2\func{Re}\left\{ \left( \rho
_{1}\alpha ^{+},\rho _{0}\alpha ^{+}\right) \right\} }{\left\vert \alpha
^{+}\right\vert ^{2}}\acute{\nu}+O\left( \acute{\nu}^{2}\right) ,  \notag \\
\rho _{0}& =Q_{0}^{-}\left[ Q_{0}^{+}\right] ^{-1},\ \rho _{1}=\left[
Q_{0}^{-}\right] ^{-1}\left\{ Q_{0}^{-}Q_{1}^{-}-Q_{0}^{+}Q_{1}^{+}\right\} %
\left[ Q_{0}^{+}\right] ^{-1}.  \notag
\end{align}

\subsection{Asymptotic analysis summary}

The final results on the reflection coefficients, transmission and flux are
formulated in Section \textquotedblleft relevant modes at degenerate
points\textquotedblright .

We reiterate that for any relevant eigenmode of a semi-infinite slab the
following fundamental relation holds for its energy flux $\left[ \Phi ,\Phi %
\right] $ and the reflection and transmission coefficients%
\begin{equation*}
t^{2}\left( \alpha ^{+}\right) =1-r^{2}\left( \alpha ^{+}\right) =\frac{%
\left[ \Phi \left( \alpha ^{+}\right) ,\Phi \left( \alpha ^{+}\right) \right]
}{\left\vert \alpha ^{+}\right\vert ^{2}}
\end{equation*}%
where the two-dimensional vector $\alpha ^{+}$ describes the incident wave
in a properly chosen basis and $\Phi \left( \alpha ^{+}\right) $ is the
corresponding EM field at the surface of the slab.

One the most important quantitative results of the analysis of the
reflection coefficient in a vicinity of band edges and inflection points is
summarized by the following formulae for the reflection coefficient $r$ as $%
\nu =\omega -\omega _{0}\rightarrow 0$ 
\begin{align*}
\text{inflection point }n& =3:r^{2}=r_{0}^{2}+c_{\limfunc{sign}\nu
}\left\vert \nu \right\vert ^{1/3},\ 0\leq r_{0}<1; \\
\text{regular band edge }n& =2:r^{2}=1-c_{0}\left\vert \nu \right\vert
^{1/2},\ \left\{ 
\begin{array}{cc}
\nu \geq 0 & \text{for a lower edge} \\ 
\nu \leq 0 & \text{for an upper edge}%
\end{array}%
\right. ; \\
\text{degenerate band edge }n& =4:r^{2}=1-c_{0}\left\vert \nu \right\vert
^{1/4},\ \left\{ 
\begin{array}{cc}
\nu \geq 0 & \text{for a lower edge} \\ 
\nu \leq 0 & \text{for an upper edge}%
\end{array}%
\right. .
\end{align*}%
where $c_{\limfunc{sign}\nu }$ denotes one of the constants $c_{\pm }$
corresponding to the sign of $\nu $.

The above formulae for the reflection coefficient indicate clearly that on
approach to a band edge the reflection coefficient approaches 1. In
contrast, in a vicinity of an inflection point the reflection coefficient
approaches a less than $1$ number $r_{0}$ and can be made arbitrarily small
for properly designed structures.

The table below shows the asymptotic behavior of the slow mode group
velocity, the saturation amplitude, and the semi-infinite slab transmittance
as the frequency approaches the respective stationary point.\bigskip

\begin{tabular}{|l|l|l|l|}
\hline
Rank of degeneracy & Group velocity & Saturation amplitude & Transmittance
\\ \hline
$2$ (regular band edge) & $\left\vert \nu\right\vert ^{1/2}$ & $1$ & $%
\left\vert \nu\right\vert ^{1/2}$ \\ \hline
$3$ (stationary inflection pint) & $\left\vert \nu\right\vert ^{2/3}$ & $%
\left\vert \nu\right\vert ^{-1/3}$ & $1$ \\ \hline
$4$ (degenerate band edge) & $\left\vert \nu\right\vert ^{3/4}$ & $%
\left\vert \nu\right\vert ^{-1/4}$ & $\left\vert \nu\right\vert ^{1/4}$ \\ 
\hline
\end{tabular}
\bigskip

Using this table we summarize the basic properties of the eigenmodes at
frequencies close to the band edges and inflection points as follows.

1. For a regular band edge there are no energy relevant Floquet modes, and
as $\nu \rightarrow 0$ the group velocity and the maximal flux vanish as $%
\left\vert \nu \right\vert ^{1/2}$, whereas the saturation amplitude remains
finite. The light does slow down in the vicinity of a regular band edge, but
only a vanishing fraction enters the photonic slab, while the rest is
reflected back to space.

2. For a stationary inflection point there is a relevant Floquet mode at $%
\omega_{0}$, and as $\nu=\omega-\omega_{0}\rightarrow0$ the group velocity
vanishes at the rate $\left\vert \nu\right\vert ^{2/3}$ with the maximal
flux remaining finite, and the saturation amplitude diverging as $\left\vert
\nu\right\vert ^{-1/3}$. The slab transmittance at $\omega=\omega_{0}$
remains finite and can even be close to unity.

3. For\ a 4-fold degenerate band edge, there is a relevant non-Bloch Floquet
mode. As $\nu \rightarrow 0$, the respective slow mode group velocity
vanishes as $\left\vert \nu \right\vert ^{3/4}$, while the saturation
amplitude diverges as $\left\vert \nu \right\vert ^{-1/4}$. The transmitted
energy flux, along with the slab transmittance, vanishes as $\left\vert \nu
\right\vert ^{1/4}$.

\section{Summary}

Although the existence of slow electromagnetic modes in photonic crystals is
quite obvious, the next question is whether and how such modes can be
excited by incident light. In other words, we need to know whether or not a
significant fraction of the incident light energy can be converted into a
slow mode with virtually zero group velocity in a semi-infinite photonic
crystal. We have shown that it can be done, but only if the slow mode is
associated with a stationary inflection point of the dispersion relation $%
\omega \left( k\right) $. In this special case, the incident light with the
proper frequency, polarization, and direction of incidence is completely
converted into the slow frozen mode with huge amplitude and vanishingly
small group velocity. Such a phenomenon constitute the frozen mode regime.
By contrast, if a slow electromagnetic mode relates to a photonic band edge,
the incident wave will be reflected back to space. Not every photonic
crystal can have the dispersion relation with a stationary inflection point
and, thereby, support the frozen mode regime. For instance, one-dimensional
periodic arrays (periodic layered structures) must include specially
oriented anisotropic layers, in order to support the proper dispersion
relation and the frozen mode regime. Photonic crystals with
three-dimensional periodicity are not required to have anisotropic
constitutive components.

Generally, the possibility of conversion of an incident wave into a slow
mode appears to be directly related to the character of the respective Bloch
dispersion relation $\omega \left( k\right) $ of the periodic structure.
This fundamental relation exists regardless of the specific physical
realization of the periodic structure supporting such a dispersion relation.
For instance, although neither periodically modulated waveguides, nor
periodic arrays of coupled resonators are formally photonic crystals, still,
as soon as the respective Bloch electromagnetic dispersion relation develops
a stationary inflection point, one can expect the frozen mode regime at the
respective frequency.

Not every periodic array can have the dispersion relation with a stationary
inflection point. Symmetry-based considerations similar to those applied
above to the case of periodic layered arrays, can provide a meaningful
guidance on how to find the proper structure.

\bigskip

\textbf{Acknowledgment and Disclaimer:} Effort of A. Figotin and I.
Vitebskiy is sponsored by the Air Force Office of Scientific Research, Air
Force Materials Command, USAF, under grant number FA9550-04-1-0359.\bigskip

{\large Appendix 1: basic properties of the transfer matrix}\bigskip

According to (\ref{JU}), the $4\times 4$ matrix $\mathcal{T}\left( \nu
\right) $, $\nu =\omega -\omega _{0}$ satisfies the following identity 
\begin{equation}
\mathcal{T}^{-1}\left( \nu \right) =J\mathcal{T}^{\ast }\left( \nu \right) J.
\label{Tru1}
\end{equation}%
The identity implies, in particular that%
\begin{equation}
\left\vert \det \mathcal{T}\left( \nu \right) \right\vert =1  \label{Tru2}
\end{equation}%
\begin{equation}
\text{if }\zeta \text{ is an eigenvalues of }\mathcal{T}\left( \nu \right) 
\text{ then }1/\zeta ^{\ast }\text{ is also an eigenvalue.}  \label{Tru3}
\end{equation}%
In other words, the statement (\ref{Tru3}) yields that if $\zeta =\rho e^{%
\mathrm{i}\phi }$ is the polar form of an eigenvalue $\zeta $ and $\rho \neq
1$ then 
\begin{equation}
\zeta =\rho e^{\mathrm{i}\phi }\text{ and }\zeta =\rho ^{-1}e^{\mathrm{i}%
\phi }\text{ are both eigenvalues of }\mathcal{T}\left( \nu \right)
\label{Tru4}
\end{equation}%
and that if $\zeta $ is an eigenvalue of $\mathcal{T}\left( \nu \right) $
then $1/\zeta ^{\ast }$ is an eigenvalue too.

The above properties of eigenvalues of $\mathcal{T}\left( \nu \right) $
imply the following statements.

1. Suppose that $\zeta \left( \nu \right) $ is an eigenvalue of $\mathcal{T}%
\left( \nu \right) $ depending on $\nu $ continuously. Suppose also that $%
\zeta \left( 0\right) $ has multiplicity $1$ and $\left\vert \zeta \left(
0\right) \right\vert =1$. Then there exists a sufficiently small $\delta >0$
such that%
\begin{equation}
\left\vert \zeta \left( \nu \right) \right\vert =1\text{ for any }\left\vert
\nu \right\vert \leq \delta .  \label{zen1}
\end{equation}%
To show (\ref{zen1}) we need the following elementary implication:%
\begin{equation}
\text{if }\left\vert \zeta _{0}\right\vert =1\text{ and }\left\vert \zeta
-\zeta _{0}\right\vert \leq \epsilon \leq \frac{1}{2}\text{ then }\left\vert 
\frac{1}{\zeta ^{\ast }}-\zeta _{0}\right\vert \leq 6\epsilon .  \label{zen2}
\end{equation}%
Assume now that the assumption of (\ref{zen2}) holds. Notice that%
\begin{equation}
\frac{1}{\zeta ^{\ast }}\equiv \frac{\zeta }{\left\vert \zeta \right\vert
^{2}}  \label{zen3}
\end{equation}%
and%
\begin{equation}
\left\vert \left\vert \zeta \right\vert -1\right\vert =\left\vert \left\vert
\zeta \right\vert -\left\vert \zeta _{0}\right\vert \right\vert \leq
\left\vert \zeta -\zeta _{0}\right\vert \leq \epsilon \leq \frac{1}{2}.
\label{zen3a}
\end{equation}%
with then implies that%
\begin{equation}
1-\epsilon \leq \left\vert \zeta \right\vert \leq 1+\epsilon .  \label{zen4}
\end{equation}%
Now using (\ref{zen3}) we get%
\begin{gather}
\left\vert \frac{1}{\zeta ^{\ast }}-\zeta _{0}\right\vert =\left\vert \frac{%
\zeta }{\left\vert \zeta \right\vert ^{2}}-\zeta _{0}\right\vert =\left\vert
\left( \frac{1}{\left\vert \zeta \right\vert ^{2}}-1\right) \zeta +\zeta
-\zeta _{0}\right\vert  \label{zen5} \\
\leq \left\vert \left( \frac{1}{\left\vert \zeta \right\vert ^{2}}-1\right)
\zeta \right\vert +\left\vert \zeta -\zeta _{0}\right\vert =\frac{%
1-\left\vert \zeta \right\vert ^{2}}{\left\vert \zeta \right\vert }%
+\left\vert \zeta -\zeta _{0}\right\vert  \notag
\end{gather}%
The inequalities (\ref{zen3a}), (\ref{zen5}) together with $\epsilon \leq 
\frac{1}{2}$ yield%
\begin{equation}
\left\vert \frac{1}{\zeta ^{\ast }}-\zeta _{0}\right\vert \leq \frac{%
1-\left( 1-\epsilon \right) ^{2}}{1-\epsilon }+\epsilon =\epsilon \frac{%
3-2\epsilon }{1-\epsilon }\leq 6\epsilon ,  \label{zen5a}
\end{equation}%
which is the desired inequality (\ref{zen2}).

Using the fact the $\zeta \left( 0\right) $ has multiplicity one and
applying the standard perturbation theory arguments we can always find $%
0<\epsilon _{0}<1$ and $\delta _{0}>0$ such that%
\begin{equation}
\text{for }\left\vert \nu \right\vert \leq \delta _{0}\text{ the eigenvalue }%
\zeta \left( \nu \right) \text{ is the only one in the circle }\left\vert
\zeta -\zeta \left( 0\right) \right\vert \leq \epsilon _{0}.  \label{zen6}
\end{equation}%
Now using the continuity of $\zeta \left( \nu \right) $ we can always find a
positive $\delta <\delta _{0}$ such that%
\begin{equation}
\text{for }\left\vert \nu \right\vert \leq \delta <\delta _{0}\text{ we have 
}\left\vert \zeta \left( \nu \right) -\zeta \left( 0\right) \right\vert \leq 
\frac{\epsilon _{0}}{6}\leq \frac{1}{6}.  \label{zen7}
\end{equation}%
Observe that (\ref{zen5a}) and (\ref{zen7}) imply that%
\begin{equation}
\left\vert \frac{1}{\zeta ^{\ast }\left( \nu \right) }-\zeta \left( 0\right)
\right\vert \leq \epsilon _{0}\text{ for }\left\vert \nu \right\vert \leq
\delta <\delta _{0}.  \label{zen8}
\end{equation}%
Assume for the sake of the argument that for some $\left\vert \nu
\right\vert \leq \delta $ we have $\left\vert \zeta \left( \nu \right)
\right\vert \neq 1$. Then based on (\ref{zen8}) and general properties of $%
\mathcal{T}\left( \nu \right) $ we have to conclude that $\frac{1}{\zeta
^{\ast }\left( \nu \right) }\neq \zeta \left( \nu \right) $ is another
eigenvalue of $\mathcal{T}\left( \nu \right) $ residing in the circle $%
\left\vert \zeta -\zeta \left( 0\right) \right\vert \leq \epsilon _{0}$. But
this clearly contradicts to (\ref{zen6}) implying the desired relation (\ref%
{zen1}).

2. Suppose that for $0<\left\vert \nu \right\vert <\delta $ the matrix $%
\mathcal{T}\left( \nu \right) $ has four different eigenvalues $\zeta
_{j}\left( \nu \right) $, $j=1,\ldots 4$ each continuously depending on $\nu 
$ and having the following properties:%
\begin{equation}
\text{There exists a }\zeta _{0}\text{ such that }\lim_{\nu \rightarrow
0}\zeta _{j}\left( \nu \right) =\zeta _{0},\ j=1,2,3\text{, and }\zeta
_{4}\left( 0\right) \neq \zeta _{0}.  \label{trip1}
\end{equation}%
In other words, for small $\left\vert \nu \right\vert $ the eigenvalue $%
\zeta _{4}\left( \nu \right) $ has multiplicity one and is well separated
from the other $3$ different eigenvalues $\zeta _{j}\left( \nu \right) $, $%
j=1,2,3$ which converge as $\nu \rightarrow 0$ to a $\zeta _{0}$.

Then we claim that%
\begin{equation}
\left\vert \zeta _{0}\right\vert =1\text{ and }\left\vert \zeta _{4}\left(
0\right) \right\vert =1,  \label{trip2}
\end{equation}%
and there exists a $\delta >0$ such that 
\begin{equation}
\left\vert \zeta _{j}\left( \nu \right) \right\vert =1\text{ for at least
one }j=1,2,3\text{ and }\left\vert \nu \right\vert <\delta .  \label{trip3}
\end{equation}%
\begin{equation}
\left\vert \zeta _{4}\left( \nu \right) \right\vert =1\text{ for }\left\vert
\nu \right\vert <\delta .  \label{trip4}
\end{equation}%
Observe, first, that the first equality in (\ref{trip1}) follows from the
conditions of (\ref{trip1}) and general properties of $\mathcal{T}\left( \nu
\right) $, since if $\left\vert \zeta _{0}\right\vert \neq 1$, we would have
three more eigenvalues $\frac{1}{\zeta _{j}^{\ast }\left( \nu \right) }$, $%
j=1,2,3$ with the total number of eigenvalues $6$. That is, of course,
impossible for $4\times 4$ matrix implying that the first equality in (\ref%
{trip1}) holds. As to the second, it follows from proving that $\left\vert
\zeta _{0}\right\vert =1$ and the identity (\ref{Tru2}), since then we must
have $\left\vert \zeta _{0}\right\vert ^{3}\left\vert \zeta _{4}\left(
0\right) \right\vert =1$.

To show (\ref{trip3}) we use the limit conditions in (\ref{trip1}) and the
fact that $\frac{1}{\zeta _{j}^{\ast }\left( \nu \right) }$ is also an
eigenvalue. Indeed,\ if, for the sake of the argument, we assume that (\ref%
{trip1}) does not hold we have to conclude that in a infinitesimally small
vicinity of $\zeta _{0}$ there will be at least four different eigenvalues
which is impossible in view of the second condition in (\ref{trip1}). This
completes the proof of (\ref{trip3}). As to the proof of (\ref{trip4}) it
follows from the statement (\ref{zen1}).\bigskip

{\large Appendix 2: perturbation theory for a diagonal matrix}\bigskip

Particular constructions of the perturbation theory we discuss here follow
from \cite{BPov} and \cite{FigGod}. Suppose that

\begin{equation}
W\left( \nu \right) =W_{0}+\nu W_{1}+\nu ^{2}W_{2}+\cdots  \label{tdg1}
\end{equation}%
and $T_{0}$ is a diagonal matrix with distinct elements, i.e.%
\begin{equation}
W_{0}=\left[ 
\begin{array}{cccc}
w_{1} & 0 & \ldots & 0 \\ 
0 & w_{2} & \ddots & \vdots \\ 
\vdots & \ddots & \ddots & 0 \\ 
0 & \ldots & 0 & w_{n}%
\end{array}%
\right] ,\ \text{where }w_{m}\neq w_{j}\text{ if }m\neq j.  \label{tdg2}
\end{equation}%
To diagonalize $T\left( \nu \right) $ we use the approach outlined in \cite%
{BPov} and used in \cite{FigGod}. Namely, there exists the following
representation for the diagonal form $X$ of $T\left( \nu \right) $ 
\begin{equation}
\zeta =e^{-S\left( \nu \right) }We^{S\left( \nu \right) }=W_{0}+\nu \zeta
_{1}+\nu ^{2}\zeta _{2}+\cdots ,\ S\left( \nu \right) =\nu S_{1}+\nu
^{2}S_{2}+\cdots ,  \label{tdg3}
\end{equation}%
where the matrices $S_{1},S_{2},\ldots $ do not depend on $\nu $ and $%
X_{1},X_{2},\ldots $ are diagonal. To find $S_{j}$ and $X_{j}$ we use the
Hausdorf's representation%
\begin{equation}
X=e^{-S}We^{S}=W+\left[ W,S\right] +\frac{1}{2!}\left[ \left[ W,S\right] ,S%
\right] +\cdots ,\ \text{where }  \label{tdg4}
\end{equation}%
where the brackets denote the commutator of two matrices%
\begin{equation*}
\left[ A,B\right] =AB-BA.
\end{equation*}%
Substituting (\ref{tdg3}) into (\ref{tdg4}) and equating the terms of like
powers in $\nu $, we obtain the following expressions for the matrices $%
X_{j} $%
\begin{equation}
X_{1}=\left[ T_{0},S_{1}\right] +T_{1},\ X_{2}=\left[ T_{0},S_{2}\right]
+T_{2}+\left[ T_{1},S_{1}\right] +\frac{1}{2}\left[ \left[ T_{0},S_{1}\right]
,S_{1}\right] ,\ldots .  \label{tdg4a}
\end{equation}%
To find $X_{j}$ we introduce for a matrix $Y$ its representation as the sum
of its diagonal $\bar{Y}$ part and the remaining part $\mathring{Y}$ with
zero diagonal elements (so called integrable matrix \cite{BPov})%
\begin{equation}
Y=\bar{Y}+\mathring{Y},\ \bar{Y}=\limfunc{diag}\left( Y\right) ,\ \mathring{Y%
}=Y-\limfunc{diag}\left( Y\right) .  \label{tdg5}
\end{equation}%
Then%
\begin{equation}
X_{1}=\left[ W_{0},S_{1}\right] +\bar{W}_{1}+\mathring{W}_{1},  \label{tdg6}
\end{equation}%
and to get rid of the integrable part $\mathring{W}_{1}$ of $W_{1}$ we take $%
S_{1}$ to be the solution of the equation%
\begin{equation}
\left[ W_{0},S_{1}\right] =-\mathring{W}_{1}.  \label{tdg7}
\end{equation}%
The solution to this equation is%
\begin{equation}
\left[ S_{1}\right] _{jm}=\frac{1}{w_{m}-w_{j}}\left[ \mathring{W}_{1}\right]
_{jm},\ j\neq m;\ \left[ S_{1}\right] _{jj}=0.  \label{tdg8}
\end{equation}%
Consequently, 
\begin{equation}
X_{1}=\limfunc{diag}\left( W_{1}\right) =\bar{W}_{1}.  \label{tdg9}
\end{equation}%
To find $X_{2}$ we recast the equation (\ref{tdg4a}) as 
\begin{equation}
X_{2}=\left[ W_{0},S_{2}\right] +\bar{Y}_{2}+\mathring{Y}_{2},\ Y_{2}=W_{2}+%
\left[ W_{1},S_{1}\right] +\frac{1}{2}\left[ \left[ W_{0},S_{1}\right] ,S_{1}%
\right] .  \label{tdg10}
\end{equation}%
Applying to this equation the same approach as for (\ref{tdg6}) we get%
\begin{align}
X_{2}& =\limfunc{diag}\left( Y_{2}\right) =\bar{Y}_{2},\ Y_{2}=W_{2}+\left[
W_{1},S_{1}\right] +\frac{1}{2}\left[ \left[ W_{0},S_{1}\right] ,S_{1}\right]
.  \label{tdg11} \\
\left[ S_{2}\right] _{jm}& =\frac{1}{w_{m}-w_{j}}\left[ \mathring{Y}_{2}%
\right] _{jm},\ j\neq m;\ \left[ S_{2}\right] _{jj}=0.  \notag
\end{align}%
Using (\ref{tdg7})-(\ref{tdg10}) we can recast (\ref{tdg11}) as 
\begin{align}
X_{2}& =\limfunc{diag}\left( W_{2}+\frac{1}{2}\left[ \mathring{W}_{1},S_{1}%
\right] \right) ,  \label{tdg12} \\
\left[ S_{2}\right] _{jm}& =\frac{1}{w_{m}-w_{j}}\left[ \mathring{Z}_{2}%
\right] _{jm},\ j\neq m;\ \left[ S_{2}\right] _{jj}=0;\ Z_{2}=W_{2}+\left[
W_{1}-\frac{1}{2}\mathring{W}_{1},S_{1}\right] .  \notag
\end{align}%

\end{document}